\documentclass[aps,prd,twocolumn,showpacs,10pt,superscriptaddress, footinbib]{revtex4-1}

\usepackage{amsmath, amsthm, amssymb,bm,relsize}
\usepackage{amsmath,autobreak}
\usepackage{ragged2e}
\usepackage{color}
\usepackage[table]{xcolor}
\usepackage{comment}

\usepackage{dsfont}
\newcommand\BbbGamma{\reflectbox{\rotatebox[origin=c]{180}{$\mathds L$}}}

\usepackage{float}

\usepackage{tabularx}
\newcolumntype{P}[1]{>{\centering\arraybackslash}p{#1}}
\usepackage{multirow, booktabs}
\usepackage{slashed}
\usepackage{array}
\allowdisplaybreaks
\let\footnote=\endnote

\usepackage{graphicx}
\usepackage[caption=false]{subfig}
\definecolor{urlblue}{rgb}{0.2,0.4,0.7}
\definecolor{citegreen}{rgb}{0,0.6,0.2}
\definecolor{linkred}{rgb}{0.9,0.2,0.1}
\usepackage{hyperref}
\hypersetup{colorlinks=true, citecolor=citegreen, linkcolor=blue,urlcolor=urlblue}

\def\la{\leftarrow}

\def\EuGa{\gamma_{E}}
\allowdisplaybreaks

\excludecomment{oldtext}
\includecomment{newtext}

\excludecomment{newtextv1}
\includecomment{newtextv2}
\usepackage[normalem]{ulem}

\begin{document}

\begin{flushright}
{\mbox{DESY 25-079 } }
\end{flushright}

\title{Soft and virtual corrections to semi-inclusive DIS up to four loops in QCD}

\author{Saurav Goyal}
\email{sauravg@imsc.res.in}
\affiliation{The Institute of Mathematical Sciences, Taramani, 600113 Chennai, India}
\affiliation{Homi Bhabha National Institute, Training School Complex, Anushakti Nagar, Mumbai 400094, India}
\author{Sven-Olaf Moch}
\email{sven-olaf.moch@desy.de}
\affiliation{II. Institute for Theoretical Physics, Hamburg University, D-22761 Hamburg, Germany} 
\author{Vaibhav Pathak}
\email{vaibhavp@imsc.res.in}
\affiliation{The Institute of Mathematical Sciences, Taramani, 600113 Chennai, India}
\affiliation{Homi Bhabha National Institute, Training School Complex, Anushakti Nagar, Mumbai 400094, India}
\author{Narayan Rana}
\email{narayan.rana@niser.ac.in}
\affiliation{School of Physical Sciences, National Institute of Science Education and Research, 752050 Jatni, India}
\affiliation{Homi Bhabha National Institute, Training School Complex, Anushakti Nagar, Mumbai 400094, India}
\author{V. Ravindran}
\email{ravindra@imsc.res.in}
\affiliation{The Institute of Mathematical Sciences, Taramani, 600113 Chennai, India}
\affiliation{Homi Bhabha National Institute, Training School Complex, Anushakti Nagar, Mumbai 400094, India}

\date{\today}


\begin{abstract}

We apply the threshold resummation formalism for semi-inclusive deep-inelastic scattering (SIDIS) to derive the soft and virtual corrections for the SIDIS cross section up to four loops in QCD.
Using the recently computed next-to-next-to-leading order QCD corrections for the SIDIS cross section together with known results for the form factor and splitting functions in QCD up to four loops, 
we derive the complete soft and collinear contributions to the SIDIS coefficient functions at four-loop order.
We also include systematically the next-to-leading power corrections, which are suppressed near threshold. 
The numerical analysis of the new four-loop corrections shows a small effect on the cross section underpinning the very good perturbative stability of the SIDIS process at that order in perturbation theory, including the reduced dependence on the renormalization and factorization scales $\mu_R$ and $\mu_F$.

\end{abstract}


\maketitle


\section{Introduction}

The semi-inclusive deep-inelastic scattering (SIDIS) process~\cite{Aschenauer:2019kzf}, which involves the scattering of a lepton off an initial hadron P and the production of an identified hadron H$'$
in the final state, is a vital experimental tool in probing the internal structure of hadrons. This process is particularly important for studying parton distribution functions (PDFs) and fragmentation functions (FFs) through Quantum Chromodynamics (QCD). 

On the theoretical side,
the earliest next-to-leading order (NLO) QCD results  for SIDIS are from a seminal paper ~\cite{Altarelli:1979kv}, see also
\cite{Nason:1993xx,Furmanski:1981cw,Graudenz:1994dq,deFlorian:2012wk,deFlorian:2012wk}.  
Recently, the QCD results at next-to-next-to-leading order (NNLO) accuracy for both spin independent and spin dependent SIDIS cross sections were obtained independently by two groups \cite{Goyal:2023zdi,Bonino:2024qbh,Goyal:2024tmo,Bonino:2024wgg,Goyal:2024emo,Bonino:2025qta}.  
In QCD factorization, which provides the backbone of the perturbative framework, the fixed order results for the coefficient functions (CFs), parameterizing the hard scattering, often provide reliable predictions for wide range of kinematic variables.  
However, they fail in the threshold regions, because of 
the presence of large logarithms.  In the case of SIDIS, these logarithms originate from the partonic configurations where partons are soft and/or collinear to each other. 
Although the CFs in SIDIS are infrared-safe, meaning that soft and collinear divergences cancel out, these cancellations lead to certain distributions in CFs represented by 
distributions such as $\delta(1-\xi)$
and 
\begin{eqnarray}
\label{eq:Dxi}
{\cal D}^j_{\overline{\xi}} =\Bigg[\frac{\ln^j(1-\xi)}{1-\xi}\Bigg]_{+}\, , \quad j=0,1,\cdots
\end{eqnarray}
where $\xi=x',z'$ are appropriate scaling variables. 
We call them soft plus virtual (SV) contributions, and their importance was recognized early on, leading to several improvements in QCD perturbation theory, see~\cite{Daleo:2003jf,Daleo:2003xg,Anderle:2012rq,Anderle:2016kwa}. 
Currently, approximations at (next-to-)$^3$-leading order (N$^3$LO) accuracy taking into account only threshold contributions, also called SV and threshold resummation to (next-to-)$^3$-leading logarithmic (N$^{3}$LL) accuracy are available~\cite{Abele:2021nyo,Abele:2022wuy,Anderle:2012rq}, 
improving upon earlier results~\cite{Cacciari:2001cw,Anderle:2013lka,Sterman:2006hu} at next-to-leading logarithmic (NLL) accuracy.
These achievements were possible thanks to an intriguing connection between SIDIS and the rapidity distributions of pair of leptons in the Drell-Yan (DY) process, \cite{Anderle:2012rq,Westmark:2017uig,Banerjee:2018vvb}. 
Beyond the SV contributions, one encounters terms with logarithmic enhancement of the type $\ln^j(1-\xi)$, for $j =0,1,\cdots$, which are classified as next-to-leading power (NLP) corrections or next-to-SV (NSV) terms. 
Threshold resummation provides a framework to resum these large logarithms, i.e. the SV terms in eq.~\eqref{eq:Dxi} and NSV terms, to all order orders  in perturbation theory to obtain improved phenomenological predictions.
In this paper, we push threshold resummation for the SIDIS CFs to N$^{4}$LL accuracy and provide their approximations at N$^{4}$LO in QCD, accounting for both SV as well as NSV terms in a systematic procedure.

The outline of the article is as follows: 
In Sec.~\ref{sec:theory}, we present the theory framework for the SIDIS process, recall the basics of mass factorization in QCD and set-up the formalism, that allows to resum the threshold enhanced SV and NSV contributions.
The new fixed-order approximate SV and NSV results for the coefficient functions at N$^3$LO and N$^4$LO and 
are presented in Sec.~\ref{sec:DNSV}.
The explicit form of the exponential for the resummation of the threshold logarithms is presented in 
Sec.~\ref{sec:resum} in Mellin space as a function of complex valued $N_i,i=1,2$, conjugate to $x', z'$ respectively.
An alternative approach to the threshold expansion through the physical evolution kernel is presented in 
Sec.~\ref{sec:pek}, while phenomenological studies are performed with Sec.~\ref{sec:pheno}.
We summarize and present our conclusions in Sec.~\ref{sec:concl}.
Relevant formulae for the threshold resummation are collected in the Appendices.



\section{Theory}
\label{sec:theory}
The SIDIS process is given by
\begin{eqnarray}
l(k_l) + \text{P}(P) \rightarrow l({k}'_l)+\text{H}(P_H) + \text{X}\, ,
\end{eqnarray}
where the incoming (outgoing) lepton $l$ has   momentum $k_l$(${k}'_l$), 
and \text{P}(\text{H}) is the incoming (outgoing) hadron with momentum $P$($P_H$). 
The remaining set of final state particles is denoted by $\text{X}$.

We consider two types of differential observables namely, the spin independent  cross section
where
we average over the spins of incoming particles and sum the spins of final state particles:  
\begin{eqnarray}
d \sigma &=& \frac{1}{4}\sum_{s_l,S,s'_l,S_H} d\sigma_{s_l ,S}^{s'_l,S_H}
\, ,
\end{eqnarray}
and the corresponding spin-dependent cross section, 
%
\begin{eqnarray}
d \Delta \sigma = {1 \over 2 } \sum_{s'_l,S_H}
  \left(d\sigma_{s_l=\frac{1}{2},S=\frac{1}{2}}^{s'_l,S_H}
-d\sigma_{s_l=\frac{1}{2},S=-\frac{1}{2}}^{s'_l,S_H} \right)
\, ,
\end{eqnarray}
where $s_l$, $S$ are the spins of incoming lepton and hadron, 
and $s'_l$, $S_H$ the spins of the outgoing ones, respectively.
The cross section depends on 
the space-like momentum transfer $Q^2=-q^2$ where $q={k}_l-{k}'_l$,
the Bjorken variable $x=Q^2/2 P\cdot q$, 
the inelasticity $y={P\cdot q}/{P\cdot k_l}$, 
and the scaling variable $z={P\cdot P_H}/{P\cdot q}$, the fraction of the initial energy transferred to the final-state hadron.  

We restrict ourselves to scattering events where there is only photon exchange between the incoming lepton and the hadron.   We find that
the SIDIS cross section factorizes into a leptonic part and a hadronic part.  
The leptonic part is simple to calculate, whereas the hadronic part encapsulates the physics of the strong interactions and is challenging due to the compositeness of both the incoming and outgoing hadrons. 
However, using parity and time reversal invariance  and electromagnetic current conservation, we can parametrize the hadronic
part of the SIDIS cross section in terms of  structure functions. 
For example, for spin averaged case, we have two structure functions, $F_J,J=1,2$.  The structure functions are Lorentz invariant, depend on the hadronic scaling variables $x,z$ and, when higher order quantum corrections are included, also on the scale $Q^2$ of the virtual momemtum transferred. 
We find
\begin{equation}
    \frac{d^3\sigma}{dxdydz}=\frac{4\pi  \alpha_e^{2}}{Q^2} \left[y F_1(x,z,Q^2) + \frac{(1-y)}{y} F_2(x,z,Q^2)\right]
    \,\! ,
\end{equation}
where $\alpha_e=e^2/(4\pi)$ is the fine structure constant and $e$ is the electric charge.  In the literature, one often parametrises the hadronic tensor in terms of structure functions denoted by $W_J$ $(J=1,2)$ and
the $F_J$ are related to $W_J$ through
$W_1 = F_1/M$ and $W_2 = F_2/(y E)$ with 
the energy $E$ of the incoming lepton and $M$ being mass of incoming hadron.
The spin-dependent cross section for longitudinally polarized hadrons in the initial state depends only on one structure function, $g_1$,
\begin{equation}
\frac{d^3\Delta \sigma}{dxdydz} = \frac{4\pi\alpha_e^2}{Q^2} \big(2-y\big)g_1 (x,z,Q^2)\, .
\end{equation}

In perturbative QCD, the structure functions factorise as a convolution of (spin-dependent) PDFs, $(\Delta) f_{a/\text{P}}(x_1,\mu_F^2)$ and FFs, $D_{\text{H}'/b}(z_1,\mu_F^2)$ 
with  (spin-dependent) CFs, $(\Delta ){\cal C}_{J,ab}(x/x_1,z/z_1,Q^2,\mu_F^2)$, the latter being calculable in perturbation theory. 
\begin{widetext}
\begin{eqnarray}\label{eq:StrucCoeff}
\big(g_{1}\big) F_J (x,z,Q^2)
&=& \sum_{a,b = q,\overline{q},g}
\int_x^1 \frac{dx_1}{x_1} (\Delta )f_{a/\text{P}}\left( x_1,\mu_F^2\right)
\int_z^1 \frac{dz_1}{z_1}
  D_{\text{H}'/b}\left( z_1,\mu_F^2\right)
  (\Delta) {\cal C}_{J,ab}\left(\frac{x}{x_1},\frac{z}{z_1},Q^2,\mu_F^2\right)\, .
\end{eqnarray}
\end{widetext}
Here, $\mu_F$ is the factorisation scale, $x_1$ is the incoming parton's momentum fraction  with respect to the hadron P, i.e.\ $x_1= p_a/P$ 
and $z_1$ is the momentum fraction of the final state parton $b$  carried away by the outgoing hadron H, i.e.\ $z_1 = P_H/p_b$. 
Note also, that the spin-dependent PDFs are given by
$\Delta f_{a/\text{P}} = f_{a(\uparrow)/\text{P}(\uparrow)}
-f_{a(\downarrow)/\text{P}(\uparrow)}$, where arrows denote the polarisations of respective particles with respect to beam director.

SIDIS coefficient functions (CFs) are derived from parton subprocesses via mass factorization, neglecting quark mass effects. Collinear divergences arise when massless partons become collinear, such as an incoming or outgoing parton with others produced in the hard scattering. Thanks to QCD factorization, the bare (spin-dependent) CFs $(\Delta) \hat {\cal C}_{J,ab}(x',z',\varepsilon)$
in dimensional regularization $(d=4+\varepsilon)$  factorize through Altarelli-Parisi kernels at the scale $\mu_F$. 
\begin{widetext}
\begin{eqnarray}
\label{massfactC}
(\Delta) \hat {\cal C}_{J,ab}(x',z',\varepsilon) &= & (\Delta) \Gamma_{c \leftarrow a}(x',\mu_F^2,\varepsilon)
 \otimes (\Delta) {\cal C}_{J,cd}(x',z',\mu_F^2,\varepsilon) \tilde \otimes \tilde \Gamma_{b \leftarrow d}(z',\mu_F^2,\varepsilon)
\end{eqnarray}   
\end{widetext}
The kernels $\Gamma_{c\la a} $ and $\tilde \Gamma_{b\la d}$ are given in terms of space- and time-like splitting functions and summarize the collinear divergences through poles in $\varepsilon$, see, e.g.~\cite{Goyal:2024emo}.
The partonic scaling variables are $x' = x/x_1$ and $z'=z/z_1$ and the convolutions $\otimes$ and $\tilde \otimes$ between the functions $A(x')$, $C(x',z')$ and $B(z')$ are defined as follows:
\begin{eqnarray}
\nonumber
{\lefteqn{
A(x')\otimes C(x',z')
\tilde \otimes B(z') \,=\,}}
\\ 
&&\int_{x'}^1 \frac{dx_1}{x_1} 
\int_{z'}^1 \frac{dz_1}{z_1} A(x_1) 
 C\left(\frac{x'}{x_1},\frac{z'}{z_1}\right) B(z_1)\, .
\end{eqnarray}
The collinear finite CFs $(\Delta) {\cal C}_{J,ab}$ in eq.~\eqref{massfactC} are expanded in powers of strong coupling constant $a_s=\alpha_s/(4\pi)$ at the renormalisation scale $\mu_R$,
\begin{eqnarray}
\label{CFexp}
{\lefteqn{
(\Delta) {\cal C}_{J,ab}(x',z',\mu_F^2) 
\,=\, \delta_{ab}\delta(1-x') \delta(1-z') 
}}
\nonumber \\
&& + \sum_{i=1}^{\infty}a_s^i(\mu_R^2) (\Delta) {\cal C}^{(i)}_{J,ab}(x',z',\mu_F^2,\mu_R^2)\, .
\end{eqnarray}
The CFs $(\Delta) {\cal C}^{(i)}_{J,ab}$ in general contain distributions $\delta(1-\xi)$ and ${\cal D}^i_{\overline \xi}$, cf. eq.~\eqref{eq:Dxi}, where $\xi=x',z'$ and regular functions of $x',z'$. 
We can write $(\Delta) {\cal C}_{J,ab} $ as,
\begin{widetext}
\begin{align}
\label{coefffinal}
(\Delta) {\cal C}_{J,ab} = 
\sum_r (\Delta) C_{J,ab}^{r} h_r(x',z') 
+ \sum_\beta\left((\Delta) C_{J,ab,x'}^{\beta}(z') X_\beta(x') + (\Delta) C_{J,ab,z'}^{\beta}(x') Z_\beta(z') \right)+ (\Delta) R_{J,ab}(x',z')\, ,
\end{align}
\end{widetext}
where $r=\{\delta_x \delta_z, \delta_x j, j \delta_z,jk\}$ with $j,k=0,1,2$ etc and
\begin{align}
\label{hdist}
h_{\delta_x \delta_z} &= \delta(1-x') \delta(1-z')\,,&
h_{\delta_x j}&=\delta(1-x') {\cal D}^j_{\overline{z}'},
\nonumber \\
h_{j \delta_z}&={\cal D}^j_{\overline x'} \delta(1-z')\,,&
h_{j k}&={\cal D}^j_{\overline x'} {\cal D}^k_{\overline z'} 
\end{align}
and $\beta=\{\delta_x,\delta_z,j\}$ with $j=0,1,2$ etc
\begin{align}
 X_{\delta_x} &= \delta(1-x')\,,& 
 Z_{\delta_x} &= \delta(1-z')\, ,\nonumber\\
 X_{j}&={\cal D}^j_{\overline x'}\,,&
 Z_{j}&={\cal D}^j_{\overline z'}\, .
\end{align} 
The functions $(\Delta) R_{J,ab}(x',z')$ contain all remaining terms that are regular (no distributions).  
Furthermore, if we expand the CFs $(\Delta) C^\beta_{J,ab,x'}$ and $(\Delta) C^\beta_{J,ab,z'}$ in eq.~\eqref{coefffinal} around $x'=1$ and $z'=1$,  
we obtain a series of the form $(1-\xi)^i \ln^j(1-\xi), \xi=x',z'$ for  $i,j=0,1$ etc. 
The leading terms in this series ($i=0$) are the so-called NSV terms, while the rest with powers $i>0$ are denoted as beyond-NSV (BNSV) terms.
In this notation, the CFs in eq.~\eqref{coefffinal} read
\begin{eqnarray}
(\Delta) {\cal C}_{J,ab} = \delta_{ab}\Big((\Delta){\cal C}_{J}^{\text{SV}}\Big) + (\Delta){\cal C}_{J,ab }^{\text{NSV}} +(\Delta){\cal C}_{J,ab }^{\text{BNSV}}\, ,
\end{eqnarray}
where the SV term does not receive any contributions if $a=b=g$. 
Each term in the above equation can be expanded in powers series of $a_s$ as,
\begin{widetext}
\begin{eqnarray}
\label{eq:svnsvbnsv}
(\Delta){\cal C}_{J}^{\text{SV}} &=& (\Delta) C_{J,qq}^{(0),\delta_{x}\delta_{z}} h_{\delta_x \delta_z} \nonumber\\
&&+ \sum_{n=1}^\infty a_s^n \Bigg[
(\Delta) C_{J,qq}^{(n),\delta_{x}\delta_{z}} h_{\delta_x \delta_z}
+
\sum_{j=0}^{2 n-1}(\Delta) C_{J,qq}^{(n),\delta_{x}j}  h_{\delta_{x}j}
+\sum_{j=0}^{2 n-1}(\Delta) C_{J,qq}^{(n),j\delta_{z}}  h_{j\delta_{z}}
+\sum_{j=0}^{2 n-2}\sum_{k=0}^{2n-2-j}(\Delta) C_{J,qq}^{(n),jk}  h_{jk}
\Bigg]\,,
\nonumber\\
(\Delta){\cal C}_{J}^{\text{NSV}} &=&\sum_{n=1}^\infty a_s^n \Bigg[\sum_{j=0}^{2 n-1}\Big( (\Delta)  C_{J}^{(n),\delta_{x}j}  X_{\delta_{x}} L_j(z') +
(\Delta) C_{J}^{(n),j\delta_{z}}  L_j(x') Z_{\delta_{z}}\Big)
\nonumber\\
&&+\sum_{j=0}^{2 n-2}\sum_{k=0}^{2 n-2-j}
\Big((\Delta) C_{J}^{(n),jk}  X_j L_k(z') +
(\Delta) C_{J}^{(n),jk} L_j(x') Z_k\Big)\Bigg]\, ,
\end{eqnarray}
\end{widetext}
where $L_j(\xi)=\ln^j(1-\xi), \xi=x',z'$.
Schematically, $(\Delta){\cal C}_{J}^{\text{SV}}$ contains double distributions ($h_r$) given in eq.~(\ref{hdist}), $(\Delta){\cal C}_{J,ab}^{\text{NSV}}$ contains single distribution in $x'$ or $z'$ ($X_{\beta}, Z_{\beta}$) each multiplied by powers of $\ln^j(1-z')$ or $\ln^j(1-x')$   
respectively, where $j=0,1,2$ etc. 
Any regular terms in $x',z'$ that do not contain any distributions fall within the BNSV categrory. 
The NNLO results in \cite{Goyal:2023zdi,Bonino:2024qbh,Goyal:2024tmo,Bonino:2024wgg,Goyal:2024emo,Bonino:2025qta} for the CFs 
$(\Delta){\cal C}_{J,ab}^{\text{SV}}$ are in complete agreement with SV predictions for unpolarised structure functions $F_{1,2}$ \cite{Abele:2021nyo}, see also \cite{Ravindran:2006bu,Ahmed:2014uya}. 
The leading terms of the NSV contributions agree as well with ref.~\cite{Abele:2021nyo}, where the dominant NLP terms have been presented.

\if{1=0}
In the following, we discuss the origin of various contributions to CFs and their logarithmic structure in detail.
In the perturbative framework, the CFs of SIDIS cross section receive contributions from different types of logarithms originating from loop and phase space integrals over square of partonic scattering amplitudes.
We can classify these contributions into three parts, namely pure virtual (VV), real virtual (RV) emission and pure real (RR) emission sub processes. 
The VV part has Born kinematics, however one requires to perform integration over loop momenta as it contains virtual loops.  The RV contains interference of sub processes in which atleast one of them should contain real emission process.  
Beyond leading order in perturbation theory,  VV, RV contain ultraviolet (UV) divergence while RR is UV finite. However, all of them encounter infrared (IR) divergences namely from soft gluons and collinear  partons  at the intermediate stages.
In order to deal with UV and IR divergences, we use dimensional regularisation which allows us to  regulate both UV and IR divergences present in the loop and phase space integrals by modifying the space time dimension from $4$ to $4+\varepsilon$. 
The UV divergences  are removed by renormalisation of $a_s$.   The renormalisation is performed at the renormalisation scale $\mu_R$.   The soft divergences arise when momenta of gluons   in the loops of VV and RV terms become soft (very close to zero) and similary real gluons in the gluon emission
processes  in RV and RR become soft. The collinear divergences are present in all the sub processes as we drop masses for the quarks and anti quarks and gluons are naturally massless. Thanks to
KLN theorem \cite{Kinoshita:1962ur,Lee:1964is}, soft and collinear divergences go away when the degenerate states
that are responsible for the divergences are summed. For SIDIS, at the parton level, the incoming parton states from the hadron and the partonic states that fragments to final state hadron  are not sumed over with respect to their momenta and other quantum numbers.  These partonic states
can have  
degenerate configurations with the remaining partons leading to  
collinear divergences.  However, if they are summed through convoluting them with bare PDFs and FFs, these singularities are expected to cancel against those from bare PDFs and bare FFs.  In practice,
the initial state collinear divergences are factored out from the partonic sub processes and then absorbed into the bare PDFs. Similarly those from final states are factored and then absorbed into bare FFs.   This process, called mass factorisation, is done at the factorisation scale $\mu_F$ and it renders all the partonic channels IR finite.
The collinear singular part resulting from the incoming parton and the corresponding one from the final state parton are called  the Space-like AP kernels $\Gamma(x',\mu_F^2,\varepsilon)$ and Time-like AP kernels $\tilde \Gamma(z',\mu_F^2,\varepsilon)$ respectively and the factorisation theorem ensures that they are the process independent functions.   

In summary, the finite CFs are obtained after performing UV renormalisation and mass factorisation.  Cancellation of UV singularies through renormalisation leaves logarithms of $\mu_R^2$ and of collinear ones gives logarithms of $\mu_F^2$.  The structure of these logarithms in the perturbative series is governed by the following renormalisation group equations, one for the UV case and the other for the collinear ones.
\begin{widetext}
\begin{eqnarray}
\mu_R^2 {d \over d \mu_R^2}(\Delta) \mathcal{C}_{J,ab}(\mu_F^2) &=& 0\,,
\\
\mu_F^2 {d \over d \mu_F^2} (\Delta) \mathcal{C}_{J,ab}(\mu_F^2) &=& -{1 \over 2} \Bigg(
(\Delta) P(x',\mu_F^2) \otimes  (\Delta)\mathcal{C}_J(\mu_F^2) 
 + (\Delta)   \mathcal{C}_J(\mu_F^2)\tilde \otimes  \tilde P(z',\mu_F^2)\Bigg)_{ab}
\end{eqnarray}
\end{widetext}
where the AP splitting functions $\mathds{P}=\{P,\Delta P,\tilde P\}$ are known in powers of $a_s$.  
These differential equations ensure that logarithms of $\mu_R$ and $\mu_F$ at any perturbative order $n$ in $a_s$ are completely determined from the knowledge of the QCD beta function coefficients $\beta$, CFs $(\Delta) \mathcal{C}_J$ and  of AP splitting functions $\mathds{P}$ known to previous orders.  This remarkable feature is due to factorisation of UV and collinear singularities as well as all order structure of perturbative series of QCD $\beta$ function and of the AP splitting  functions, see \cite{Moch:2005ky,Ravindran:2005vv,Ravindran:2006cg,deFlorian:2012za,Ahmed:2014cha, Kumar:2014uwa,Ahmed:2014cla,Catani:2014uta,Li:2014bfa}.

Note that the CFs also contain logarithms that are functions of the scaling variables $x',z'$ and it is important to understand their structure like the way we understand those from $\mu_R$ and $\mu_F$.  In the following we would like to investigate the structure of these logarithms in a particular set of limits say, at large $x'$ and $,z'$, namely $x'\rightarrow 1$ and $z'\rightarrow 1$.  These limits are called threshold limits.  In inclusive  cross sections such as DIS, DY and Higgs boson or in genernal n-colorless particle productions, one finds that the corresponding CFs demonstrate a peculiar logarithmic structure at every order in perturbative series allowing one to resum them to all orders in a systematic fashion, see
\cite{Sterman:1986aj,Catani:1989ne,
Bonvini:2016fgf, Moch:2005ba,Ravindran:2006cg,Ravindran:2005vv,
Bonvini:2010ny,Bonvini:2012sh}.     Similarly for rapidity distribution of DY pairs or Higgs boson or of n-colorless state in the double threshold limit, the logarithms from two different scaling variables were   systematically resumed to all orders in perturbation theory, see \cite{
Laenen:2008ux,
Laenen:2010uz,
Bonocore:2014wua,
Bonocore:2015esa,
Bonocore:2016awd,
DelDuca:2017twk,
Bahjat-Abbas:2019fqa,
Soar:2009yh,
Moch:2009hr,deFlorian:2014vta,
Beneke:2018gvs,
Beneke:2019mua,
Beneke:2019oqx}.  In addition,
both single as well as double variable resummations were extended for flavour diagonal channels to include next to threshold contributions, also called NSV terms to all orders in perturbation theory  \cite{Catani:1989ne,Ravindran:2006bu,Ravindran:2007sv,Banerjee:2017cfc,Banerjee:2018vvb,Ahmed:2014uya}.   In the following we develop a formalism that can resum both SV as well as NSV logarithms of flavour diagonal contribution, namely $(\Delta) \mathcal{C}_{J,qq}$ to all orders both in $(x',z')$ space as well in their corresponding Mellin space $(N_1,N_2)$.  The reason to consider only flavour diagonal channels is because of their simple collinear factorisation structure.  We elaborate this below.
Consider the diagonal channel in the mass factorised form (writing out the implicit sum over the parton channels in eq.~\eqref{massfactC}),
\fi

The first principle's derivation of the threshold enhanced logarithms in eq.~\eqref{eq:svnsvbnsv} exploits the renormalisation group equations and the factorization of soft and collinear singularities to all orders in QCD, building on well-known developments for inclusive cross sections such as DIS, DY and Higgs boson production (or, more generally, colorless $n$-particle production), see e.g.~\cite{Sterman:1986aj,Catani:1989ne,
Moch:2005ky,Moch:2005ba,Ravindran:2005vv,Ravindran:2006cg,deFlorian:2012za,Bonvini:2012sh,Ahmed:2014cha, Kumar:2014uwa,Ahmed:2014cla,Catani:2014uta}.
For observables in differential kinematics, where the threshold limit is approached in terms of two different scaling variables,
such as the rapidity distributions of lepton pairs in the DY process or Higgs boson production, threshold logarithms have also been systematically resumed to all orders in perturbation theory, see~\cite{
Laenen:2008ux,
Laenen:2010uz,
Bonocore:2014wua,
Bonocore:2015esa,
Bonocore:2016awd,
DelDuca:2017twk,
Bahjat-Abbas:2019fqa,
Soar:2009yh,
Moch:2009hr,deFlorian:2014vta,
Beneke:2018gvs,
Beneke:2019mua,
Beneke:2019oqx}. 
In addition, both single as well as double variable resummations were extended for flavour diagonal channels to include NSV contributions~\cite{Catani:1989ne,Ravindran:2006bu,Ravindran:2007sv,Banerjee:2017cfc,Banerjee:2018vvb,Ahmed:2014uya}.

The  kinematics in these cases, is similar to the SIDIS process, where the limits $x'\rightarrow 1$ and $z'\rightarrow 1$ are considered. 
In the following we develop the formalism to resum both SV as well as NSV logarithms of flavour diagonal contributions, namely $(\Delta) \mathcal{C}_{J,qq}$ to all orders both in $(x',z')$ space as well in their corresponding Mellin space $(N_1,N_2)$.  
The reason to consider only flavour diagonal channels is because of their simple collinear factorisation structure. 
The starting point is the mass factorised form of the bare CFs in eq.~\eqref{massfactC} in $d=4+\varepsilon$ dimensions, 
expanding the sum over the parton channels,
\begin{widetext}
\begin{align}\label{massfactDY}
(\Delta) \hat {\cal C}_{J,qq}(x',z',\varepsilon) = & ~(\Delta) \Gamma_{q \leftarrow q} \otimes (\Delta) {\cal C}_{J,qq} \tilde \otimes \tilde \Gamma_{q \leftarrow q} + (\Delta) \Gamma_{q\leftarrow  q} \otimes (\Delta) {\cal C}_{J,q\overline{q}} \tilde \otimes \tilde \Gamma_{q \leftarrow\overline{q}} +(\Delta) \Gamma_{q\leftarrow  q} \otimes (\Delta) {\cal C}_{J,qg} \tilde \otimes \tilde \Gamma_{q\leftarrow g}\\ \nonumber
& + (\Delta) \Gamma_{\overline{q}\leftarrow q} \otimes (\Delta) {\cal C}_{J,\overline{q}q} \tilde \otimes \tilde \Gamma_{q \leftarrow q} + (\Delta) \Gamma_{\overline{q}\leftarrow  q} \otimes (\Delta) {\cal C}_{J,\overline{q}\overline{q}} \tilde \otimes \tilde \Gamma_{q\leftarrow \overline{q}} +(\Delta) \Gamma_{\overline{q}\leftarrow  q} \otimes (\Delta) {\cal C}_{J,\overline{q}g} \tilde \otimes \tilde \Gamma_{q\leftarrow g}\\ \nonumber 
& + (\Delta) \Gamma_{g\leftarrow q} \otimes (\Delta) {\cal C}_{J,gq} \tilde \otimes \tilde \Gamma_{q\leftarrow q} + (\Delta) \Gamma_{g\leftarrow q} \otimes (\Delta) {\cal C}_{J,g\overline{q}} \tilde \otimes \tilde \Gamma_{q\leftarrow \overline{q}} +(\Delta) \Gamma_{g \leftarrow q} \otimes (\Delta) {\cal C}_{J,gg} \tilde \otimes \tilde \Gamma_{q\leftarrow g}
\, .
\end{align}
\end{widetext}
The convolutions with the space- and time-like AP kernels $\Gamma_{a\leftarrow b}$ and $\tilde \Gamma_{a\leftarrow b}$ can be distinguished as follows: i) terms involving only diagonal combinations, such as $ (\Delta) \Gamma_{q\leftarrow  q} \otimes (\Delta) {\cal C}_{J,qq} \tilde \otimes \tilde \Gamma_{q\leftarrow q}$; 
ii) contributions involving a product of one diagonal with a pair of non-diagonal terms, for example $ (\Delta) \Gamma_{q \leftarrow q} \otimes (\Delta) {\cal C}_{J,qg} \tilde \otimes \tilde \Gamma_{q\leftarrow g}$.
The terms in i) contain SV plus NSV terms when the convolutions have been performed 
while the latter ones in case ii) will be composed beyond NSV terms only. 
Note that the diagonal AP kernels $(\Delta)\Gamma_{q\leftarrow q}$ and $\tilde \Gamma_{q\leftarrow q}$ contain convolutions among diagonal splitting functions, $(\Delta)P_{qq}$ and $\tilde P_{qq}$ respectively,
or one diagonal and a pair of off-diagonal ones, $(\Delta)P_{ab}, \tilde P_{ab}, a\not = b$.
We drop those terms in $(\Delta)\Gamma_{q\leftarrow q}$ ($\tilde \Gamma_{q\leftarrow q}$), that contain a pair of off-diagonal splitting functions $(\Delta)P_{ab}$ ($\tilde P_{ab}$), as they contribute only beyond NSV accuracy. 
In addition, we drop all the BNSV terms in the diagonal splitting functions as well.
This leads to
\begin{eqnarray}
\label{Massfactqq}
\!(\Delta) \hat {\cal C}_{J,qq}^{\text{SV+NSV}} \!\!=\!(\Delta) \Gamma_{q\leftarrow q}^{\mathrm{SV+NSV}} \!\! \otimes \! (\Delta)\mathcal{C}_{J,qq}^{\mathrm{SV+NSV}}    \tilde{\otimes}  \tilde  \Gamma_{q\leftarrow q}^{\mathrm{SV+NSV}}\, .
\nonumber \\
\end{eqnarray}
Here, the superscript on the AP kernels indicate their truncation to SV and NSV terms only. 
In this way we can write the bare CFs in eq.~(\ref{massfactC}) are given in terms of only diagonal quantities, i.e., $(\Delta)\hat{\mathcal{C}}_{J,qq}$, $(\Delta){\mathcal{C}}_{J,qq}$ and AP kernels $(\Delta) \Gamma_{q\leftarrow q}$, $\tilde \Gamma_{q\leftarrow q}$, where BNSV terms and the sum $ab$ over the parton channels are dropped. 

The bare flavor off-diagonal CFs instead contain both diagonal and off-diagonal terms from $(\Delta) {\cal C}_{J,ab}$ and AP kernels.  
For instance $(\Delta) \hat {\cal C}_{J,qg}$ reads
\begin{widetext}
\begin{align}\label{massfactqg}
(\Delta) \hat {\cal C}_{J,qg}(x',z',\varepsilon) = &~(\Delta) \Gamma_{q \leftarrow q} \otimes (\Delta) {\cal C}_{J,q q} \tilde \otimes \tilde \Gamma_{g \leftarrow q}+ (\Delta) \Gamma_{q \leftarrow q} \otimes (\Delta) {\cal C}_{J,q\overline{q}} \tilde \otimes \tilde \Gamma_{g \leftarrow \overline{q}}+
(\Delta) \Gamma_{q \leftarrow q} \otimes (\Delta) {\cal C}_{J,qg} \tilde \otimes \tilde \Gamma_{g \leftarrow g}\\\nonumber
&+
(\Delta) \Gamma_{\overline{q} \leftarrow q} \otimes (\Delta) {\cal C}_{J,\overline{q} q} \tilde \otimes \tilde \Gamma_{g \leftarrow q}+ (\Delta) \Gamma_{\overline{q} \leftarrow q} \otimes (\Delta) {\cal C}_{J,\overline{q}\overline{q}} \tilde \otimes \tilde \Gamma_{g \leftarrow \overline{q}}+
(\Delta) \Gamma_{\overline{q} \leftarrow q} \otimes (\Delta) {\cal C}_{J,\overline{q}g} \tilde \otimes \tilde \Gamma_{g \leftarrow g}\\\nonumber
&+
(\Delta) \Gamma_{g \leftarrow q} \otimes (\Delta) {\cal C}_{J,g q} \tilde \otimes \tilde \Gamma_{g \leftarrow q}+ (\Delta) \Gamma_{g \leftarrow q} \otimes (\Delta) {\cal C}_{J,g\overline{q}} \tilde \otimes \tilde \Gamma_{g \leftarrow \overline{q}}+
(\Delta) \Gamma_{g \leftarrow q} \otimes (\Delta) {\cal C}_{J,gg} \tilde \otimes \tilde \Gamma_{g \leftarrow g}
\, .
\end{align}
\end{widetext}
Unlike the diagonal case in eq.~\eqref{massfactDY}, no single term provides pure SV contributions in the threshold limit. 
Hence, the mass factorised result for the off-diagonal channel contains NSV terms only and those beyond, where the former ones come from contributions that contain at least two diagonal quantities, either from $(\Delta) {\cal C}_{J,ab}$ or $(\Delta) \Gamma_{a\leftarrow b},\tilde \Gamma_{a\leftarrow b}$.  
Dropping all terms that contain more than two off-diagonal quantities in the mass factorisation formula, we obtain
\begin{eqnarray}
(\Delta) \hat {\cal C}^{\text{NSV}}_{J,qg}  &=&(\Delta) \Gamma_{q q}^{\mathrm{SV+NSV}}\otimes (\Delta){\cal C}_{J,q  q}^{\mathrm{SV+NSV}} \tilde{\otimes} \tilde \Gamma_{\overline q g}^{\mathrm{NSV}} 
\nonumber \\
&+&(\Delta) \Gamma_{q q}^{\mathrm{SV+NSV}} \!\otimes \! (\Delta){\cal C}_{J,qg}^{\mathrm{NSV}}\! \tilde{\otimes} \tilde \Gamma_{gg}^{\mathrm{SV+NSV}}\, ,
\qquad
\end{eqnarray}
which receives contribution from both CFs, $(\Delta) {\cal{C}}_{J,qg}$ and $(\Delta){\cal C}_{J,q  q}$, 
unlike the diagonal channel in eq.~\eqref{Massfactqq} which is directly proportional to $(\Delta) \hat{\cal C}_{J,q q}$.
%

The simple structure in the diagonal channel makes it easy for us to study its all order structure of SV and NSV logarithms.  
We achieve this by setting up a  homogeneous differential equations for CFs in terms of the soft and collinear anomalous dimensions.  
The presence of inhomogeneous terms in corresponding differential equations for  the off-diagonal channel poses
technical challenges when solving them.   Hence, in the rest of the paper, we will focuss only to diagonal partonic channels.  

We begin with the eq.~\eqref{Massfactqq} for the diagonal CFs and rewrite as
\begin{widetext}
\begin{align}
\label{MassFactRe}
(\Delta)\mathcal{C}_{J,qq}^{\rm {SV+NSV}}   = 
 \left | \hat F_q (Q^2,\varepsilon)\right|^2    \Big((\Delta)\Gamma\Big)^{-1}_{q\leftarrow q}(x',\mu_F^2,\varepsilon) &\otimes 
(\Delta) \hat {\cal S}_{J,q} (Q^2,x',z',\varepsilon)  
 \tilde{\otimes} \Big(\tilde \Gamma\Big)^{-1}_{ q \leftarrow  q}(z',\mu_F^2,\varepsilon) 
\end{align}
\end{widetext}
where $\hat F_q(Q^2,\varepsilon)$ is the massles quark form factor (FF) in dimensional regularization 
and $(\Delta)\hat {\cal S}_{J,q}(Q^2,x',z',\varepsilon)$, called soft function, is defined by
\begin{eqnarray}
(\Delta)\hat {\cal S}_{J,q}(Q^2,x',z',\varepsilon) = \frac{(\Delta)\hat {\cal C}_{J,qq}(Q^2,x',z',\varepsilon)}{\left| \hat F_{q}(Q^2,\varepsilon)\right|^2}
\end{eqnarray}
Note that in the absence of any real radiation, the soft functions is constrained by kinematics and takes the simple form: $(\Delta)\hat  {\cal S}_{J,q}|_{\text  {no radiation}} = \delta(1-x') \delta(1-z')$.

The vector form factor $\hat F$ has been subject to extensive studies, see, e.g.~\cite{Sen:1981sd,Collins:1989bt,Magnea:1990zb,Magnea:2000ss,Sterman:2002qn,
Moch:2005id,Ravindran:2005vv}.
In massless QCD and within dimensional regularization it exhibits IR and UV singularities as poles in $\varepsilon$. 
The UV divergences can be removed by renormalisation of the strong coupling, while the soft and collinear, i.e., IR
sigularities, factorise as 
\begin{eqnarray}
\hat F_q(Q^2,\varepsilon)
= Z_{\hat F_q}(Q^2,\mu_s^2,\varepsilon) F_{q,fin}(Q^2,\mu_s^2,\varepsilon)\, ,
\end{eqnarray}
where $Z_{\hat F_q}$ is IR singular and $F_{q,fin}$ is IR finite, the scale $\mu_s$ is IR factorisation scale.
Renormalisation group invariance implies that 
\begin{eqnarray}
\label{RG-ZF}
\mu_s^2 \frac{d}{d\mu_s^2} \ln Z_{\hat F_q} (\mu_s^2)= \gamma_{\hat F_q} (\mu_s^2)\, ,
\end{eqnarray}
where $\gamma_{\hat F_q}$ to all order in $a_s$ takes the form
\begin{align}
\gamma_{\hat F_q} = {1 \over 2} 
\Bigg(-A^q(\mu_s^2) \ln \Bigg(\frac{Q^2} {\mu_s^2} \Bigg) + 2 B^q(\mu_s^2)+ f^q(\mu_s^2) \Bigg)
\, .
\end{align}
Here, $A^q, B^q, f^q$ denote the cusp ($A^q$), the quark virtual $B^q$ and the eikomal anomalous dimensions, which all have expansions in powers of the strong coupling as $A^q(\mu_s^2)=\sum_i^\infty a_s^{i}(\mu_s^2) A^q_i$ etc.
They are currently known to four-loop accuracy in QCD~\cite{Henn:2019swt,vonManteuffel:2020vjv,Das:2019btv,Das:2020adl,Agarwal:2021zft,Lee:2021uqq,Moch:2023tdj,Kniehl:2025ttz}.
Thus, the solution to eq.~\eqref{RG-ZF} for $Z_{\hat F_q}$
in the $\overline {\text{MS}}$ scheme contains only poles in $\varepsilon$ whose coefficients are completely controlled by universal anomalous dimensions $A^q, B^q$ and $f^q$ and the coefficients of QCD $\beta$ function, the latter being known even to five loops~\cite{Baikov:2016tgj,Herzog:2017ohr,Luthe:2017ttg,Chetyrkin:2017bjc}.

The finite part $F_{q,fin}$ of the form factor is subject to a differential equation with respect to $Q^2$,  
\begin{eqnarray}
\label{KGeqF}
Q^2 \frac{d}{d Q^2}\ln \hat F_q(Q^2,\varepsilon)= \Gamma_{\hat F_q}(Q^2,
\varepsilon)\, ,
\end{eqnarray}
where
\begin{eqnarray}
\label{Fker}
\Gamma_{\hat F_q}(Q^2,
\varepsilon)=\frac{1}{2} \bigg(K_d^{q}(\mu_s^2,\varepsilon)+G_d^{q}(Q^2,\mu_s^2,\varepsilon)\bigg) 
\, .
\end{eqnarray}
The kernel $K_d^{q} = 2 d \ln Z_{\hat F_q}/d\ln(Q^2)$ contains entire IR singular terms while the function $G_d^{q}=2 d \ln F_{q,fin}/d\ln(Q^2)$ is IR finite by definition.  
Since, we have defined the form factor in dimensional regularisation, the solution to the $K+G$ equation
takes the simple form
\begin{eqnarray}
\label{Fcsol}
\hat F_q(Q^2,\varepsilon) = \exp\left(\int_0^{Q^2} {d \lambda^2 \over \lambda^2} \Gamma_{\hat F_q}(\lambda^2,\varepsilon)\right)\, ,
\end{eqnarray}
with $\hat F_q(Q^2=0,\varepsilon)=1$ as boundary condition.   
Since all IR poles in $\varepsilon$ are contained in $K_d^{q}$ only and $G_d^{q}$ is finite, the solution reveals the remarkable structure for the form factor, namely, at given order $a_s^i$, all the poles except
the single pole in $\varepsilon$ can be predicted from knowledge of $K_d^{q}$ and $G_d^{q}$ at previous orders up to $a_s^{i-1}$. 
Explicit results for $K_d^{q}$ and $G_d^{q}$ are presented in \cite{Moch:2005ky,Ravindran:2005vv} and also in App.~\ref{FKG}.

Next, we briefly review the structure of the AP kernels in eq.~\eqref{MassFactRe} and the associated splitting functions, controlling their evolution as a function of the factorisation scale $\mu_F$. 
We denote them collectively by $\BbbGamma_{q \la q}$ and $\mathds{P}_{qq}$ for (polarized) space-like and time-like evolution, i.e., $(\Delta)\Gamma_{q \la q}$ and $\tilde \Gamma_{q \la q}$.
In the threshold limit only the diagonal slitting functions $\mathds{P}_{c\la c} , c=q,\overline q $ contribute,
\begin{align}
\label{eq:APevolution}
\mu_F^2 \frac{d }{d\mu_F^2}\BbbGamma_{c\la c}(\mu_F^2,\xi) = \frac{1}{2}
\mathds{P}_{c\la c}(\mu_F^2,\xi)\otimes \BbbGamma_{c\la c} (\mu_F^2,\xi)
\, ,
\end{align}
where $\xi=x',z'$. 
The splitting functions are expanded in powers of $a_s(\mu_F^2)$ as $\mathds{P}_{c \la c}(\mu_F^2) =  \sum_{i=1}^{\infty} a_s^{i}(\mu_F^2) \mathds{P}^c_{i-1}$ and 
we have suppressed any dependence on $\varepsilon$ in both ${\BbbGamma_{c\la c}}$ and $\mathds{P}_{c\la c}$.
Due to its simple structure the all order solution of eq.~\eqref{eq:APevolution} takes the simple form
\begin{eqnarray}
\label{Gcsol}
\BbbGamma_{c\la c}(\mu_F^2,\xi)={\cal C} \exp\left({1\over 2} \int_0^{\mu_F^2}
{d \lambda^2 \over \lambda^2} \mathds{P}_{c\la c}(\lambda^2,\xi)\right)
\, ,
\end{eqnarray}
where the symbol ${\cal C}$ 
denotes an ordered exponential, appropriately defined through iterated convolutions with respect to $x'$ or $z'$, see \cite{Ravindran:2005vv}. 
The relevant splitting functions are known to third order in perturbative QCD~\cite{Moch:2004pa,Vogt:2004mw,Moch:2014sna,Almasy:2011eq,Chen:2020uvt} and their expansion in the soft and collinear limit reads
\begin{eqnarray}
\label{eq:APspSVNSV}
\mathds{P}_{c\la c}(\mu_F^2,\xi) &=& 2 \Bigg(
\frac{A^c(\mu_F^2)}{(1-\xi)_{+}} + B^c(\mu_F^2)\delta(1-\xi)
\nonumber \\
&&+ \mathds{C}^c(\mu_F^2) \ln(1-\xi)
+\mathds{D}^c(\mu_F^2)
\Bigg)\, ,
\end{eqnarray}
in terms of the cusp and the virtual anomalous dimensions $A^c$ and $B^c$ already encountered in the form factor above.
The NSV terms, denoted as $\mathds{C}^c=C^q,$ and $\mathds{D}^c=D^q$, are determined from $A^c$, $B^c$ and coefficients of the QCD $\beta$-function at lower orders.
There is distinction between space- and time-like kinematics, i.e.\ $\mathds{C}^c=C^q,\mathds{D}^c=D^q$ for space-like and $\mathds{C}^c=\tilde C^q,\mathds{D}^c=\tilde D^q$ for time-like splitting functions.
Explicit results are isted in App.~\ref{App:ano}.

The final ingredient in the threshold factorization formulae in eq.~\eqref{MassFactRe} is the soft function $\hat S_q$.
A differential equation for $\hat S_q$ in dimensional regularisation, organizing its all-order perturbative structure, can be obtained with eq.~\eqref{MassFactRe} and the differential equation for the form factor, see eq.~\eqref{KGeqF}.  
\begin{eqnarray}
\label{KGeqS}
Q^2{d\hat { \cal S}_q(Q^2,\varepsilon) \over dQ^2} = \Gamma_{\hat {\cal S}_q}(Q^2,\varepsilon)\otimes \hat { \cal S}_q(Q^2,\varepsilon)\, ,
\end{eqnarray}
where we have suppressed the dependence on $x',z'$ and 
\begin{eqnarray}
\label{Scker}
\Gamma_{\hat {\cal S}_q}(Q^2,\varepsilon)\!\!&=&\!\!Q^2{d\over dQ^2}\Big({\cal C} \ln (\Delta) {\cal C}_{J,qq}(Q^2,\mu_R^2,\mu_F^2,x',z',\varepsilon) ) \nonumber\\
&&-\ln|\hat F_q(Q^2,\varepsilon)|^2 \delta(1-x')\delta(1-z')\Big)\, .
\end{eqnarray}
The fact that $\hat {\cal S}_q$ is renormalization group invariant with respect to $\mu_R$ and $\mu_F$ implies that the $Q^2$ derivative of $(\Delta) {\cal C}_{J,qq}$ in the first term in eq.~\eqref{Scker} has to be a function of only $Q^2$ and $x',z'$.  This implies that $\Gamma_{\hat {\cal S}_q}$ is  independent of $\mu_R$ and $\mu_F$. 
Observe that the first term is finite and the second  term contains both  singular and finite parts through $K_{d}^{q}$ and $G_{d}^{q}$ via $K+G$ equation of $\Gamma_{\hat F_q}$, see eq.~\eqref{KGeqF}. 
This means that we can express the kernel $\Gamma_{\hat {\cal S}_q}$ in eq.~\eqref{Scker} as a sum of singular term $\overline K_{d}^{q}$ and finite $\overline G_{d}^{q}$ pieces to all orders in perturbation theory, similar to the $Q^2$ evolution equation of the form factor, i.e., $\Gamma_{\hat {\cal S}_q}=   (\overline K_{d}^{q}(\mu_s^2,x',z') +\overline G_{d}^{q}(Q^2,\mu_s^2,x',z'))/2$. 
Since the singular part of the kernel comes from $K_{d}^{q}$ of the form factor, the term  $\overline K_{d}^{q}$ can depend only on $\mu_s$ and process independent anomalous dimension, $A^q$. 
The finite $\overline G_q(Q^2,\mu_s^2,z)$ will contain $G_{d}^{q}$ from the form factor and the other process dependent terms from CF $(\Delta) C_{J,qq}$. 

The fact that the kernel admits the decomposition of singular and finite parts as it happens for the form factor, implies that  the $\hat {\cal S}_q$ is also factorisable, i.e., we can  write $\hat {\cal S}_q(Q^2,x',z') = Z_q(Q^2,\mu_s^2,x',z')\otimes  {\cal S}_{q,{\rm fin}}(Q^2,\mu_s^2,x',z')$, where ${\cal S}_{q,{\rm fin}}$ is IR finite, $Z_q$ is IR singular and identify the IR singular $\overline K_{d}^{q}=2 d\ln Z_q/d\ln(Q^2)$ and IR finite $\overline G_{d}^{q}=2 d\ln  {\cal S}_{q,fin}/d\ln(Q^2)$. 
Since $\overline K_{d}^{q}(\mu_s^2)$ is independent of $Q^2$, we can fix only the $\ln(Q^2)$ terms in $Z_q$.  The factorisation of $\hat {\cal S}_{q}$ in terms of $Z_q$ and $\hat {\cal S}_{q,\text{fin}}$ at arbitrary scale $\mu_s$ leads to the renormalisation group equation,
\begin{eqnarray}
\label{RG-ZS}
\mu_s^2 {d Z_q(Q^2,\mu_s^2 )\over d \mu_s^2}  &=&\gamma_{\hat {\cal S}_q} (
Q^2,\mu_s^2)\otimes Z_q(Q^2,\mu_s^2)\, ,
\nonumber
\end{eqnarray}
where we have suppressed any dependence on $x',z'$ in $Z_q$ and $\gamma_{\hat {\cal S}_q}$. 
From our two-loop results we find that $\gamma_{\hat {\cal S}_q}$ takes the remarkable structure, namely $\gamma_{\hat {\cal S}_q}(Q^2,\mu_s^2)=\xi_1(\mu_s^2,x',z') \ln(Q^2/\mu_s^2) +
\xi_2(\mu_s^2,x',z')$. 
This is expected to continue to all orders in perturbation theory because it results from the IR  pole structure of $\hat {\cal S}_q$. 
In other words, the singular part of $\hat {\cal S}_q$, namely, $Z_q$ has to contain precisley those IR poles to cancel against the ones from the form factor and AP kernels leaving $(\Delta){\cal C}_{J,qq}$ finite and this pole structure determines the structure of $\gamma_{\hat {\cal S}_q}$.     
Since the poles of the form factor and AP kernels are controlled by the cusp, virtual and eikonal anomalous dimensions, $\gamma_{\hat {\cal S}_q}$ can be expressed as follows:
\begin{align}
\label{scanom}
\gamma_{\hat {\cal S}_q}\!\!&= \!\delta(\bar x')\delta(\bar z') A^q(\mu_s^2) \ln\Big({Q^2\over \mu_s^2}\Big)
+P'_{q\la q}(\mu_s,\overline{x}',\overline{z}')
\, ,
\end{align}
where
\begin{align}
P'_{q\la q}=&
\bigg[-{1 \over 2} \delta(\bar x')\delta(\bar z')   {f^q(\mu_s^2)} 
+\delta(\bar z'){A^q(\mu_s^2)\over (\bar x')_+} 
\nonumber\\&
+\delta(\bar z')\Big( C^q(\mu_s^2)\ln(\bar x') 
+ D^q(\mu_s^2)\Big)\bigg]\nonumber\\&
+\Big(\overline{x}'\leftrightarrow \overline{z}', C^q \leftrightarrow \tilde C^q, D^q \leftrightarrow \tilde D^q \Big) 
\, ,
\end{align}
and we use the shorthand notation $\bar x'=1-x',\bar z'=1-z'$.

With the boundary $Q^2=0$, $\hat {\cal S}_{q}(Q^2=0,x',z',\varepsilon) = \delta(1-x') \delta(1-z')$ to all orders in $a_s$ we obtain the solution to eq.~\eqref{KGeqS} as
\begin{eqnarray}
\label{eq:calScsol}
\hat {\cal S}_q(Q^2,x',z',\varepsilon) &= &{\cal C} \exp\left(\int_0^{Q^2} {d \lambda^2 \over \lambda^2} \Gamma_{\hat {\cal S}_q}(\lambda^2,x',z',\varepsilon)\right)
\nonumber\\
&=&{\cal C} \exp\left(2 \Phi_{d,q}(Q^2,x',z',\varepsilon)\right)
\, .
\end{eqnarray}
The exponent $\Phi_{d,q}$ can be decomposed into an IR singular part $\Phi_{d,q,s}$ and a finite part $\Phi_{d,q,f}$, i.e., $\Phi_{d,q}=\Phi_{d,q,s}+\Phi_{d,q,f}$.
Using this decomposition, we obtain $Z_q= {\cal C} \exp(2 \Phi_{d,q,s})$ and ${\cal S}_{q,fin}={\cal C} \exp(2 \Phi_{d,q,f})$.
The singular part is controlled by $\gamma_{\hat {\cal S}_q}$ given in eq.~\eqref{RG-ZS}.  A careful inspection of the one- and two-loop results
for $\hat {\cal S}_q$, i.e., using the results presented in \cite{Goyal:2023zdi,Goyal:2024tmo}, provides hints to its all order structure.  
However, before we proceed, we split $\Phi_{d,q}$ into the sum of two parts.
The SV part denoted by $\Phi_{d,q}^{\text{SV}}$ contains only distributions such as $\delta(1-\xi)$ and ${\cal D}_{\xi,i}$, while NSV part $\Phi_{d,q}^{\text{NSV}}$ summarizes all terms that are products of distributions and $\ln^i(1-\xi)$ with $\xi = x',z'$.  
We find that $\Phi_{d,q}^{\text{SV}}$ is symmetric under the interchange of $x'$ and $z'$ while $\Phi_{d,q}^{\text{NSV}}$ is not.
Our observations for $\Phi_{d,q}$ read as follows:
\begin{itemize}
\item  For the SV terms, at order $\hat a_s$, the leading pole in $\varepsilon$ starts at order two, at $\hat a_s^2$ it is three and it is expected to increase by one unit at every order in $\hat a_s$.
Hence, at order $\hat a_s^i$, the leading pole in $\varepsilon$
will be of order $i+1$.  As we had already mentioned, this pole structure is completely dictated by pole structure of $\ln |\hat F_{q}|^2$ and the AP kernels.  For the NSV terms, we find that at order $\hat a_s$ the leading pole in $\varepsilon$ is of order one and at two loops, it is two etc.  This pattern indicates that the increment of one unit for the leading poles is expected to continue
with the order of perturbation and hence at order $\hat a_s^i$, the leading pole
will be of order $i$.
\item   
At every order in $\hat a_s$, for a given color factor, the combination of $\varepsilon$
and the leading logarithm shows uniform transcendentality weight. If we
assign $n_\varepsilon$ weight for $\varepsilon^{-n}$ and $n_{{\cal D}}$ for ${\cal D}_{\overline \xi}^{n}$, $n_L$ for 
$\ln^n (1-\xi)$ and $n_\zeta$ for $\zeta_n$ then the highest weight at 
every order in $\varepsilon$
shows uniform transcendentality $\omega =n_\varepsilon + n_{{\cal D}} + n_L + n_\zeta$. For example, in SV, at one loop, we find $\omega = 2$  at
every order of $\varepsilon$ and at two loops it is three ($\omega = 3$) etc.  For NSV, $\omega =1 $ at one loop and $\omega=2$ at two loops at every order in 
$\varepsilon$ etc.
\end{itemize}
The above observations on the fixed order results and the resemblance of eq.~(\ref{KGeqF}) with eq.~(\ref{KGeqS}), the similarity in the pole structure of the respective kernels eqs.~(\ref{Fker}) and (\ref{Scker}) and the renormalization group invariance reveal an all order structure for $\phi^{\text{SV},(i)}_{d,q}$ of the form:
\begin{widetext}  
\begin{align}
\label{eq:PhiSV}
\mathrm{\Phi}^{\text{SV}}_{q} =& \sum_{i=1}^\infty \hat a_s^i \left(Q^2 (1-x') (1-z') \over \mu^2\right)^{i{\varepsilon \over 2}} S_\varepsilon^i
\Bigg[ {(i \varepsilon)^2 \over 4 (1-x') (1-z') } \hat \phi_{d,q}^{\text{SV},(i)}(\varepsilon)\Bigg] \,,
\end{align}
where
\begin{eqnarray}
\hat{\phi}_{d,q}^{\text{SV},(i)}(\varepsilon) = \frac{1}{i\epsilon}\Big[ \overline{K}_{d}^{q(i)}(\varepsilon) + \overline{G}_{d,\text{SV}}^{q(i)}(\varepsilon) \Big]\, ,
\end{eqnarray} 
and for NSV part
\begin{eqnarray}
\label{eq:PhiNSV}
\mathrm{\Phi}^{\text{NSV}}_{q} &=& \sum_{i=1}^\infty \hat a_s^i \left(Q^2 (1-x') (1-z') \over \mu^2\right)^{i{\varepsilon \over 2}} S_\varepsilon^i
\Bigg[ {i\varepsilon \over 4(1-x') }\hat \varphi_{d,z',q}^{\text{NSV},(i)} ( z',\varepsilon) 
+ {i\varepsilon \over 4(1-z') }\hat \varphi_{d,x',q}^{\text{NSV},(i)} (x',\varepsilon) \Bigg]
\, .
\end{eqnarray}
\end{widetext}
For reader's convenience, we have listed the results of $\overline{K}_{d}^{q(i)}$ and $\overline{G}_{d,\text{SV}}^{q(i)}$ in the App.~\ref{AppendixKbGb}. 
After substituting them the constants $\hat \phi_{d,q}^{\text{SV},(i)}(\varepsilon)$ are found to be
\begin{widetext}
\begin{eqnarray}
\label{phisvi}
\hat{\phi}^{\text{SV},(1)}_{d,q}(\varepsilon)&=&{1\over \varepsilon^2} \Bigg(2 A_1^q\Bigg) 
              +{1 \over \varepsilon} \Bigg({\cal \overline{G}}_{d,1}^q(\varepsilon)\Bigg)\, ,
\nonumber\\[2ex]
\hat{\phi}_{d,q}^{\text{SV},(2)}(\varepsilon)&=&{1\over \varepsilon^3} \Bigg(-\beta_0 A_1^q\Bigg) 
                  +{1\over \varepsilon^2} \Bigg({1 \over 2} A_2^q 
                  - \beta_0  {\cal \overline{G}}_{d,1}^q(\varepsilon)\Bigg)
                  +{1 \over 2 \varepsilon} {\cal \overline{G}}_{d,2}^q(\varepsilon)\, ,
\nonumber\\[2ex]
\hat{\phi}_{d,q}^{\text{SV},(3)}(\varepsilon)&=& {1\over \varepsilon^4} \Bigg({8 \over 9}\beta_0^2 A_1^q\Bigg) 
                  + {1\over \varepsilon^3} \Bigg(-{2 \over 9} \beta_1 A_1^q
                    -{8 \over 9} \beta_0 A_2^q -{4 \over 3} 
                     \beta_0^2 {\cal \overline{G}}_{d,1}^q(\varepsilon)\Bigg) 
+{1\over \varepsilon^2} \Bigg({2 \over 9} A_3^c 
                   -{1 \over 3} \beta_1 {\cal \overline{G}}_{d,1}^q(\varepsilon) 
                   -{4 \over 3}\beta_0 {\cal \overline{G}}_{d,2}^q(\varepsilon)\Bigg)
\nonumber\\[2ex]
&&                  +{1 \over \varepsilon}\Bigg({1 \over 3} {\cal \overline{G}}_{d,3}^q(\varepsilon)\Bigg)\, ,
\nonumber\\[2ex]
\hat{\phi}_{d,q}^{\text{SV},(4)}(\varepsilon)&=& {1\over \varepsilon^5} \Bigg(-\beta_0^3 A_1^q\Bigg) 
                  +{1 \over \varepsilon^4} \Bigg({2 \over 3} \beta_0 \beta_1 A_1^q
                   +{3 \over 2}\beta_0^2 A_2^q -2 \beta_0^3 {\cal \overline{G}}_{d,1}^q(\varepsilon)\Bigg)               
-{1 \over \varepsilon^3} \Bigg({1 \over 12} \beta_2 A_1^q 
 -{1 \over 4} \beta_1 A_2^q
                    - {3 \over 4}\beta_0 A_3^q 
\nonumber\\[2ex]
&&
                    +{4 \over 3} 
                       \beta_0 \beta_1 {\cal \overline{G}}_{d,1}^q(\varepsilon)
                    +3\beta_0^2 {\cal \overline{G}}_{d,2}^q(\varepsilon)\Bigg)
+{1 \over \varepsilon^2} \Bigg({1 \over 8} A_4^q
                       -{1 \over 6} \beta_2 {\cal \overline{G}}_{d,1}^q(\varepsilon)
                       -{1 \over 2} \beta_1 {\cal \overline{G}}_{d,2}^q(\varepsilon) 
                    -{3\over 2} \beta_0 
            {\cal \overline{G}}_{d,3}^q(\varepsilon)\Bigg)
                  +{1 \over \varepsilon} \Bigg({1 \over 4} {\cal \overline{G}}_{d,4}^q(\varepsilon)\Bigg)\, .
\nonumber\\[2ex]
&&                  
\end{eqnarray}
\end{widetext}
The cusp anomalous dimensions $A^q_i$ in the above equations carry the right sign to cancel all the leading singularities against the form factor contributions in CFs.  
The $\varepsilon$ dependent terms $\overline{\cal G}^q_{d,i}$ originate from  {$|{ \hat{F}}_q|^2$} as well as the CFs $(\Delta)\mathcal{C}_{J,qq}$ in eq.~(\ref{Scker}).  
Using renormalization group invariance and demanding the cancellation of single poles in eq.~(\ref{phisvi}) against the form factor and the AP kernels, we can parameterise ${\overline {\cal G}}^q_{d,i}$ as,
\begin{widetext}
\begin{align}\label{eq:gbarsv}
\overline {\cal G}^{q}_{d,i}(\varepsilon)=- f_i^q+ \overline \chi_{d,i}^q +
\sum_{j=1}^\infty \varepsilon^j \overline {\cal G}^{q,(j)}_{d,i}\,,
\end{align}
where
\begin{align}
  \label{eq:gbarsvi}
  \overline \chi^{q}_{d,1} &= 0\, ,
  \overline\chi^{q}_{d,2} = - 2 \beta_{0} \overline {\cal G}^{q,(1)}_{d,1}\, ,
 \overline\chi^{q}_{d,3} = - 2 \beta_{1}\overline {\cal G}^{q,(1)}_{d,1} - 2
                \beta_{0} \left(\overline {\cal G}^{q,(1)}_{d,2}  + 2 \beta_{0}\overline {\cal G}^{q,(2)}_{d,1}\right)
                \nonumber\\
\overline\chi^{q}_{d,4} &= - 2 \beta_{2}\overline {\cal G}^{q,(1)}_{d,1} - 2
                \beta_{1} \left(\overline {\cal G}^{q,(1)}_{d,2}  + 4 \beta_{0}\overline {\cal G}^{q,(2)}_{d,1}\right)
                - 2\beta_{0} 
                \left(\overline {\cal G}^{q,(1)}_{d,3}  + 2 \beta_{0}\overline {\cal G}^{q,(2)}_{d,2} + 4 \beta_{0}^2\overline {\cal G}^{q,(3)}_{d,1}\right)
                \,.
\end{align}
\begin{align}\label{app:SVGij}
\overline{\cal G}^{q,(1)}_{d,1}
&=C_F   \Bigg\{  - \zeta_2 \Bigg\}\,,
\qquad \qquad
\overline{\cal G}^{q,(2)}_{d,1}
= C_F   \Bigg\{ \frac{1}{3} \zeta_3 \Bigg\}\,,
\nonumber\\
\overline{\cal G}^{q,(3)}_{d,1}
&= C_F   \Bigg\{ \frac{1}{80} \zeta_2^2 \Bigg\}\,,
\qquad \qquad
\overline{\cal G}^{q,(4)}_{d,1} =  C_F   \Bigg\{ \frac{1}{20} \zeta_5 - \frac{1}{24}    \zeta_2 \zeta_3 \Bigg\}\,,
\nonumber\\
\overline{\cal G}^{q,(5)}_{d,1}
&= C_F   \Bigg\{ \frac{1}{144} \zeta_3^2 -              \frac{1}{896} \zeta_2^3 \Bigg\}\,,
\nonumber\\
\overline{\cal G}^{q,(1)}_{d,2}
&=  C_F n_f   \Bigg\{  - \frac{328}{81} + \frac{8}{3}    \zeta_3 + \frac{10}{3} \zeta_2 \Bigg\}
       + C_F C_A   \Bigg\{ \frac{2428}{81} - \frac{44}{3}     \zeta_3 - \frac{67}{3} \zeta_2 - 4 \zeta_2^2 \Bigg\}\,,
            \nonumber    \\
\overline{\cal G}^{q,(2)}_{d,2} &=
 C_F n_f   \Bigg\{ \frac{976}{243} - \frac{70}{27}    \zeta_3 - \frac{28}{9} \zeta_2 + \frac{29}{60} \zeta_2^2      \Bigg\}
       - C_F C_A   \Bigg\{   \frac{7288}{243} - 43 \zeta_5 - \frac{469}{27} \zeta_3 - \frac{202}{9} \zeta_2 + \frac{71}{3}
         \zeta_2 \zeta_3 + \frac{319}{120} \zeta_2^2 \Bigg\}\,,
   \nonumber\\
 \overline{\cal G}^{q,(1)}_{d,3} &= 
C_F n_f^2   \Bigg\{ \frac{11584}{2187} -   \frac{320}{81} \zeta_3 - \frac{316}{27} \zeta_2 +             \frac{152}{45} \zeta_2^2
          \Bigg\}
- C_F^2 n_f   \Bigg\{   \frac{42727}{324} -          \frac{112}{3} \zeta_5 - \frac{1672}{27} \zeta_3 -     \frac{275}{6} \zeta_2
+ 40 \zeta_2 \zeta_3 - \frac{152}{15} \zeta_2^2  \Bigg\}   
    \nonumber \\
&+ C_F C_A^2   \Bigg\{ \frac{7135981}{8748} -           \frac{1430}{3} \zeta_5 - 936 \zeta_3 + \frac{536}{3} \zeta_3^2
      - \frac{379417}{486} \zeta_2 + \frac{4136}{9}       \zeta_2 \zeta_3 + \frac{1538}{45} \zeta_2^2 + \frac{17392}{315} \zeta_2^3 \Bigg\}\nonumber\\
& - C_F C_A n_f  \Bigg\{   \frac{716509}{4374} -       \frac{148}{3} \zeta_5 - \frac{12356}{81} \zeta_3 - 
   \frac{51053}{243} \zeta_2 + \frac{392}{9} \zeta_2 \zeta_3 + \frac{1372}{45} \zeta_2^2 \Bigg\}\,.
\end{align}
\end{widetext}
Here, $C_A=N_c$ and $C_F \equiv \frac{N_c^2-1}{2N_c}$ are the quadradic Casimirs of the adjoint and fundamental representations of $SU(N_c)$, respectively, 
while $n_f$ denoted the number of quark flavors.
Note, that the constants in eq.~\eqref{app:SVGij} are also related to the threshold exponent $\textbf{D}^q\big(a_s(q^2(1-z)^2)\big)$ via eq.~(46) of \cite{Ravindran:2006cg}. 

The results for $\overline{\cal G}^{q,(j)}_{d,i}$ for $i=1,2$ have been obtained from the NLO and NNLO CFs expanded to sufficient order in $\varepsilon$ \cite{Altarelli:1979kv,Goyal:2023zdi,Bonino:2024qbh}. 
We find that these results are identical to the corresponding ones for the rapidity distribution of pairs of leptons in DY production~\cite{Ravindran:2006bu,Ravindran:2007sv,Ahmed:2014uya,Ahmed:2014era,Banerjee:2017cfc,Banerjee:2018vvb,AH:2020qoa,Ahmed:2020amh,AH:2021vhf,Ravindran:2022aqr,Ravindran:2023qae}, see also \cite{Anderle:2012rq,Westmark:2017uig,Banerjee:2018vvb}, 
where a conjecture on the relation between these two processes in the threshold limit had been proposed. 
Assuming that this conjecture holds true beyond NNLO, we have used $\overline{\cal G}^{q,(1)}_{d,3}$ of the rapidity distribution for the DY process also for SIDIS at N${}^3$LO.

The all order result given in eq.~\eqref{eq:PhiSV} is useful to obtain predictions for the SV part of the CFs order by order $a_s$.  
However, to study the numerical impact of the all order result, we need to perform a resummation in Mellin space.  
This can be achieved by expressing the summation in the integral form. The integral representation for $\mathrm{\Phi}_{q}^{\text{SV}}$ for the DY process is given in~\cite{Ravindran:2006bu} and the same can be 
used for SIDIS in eq.~\eqref{eq:PhiSV}: 
\begin{widetext}
\begin{eqnarray}
\label{eq:SVintrep}
\Phi^{\text{SV}}_{q}(\hat a_s,Q^2,\mu^2,x',z',\varepsilon)&=&
{1 \over 2}\delta(\bar z') \Bigg( {1 \over \bar x'} \Bigg\{
\int_{\mu_F^2}^{Q^2 \bar x'} {d \lambda^2 \over \lambda^2}
A^q \left(a_s(\lambda^2)\right) 
+ \overline G^{q}_d \left(
a_s\left(Q^2 \bar x'\right),\varepsilon\right)\Bigg\} \Bigg)_+
\nonumber\\[2ex]
&&+Q^2 {d \over dQ^2} \Bigg[
\Bigg({1 \over 4 \bar x' \bar z'}
\Bigg\{\int_{\mu_F^2}^{Q^2 \bar x' \bar z'} {d \lambda^2 \over \lambda^2} 
A^q\left(a_s(\lambda^2)\right)
+\overline G^q_d\left(a_s\left(Q^2 \bar x' \bar z'\right),\varepsilon\right)\Bigg\}\Bigg)_+\Bigg]
\nonumber\\[2ex]
&&+{1 \over 2} \delta(\bar x')\delta(\bar z') \sum_{i=1}^\infty \hat a_s^i
\left({Q^2 \over \mu^2}\right)^{i {\varepsilon \over 2}}
S_{\varepsilon}^i~
\hat \phi^{\text{SV},(i)}_{d,q}(\varepsilon)
+{1 \over 2} \delta(\bar z')
\left({1 \over \bar x'}\right)_+ \sum_{i=1}^\infty
\hat a_s^i \left({\mu_F^2 \over \mu^2}\right)^{i {\varepsilon \over 2}}
S_{\varepsilon}^i~
\overline K^{q,(i)}_{d}(\varepsilon)
\nonumber\\[2ex]
&&+ (x' \leftrightarrow z')\,,
\label{resum}
\end{eqnarray}
\end{widetext}
In complete analogy to the SV part, using the renormalization group, the coefficients $\varphi^{\text{NSV},(i)}_{d,\xi,q}$ in the NSV solution $\Phi^{\text{NSV}}_q$, eq. (\ref{eq:PhiNSV}) can be expanded as follows,
\begin{widetext}
\begin{align}
\label{phifc}
\hat \varphi_{d,\xi,q}^{\text{NSV},(1)}(\xi,\varepsilon) &= 
\frac{1}{\varepsilon}   
\overline {\mathcal{G}}_{d,\xi,1}^q
(\xi,\varepsilon) \,, \nonumber \\
{ \hat \varphi_{d,\xi,q}^{\text{NSV},(2)}(\xi,\varepsilon)} &= \frac{1}{\varepsilon^2}(-\beta_0 
\overline {\mathcal{G}}_{d,\xi,1}^q(\xi,\varepsilon)) + \frac{1}{2\varepsilon} 
\overline {\mathcal{G}}_{d,\xi,2}^q(\xi,\varepsilon) \,, \nonumber \\
{\hat  \varphi_{d,\xi,q}^{\text{NSV},(3)}(\xi,\varepsilon) }&=  \frac{1}{\varepsilon^3} \bigg(\frac{4}{3} \beta_0^2  
\overline {\mathcal{G}}_{d,\xi,1}^q(\xi,\varepsilon)\bigg)
 + \frac{1}{\varepsilon^2} \bigg(-\frac{1}{3} \beta_1  \overline {\mathcal{G}}_{d,\xi,1}^q(\xi,\varepsilon) - \frac{4}{3} \beta_0 
 \overline{\mathcal{G}}_{d,\xi,2}^q(\xi,\varepsilon) \bigg)
                     + \frac{1}{3\varepsilon}  
 \overline{ \mathcal{G}}_{d,\xi,3}^q(\xi,\varepsilon) \,, \nonumber \\
{\color{black} \hat \varphi_{d,\xi,q}^{\text{NSV},(4)}(\xi,\varepsilon)} &= \frac{1}{\varepsilon^4} (-2\beta_0^3  
\overline {\mathcal{G}}_{d,\xi,1}^q(\xi,\varepsilon))
 + \frac{1}{\varepsilon^3} \bigg( \frac{4}{3} \beta_0 \beta_1  
 \overline {\mathcal{G}}_{d,\xi,1}^q(\xi,\varepsilon) + 3 \beta_0^2 
 \overline {\mathcal{G}}_{d,\xi,2}^q(\xi,\varepsilon) \bigg)
                   \nonumber \\
& + \frac{1}{\varepsilon^2} \bigg( -\frac{1}{6} \beta_2  \overline {\mathcal{G}}_{d,\xi,1}^q(\xi,\varepsilon) 
-\frac{1}{2} \beta_1  
\overline {\mathcal{G}}_{d,\xi,2}^q(\xi,\varepsilon) -\frac{3}{2} \beta_0  
\overline {\mathcal{G}}_{d,\xi,3}^q(\xi,\varepsilon) \bigg)
                   + \frac{1}{4\varepsilon} 
\overline {\mathcal{G}}_{d,\xi,4}^q(\xi,\varepsilon) \,.
\end{align}
\end{widetext}
In order to cancel the collinear poles from the AP kernels, the quantities $\overline {\mathcal{G}}_{d,\xi,i}^q(\xi,\varepsilon)$ can be further decomposed as
\begin{eqnarray}
\label{eq:gbxzi}
    \overline {\mathcal{G}}_{d,x',i}^q(x',\varepsilon) &=&  2 L^q_i(x') + \overline \chi_{d,x',i}^q(x') + \sum_{j=1}^{\infty}\varepsilon^j \overline{\mathcal{G}}_{d,x',i}^{q,(j)}(x')\,,
    \nonumber \\
    \overline {\mathcal{G}}_{d,z',i}^q(z',\varepsilon) &=&  2 \tilde L^q_i(z') + \overline \chi_{d,z',i}^q(z') + \sum_{j=1}^{\infty}\varepsilon^j \overline{\mathcal{G}}_{d,z',i}^{q,(j)}(z')
    \nonumber \\
\end{eqnarray}
with $L_i^q(x') = C_i^q \ln(1-x') + D_i^q$ and $\tilde L_i^q(z') = \tilde C_i^q \ln(1-z') + \tilde D_i^q$  and 
\begin{align}
\label{phisc}
   \overline \chi_{d,\xi,i}^q(\xi) =  \overline \chi_{d,i}^q\Big|_{\left(\overline {\cal G}^{q,(j)}_{d,i} \rightarrow \overline{\mathcal{G}}_{d,\xi,i}^{q,(j)}(\xi)\right)} \,,
\end{align}
where $\overline \chi^q_{d,i}$ is given in eq.~\eqref{eq:gbarsvi}. 
The coefficients $C^q_i (\tilde C^q_i)$ and $D^q_i(\tilde D^q_i)$ are related to the cusp $A^q_i$ and virtual $B^q_i$ anomalous dimensions in the following way~\cite{Moch:2004pa,Dokshitzer:2005bf}.
For space-like splitting functions one has
\begin{align}
     D^q_1=&  -A_1^q\,, \qquad
 D_2^q = -A_2^q + A_1^q\left(B_1^q-\beta_0\right)\,,
 \nonumber\\
 D_3^q =&-A_3^q-A_1^q\left(-B_2^q+\beta_1\right)-A_2^q\left(-B_1^q+\beta_0\right)\,,
 \nonumber\\
 C_1^q =& 0\,, \qquad
 C_2^q = \Big(A_1^q\Big)^2\,, \qquad
 C_3^q = 2 A_1^q A_2^q\,,
\end{align}
and for the time-like ones
\begin{align}
     \tilde D^q_1=&  -A_1^q\,, \qquad
 \tilde D_2^q = -A_2^q + A_1^q\left(-B_1^q-\beta_0\right)\,,
 \nonumber\\
 \tilde D_3^q =&-A_3^q-A_1^q\left(B_2^q+\beta_1\right)-A_2^q\left(B_1^q+\beta_0\right)\,,
 \nonumber\\
 \tilde C_1^q =& 0\,, \qquad
 \tilde C_2^q = -\Big(A_1^q\Big)^2\,, \qquad
 \tilde C_3^q = -2 A_1^q A_2^q\,.
\end{align}
The renormalized coefficients  $\overline{\mathcal{G}}_{d,\xi,i}^{q,(j)}(\xi)$ in the above equations are parametrised in terms of
$\ln^k(1-\xi), k=0,1,\cdots$ and all the terms vanishing as $\xi\rightarrow 1$ are dropped
\begin{eqnarray}
\label{GikLj}
\overline{\mathcal{G}}_{d,\xi,i}^{q,(j)}(\xi) = \sum_{k=0}^{i+j-1} \overline{\mathcal{G}}_{d,\xi,i}^{q,(j,k)} \ln^k(1-\xi) \,.
\end{eqnarray}
The leading term in $\ln(1-\xi)$ at every order in $a_s$ depends on the power of $a_s$, but also the power of $\varepsilon$.  
The leading power of $\ln(1-\xi)$ can be determined by expanding the partonic cross sections to high accuracy in $\varepsilon$ at every order in $a_s$. 
One can also use the mass factorised results such as the CFs to determine them. 
In the former approach, we havev computed the partonic cross section at NLO to an accuracy of $\varepsilon$ up to third order and at NNLO to
zeroth power in $\varepsilon$.  
We expect that the observed pattern for those logarithms persists at higher orders and we can extrapolate to all orders in $\hat a_s$ and $\varepsilon$ by fixing the highest power for $\ln(1-\xi)$ to be $i+j-1$ in eq.~\eqref{GikLj}.

For a resummation to all orders an integral form for $\Phi^{\rm{NSV}}_{q}$ is needed to study its numerical impact. 
In order to do this, we decompose $\Phi^{\rm{NSV}}_{q}$ as a sum of a singular part and a finite part.  
The singular part can be obtained by setting all the terms to zero in eq.~\eqref{eq:gbxzi} except the first one, $2 L_i^q$ and $2\tilde  L_i^q$. 
It contains the correct terms that will cancel against NSV terms of the AP kernels. 
We write $\Phi^{{\text{NSV}}}_{q} = \Phi^{{\text{NSV}}}_{q,s} + \Phi^{{\text{NSV}}}_{q,fin}$, where the first term is singular as $\varepsilon \rightarrow 0$  and contains only $L_i^q$ as well as $\tilde L_i^q$ and the second one contains the finite remainder. 
This decomposition facilitates to express $\Phi^{\rm{NSV}}_{q}$ in an integral form as follows:
\begin{widetext}
\begin{eqnarray}
\label{eq:NSVintrep}
   \mathrm{\Phi}_{q}^{\text{NSV}}  &= & 
   {1 \over 2}\delta(\bar x')\int_{\mu_F^2}^{Q^2 \bar z'} \frac{d\lambda^2}{\lambda^2}\tilde L^q\big(a_s(\lambda^2),z'\big) 
   + {1\over 2} \Bigg({1 \over \bar x' }\tilde  L^q \big(a_s(Q^2 \bar x' \bar z'), z'\big) \Bigg)_+
   +{1 \over 2} \delta(\bar x') \varphi_{d,z',f}^q\big(a_s(Q^2 \bar z'),z'\big)\nonumber\\ 
    &&
   +{1 \over 2} \Bigg({1 \over \bar x'} Q^2 {d \over d Q^2} \varphi_{d,z',f}^q \big(a_s(Q^2 \bar x' \bar z'),z'\big)\Bigg)_+
   +  {1 \over 2}\delta(\bar x') \sum_{i=1}^\infty \hat a_s^i \left({\mu_F^2 \over \mu^2}\right)^{i {\varepsilon \over 2}}  S_\varepsilon^i \varphi^{(i)}_{d,z',s}(z',\varepsilon) 
   \nonumber \\
   && +\left( \tilde L^q \leftrightarrow L^q ,  x'\leftrightarrow z'\right) \,.
\end{eqnarray}
\end{widetext}
In the above equation, the term containg $\varphi_{d,\xi,s}^{(i)}$ is singular while the remaining ones are all finite as $\varepsilon \rightarrow 0$.  
The fact that $\mathrm{\Phi}_{q}^{\text{NSV}}$ is renormalization group invariant implies that $\varphi^q_{d,\xi,f}$ satisfies the equations
\begin{eqnarray}
\label{RGphis}
\mu_F^2 {d \over d\mu_F^2}  \varphi^q_{d,x',f}(a_s(\mu_F^2),x') &=& L^q (a_s(\mu_F^2),x')\, ,
\nonumber \\
\mu_F^2 {d \over d\mu_F^2}  \varphi^q_{d,z',f}(a_s(\mu_F^2),z') &=& \tilde L^q (a_s(\mu_F^2),z')\,,
\end{eqnarray}
where
\begin{align}
L^q (a_s(\mu_F^2),x') &= \sum_{i=1}^{\infty}a_s^i(\mu_F^2)L_i^q(x')\,,
\nonumber \\
\tilde L^q (a_s(\mu_F^2),z') &= \sum_{i=1}^{\infty}a_s^i(\mu_F^2)\tilde L_i^q(z')\,.
\end{align}  
Having fixed the divergent part of 
$\mathrm{\Phi}_{q}^{\rm{NSV}}$ completely, we need to understand all order structure of the finite piece $\varphi^q_{d,\xi,f}$. 
Using eq.~\eqref{GikLj}, we can expand this functions in powers of renormalised coupling $a_s$:
\begin{align}
\label{varphiexp}
\varphi_{d,\xi,f}^q(a_s(\lambda^2),\xi) =&
\sum_{i=1}^\infty a_s^i(\lambda^2) \sum_{k=0}^i \overline \varphi_{q,\xi,i}^{(k)} \ln^k(1-\xi) \,,
\end{align}
where the highest power of $\ln(1-\xi)$ aligns with that in eq.~\eqref{GikLj}.  
The coefficients $\overline{\varphi}_{q,\xi,i}^{(k)}$ can be expressed in terms of their unrenormalised counter part $\overline{\cal G}^{q,(j,k)}_{d,\xi,i}$'s in eq.~\eqref{GikLj} as:
\begin{widetext}
\begin{eqnarray}\label{eq:varphi}
\overline \varphi_{q,\xi,1}^{(k)} &=& \overline {\mathcal{G}}_{d,\xi,1}^{q,(1,k)}, \quad\quad k=0,1\nonumber\\
\overline \varphi_{q,\xi,2}^{(k)} &=&  \bigg(\frac{1}{2}
\overline {\mathcal{G}}_{d,\xi,2}^{q,(1,k)} + \beta_0
\overline {\mathcal{G}}_{d,\xi,1}^{q,(2,k)}\bigg), 
\quad\quad k = 0,1,2\nonumber\\
\overline \varphi_{q,\xi,3}^{(k)} &=&  \bigg(\frac{1}{3}
\overline{\mathcal{G}}_{d,\xi,3}^{q,(1,k)} + \frac{2}{3}\beta_1 \overline{\mathcal{G}}_{d,\xi,1}^{q,(2,k)} + \frac{2}{3}\beta_0\overline{\mathcal{G}}_{d,\xi,2}^{q,(2,k)} + \frac{4}{3}
         \beta_0^2\overline{\mathcal{G}}_{d,\xi,1}^{q,(3,k)}\bigg), 
         \quad\quad k=0,1,2,3\nonumber\\
\overline \varphi_{q,\xi,4}^{(k)} &=& \bigg( \frac{1}{4}\overline{\mathcal{G}}_{d,\xi,4}^{q,(1,k)} + \frac{1}{2}\beta_2\overline{\mathcal{G}}_{d,\xi,1}^{q,(2,k)} + \frac{1}{2}\beta_1\overline{\mathcal{G}}_{d,\xi,2}^{q,(2,k)} + \frac{1}{2}
         \beta_0\overline{\mathcal{G}}_{d,\xi,3}^{q,(2,k)} 
         + 2\beta_0\beta_1\overline{\mathcal{G}}_{d,\xi,1}^{q,(3,k)} 
         + \beta_0^2\overline{\mathcal{G}}_{d,\xi,2}^{q,(3,k)}+ 
         2\beta_0^3\overline{\mathcal{G}}_{d,\xi,1}^{q,(4,k)}\bigg), \nonumber\\
         && \quad\quad \hspace{0.6\textwidth} k=0,1,2,3,4
\end{eqnarray}
\end{widetext}
 The structure of divergent and finite pieces of 
 ${\rm \Phi}_{q}^{\text{NSV}}$ allows us to determine the coefficients  $\overline{\cal G}^{q,(j,k)}_{d,\xi,i}$.  
 Since our final result depends on
 $\overline{\varphi}_{q,\xi,i}^{(k)}$ we can directly determine them by comparison with the fixed order results for the CFs, known to NNLO level \cite{Goyal:2023zdi,Bonino:2024qbh,Goyal:2024tmo,Bonino:2024wgg,Goyal:2024emo,Bonino:2025qta} and they are listed below:
\begin{widetext}
\begin{align}
\overline \varphi^{(0)}_{q,x',1} &= 
         C_F \bigg\{-4\bigg\} \,,
\nonumber \quad 
\overline \varphi^{(1)}_{q,x',1} = 0\,,
\nonumber \\ 
\overline \varphi^{(0)}_{q,x',2} &=
    C_F C_A   \bigg\{ \frac{1108}{27} - 28 \zeta_3 - \frac{44}{3} \zeta_2 \bigg\}
   - C_F n_f   \bigg\{ \frac{52}{27} - \frac{8}{3} \zeta_2 \bigg\}
   - C_F^2   \bigg\{   8 \zeta_2 \bigg\}\,,
\nonumber\\
\overline \varphi^{(1)}_{q,x',2} &=
       C_F C_A   \bigg\{  - 10 \bigg\}
       + C_F^2   \bigg\{ 10 \bigg\}\,,
 \quad
\overline \varphi^{(2)}_{q,x',2} =
        C_F^2   \bigg\{ 4 \bigg\}\,,
\end{align}\label{PhiqX}
\begin{align}
\overline \varphi^{(0)}_{q,z',1} &= 
         C_F \bigg\{ 4\bigg\} \,,
\nonumber \quad 
\overline \varphi^{(1)}_{q,z',1} = 0\,,
\nonumber \\ 
\overline \varphi^{(0)}_{q,z',2} &=
C_F C_A   \bigg\{ \frac{1492}{27} - 28 \zeta_3 - \frac{68}{3} \zeta_2 \bigg\}-
C_F n_f   \bigg\{   \frac{364}{27} - \frac{8}{3} \zeta_2 \bigg\}
       - C_F^2   \bigg\{ 16 \zeta_2 \bigg\} \,,\nonumber\\
\overline \varphi^{(1)}_{q,z',2} &=
       C_F C_A   \bigg\{   10 \bigg\}
       - C_F^2   \bigg\{ 10 \bigg\}\,,
 \quad
\overline \varphi^{(2)}_{q,z',2} =
        C_F^2   \bigg\{- 4 \bigg\}\,,
\end{align}\label{PhiqZ}
\end{widetext}
Note that the constants $\overline \varphi^{(k)}_{q,x',i}$ are different from $\overline \varphi^{(k)}_{q,z',i}$ unlike those that appearing in rapidity distributions of the DY process, see \cite{AH:2020qoa}.  

Having obtained the  general structure
of the soft function, the final step is the substitution of the integral representations for 
$\Phi_q^{\rm{SV}}$ in eq.~\eqref{eq:SVintrep}, 
for $\Phi_q^{\rm{NSV}}$ in eq.~\eqref{eq:NSVintrep}, 
the solutions to the form factor in eq.~\eqref{Fcsol} and the AP kernels in eq.~\eqref{Gcsol}
into eq.~\eqref{MassFactRe}. 
We find that all the soft and collinear singularities cancel among themselves leaving SV and NSV parts of the CFs IR finite.
\begin{widetext}
\begin{align}
\label{eq:MasterF}
 (\Delta){\cal C}_{J,qq}^{\rm SV+NSV}  = \mathcal{C}\exp \bigg( \Psi^q_{d}\big(Q^2,\mu_F^2,x',z',\varepsilon\big)\bigg)\bigg |_{\varepsilon=0} \,,
\end{align}
where the IR finite expression $\Psi_d^q$ takes a simple form
\begin{align}
\label{IntSVpNSV}
\mathrm{\Psi}_{d}^q =&
\frac{\delta(\overline x')}{2} \Bigg(\Bigg\{
\int_{\mu_F^2}^{Q^2 \overline z'} {d \lambda^2 \over \lambda^2}
\mathcal{\tilde{P}}^{q} \left(a_s(\lambda^2), \overline z'\right) 
+\mathcal{Q}^{q} \left(a_s(Q_2^2), \overline z'\right) \Bigg)_+
+ \frac{1}{4} \Bigg(\frac{1}{\overline x'} \Bigg\{ \mathcal{\tilde{P}}^{q} \left(a_s(Q_{12}^2), \overline z'\right)  + 2 \tilde L^{q} \left(a_s(Q_{12}^2), \overline z'\right) 
\nonumber \\ & 
 +Q^2 {\frac{d}{dQ^2} } \bigg( \mathcal{Q}^{q} \left(a_s(Q_2^2), \overline z'\right) + 2 {\varphi}^{q}_{d,z',f} \left(a_s(Q_2^2), \overline z'\right)  \bigg)    \Bigg\}\Bigg)_+
+ \frac{1}{2} \delta(\overline x')~\delta(\overline z')~ 
\text{ln}\Big(g^q_{d,0}\left(a_s(\mu_F^2) \right) \Big)
+  \Big( x' \leftrightarrow  z'\Big)
\, .
\end{align}
\end{widetext}
Here, we use the short-hands $\overline x' = (1-x')$ and $\overline z' = (1-z')$, $Q_1^2 = Q^2~\overline x' $,$Q_2^2 = Q^2~\overline z' $ and $Q^2_{12}=Q^2 ~\overline x' ~\overline z'$. 
The subscript $+$ indicates the standard plus distribution.  
In the above equation, ${\cal P}^q,\tilde {\cal P}^q$ are related diagonal splitting functions given in eq.~\eqref{eq:APspSVNSV},
\begin{align}    
 {\cal P}^{q} (a_s, \overline x')&= P_{q \la q}(\mu_s^2,\overline x') - 2 B^q(a_s) \delta(\overline x'),\nonumber\\  
 \tilde{{\cal P}}^{q} (a_s, \overline z')&= \tilde P_{q \la q}(\mu_s^2,\overline z') - 2 B^q(a_s) \delta(\overline z')\, , 
\end{align}
and $\mathcal{Q}^{q}_d(a_s,\overline z')$ = $\frac{2}{\overline z'}G^q_d(a_s)$ + 2 $\varphi^q_{d,z',f}(a_s,\overline z')$ and similar for $x'$. 
Note that $g_{d,0}^q$ is process-dependent and receives contributions only from the form factor and finite parts of $\mathrm{\Phi}^{\text{SV}}_{q}$.   

The result for $(\Delta) {\cal C}^{{\text{SV+NSV}}}_{J,qq}$ in eq.~\eqref{IntSVpNSV} is valid to all orders in $a_s$. 
The exponent $\Psi^q_d$ contains universal (process independent) quantities such as the QCD $\beta$ function and anomalous dimensions $A^q, B^q, f^q$ etc.\ 
and process dependent quantities $G^q_d,\varphi^q_{d,\xi,f}$ and $g^q_{d,0}$.  
Each of them can be expanded in powers of $a_s$, hence $\Psi^q_d$ is expanded as
\begin{eqnarray}
\Psi^q_d(Q^2,\mu_F^2) = \sum_{i=1}^\infty a_s^i(\mu_R^2) 
\Psi^{q,(i)}_d(Q^2,\mu_F^2,\mu_R^2)
\, ,
\end{eqnarray}
where we have suppressed the dependence on $x',z'$ in $\Psi^q_d$ and $\Psi^{q,(i)}_d$.   

The result in eq.~\eqref{eq:MasterF} is the main result of the paper. 
It sums up towers of SV and NSV terms to all orders in $a_s$ in those CFs, which are dominat near threshold. 
The particular logarithmic structure of $\Psi^q_d$ given in eq.~\eqref{IntSVpNSV}
allows us to predict certain SV and NSV terms to any order in $a_s$, i.e., $a_s^n$ for the diagonal CFs for all $n \geq 1$, provided $\Psi^q_d$ is known up to order $a_s^m$ with $n > m$.
The achievable accuray of those predictions for $\Psi^{q,(i)}_d$s for $i=1,2,3$ and for general $i=n$ 
is summarized in Tabs.~\ref{TableSVz} and \ref{TableNSVz}.
For example, if we know only $\Psi_{d}^{q,(1)}$, we can predict the following terms for $(\Delta){\cal C}_{J,cc}^{(2)}$, i.e., for the SV part, $\{\delta_{\overline{x}'}{\cal D}_{\overline{z}'}^{3}, \delta_{\overline{z}'}{\cal D}_{\overline{x}'}^{3}\}$  and $\{{\cal D}_{\overline{x}'}^{i}{\cal D}_{\overline{z}'}^{j}\}$ for $i+j=2$ and, likewise, for the NSV one 
$\{\delta_{\overline{x}'}{\cal L}_{\overline{z}'}^{3},\delta_{\overline{z}'}{\cal L}_{\overline{x}'}^{3}\}$, and $\{ {\cal D}_{\overline{x}'}^{i}{\cal L}_{\overline{z}'}^{j},{\cal D}_{\overline{z}'}^{i}{\cal L}_{\overline{x}'}^{j} \}$ for $i+j=2$.
If, in addition, we allow for terms present in $\Psi_{d}^{q,(2)}$ which depend only 
on information about $\Psi_{d}^{q,(1)}$, then, because of the renormalization group equations, we can predict, 
for the SV part also $\{\delta_{\overline{x}'}{\cal D}_{\overline{z}'}^{2}, \delta_{\overline{z}'}{\cal D}_{\overline{x}'}^{2}\}$ and $\{{\cal D}_{\overline{x}'}^{i}{\cal D}_{\overline{z}'}^{j}\}$ for $i+j=1$,
and for the NSV part $\{ {\cal D}_{\overline{x}'}^{1}{\cal L}_{\overline{z}'}^{0},{\cal D}_{\overline{z}'}^{1}{\cal L}_{\overline{x}'}^{0} \}$. 
Knowledge on higher-order corrections to the anomalous dimensions leads to predictions for the SV terms $\{\delta_{\overline{x}'}{\cal D}_{\overline{z}'}^{i}, \delta_{\overline{z}'}{\cal D}_{\overline{x}'}^{i}\}$ for $i = 0,1$ and $\{{\cal D}_{\overline{x}'}^{0}{\cal D}_{\overline{z}'}^{0}\}$ and for the NSV terms $\{ {\cal D}_{\overline{x}'}^{0}{\cal L}_{\overline{z}'}^{i},{\cal D}_{\overline{z}'}^{0}{\cal L}_{\overline{x}'}^{i} \}$ for $i=0,1$.
We elaborate on this further below when we make predictions at third and fourth order in $a_s$ using the $\Psi^{q,(i)}_d$s at $i=1,2$ and partial results on $\Psi^{q,(i)}_d$s for $i>2$.



\bigskip
\section{Results of the SV and NSV Coefficient functions}
\label{sec:DNSV}
In this section, we present the analytic results of the SV and NSV CFs, $(\Delta){\cal C}^{\text{SV}}_{J,cc}$ and $(\Delta){\cal C}^{\text{NSV}}_{J,cc}$ for $c=\{q,\overline{q}\}$ respectively. 
We present complete results at NLO and NNLO and partial results up to N$^4$LO in the thresold region for the CFs, expanded in $a_s$ as 
$(\Delta){\cal C}^{\text{(N)SV}}_{J,cc} = \sum_{i=0}^{\infty} a_s^{i}(\mu_R^2)(\Delta){\cal C}^{(i),\text{(N)SV}}_{J,cc}$.
Their normalization at LO reads $(\Delta){\cal C}^{(0),\text{SV}}_{J,cc} =\delta(1-x')\delta(1-z')$ and $(\Delta){\cal C}^{(0),\text{NSV}}_{J,cc} =0$, and we have set $\mu_F^2 = \mu_R^2=Q^2$ in the sequel.
\begin{widetext} 
\subsubsection{SV Coefficient functions}
\noindent
For the perturbative expansion of the SV part of the CFs, 
we define in addition to $C_A$ and $C_F$ further group invariants of the color $SU(N_c)$ gauge group, namely
We also define 
\begin{equation}
 \frac{d^{abcd}_F d^{abcd}_F}{N_c} = \frac{(N_c^2-1)(N_c^4-6N_c^2+18)}{96 N_c^3}\, , \qquad
 \frac{d^{abcd}_F d^{abcd}_A}{N_c} = \frac{(N_c^2-1)(N_c^2+6)}{48}\, , \qquad
 N_4= \frac{(N_c^2-4)}{N_c}\, .
\end{equation}
Also, below we will use $n_{fv}$, which is defined as the charge weighted sum of quark flavors.
At NLO, i.e. at ${\mathcal O}(a_s)$,
\begin{align}
(\Delta){\cal C}^{(1),\text{SV}}_{J,cc} &=  
\bm{\delta_{\overline{x}'}\mathcal{D}^{1}_{\overline{z}'}} \Bigg[
        C_F   \bigg\{ 4 \bigg\}\Bigg]
-
   \bm{\delta_{\overline{x}'}\delta_{\overline{z}'}} \Bigg[
        C_F   \bigg\{   8  \bigg\}\Bigg]
+
\bm{\mathcal{D}^{0}_{\overline{x}'}\mathcal{D}^{0}_{\overline{z}'}} \Bigg[
     C_F   \bigg\{ 2 \bigg\}\Bigg] +\bm{({x}'\leftrightarrow {z}')}      \,.
\end{align}
The NNLO CF at SV accuracy read
\begin{align}
(\Delta)\mathcal{C}^{(2),\text{SV}}_{J,cc} &= 
   \bm{\delta_{\overline{x}'}\mathcal{D}^{3}_{\overline{z}'}} \bigg[
        C_F^2   \bigg\{ 8 \bigg\}\Bigg]
-
  \bm{\delta_{\overline{x}'}\mathcal{D}^{2}_{\overline{z}'}} \Bigg[
        C_F C_A   \bigg\{   \frac{22}{3} \bigg\}
       - C_F n_f   \bigg\{ \frac{4}{3}
       \bigg\}\bigg]
+
   \bm{\mathcal{D}^{1}_{\overline{x}'}\mathcal{D}^{1}_{\overline{z}'}} \bigg[
        C_F^2   \bigg\{ 24 \bigg\}\bigg] 
+
   \bm{\mathcal{D}^{0}_{\overline{x}'}\mathcal{D}^{2}_{\overline{z}'}} \bigg[
        C_F^2   \bigg\{ 24 \bigg\}\bigg]
\nonumber\\&
-
   \bm{\mathcal{D}^{0}_{\overline{x}'}\mathcal{D}^{1}_{\overline{z}'}} \bigg[
       C_F C_A   \bigg\{   \frac{44}{3} \bigg\}
       - C_F n_f   \bigg\{ \frac{8}{3} \bigg\}\bigg]
+
   \bm{\delta_{\overline{x}'}\mathcal{D}^{1}_{\overline{z}'}} \bigg[
        C_F C_A   \bigg\{ \frac{268}{9} - 8 \zeta_2  \bigg\}
       - C_F n_f   \bigg\{   \frac{40}{9}  \bigg\}
       - C_F^2   \bigg\{   64 + 32 \zeta_2  \bigg\}\bigg] 
\nonumber\\&
-
   \bm{\delta_{\overline{x}'}\delta_{\overline{z}'}} \bigg[
        C_F C_A   \bigg\{   \frac{1535}{24} - \frac{86}{3} \zeta_3 + \frac{269}{9} \zeta_2 - \frac{34}{5} \zeta_2^2   \bigg\}
          - C_F^2   \bigg\{ \frac{511}{8} - 30 \zeta_3 + 29 \zeta_2 - \frac{28}{5} \zeta_2^2  
        \bigg\}
\nonumber\\&
          - C_F n_f   \bigg\{ \frac{127}{12} + \frac{4}{3} \zeta_3 + \frac{38}{9} \zeta_2 \bigg\}\bigg]
+
  \bm{ \mathcal{D}^{0}_{\overline{x}'}\mathcal{D}^{0}_{\overline{z}'}} \bigg[
        C_F C_A   \bigg\{ \frac{134}{9} - 4 \zeta_2 \bigg\}
       - C_F n_f   \bigg\{   \frac{20}{9} \bigg\}
       - C_F^2   \bigg\{   32 + 16 \zeta_2  \bigg\}\bigg]
\nonumber\\&
-
   \bm{\delta_{\overline{x}'}\mathcal{D}^{0}_{\overline{z}'}} \bigg[
        C_F C_A   \bigg\{   \frac{808}{27} - 28 \zeta_3 - \frac{44}{3} \zeta_2 
  \bigg\}
       - C_F n_f   \bigg\{ \frac{112}{27} - \frac{8}{3} \zeta_2  \bigg\}
       - C_F^2   \bigg\{ 32 \zeta_3  \bigg\}\bigg]
+ \bm{(x' \leftrightarrow z')}   \,.
\end{align}

%
In the following we predict the CF at third order in $a_s$ using the available information on the perturbative results.
\begin{itemize}
\item the knowledge of $\Psi_{d}^{q,(1)}$ and $\Psi_{d}^{q,(2)}$ can be used to predict
the coefficients of
$\{ \delta_{\overline{x}'}\mathcal{D}_{\overline{z}'}^{i}, \delta_{\overline{z}'}\mathcal{D}_{\overline{x}'}^{i}\}$ for $i = 5, 4, 3, 2$ 
and $\{ \mathcal{D}_{\overline{x}'}^{i}\mathcal{D}_{\overline{z}'}^{j}\}$ for $i+j=4,3,2,1$
in $(\Delta){\cal C}_{J,cc}^{(3),\text{SV}}$. 
\item the results on the cusp and collinear anomalous dimensions ($A_3^q, f_{3}^{q}$)
can be used to obtain the coefficients of $\{ \delta_{\overline{x}'}\mathcal{D}_{\overline{z}'}^{i}, \delta_{\overline{z}'}\mathcal{D}_{\overline{x}'}^{i}\}$ for $i=1,0$ and $\{ \mathcal{D}_{\overline{x}'}^{0}\mathcal{D}_{\overline{z}'}^{0}\}$ 
\item using  ${\cal G}_{d,3}^{q,(1)}$ from the three-loop form factors
and $\overline{\cal G}_{d,3}^{q,(1)}$ from the three-loop soft functions,
we can obtain the coefficient of $\delta_{\overline{x}'}\delta_{\overline{z}'}$.
\end{itemize}
The results for the SV part of the CFs at third order obtained in this way read
\begin{align}
(\Delta)\mathcal{C}^{(3),\text{SV}}_{J,cc} &= 
   \bm{\mathcal{D}^{2}_{\overline{x}'}\mathcal{D}^{2}_{\overline{z}'} }\Bigg[
        C_F^3   \bigg\{ 120 \bigg\}\Bigg] 
+
 \bm{  \delta_{\overline{x}'}\mathcal{D}^{5}_{\overline{z}'} }\Bigg[
        C_F^3   \bigg\{ 8 \bigg\}\Bigg] 
-
  \bm{ \delta_{\overline{x}'}\mathcal{D}^{4}_{\overline{z}'}} \Bigg[
        C_F^2 C_A   \bigg\{  \frac{220}{9} \bigg\}
       - C_F^2 n_f   \bigg\{ \frac{40}{9} \bigg\}\Bigg] 
\nonumber\\&
+
  \bm{ \delta_{\overline{x}'}\mathcal{D}^{3}_{\overline{z}'} }\Bigg[
        C_F C_A^2   \bigg\{ \frac{484}{27} \bigg\}
       - C_F n_f C_A   \bigg\{   \frac{176}{27} \bigg\}
       + C_F n_f^2   \bigg\{ \frac{16}{27} \bigg\}
       + C_F^2 C_A   \bigg\{ \frac{1072}{9} - 32 \zeta_2 \bigg\}
       - C_F^2 n_f   \bigg\{  \frac{160}{9} \bigg\}
\nonumber\\&       
       - C_F^3   \bigg\{   128 + 128 \zeta_2 \bigg\}\Bigg] 
-
   \bm{\delta_{\overline{x}'}\mathcal{D}^{2}_{\overline{z}'} }\Bigg[
        C_F C_A^2   \bigg\{   \frac{3560}{27} - \frac{88}{3} \zeta_2 \bigg\}
       - C_F  C_A n_f  \bigg\{ \frac{1156}{27} - \frac{16}{3} \zeta_2 \bigg\}
       + C_F n_f^2   \bigg\{   \frac{80}{27} \bigg\}
       
\nonumber\\&       
       + C_F^2 C_A   \bigg\{   \frac{560}{9} - 168 \zeta_3 - 264 \zeta_2 \bigg\}
       - C_F^2 n_f   \bigg\{ \frac{68}{9} - 48 \zeta_2 \bigg\}
       - C_F^3   \bigg\{ 320 \zeta_3 \bigg\}\Bigg] 
+
  \bm{ \mathcal{D}^{1}_{\overline{x}'}\mathcal{D}^{3}_{\overline{z}'} }\Bigg[
        C_F^3   \bigg\{ 160 \bigg\}\Bigg]   
\nonumber\\&
-
   \bm{\mathcal{D}^{1}_{\overline{x}'}\mathcal{D}^{2}_{\overline{z}'} }\Bigg[
        C_F^2 C_A   \bigg\{  \frac{880}{3} \bigg\}
       - C_F^2 n_f   \bigg\{ \frac{160}{3} \bigg\}\Bigg]
+
  \bm{ \delta_{\overline{x}'}\mathcal{D}^{1}_{\overline{z}'} }\Bigg[
        C_F C_A^2   \bigg\{ \frac{31006}{81} - 176 \zeta_3 - \frac{680}{3} \zeta_2 + \frac{176}{5} \zeta_2^2 \bigg\}
\nonumber\\&       
       - C_F  C_A n_f  \bigg\{   \frac{8204}{81} - \frac{512}{9} \zeta_2 \bigg\}  
       + C_F n_f^2   \bigg\{ \frac{400}{81} - \frac{32}{9} \zeta_2 \bigg\}
       - C_F^2 C_A   \bigg\{   \frac{8893}{9} + \frac{368}{3} \zeta_3 + \frac{5288}{9} \zeta_2 - \frac{912}{5} \zeta_2^2 \bigg\}
\nonumber\\&  
        + C_F^2 n_f   \bigg\{ \frac{1072}{9} + \frac{320}{3} \zeta_3 + \frac{944}{9} \zeta_2 \bigg\}
       + C_F^3   \bigg\{ 511 - 240 \zeta_3 + 744 \zeta_2 + \frac{288}{5} \zeta_2^2 \bigg\}\Bigg] 
+
 \bm{  \mathcal{D}^{0}_{\overline{x}'}\mathcal{D}^{4}_{\overline{z}'} }\Bigg[
        C_F^3   \bigg\{ 40 \bigg\}\Bigg] 
\nonumber\\&
+
  \bm{ \mathcal{D}^{0}_{\overline{x}'}\mathcal{D}^{3}_{\overline{z}'} }\Bigg[
       - C_F^2 C_A   \bigg\{   \frac{880}{9} \bigg\}
       + C_F^2 n_f   \bigg\{ \frac{160}{9} \bigg\}\Bigg] 
+
  \bm{ \mathcal{D}^{1}_{\overline{x}'}\mathcal{D}^{1}_{\overline{z}'} }\Bigg[
        C_F C_A^2   \bigg\{ \frac{484}{9} \bigg\}
       - C_F C_A n_f   \bigg\{   \frac{176}{9} \bigg\}
       + C_F n_f^2   \bigg\{ \frac{16}{9} \bigg\}
\nonumber\\&
       + C_F^2 C_A   \bigg\{ \frac{1072}{3} - 96 \zeta_2 \bigg\}
       - C_F^2 n_f   \bigg\{  \frac{160}{3} \bigg\}
       - C_F^3   \bigg\{   384 + 384 \zeta_2 \bigg\}\Bigg] 
       -
 \bm{  \mathcal{D}^{0}_{\overline{x}'}\mathcal{D}^{1}_{\overline{z}'} }\Bigg[
     C_F C_A^2   \bigg\{   \frac{7120}{27} - \frac{176}{3} \zeta_2 \bigg\}
      
\nonumber\\&       
 - C_F C_A n_f  \bigg\{ \frac{2312}{27} - \frac{32}{3} \zeta_2 \bigg\}
       + C_F n_f^2   \bigg\{   \frac{160}{27} \bigg\}
       + C_F^2 C_A   \bigg\{   \frac{1120}{9} - 336 \zeta_3 - 528 \zeta_2 \bigg\}
       - C_F^2 n_f   \bigg\{ \frac{136}{9} - 96 \zeta_2 \bigg\} 
\nonumber\\&  
       - C_F^3   \bigg\{ 640 \zeta_3 \bigg\}\Bigg]
-
 \bm{  \delta_{\overline{x}'}\mathcal{D}^{0}_{\overline{z}'} }\Bigg[
        C_F n_f^2   \bigg\{  \frac{1856}{729} + \frac{32}{27} \zeta_3 - \frac{160}{27} \zeta_2 \bigg\}
         + C_F^2 n_f   \bigg\{   3 + \frac{944}{9} \zeta_3 - \frac{40}{27} \zeta_2 - \frac{224}{15} \zeta_2^2 \bigg\}
\nonumber\\&         
       - C_F  C_A n_f  \bigg\{ \frac{62626}{729} - \frac{536}{9} \zeta_3 - \frac{7760}{81} \zeta_2 + \frac{208}{15} \zeta_2^2
          \bigg\}
       - C_F^2 C_A   \bigg\{ \frac{12928}{27} + \frac{256}{9} \zeta_3 + \frac{128}{27} \zeta_2 - 352 \zeta_2 \zeta_3
          - \frac{352}{3} \zeta_2^2 \bigg\}
\nonumber\\&       
       + C_F C_A^2   \bigg\{   \frac{297029}{729} + 192 \zeta_5 - \frac{14264}{27} \zeta_3 - \frac{27752}{81} 
         \zeta_2 + \frac{176}{3} \zeta_2 \zeta_3 + \frac{616}{15} \zeta_2^2 \bigg\}
      - C_F^3   \bigg\{ 384 \zeta_5 - 512 \zeta_3 - 512 \zeta_2 \zeta_3 \bigg\}\Bigg] 
\nonumber\\&
+
  \bm{ \delta_{\overline{x}'}\delta_{\overline{z}'} }\Bigg[
        C_F N_{4} n_{fv} \bigg\{ 4 - \frac{80}{3} \zeta_5 + \frac{14}{3} \zeta_3 + 10 \zeta_2 - \frac{2}{5} \zeta_2^2 \bigg\}
        - C_F n_f^2   \bigg\{   \frac{7081}{486} + \frac{152}{81} \zeta_3 + \frac{1028}{81} \zeta_2 + \frac{64}{135} 
         \zeta_2^2 \bigg\}
\nonumber\\& 
       - C_F C_A^2   \bigg\{   \frac{1505881}{1944} + 102 \zeta_5 - \frac{125105}{162} \zeta_3 + \frac{200}{3} 
         \zeta_3^2 + \frac{124895}{162} \zeta_2  - 234 \zeta_2 \zeta_3 - \frac{23539}{270} \zeta_2^2 + 
         \frac{4456}{315} \zeta_2^3 \bigg\}
\nonumber\\&         
       + C_F  C_A n_f  \bigg\{ \frac{110651}{486} - 4 \zeta_5 - \frac{9944}{81} \zeta_3 + \frac{5696}{27} \zeta_2 - \frac{8}{3} \zeta_2 \zeta_3 - \frac{1054}{135} \zeta_2^2 \bigg\}
       + C_F^2 C_A   \bigg\{ \frac{74321}{72} - \frac{3812}{9} \zeta_5 - \frac{2986}{3} \zeta_3
\nonumber\\&       
        + \frac{632}{3} \zeta_3^2 + \frac{33451}{27} \zeta_2  - \frac{2480}{9} \zeta_2 \zeta_3 + \frac{1304}{135} \zeta_2^2 - \frac{3896}{63} 
         \zeta_2^3 \bigg\}
       - C_F^2 n_f   \bigg\{  \frac{421}{6} + \frac{112}{9} \zeta_5 - 180 \zeta_3 + \frac{3844}{27} \zeta_2 
\nonumber\\&       
       + \frac{448}{9} \zeta_2 \zeta_3 + \frac{2504}{135} \zeta_2^2 \bigg\}       
       - C_F^3   \bigg\{   \frac{5599}{12} - 664 \zeta_5 + 230 \zeta_3 - \frac{368}{3} \zeta_3^2 + \frac{1598}{3} 
         \zeta_2 - 280 \zeta_2 \zeta_3 + \frac{442}{5} \zeta_2^2 - \frac{20816}{315} \zeta_2^3 \bigg\}\Bigg] 
\nonumber\\&
+
 \bm{  \mathcal{D}^{0}_{\overline{x}'}\mathcal{D}^{2}_{\overline{z}'} }\Bigg[
        C_F C_A^2   \bigg\{ \frac{484}{9} \bigg\}
       - C_F  C_A  n_f \bigg\{   \frac{176}{9} \bigg\}
       + C_F n_f^2   \bigg\{ \frac{16}{9} \bigg\}
       + C_F^2 C_A   \bigg\{ \frac{1072}{3} - 96 \zeta_2 \bigg\}
       - C_F^2 n_f   \bigg\{   \frac{160}{3} \bigg\}
\nonumber\\&
       - C_F^3   \bigg\{   384 + 384 \zeta_2 \bigg\}\Bigg] 
+
 \bm{  \mathcal{D}^{0}_{\overline{x}'}\mathcal{D}^{0}_{\overline{z}'} }\Bigg[
        C_F C_A^2   \bigg\{ \frac{15503}{81} - 88 \zeta_3 - \frac{340}{3} \zeta_2 + \frac{88}{5} \zeta_2^2 \bigg\}
       - C_F C_A n_f  \bigg\{   \frac{4102}{81} - \frac{256}{9} \zeta_2 \bigg\}
\nonumber\\&
       + C_F n_f^2   \bigg\{ \frac{200}{81} - \frac{16}{9} \zeta_2 \bigg\}
       - C_F^2 C_A   \bigg\{   \frac{8893}{18} + \frac{184}{3} \zeta_3 + \frac{2644}{9} \zeta_2 - \frac{456}{5} \zeta_2^2 \bigg\}
       + C_F^2 n_f   \bigg\{ \frac{536}{9} + \frac{160}{3} \zeta_3 + \frac{472}{9} \zeta_2 \bigg\}
\nonumber\\&       
        + C_F^3   \bigg\{ \frac{511}{2} - 120 \zeta_3 + 372 \zeta_2 + \frac{144}{5} \zeta_2^2 \bigg\}
       \Bigg]
        + \bm{(x'\leftrightarrow z')} \,.
\end{align}
%

%
Similarly at fourth order we can predict various terms as follows:
\begin{itemize}
\item
 using $\Psi_{d}^{q,(1)}$, $\Psi_{d}^{q,(2)}$, 
and the three-loop cusp and eikonal anomalous dimensions ($A_3^q, f_{3}^{q}$),
we can obtain the coefficients of
$\{ \delta_{\overline{x}'}\mathcal{D}_{\overline{z}'}^{i}, \delta_{\overline{z}'}\mathcal{D}_{\overline{x}'}^{i}\}$ for $i = 7, 6, 5, 4, 3, 2$ 
and $\{ \mathcal{D}_{\overline{x}'}^{i}\mathcal{D}_{\overline{z}'}^{j}\}$ for $i+j=6,5,4,3,2,1$,
in $(\Delta){\cal C}_{J,cc}^{(4),\text{SV}}$. 
\item using $\overline{\cal G}_{d,3}^{q,(1)}$ from the three-loop soft functions as given in eq.~(\ref{app:SVGij}), along with higher order anomalous dimensions $(A_{3}^{q}, A_{4}^{q}, B_{3}^{q},f_{3}^{q})$, we  obtain the coefficients of $\{ \delta_{\overline{x}'}\mathcal{D}_{\overline{z}'}^{1}, \delta_{\overline{z}'}\mathcal{D}_{\overline{x}'}^{1}\}$ and $\{ \mathcal{D}_{\overline{x}'}^{0}\mathcal{D}_{\overline{z}'}^{0}\}$.
\item
finally, thanks to the recent computation  \cite{Kniehl:2025ttz}  on four-loop eikonal anomalous dimension ($f_4$) the coefficients of $\{\delta_{\overline{x}'}\mathcal{D}^{0}_{\overline{z}'}, \delta_{\overline{z}'}\mathcal{D}^{0}_{\overline{x}'}\}$ 
in $(\Delta)\mathcal{C}^{(4),\text{SV}}_{J,cc}$, can now be predicted.  
\end{itemize}
However, the coefficient of $\delta_{\overline{x}'}\delta_{\overline{z}'}$ cannot be predicted with the available information. Thus, the SV part of the cFs at fourth order is given by
\begin{align}
(\Delta)\mathcal{C}^{(4),\text{SV}}_{J,cc} &=    
\bm{\mathcal{D}^{3}_{\overline{x}'}\mathcal{D}^{3}_{\overline{z}'} }\Bigg[
        C_F^4   \bigg\{ \frac{1120}{3} \bigg\}\Bigg] 
+
\bm{\mathcal{D}^{2}_{\overline{x}'}\mathcal{D}^{4}_{\overline{z}'}} \Bigg[
        C_F^4   \bigg\{ 560 \bigg\}\Bigg] 
-   \bm{\mathcal{D}^{2}_{\overline{x}'}\mathcal{D}^{3}_{\overline{z}'}} \Bigg[
        C_F^3 C_A   \bigg\{   \frac{6160}{3} \bigg\}
       - C_F^3 n_f   \bigg\{ \frac{1120}{3} \bigg\}\Bigg] 
\nonumber\\&
+
\bm{\mathcal{D}^{2}_{\overline{x}'}\mathcal{D}^{2}_{\overline{z}'}} \Bigg[
        C_F^2 C_A^2   \bigg\{ \frac{9680}{9} \bigg\}
       - C_F^2 n_f C_A   \bigg\{   \frac{3520}{9} \bigg\}
       + C_F^2 n_f^2   \bigg\{ \frac{320}{9} \bigg\}
       + C_F^3 C_A   \bigg\{ 2680 - 720 \zeta_2 \bigg\}
       - C_F^3 n_f   \bigg\{   400 \bigg\}
\nonumber\\&
       - C_F^4   \bigg\{  1920 + 2880 \zeta_2 \bigg\}\Bigg] 
+
 \bm{  \delta_{\overline{x}'}\mathcal{D}^{7}_{\overline{z}'}} \Bigg[
        C_F^4   \bigg\{ \frac{16}{3} \bigg\}\Bigg] 
-
 \bm{  \delta_{\overline{x}'}\mathcal{D}^{6}_{\overline{z}'}} \Bigg[
        C_F^3 C_A   \bigg\{   \frac{308}{9} \bigg\}
       - C_F^3 n_f   \bigg\{ \frac{56}{9} \bigg\}\Bigg] 
\nonumber\\&
+
 \bm{  \delta_{\overline{x}'}\mathcal{D}^{5}_{\overline{z}'}} \Bigg[
        C_F^2 C_A^2   \bigg\{ \frac{1936}{27} \bigg\}
       - C_F^2 n_f C_A   \bigg\{   \frac{704}{27} \bigg\}
       + C_F^2 n_f^2   \bigg\{ \frac{64}{27} \bigg\}
       + C_F^3 C_A   \bigg\{ \frac{536}{3} - 48 \zeta_2 \bigg\}
       - C_F^3 n_f   \bigg\{   \frac{80}{3} \bigg\}
\nonumber\\&       
       - C_F^4   \bigg\{   128 + 192 \zeta_2 \bigg\}\Bigg] 
+
 \bm{  \delta_{\overline{x}'}\mathcal{D}^{4}_{\overline{z}'} }\Bigg[
         C_F n_f C_A^2   \bigg\{ \frac{242}{9} \bigg\}
       - C_F C_A^3   \bigg\{   \frac{1331}{27} \bigg\}
       - C_F n_f^2 C_A   \bigg\{   \frac{44}{9} \bigg\}
       + C_F n_f^3   \bigg\{ \frac{8}{27} \bigg\}
\nonumber\\&       
       + C_F^4   \bigg\{ \frac{2240}{3} \zeta_3 \bigg\}       
       - C_F^2 C_A^2   \bigg\{   \frac{16780}{27} - \frac{440}{3} \zeta_2 \bigg\}
       + C_F^2 n_f C_A   \bigg\{ \frac{5480}{27} - \frac{80}{3} \zeta_2 \bigg\}
       - C_F^3 n_f   \bigg\{   \frac{440}{27} + \frac{1520}{9} \zeta_2 \bigg\}
\nonumber\\&       
       - C_F^2 n_f^2   \bigg\{   \frac{400}{27} \bigg\}
       + C_F^3 C_A   \bigg\{ \frac{2480}{27} + 280 \zeta_3 + \frac{8360}{9} \zeta_2 \bigg\}\Bigg] 
+
 \bm{  \delta_{\overline{x}'}\mathcal{D}^{3}_{\overline{z}'} }\Bigg[
         C_F^3 n_f   \bigg\{ \frac{2764}{9} + \frac{5696}{9} \zeta_3 + \frac{4448}{9} \zeta_2 \bigg\}
\nonumber\\&        
        - C_F n_f^3   \bigg\{   \frac{160}{81} \bigg\}
        + C_F^2 C_A^2   \bigg\{ \frac{481216}{243} - \frac{8800}{9} \zeta_3 - \frac{56080}{27} \zeta_2 + \frac{864}{5} 
         \zeta_2^2 \bigg\}
        - C_F n_f C_A^2   \bigg\{   \frac{7324}{27} - \frac{352}{9} \zeta_2 \bigg\}
\nonumber\\&       
       + C_F n_f^2 C_A   \bigg\{ \frac{1144}{27} - \frac{32}{9} \zeta_2 \bigg\}     
       - C_F^2 n_f C_A   \bigg\{   \frac{133988}{243} - \frac{448}{9} \zeta_3 - \frac{16256}{27} \zeta_2 \bigg\}
        + C_F^2 n_f^2   \bigg\{ \frac{7768}{243} - \frac{1216}{27} \zeta_2 \bigg\}
\nonumber\\&       
     - C_F^3 C_A   \bigg\{   \frac{26362}{9} + \frac{19808}{9} \zeta_3 + \frac{25424}{9} \zeta_2 - \frac{4384}{5} 
         \zeta_2^2 \bigg\}         
       + C_F^4   \bigg\{ 1022 - 480 \zeta_3 + 2512 \zeta_2 + \frac{3136}{5} \zeta_2^2 \bigg\}
\nonumber\\&       
       + C_F C_A^3   \bigg\{ \frac{43648}{81} - \frac{968}{9} \zeta_2 \bigg\}
       \Bigg] 
+
 \bm{  \delta_{\overline{x}'}\mathcal{D}^{2}_{\overline{z}'} }\Bigg[
           C_F^3 C_A   \bigg\{ \frac{34849}{18} + \frac{14696}{3} \zeta_3 - \frac{15988}{9} \zeta_2 - 4608 \zeta_2 
         \zeta_3 - \frac{36784}{15} \zeta_2^2 \bigg\}
\nonumber\\&    
        +C_F n_f C_A^2   \bigg\{ \frac{10277}{9} - 176 \zeta_3 - \frac{5096}{9} \zeta_2 + \frac{176}{5} \zeta_2^2 \bigg\}
       - C_F C_A^3   \bigg\{   \frac{206440}{81} - 968 \zeta_3 - \frac{4012}{3} \zeta_2 + \frac{968}{5} \zeta_2^2 \bigg\}       
\nonumber\\&       
       - C_F n_f^2 C_A   \bigg\{   \frac{3947}{27} - \frac{688}{9} \zeta_2 \bigg\}
       + C_F n_f^3   \bigg\{ \frac{400}{81} - \frac{32}{9} \zeta_2 \bigg\}
       + C_F^3 n_f   \bigg\{ \frac{259}{3} - \frac{4048}{3} \zeta_3 + \frac{2728}{9} \zeta_2 + \frac{6112}{15} \zeta_2^2 \bigg\}
\nonumber\\&       
       + C_F^2 n_f C_A   \bigg\{ \frac{42140}{243} - \frac{13712}{9} \zeta_3 - \frac{21376}{9} \zeta_2 + \frac{976}{3} 
         \zeta_2^2 \bigg\}
         + C_F^4   \bigg\{ 5376 \zeta_5 - 5120 \zeta_3 - 7680 \zeta_2 \zeta_3 \bigg\}
\nonumber\\&       
       - C_F^2 C_A^2   \bigg\{   \frac{356573}{486} +1152 \zeta_5 - \frac{57296}{9} \zeta_3 - \frac{202868}{27} 
         \zeta_2 + 688 \zeta_2 \zeta_3 + \frac{23672}{15} \zeta_2^2 \bigg\}     
\nonumber\\&       
        - C_F^2 n_f^2   \bigg\{   \frac{3718}{243} - \frac{320}{3} \zeta_3 - \frac{4624}{27} \zeta_2 \bigg\}\Bigg] 
+
 \bm{  \delta_{\overline{x}'}\mathcal{D}^{1}_{\overline{z}'} }\Bigg[
        \frac{d^{abcd}_Fd^{abcd}_A}{N_F}    \bigg\{ \frac{3520}{3} \zeta_5 + \frac{128}{3} \zeta_3 - 384 \zeta_3^2 - 128 \zeta_2 - \frac{7936}{35} \zeta_2^3 \bigg\}
\nonumber\\&       
       - n_f \frac{d^{abcd}_Fd^{abcd}_F}{N_F}   \bigg\{   \frac{1280}{3} \zeta_5 + \frac{256}{3} \zeta_3 - 256 \zeta_2 \bigg\}       
        + C_F^2 N_{4} n_{fv}  \bigg\{ 32 - \frac{640}{3} \zeta_5 + \frac{112}{3} \zeta_3 + 80 \zeta_2 - \frac{16}{5} 
         \zeta_2^2 \bigg\}
\nonumber\\&       
       + C_F C_A^3   \bigg\{ \frac{4520317}{729} + \frac{15400}{9} \zeta_5 - \frac{51032}{9} \zeta_3 - 16 \zeta_3^2
          - \frac{140200}{27} \zeta_2 + 528 \zeta_2 \zeta_3 + \frac{3520}{3} \zeta_2^2 - \frac{20032}{105} 
         \zeta_2^3 \bigg\}
\nonumber\\&
       - C_F n_f C_A^2   \bigg\{   \frac{571387}{243} + \frac{1360}{9} \zeta_5 - \frac{3128}{3} \zeta_3 - \frac{58000}{27}
          \zeta_2 - 32 \zeta_2 \zeta_3 + \frac{3872}{15} \zeta_2^2 \bigg\}
        - C_F n_f^3   \bigg\{   \frac{4000}{729} - \frac{320}{27} \zeta_2 \bigg\}
\nonumber\\&
       - C_F^2 C_A^2   \bigg\{   \frac{11106458}{729} + 816 \zeta_5 - \frac{6164}{81} \zeta_3 - \frac{752}{3} 
         \zeta_3^2 + \frac{1049066}{81} \zeta_2 - 7984 \zeta_2 \zeta_3 - \frac{796252}{135} \zeta_2^2 + 
         \frac{287648}{315} \zeta_2^3 \bigg\}
\nonumber\\&
       + C_F^3 C_A   \bigg\{ 12062 - \frac{110752}{9} \zeta_5 - \frac{214592}{27} \zeta_3 + \frac{15808}{3} 
         \zeta_3^2 + \frac{602990}{27} \zeta_2 + \frac{86240}{9} \zeta_2 \zeta_3 + \frac{193264}{135} \zeta_2^2
          - \frac{91648}{63} \zeta_2^3 \bigg\}  
\nonumber\\&
       - C_F^4   \bigg\{   \frac{11198}{3} - 5312 \zeta_5 + 1840 \zeta_3 - \frac{18304}{3} \zeta_3^2 + 
         \frac{25048}{3} \zeta_2 - 4160 \zeta_2 \zeta_3 + \frac{21008}{5} \zeta_2^2 - \frac{48512}{63} \zeta_2^3\bigg\}
\nonumber\\&
       + C_F^2 n_f C_A   \bigg\{ \frac{2300107}{729} + 128 \zeta_5 + \frac{234368}{81} \zeta_3 + \frac{37180}{9} 
         \zeta_2 - \frac{2560}{3} \zeta_2 \zeta_3 - \frac{182576}{135} \zeta_2^2 \bigg\}   
       - C_F^3 n_f   \bigg\{   \frac{4310}{9}
\nonumber\\& 
       -\frac{10816}{9} \zeta_5 - \frac{29456}{27} \zeta_3 + \frac{70880}{27} 
         \zeta_2 + \frac{30464}{9} \zeta_2 \zeta_3 + \frac{95872}{135} \zeta_2^2 \bigg\}                 
        + C_F n_f^2 C_A   \bigg\{ \frac{58045}{243} - \frac{208}{9} \zeta_3
       - \frac{7616}{27} \zeta_2 
\nonumber\\&       
       + \frac{64}{5} \zeta_2^2\bigg\}
        - C_F^2 n_f^2   \bigg\{   \frac{72590}{729} + \frac{30688}{81} \zeta_3         
        + \frac{22000}{81} \zeta_2 - \frac{8512}{135} \zeta_2^2 \bigg\}
          \Bigg] 
+
 \bm{  \mathcal{D}^{1}_{\overline{x}'}\mathcal{D}^{5}_{\overline{z}'} }\Bigg[
        C_F^4   \bigg\{ 224 \bigg\}\Bigg] 
\nonumber\\&
-
 \bm{  \mathcal{D}^{1}_{\overline{x}'}\mathcal{D}^{4}_{\overline{z}'} }\Bigg[
     C_F^3 C_A   \bigg\{   \frac{3080}{3} \bigg\}
     - C_F^3 n_f   \bigg\{ \frac{560}{3} \bigg\}\Bigg]    
+
  \bm{ \mathcal{D}^{1}_{\overline{x}'}\mathcal{D}^{3}_{\overline{z}'} }\Bigg[
        C_F^2 C_A^2   \bigg\{ \frac{38720}{27} \bigg\}
       + C_F^3 C_A   \bigg\{ \frac{10720}{3} - 960 \zeta_2 \bigg\}
\nonumber\\&       
       - C_F^2 n_f C_A   \bigg\{   \frac{14080}{27} \bigg\}        
       - C_F^4   \bigg\{   2560 + 3840 \zeta_2 \bigg\}
       + C_F^2 n_f^2   \bigg\{ \frac{1280}{27} \bigg\}
       - C_F^3 n_f   \bigg\{   \frac{1600}{3} \bigg\}\Bigg] 
\nonumber\\&
+
 \bm{  \mathcal{D}^{1}_{\overline{x}'}\mathcal{D}^{2}_{\overline{z}'} }\Bigg[
        C_F n_f C_A^2   \bigg\{ \frac{968}{3} \bigg\}
       - C_F C_A^3   \bigg\{   \frac{5324}{9} \bigg\}
       + C_F^2 n_f C_A   \bigg\{ \frac{21920}{9} - 320 \zeta_2 \bigg\}
       - C_F^2 C_A^2   \bigg\{   \frac{67120}{9} 
                 - 1760 \zeta_2 \bigg\}
\nonumber\\&       
        - C_F n_f^2 C_A   \bigg\{   \frac{176}{3} \bigg\} 
       + C_F n_f^3   \bigg\{ \frac{32}{9} \bigg\}
       - C_F^3 n_f   \bigg\{   \frac{1760}{9} + \frac{6080}{3} \zeta_2 \bigg\}
       + C_F^3 C_A   \bigg\{ \frac{9920}{9} + 3360 \zeta_3 + \frac{33440}{3} \zeta_2 \bigg\}
\nonumber\\&       
       - C_F^2 n_f^2   \bigg\{   \frac{1600}{9} \bigg\}
       + C_F^4   \bigg\{ 8960 \zeta_3 \bigg\}\Bigg] 
+
 \bm{  \mathcal{D}^{1}_{\overline{x}'}\mathcal{D}^{1}_{\overline{z}'} }\Bigg[
        C_F C_A^3   \bigg\{ \frac{43648}{27} - \frac{968}{3} \zeta_2 \bigg\}
      - C_F n_f C_A^2   \bigg\{   \frac{7324}{9} 
            -  \frac{352}{3} \zeta_2 \bigg\}
\nonumber\\&
      - C_F^2 n_f C_A   \bigg\{   \frac{133988}{81} - \frac{448}{3} \zeta_3 - \frac{16256}{9} \zeta_2 \bigg\}
       + C_F^2 C_A^2   \bigg\{ \frac{481216}{81} - \frac{8800}{3} \zeta_3 - \frac{56080}{9} \zeta_2 + \frac{2592}{5}   \zeta_2^2 \bigg\}
\nonumber\\&
       - C_F n_f^3   \bigg\{   \frac{160}{27} \bigg\}
       + C_F^2 n_f^2   \bigg\{ \frac{7768}{81} - \frac{1216}{9} \zeta_2 \bigg\}
       - C_F^3 C_A   \bigg\{   \frac{26362}{3} + \frac{19808}{3} \zeta_3 + \frac{25424}{3} \zeta_2 - \frac{13152}{5} 
         \zeta_2^2 \bigg\}
\nonumber\\&     
      + C_F^3 n_f   \bigg\{ \frac{2764}{3} + \frac{5696}{3} \zeta_3 + \frac{4448}{3} \zeta_2 \bigg\}         
      + C_F^4   \bigg\{ 3066 - 1440 \zeta_3 + 7536 \zeta_2 + \frac{9408}{5} \zeta_2^2 \bigg\}
\nonumber\\&         
       + C_F n_f^2 C_A   \bigg\{ \frac{1144}{9} - \frac{32}{3} \zeta_2 \bigg\}
       \Bigg] 
+
\bm{   \mathcal{D}^{0}_{\overline{x}'}\mathcal{D}^{6}_{\overline{z}'} }\Bigg[
        C_F^4   \bigg\{ \frac{112}{3} \bigg\}\Bigg]        
-
  \bm{ \mathcal{D}^{0}_{\overline{x}'}\mathcal{D}^{5}_{\overline{z}'} }\Bigg[
        C_F^3 C_A   \bigg\{  \frac{616}{3} \bigg\}
       - C_F^3 n_f   \bigg\{ \frac{112}{3} \bigg\}\Bigg]         
\nonumber\\&
+
 \bm{  \mathcal{D}^{0}_{\overline{x}'}\mathcal{D}^{4}_{\overline{z}'} }\Bigg[
        C_F^2 C_A^2   \bigg\{ \frac{9680}{27} \bigg\}
       - C_F^2 n_f C_A   \bigg\{   \frac{3520}{27} \bigg\}
       + C_F^3 C_A   \bigg\{ \frac{2680}{3} - 240 \zeta_2 \bigg\}
       - C_F^4   \bigg\{   640 + 960 \zeta_2 \bigg\}
\nonumber\\& 
        + C_F^2 n_f^2   \bigg\{ \frac{320}{27} \bigg\}
       - C_F^3 n_f   \bigg\{   \frac{400}{3} \bigg\} \Bigg]   
+   
 \bm{   \mathcal{D}^{0}_{\overline{x}'}\mathcal{D}^{3}_{\overline{z}'} }\Bigg[
         C_F^2 n_f C_A   \bigg\{ \frac{21920}{27} - \frac{320}{3} \zeta_2 \bigg\}
         - C_F^2 C_A^2   \bigg\{   \frac{67120}{27}       - \frac{1760}{3} \zeta_2 \bigg\} 
\nonumber\\&        
       + C_F n_f C_A^2   \bigg\{ \frac{968}{9} \bigg\}
       - C_F C_A^3   \bigg\{   \frac{5324}{27} \bigg\}
       - C_F n_f^2 C_A   \bigg\{   \frac{176}{9} \bigg\}
       + C_F n_f^3   \bigg\{ \frac{32}{27} \bigg\}
       - C_F^3 n_f   \bigg\{   \frac{1760}{27} + \frac{6080}{9} \zeta_2 \bigg\} 
\nonumber\\& 
       - C_F^2 n_f^2   \bigg\{  \frac{1600}{27} \bigg\}
        + C_F^3 C_A   \bigg\{ \frac{9920}{27} + 1120 \zeta_3 + \frac{33440}{9} \zeta_2 \bigg\}
       + C_F^4   \bigg\{ \frac{8960}{3} \zeta_3 \bigg\}\Bigg] 
\nonumber\\&
+
 \bm{  \mathcal{D}^{0}_{\overline{x}'}\mathcal{D}^{2}_{\overline{z}'} }\Bigg[
        C_F C_A^3   \bigg\{ \frac{43648}{27} - \frac{968}{3} \zeta_2 \bigg\}
        + C_F^2 C_A^2   \bigg\{ \frac{481216}{81} - \frac{8800}{3} \zeta_3 - \frac{56080}{9} \zeta_2 + \frac{2592}{5} 
         \zeta_2^2 \bigg\}
       - C_F n_f^3   \bigg\{   \frac{160}{27} \bigg\}
\nonumber\\&       
        - C_F n_f C_A^2   \bigg\{  \frac{7324}{9} - \frac{352}{3} \zeta_2 \bigg\}
       + C_F n_f^2 C_A   \bigg\{ \frac{1144}{9} - \frac{32}{3} \zeta_2 \bigg\}
       - C_F^2 n_f C_A   \bigg\{   \frac{133988}{81} - \frac{448}{3} \zeta_3 - \frac{16256}{9} \zeta_2 \bigg\}
\nonumber\\&       
       - C_F^3 C_A   \bigg\{   \frac{26362}{3} + \frac{19808}{3} \zeta_3 + \frac{25424}{3} \zeta_2 - \frac{13152}{5} 
         \zeta_2^2 \bigg\}   
       + C_F^3 n_f   \bigg\{ \frac{2764}{3} + \frac{5696}{3} \zeta_3 + \frac{4448}{3} \zeta_2 \bigg\}
\nonumber\\&         
       + C_F^2 n_f^2   \bigg\{ \frac{7768}{81} - \frac{1216}{9} \zeta_2 \bigg\}
       + C_F^4   \bigg\{ 3066 - 1440 \zeta_3 + 7536 \zeta_2 + \frac{9408}{5} \zeta_2^2 \bigg\}\Bigg] 
+
 \bm{  \mathcal{D}^{0}_{\overline{x}'}\mathcal{D}^{1}_{\overline{z}'} }\Bigg[
        C_F n_f^3   \bigg\{ \frac{800}{81} - \frac{64}{9} \zeta_2 \bigg\}
\nonumber\\&       
       - C_F C_A^3   \bigg\{   \frac{412880}{81} - 1936 \zeta_3 - \frac{8024}{3} \zeta_2 + \frac{1936}{5} \zeta_2^2
          \bigg\}
       + C_F n_f C_A^2   \bigg\{ \frac{20554}{9} - 352 \zeta_3 - \frac{10192}{9} \zeta_2 + \frac{352}{5} \zeta_2^2 \bigg\}
\nonumber\\&
       - C_F n_f^2 C_A   \bigg\{   \frac{7894}{27} - \frac{1376}{9} \zeta_2 \bigg\}
              + C_F^3 C_A   \bigg\{ \frac{34849}{9} + \frac{29392}{3} \zeta_3 - \frac{31976}{9} \zeta_2 - 9216 \zeta_2 
         \zeta_3 - \frac{73568}{15} \zeta_2^2 \bigg\}
\nonumber\\&
       - C_F^2 C_A^2   \bigg\{   \frac{356573}{243} + 2304 \zeta_5 - \frac{114592}{9} \zeta_3 - \frac{405736}{27}
          \zeta_2 + 1376 \zeta_2 \zeta_3 + \frac{47344}{15} \zeta_2^2 \bigg\}
\nonumber\\&
       + C_F^2 n_f C_A   \bigg\{ \frac{84280}{243} - \frac{27424}{9} \zeta_3 - \frac{42752}{9} \zeta_2 + \frac{1952}{3} 
         \zeta_2^2 \bigg\}
        + C_F^3 n_f   \bigg\{ \frac{518}{3} - \frac{8096}{3} \zeta_3 + \frac{5456}{9} \zeta_2 + \frac{12224}{15} \zeta_2^2 \bigg\}
\nonumber\\&
        - C_F^2 n_f^2   \bigg\{   \frac{7436}{243} - \frac{640}{3} \zeta_3 - \frac{9248}{27} \zeta_2 \bigg\}
       + C_F^4   \bigg\{ 10752 \zeta_5 - 10240 \zeta_3 - 15360 \zeta_2 \zeta_3 \bigg\}\Bigg] 
\nonumber\\&
+
  \bm{ \mathcal{D}^{0}_{\overline{x}'}\mathcal{D}^{0}_{\overline{z}'} }\Bigg[
        \frac{d^{abcd}_Fd^{abcd}_A}{N_F}   \bigg\{ \frac{1760}{3} \zeta_5 + \frac{64}{3} \zeta_3 - 192 \zeta_3^2 - 64 \zeta_2 - \frac{3968}{35} 
         \zeta_2^3 \bigg\}
       - n_f \frac{d^{abcd}_Fd^{abcd}_F}{N_F}  \bigg\{   \frac{640}{3} \zeta_5 + \frac{128}{3} \zeta_3 
\nonumber\\&       
       - 128 \zeta_2 \bigg\}
       + C_F C_A^3   \bigg\{ \frac{4520317}{1458} + \frac{7700}{9} \zeta_5 - \frac{25516}{9} \zeta_3 - 8 \zeta_3^2
          - \frac{70100}{27} \zeta_2 + 264 \zeta_2 \zeta_3 + \frac{1760}{3} \zeta_2^2 - \frac{10016}{105} 
         \zeta_2^3 \bigg\}
\nonumber\\&
       - C_F n_f C_A^2   \bigg\{   \frac{571387}{486} + \frac{680}{9} \zeta_5 - \frac{1564}{3} \zeta_3 - \frac{29000}{27} 
         \zeta_2 - 16 \zeta_2 \zeta_3 + \frac{1936}{15} \zeta_2^2 \bigg\}
       - C_F n_f^3   \bigg\{   \frac{2000}{729} - \frac{160}{27} \zeta_2 \bigg\}
\nonumber\\&
       + C_F n_f^2 C_A   \bigg\{ \frac{58045}{486} - \frac{104}{9} \zeta_3 - \frac{3808}{27} \zeta_2 + \frac{32}{5} \zeta_2^2
          \bigg\}
       + C_F^2 N_{4} n_{fv}  \bigg\{ 16 - \frac{320}{3} \zeta_5 + \frac{56}{3} \zeta_3 + 40 \zeta_2 - \frac{8}{5} 
         \zeta_2^2 \bigg\}
\nonumber\\&
       - C_F^2 C_A^2   \bigg\{   \frac{5553229}{729} + 408 \zeta_5 - \frac{3082}{81} \zeta_3 - \frac{376}{3} 
         \zeta_3^2 + \frac{524533}{81} \zeta_2 - 3992 \zeta_2 \zeta_3 - \frac{398126}{135} \zeta_2^2 + 
         \frac{143824}{315} \zeta_2^3 \bigg\}
\nonumber\\&
       + C_F^3 C_A   \bigg\{ 6031 - \frac{55376}{9} \zeta_5 - \frac{107296}{27} \zeta_3 + \frac{7904}{3} \zeta_3^2
          + \frac{301495}{27} \zeta_2 + \frac{43120}{9} \zeta_2 \zeta_3 + \frac{96632}{135} \zeta_2^2 - 45824/
         63 \zeta_2^3 \bigg\}
\nonumber\\&
       + C_F^2 n_f C_A   \bigg\{ \frac{2300107}{1458} + 64 \zeta_5 + \frac{117184}{81} \zeta_3 + \frac{18590}{9} 
         \zeta_2 - \frac{1280}{3} \zeta_2 \zeta_3 - \frac{91288}{135} \zeta_2^2 \bigg\}
       - C_F^2 n_f^2   \bigg\{   \frac{36295}{729} 
\nonumber\\&       
       + \frac{15344}{81} \zeta_3 + \frac{11000}{81} \zeta_2 - \frac{4256}{135} \zeta_2^2 \bigg\}         
       - C_F^3 n_f   \bigg\{   \frac{2155}{9} - \frac{5408}{9} \zeta_5 - \frac{14728}{27} \zeta_3 + \frac{35440}{27} 
         \zeta_2 + \frac{15232}{9} \zeta_2 \zeta_3 
\nonumber\\&         
         + \frac{47936}{135} \zeta_2^2 \bigg\}
       - C_F^4   \bigg\{   \frac{5599}{3} - 2656 \zeta_5 + 920 \zeta_3 - \frac{9152}{3} \zeta_3^2 + \frac{12524}{3} \zeta_2 - 2080 \zeta_2 \zeta_3 + \frac{10504}{5} \zeta_2^2 - \frac{24256}{63} \zeta_2^3 \bigg\}\Bigg] 
\nonumber\\&
+
   \bm{\delta_{x}\mathcal{D}^{0}_{z}} \Bigg[
 n_f \frac{d^{abcd}_Fd^{abcd}_F}{N_F} 
        \bigg\{ \frac{1600}{9} \zeta_5 - \frac{640}{9} \zeta_3 + \frac{320}{3} \zeta_3^2 - 256
         \zeta_2 + \frac{64}{5} \zeta_2^2 + \frac{1280}{21} \zeta_2^3 \bigg\}
       - 
       \frac{d^{abcd}_Fd^{abcd}_A}{N_F} \bigg\{   192 
\nonumber\\&       
 -2~{\color{blue}b^{q}_{4,FA}}
- \frac{1840}{9} \zeta_5 + 3484 \zeta_7 -
         \frac{7808}{9} \zeta_3 - \frac{3344}{3} \zeta_3^2 - \frac{2176}{3} \zeta_2 + 1024 \zeta_2 \zeta_5 -
         1792 \zeta_2 \zeta_3 + \frac{224}{15} \zeta_2^2 
\nonumber\\&         
         - \frac{736}{5} \zeta_2^2 \zeta_3 + \frac{27808}{315}
         \zeta_2^3 \bigg\}
       - C_F C_A^3   \bigg\{   \frac{28290079}{4374} 
         + \frac{1}{12} {\color{blue}b^{q}_{4,FA}}   
       + \frac{74990}{27} \zeta_5 -
         \frac{11071}{6} \zeta_7 - \frac{258224}{27} \zeta_3 
\nonumber\\&         
         + \frac{7414}{9} \zeta_3^2 - \frac{1634851}{243} \zeta_2
          - \frac{1376}{3} \zeta_2 \zeta_5 + \frac{31444}{9} \zeta_2 \zeta_3 + \frac{15632}{9} \zeta_2^2 - \frac{4228}{15} \zeta_2^2 \zeta_3 - \frac{27808}{189} \zeta_2^3 \bigg\}
\nonumber\\&       
       + C_F n_f C_A^2   \bigg\{ \frac{11551831}{5832} 
       - \frac{3532}{27} \zeta_5 - \frac{150988}{81} \zeta_3 - \frac{2276}{9} \zeta_3^2 - \frac{645476}{243} \zeta_2 + \frac{4328}{9} \zeta_2 \zeta_3 + \frac{29624}{45} \zeta_2^2
          - \frac{7288}{105} \zeta_2^3 \bigg\}
\nonumber\\&          
       - C_F n_f^2 C_A   \bigg\{  \frac{898033}{5832} - \frac{304}{3} \zeta_5 - \frac{2456}{81} \zeta_3 - \frac{75718}{
         243} \zeta_2 - \frac{80}{9} \zeta_2 \zeta_3 + \frac{3104}{45} \zeta_2^2 \bigg\}
       + C_F n_f^3   \bigg\{ \frac{5216}{2187} + \frac{80}{81} \zeta_3 - \frac{800}{81} \zeta_2 
\nonumber\\&       
       + \frac{16}{9} \zeta_2^2 \bigg\}
       + C_F^2 C_A^2   \bigg\{ \frac{7543094}{729} + \frac{58624}{9} \zeta_5 - \frac{134305}{27} \zeta_3 - 2032
         \zeta_3^2 - \frac{384365}{729} \zeta_2 + 1536 \zeta_2 \zeta_5 - \frac{361240}{27} \zeta_2 \zeta_3
\nonumber\\&         
          - \frac{382312}{81} \zeta_2^2 + \frac{5968}{3} \zeta_2^2 \zeta_3 + \frac{2288}{3} \zeta_2^3 \bigg\}
       - C_F^2 n_f C_A   \bigg\{   \frac{1792393}{2916} + \frac{15176}{9} \zeta_5 + \frac{55924}{27} \zeta_3 - 536
          \zeta_3^2 + \frac{202685}{729} \zeta_2 
\nonumber\\&          
          - \frac{31040}{9} \zeta_2 \zeta_3 - \frac{96664}{81} \zeta_2^2
          + \frac{4064}{21} \zeta_2^3 \bigg\}
       - C_F^2 n_f^2   \bigg\{  \frac{142769}{1458} - \frac{1168}{9} \zeta_5 - \frac{8552}{27} \zeta_3 - \frac{26980}{729} \zeta_2 + \frac{5888}{27} \zeta_2 \zeta_3 
\nonumber\\&       
       + \frac{2240}{27} \zeta_2^2 \bigg\}
       - C_F^3 C_A   \bigg\{   \frac{103222}{27} - 8576 \zeta_5 + \frac{57095}{9} \zeta_3 + 5008 \zeta_3^2
          + \frac{33233}{9} \zeta_2 + 2304 \zeta_2 \zeta_5 + \frac{62744}{9} \zeta_2 \zeta_3 
\nonumber\\&          
          - \frac{34456}{15} \zeta_2^2 - \frac{19552}{5} \zeta_2^2 \zeta_3 - \frac{1056}{5} \zeta_2^3 \bigg\}
       - C_F^3 n_f   \bigg\{   \frac{73309}{108} + \frac{2240}{3} \zeta_5 - \frac{9256}{9} \zeta_3 - 1616 \zeta_3^2
          - \frac{1510}{27} \zeta_2 - \frac{21664}{9} \zeta_2 \zeta_3 
\nonumber\\&          
          + \frac{4492}{15} \zeta_2^2 - \frac{2048}{35}
         \zeta_2^3 \bigg\}
       - C_F^4   \bigg\{ 6144 \zeta_5 - 7680 \zeta_7 - 4088 \zeta_3 + 1920 \zeta_3^2 +
         9216 \zeta_2 \zeta_5 - 10048 \zeta_2 \zeta_3 - \frac{12544}{5} \zeta_2^2 \zeta_3 \bigg\}\Bigg]
\nonumber\\&
+\bm{(x' \leftrightarrow z')} \,.
\end{align}  
In the above equation, the constant,  $b^{q}_{4,FA} =-998.02\pm  0.02$, \cite{Kniehl:2025ttz,Moch:2023tdj}.


\subsubsection{NSV Coefficient functions}
\noindent
The NSV contributions to the CFs are next. At NLO, i.e at ${\mathcal O} (a_s)$, we have
\begin{align}
(\Delta){\cal C}^{(1),\text{NSV}}_{J,cc} &=  
  \bm{ \delta_{\overline{z}'}\mathcal{L}^{0}_{\overline{x}'} }\Bigg[
        C_F   \bigg\{ 4 \bigg\}\Bigg]
- 
   \bm{\delta_{\overline{x}'}\mathcal{L}^{0}_{\overline{z}'}} \Bigg[
        C_F   \bigg\{   4 \bigg\}\Bigg]
-
   \bm{\mathcal{D}^{0}_{\overline{x}'}\mathcal{L}^{0}_{\overline{z}'}} \Bigg[
        C_F   \bigg\{   4 \bigg\}\Bigg]
-
   \bm{\mathcal{D}^{0}_{\overline{z}'}\mathcal{L}^{0}_{\overline{x}'} }\Bigg[
        C_F   \bigg\{   4 \bigg\}\Bigg]
-
   \bm{\delta_{\overline{x}'}\mathcal{L}^{1}_{\overline{z}'} }\Bigg[
        C_F   \bigg\{   4 \bigg\}\Bigg]
\nonumber\\&
-
   \bm{\delta_{\overline{z}'}\mathcal{L}^{1}_{\overline{x}'} }\Bigg[
        C_F   \bigg\{   4 \bigg\}\Bigg]
\, .
\end{align}
The NSV terms of the CFs at NNLO are given by
\begin{align}
(\Delta)\mathcal{C}^{(2),\text{NSV}}_{J,cc} &=
   \bm{\delta_{\overline{x}'}\mathcal{L}^{0}_{\overline{z}'} }\Bigg[
        C_F C_A   \bigg\{ \frac{1108}{27} - 28 \zeta_3 - \frac{44}{3} \zeta_2 \bigg\}
       - C_F n_f   \bigg\{   \frac{52}{27} - \frac{8}{3} \zeta_2 \bigg\}
       + C_F^2   \bigg\{ 64 - 32 \zeta_3 - 24 \zeta_2 \bigg\}\Bigg] 
\nonumber\\&
+
   \bm{\delta_{\overline{z}'}\mathcal{L}^{0}_{\overline{x}'} }\Bigg[
        C_F C_A   \bigg\{ \frac{1492}{27} - 28 \zeta_3 - \frac{68}{3} \zeta_2 \bigg\}
       - C_F n_f   \bigg\{   \frac{364}{27} - \frac{8}{3} \zeta_2 \bigg\}
       - C_F^2   \bigg\{   64 + 32 \zeta_3 + 32 \zeta_2 \bigg\}\Bigg] 
\nonumber\\&
-
   \bm{\mathcal{D}^{0}_{\overline{x}'}\mathcal{L}^{0}_{\overline{z}'} }\Bigg[
        C_F C_A   \bigg\{   \frac{268}{9} - 8 \zeta_2 \bigg\}
       - C_F n_f   \bigg\{ \frac{40}{9} \bigg\}
       - C_F^2   \bigg\{ 52 + 32 \zeta_2 \bigg\}\Bigg] 
-
   \bm{\mathcal{D}^{0}_{\overline{z}'}\mathcal{L}^{0}_{\overline{x}'} }\Bigg[
        C_F C_A   \bigg\{   \frac{532}{9} - 8 \zeta_2 \bigg\}
\nonumber\\&        
       - C_F n_f   \bigg\{ \frac{88}{9} \bigg\}
       - C_F^2   \bigg\{ 76 + 32 \zeta_2 \bigg\}\Bigg] 
+
   \bm{\mathcal{D}^{0}_{\overline{x}'}\mathcal{L}^{1}_{\overline{z}'} }\Bigg[
        C_F C_A   \bigg\{ \frac{44}{3} \bigg\}
       - C_F n_f   \bigg\{   \frac{8}{3} \bigg\}
       - C_F^2   \bigg\{   16 \bigg\}\Bigg] 
\nonumber\\&
+
   \bm{\mathcal{D}^{0}_{\overline{z}'}\mathcal{L}^{1}_{\overline{x}'} }\Bigg[
        C_F C_A   \bigg\{ \frac{44}{3} \bigg\}
       - C_F n_f   \bigg\{   \frac{8}{3} \bigg\}
       + C_F^2   \bigg\{ 48 \bigg\}\Bigg] 
-
   \bm{\mathcal{D}^{0}_{\overline{x}'}\mathcal{L}^{2}_{\overline{z}'}} \Bigg[
        C_F^2   \bigg\{   24 \bigg\}\Bigg] 
-
   \bm{\mathcal{D}^{0}_{\overline{z}'}\mathcal{L}^{2}_{\overline{x}'}} \Bigg[
        C_F^2   \bigg\{   24 \bigg\}\Bigg] 
\nonumber\\&
-
   \bm{\mathcal{D}^{1}_{\overline{x}'}\mathcal{L}^{1}_{\overline{z}'} }\Bigg[
        C_F^2   \bigg\{   48 \bigg\}\Bigg] 
-
   \bm{\mathcal{D}^{1}_{\overline{z}'}\mathcal{L}^{1}_{\overline{x}'}} \Bigg[
        C_F^2   \bigg\{   48 \bigg\}\Bigg]   
+
   \bm{\mathcal{D}^{1}_{\overline{x}'}\mathcal{L}^{0}_{\overline{z}'}} \Bigg[
        C_F C_A   \bigg\{ \frac{44}{3} \bigg\}
       - C_F n_f   \bigg\{   \frac{8}{3} \bigg\}\Bigg]         
\nonumber\\&
+
   \bm{\mathcal{D}^{1}_{\overline{z}'}\mathcal{L}^{0}_{\overline{x}'} }\Bigg[
        C_F C_A   \bigg\{ \frac{44}{3} \bigg\}
       - C_F n_f   \bigg\{   \frac{8}{3} \bigg\}
       + C_F^2   \bigg\{ 32 \bigg\}\Bigg]  
-
   \bm{\delta_{\overline{x}'}\mathcal{L}^{1}_{\overline{z}'}} \Bigg[
        C_F C_A   \bigg\{   \frac{358}{9} - 8 \zeta_2 \bigg\}
       - C_F n_f   \bigg\{ \frac{40}{9} \bigg\}
\nonumber\\&       
       - C_F^2   \bigg\{ 62 + 32 \zeta_2 \bigg\}\Bigg] 
-
   \bm{\delta_{\overline{z}'}\mathcal{L}^{1}_{\overline{x}'} }\Bigg[
        C_F C_A   \bigg\{   \frac{442}{9} - 8 \zeta_2 \bigg\}
       - C_F n_f   \bigg\{ \frac{88}{9} \bigg\}
       - C_F^2   \bigg\{ 66 + 32 \zeta_2 \bigg\}\Bigg] 
\nonumber\\&
+
   \bm{\delta_{\overline{x}'}\mathcal{L}^{2}_{\overline{z}'} }\Bigg[
        C_F C_A   \bigg\{ \frac{22}{3} \bigg\}
       - C_F n_f   \bigg\{   \frac{4}{3} \bigg\}
       - C_F^2   \bigg\{   12 \bigg\}\Bigg] 
+
   \bm{\delta_{\overline{z}'}\mathcal{L}^{2}_{\overline{x}'}} \Bigg[
        C_F C_A   \bigg\{ \frac{22}{3} \bigg\}
       - C_F n_f   \bigg\{  \frac{4}{3} \bigg\}
       + C_F^2   \bigg\{ 28 \bigg\}\Bigg] 
\nonumber\\&
-
   \bm{\mathcal{D}^{2}_{\overline{x}'}\mathcal{L}^{0}_{\overline{z}'} }\Bigg[
        C_F^2   \bigg\{   24 \bigg\}\Bigg] 
-
   \bm{\mathcal{D}^{2}_{\overline{z}'}\mathcal{L}^{0}_{\overline{x}'}} \Bigg[
        C_F^2   \bigg\{   24 \bigg\}\Bigg] 
- 
     \bm{\delta_{\overline{x}'}\mathcal{L}^{3}_{\overline{z}'} }\Bigg[
        C_F^2   \bigg\{   8 \bigg\}\Bigg] 
-
  \bm{ \delta_{\overline{z}'}\mathcal{L}^{3}_{\overline{x}'} }\Bigg[
        C_F^2   \bigg\{  8 \bigg\}\Bigg] 
\, .
\end{align}

\noindent
At third order, only partial result for NSV contributions can be obtained with the the presently known ingredients, unlike the SV case. 
Using the complete second order results for $\Psi_{d}^{q,(1)}$ and $\Psi_{d}^{q,(2)}$ along with third order results ($A_{3}^{q}, C_{3}^{q}, \tilde{C}_{3}^{q}, D_{3}^{q},\tilde{D}_{3}^{q}$) and an 
intriquing relation $\overline{\varphi}_{q,z',3}^{(3)}=-\overline{\varphi}_{q,x',3}^{(3)} = -16/9 \beta_0 C_F^2$, see \cite{Das:2024pac},
we can predict the coefficients of $\{\delta_{\overline{x}'}\mathcal{L}^{i}_{\overline{z}'},\delta_{\overline{z}'}\mathcal{L}^{i}_{\overline{x}'}\}$ for $i=5,4,3$ and of $\{\mathcal{D}_{\overline{x}'}^{i}\mathcal{L}^{j}_{\overline{z}'},\mathcal{D}_{\overline{z}'}^{i}\mathcal{L}^{j}_{\overline{x}'}\}$ for $i+j=4,3,2,1,0$. 
The coefficients of $\{\delta_{\overline{x}'}\mathcal{L}^{i}_{\overline{z}'},\delta_{\overline{z}'}\mathcal{L}^{i}_{\overline{x}'}\}$ for $i=2,1,0$ cannot be predicted at third order, as they require explicit information at that order, namely $\overline{\varphi}_{q,\xi,3}^{(i)}$ for $i=0,1,2$ and $\xi = x', z'$.
Thus, the available partial NSV results at ${\mathcal O} (a_s^3)$ are
%
\begin{align}
(\Delta)\mathcal{C}^{(3),\text{NSV}}_{J,cc} &=
   \bm{\mathcal{D}^{0}_{\overline{x}'}\mathcal{L}^{0}_{\overline{z}'} }\Bigg[
        C_F n_f C_A   \bigg\{ \frac{11012}{81} - \frac{560}{9} \zeta_2 \bigg\}
       - C_F n_f^2   \bigg\{   \frac{400}{81} - \frac{32}{9} \zeta_2 \bigg\}
         - C_F^2 n_f   \bigg\{  \frac{940}{9} + \frac{320}{3} \zeta_3 + \frac{1136}{9} \zeta_2 \bigg\}
\nonumber\\&       
       - C_F C_A^2   \bigg\{   \frac{46450}{81} - 176 \zeta_3 - 256 \zeta_2 + \frac{176}{5} \zeta_2^2 \bigg\}
       + C_F^2 C_A   \bigg\{ \frac{7987}{9} + \frac{512}{3} \zeta_3 + \frac{6920}{9} \zeta_2 - \frac{912}{5} \zeta_2^2 \bigg\}
\nonumber\\&       
       - C_F^3   \bigg\{   325 - 112 \zeta_3 + 640 \zeta_2 + \frac{288}{5} \zeta_2^2 \bigg\}\Bigg] 
+
   \bm{\mathcal{D}^{0}_{\overline{z}'}\mathcal{L}^{0}_{\overline{x}'}} \Bigg[
        C_F^2 C_A   \bigg\{ \frac{35605}{27} + \frac{896}{3} \zeta_3 + \frac{10520}{9} \zeta_2 - \frac{912}{5} \zeta_2^2 \bigg\} 
\nonumber\\&
        + C_F n_f C_A   \bigg\{ \frac{21572}{81} - \frac{656}{9} \zeta_2 \bigg\}
       - C_F n_f^2   \bigg\{   \frac{1648}{81} - \frac{32}{9} \zeta_2 \bigg\}
        - C_F^2 n_f   \bigg\{   \frac{4588}{27} + \frac{320}{3} \zeta_3 + \frac{2000}{9} \zeta_2 \bigg\}
\nonumber\\&       
       - C_F C_A^2   \bigg\{   \frac{62242}{81} - 176 \zeta_3 - \frac{944}{3} \zeta_2 + \frac{176}{5} \zeta_2^2 \bigg\} 
       - C_F^3   \bigg\{   697 - 880 \zeta_3 + 848 \zeta_2 + \frac{288}{5} \zeta_2^2 \bigg\}\Bigg] 
\nonumber\\&
+
   \bm{\mathcal{D}^{0}_{\overline{x}'}\mathcal{L}^{1}_{\overline{z}'} }\Bigg[
        C_F C_A^2   \bigg\{ \frac{9100}{27} - \frac{176}{3} \zeta_2 \bigg\}
       - C_F n_f C_A   \bigg\{   \frac{2672}{27} - \frac{32}{3} \zeta_2 \bigg\}
       + C_F^2 C_A   \bigg\{ \frac{580}{9} - 336 \zeta_3 - 496 \zeta_2 \bigg\}
\nonumber\\&       
       + C_F n_f^2   \bigg\{ \frac{160}{27} \bigg\}
       + C_F^2 n_f   \bigg\{ \frac{80}{9} + 96 \zeta_2 \bigg\}
       + C_F^3   \bigg\{ 256 - 640 \zeta_3 \bigg\}\Bigg] 
+
   \bm{\mathcal{D}^{0}_{\overline{z}'}\mathcal{L}^{1}_{\overline{x}'} }\Bigg[
        C_F C_A^2   \bigg\{ \frac{10948}{27} - \frac{176}{3} \zeta_2 \bigg\}
\nonumber\\&       
       - C_F n_f C_A   \bigg\{   \frac{4064}{27} - \frac{32}{3} \zeta_2 \bigg\}
       + C_F^2 C_A   \bigg\{ \frac{2420}{3} - 336 \zeta_3 - 720 \zeta_2 \bigg\}
       - C_F^3   \bigg\{  768 + 640 \zeta_3 + 736 \zeta_2 \bigg\}       
\nonumber\\&       
       + C_F n_f^2   \bigg\{ \frac{352}{27} \bigg\}
       - C_F^2 n_f   \bigg\{   \frac{416}{3} - 96 \zeta_2 \bigg\}\Bigg] 
+
   \bm{\mathcal{D}^{0}_{\overline{x}'}\mathcal{L}^{2}_{\overline{z}'} }\Bigg[
        C_F n_f C_A   \bigg\{ \frac{176}{9} \bigg\}
       - C_F C_A^2   \bigg\{   \frac{484}{9} \bigg\}
       - C_F n_f^2   \bigg\{   \frac{16}{9} \bigg\}
\nonumber\\&       
       - C_F^2 C_A   \bigg\{   368 - 96 \zeta_2 \bigg\}
       + C_F^2 n_f   \bigg\{ 48 \bigg\}       
       + C_F^3   \bigg\{ 352 + 384 \zeta_2 \bigg\}\Bigg] 
+
   \bm{\mathcal{D}^{0}_{\overline{z}'}\mathcal{L}^{2}_{\overline{x}'} }\Bigg[
        C_F n_f C_A   \bigg\{ \frac{176}{9} \bigg\}
        - C_F n_f^2   \bigg\{   \frac{16}{9} \bigg\}
\nonumber\\&       
       - C_F C_A^2   \bigg\{   \frac{484}{9} \bigg\}
       - C_F^2 C_A   \bigg\{   \frac{2096}{3} - 96 \zeta_2 \bigg\}
       + C_F^2 n_f   \bigg\{ \frac{368}{3} \bigg\}       
       + C_F^3   \bigg\{ 416 + 384 \zeta_2 \bigg\}\Bigg] 
\nonumber\\&
+
   \bm{\mathcal{D}^{0}_{\overline{x}'}\mathcal{L}^{3}_{\overline{z}'} }\Bigg[
        C_F^2 C_A   \bigg\{ \frac{880}{9} \bigg\}
       - C_F^2 n_f   \bigg\{   \frac{160}{9} \bigg\}
       - C_F^3   \bigg\{   48 \bigg\}\Bigg] 
+
   \bm{\mathcal{D}^{0}_{\overline{z}'}\mathcal{L}^{3}_{\overline{x}'} }\Bigg[
        C_F^2 C_A   \bigg\{ \frac{880}{9} \bigg\}
       - C_F^2 n_f   \bigg\{   \frac{160}{9} \bigg\}
\nonumber\\&       
       + C_F^3   \bigg\{ 176 \bigg\}\Bigg] 
-
   \bm{\mathcal{D}^{0}_{\overline{x}'}\mathcal{L}^{4}_{\overline{z}'} }\Bigg[
        C_F^3   \bigg\{   40 \bigg\}\Bigg] 
-
   \bm{\mathcal{D}^{0}_{\overline{z}'}\mathcal{L}^{4}_{\overline{x}'}} \Bigg[
        C_F^3   \bigg\{   40 \bigg\}\Bigg] 
+
   \bm{\mathcal{D}^{1}_{\overline{x}'}\mathcal{L}^{0}_{\overline{z}'} }\Bigg[
        C_F C_A^2   \bigg\{ \frac{7120}{27} - \frac{176}{3} \zeta_2 \bigg\}
\nonumber\\&
       + C_F n_f^2   \bigg\{ \frac{160}{27} \bigg\} 
       - C_F n_f C_A   \bigg\{   \frac{2312}{27} - \frac{32}{3} \zeta_2 \bigg\}
       + C_F^2 C_A   \bigg\{ 376 - 336 \zeta_3 - 560 \zeta_2 \bigg\}
       - C_F^3   \bigg\{   640 \zeta_3 + 96 \zeta_2 \bigg\}
\nonumber\\&       
        - C_F^2 n_f   \bigg\{   40 - 96 \zeta_2 \bigg\}    \Bigg] 
+
   \bm{\mathcal{D}^{1}_{\overline{z}'}\mathcal{L}^{0}_{\overline{x}'}} \Bigg[
        C_F C_A^2   \bigg\{ \frac{12928}{27} - \frac{176}{3} \zeta_2 \bigg\}
       + C_F n_f^2   \bigg\{ \frac{352}{27} \bigg\}        
       - C_F n_f C_A   \bigg\{   \frac{4424}{27} - \frac{32}{3} \zeta_2 \bigg\}
\nonumber\\&
       + C_F^2 C_A   \bigg\{ \frac{4456}{9} - 336 \zeta_3 - 656 \zeta_2 \bigg\}
       - C_F^2 n_f   \bigg\{   \frac{808}{9} - 96 \zeta_2 \bigg\}
       - C_F^3   \bigg\{   512 + 640 \zeta_3 + 640 \zeta_2 \bigg\}\Bigg] 
\nonumber\\&
+
   \bm{\mathcal{D}^{1}_{\overline{x}'}\mathcal{L}^{1}_{\overline{z}'} }\Bigg[
        C_F n_f C_A   \bigg\{ \frac{352}{9} \bigg\}       
       - C_F C_A^2   \bigg\{   \frac{968}{9} \bigg\}
       - C_F n_f^2   \bigg\{   \frac{32}{9} \bigg\}
       - C_F^2 C_A   \bigg\{   \frac{2264}{3} - 192 \zeta_2 \bigg\}
       + C_F^2 n_f   \bigg\{ \frac{320}{3} \bigg\}
\nonumber\\&       
       + C_F^3   \bigg\{ 664 + 768 \zeta_2 \bigg\}\Bigg] 
+
   \bm{\mathcal{D}^{1}_{\overline{z}'}\mathcal{L}^{1}_{\overline{x}'}} \Bigg[
        C_F n_f C_A   \bigg\{ \frac{352}{9} \bigg\}       
       - C_F C_A^2   \bigg\{  \frac{968}{9} \bigg\}
       - C_F n_f^2   \bigg\{   \frac{32}{9} \bigg\}
       + C_F^2 n_f   \bigg\{ \frac{704}{3} \bigg\}       
\nonumber\\&       
       - C_F^2 C_A   \bigg\{   \frac{4136}{3} - 192 \zeta_2 \bigg\}       
       + C_F^3   \bigg\{ 872 + 768 \zeta_2 \bigg\}\Bigg] 
+
   \bm{\mathcal{D}^{1}_{\overline{x}'}\mathcal{L}^{2}_{\overline{z}'} }\Bigg[
        C_F^2 C_A   \bigg\{ \frac{880}{3} \bigg\}
       - C_F^2 n_f   \bigg\{   \frac{160}{3} \bigg\}
       - C_F^3   \bigg\{   80 \bigg\}\Bigg] 
\nonumber\\&
+
   \bm{\mathcal{D}^{1}_{\overline{z}'}\mathcal{L}^{2}_{\overline{x}'}} \Bigg[
        C_F^2 C_A   \bigg\{ \frac{880}{3} \bigg\}
       - C_F^2 n_f   \bigg\{   \frac{160}{3} \bigg\}
       + C_F^3   \bigg\{ 464 \bigg\}\Bigg] 
-
   \bm{\mathcal{D}^{1}_{\overline{x}'}\mathcal{L}^{3}_{\overline{z}'} }\Bigg[
        C_F^3   \bigg\{   160 \bigg\}\Bigg] 
-
   \bm{\mathcal{D}^{1}_{\overline{z}'}\mathcal{L}^{3}_{\overline{x}'} }\Bigg[
        C_F^3   \bigg\{   160 \bigg\}\Bigg] 
\nonumber\\&
+
   \bm{\mathcal{D}^{2}_{\overline{x}'}\mathcal{L}^{0}_{\overline{z}'} }\Bigg[
        C_F n_f C_A   \bigg\{ \frac{176}{9} \bigg\}       
       - C_F C_A^2   \bigg\{   \frac{484}{9} \bigg\}
       - C_F n_f^2   \bigg\{   \frac{16}{9} \bigg\}
       - C_F^2 C_A   \bigg\{   416 - 96 \zeta_2 \bigg\}
       + C_F^2 n_f   \bigg\{ 64 \bigg\}
\nonumber\\&       
       + C_F^3   \bigg\{ 312 + 384 \zeta_2 \bigg\}\Bigg] 
+
   \bm{\mathcal{D}^{2}_{\overline{z}'}\mathcal{L}^{0}_{\overline{x}'}} \Bigg[
        C_F n_f C_A   \bigg\{ \frac{176}{9} \bigg\}       
       - C_F C_A^2   \bigg\{   \frac{484}{9} \bigg\}
       - C_F n_f^2   \bigg\{   \frac{16}{9} \bigg\}
      + C_F^2 n_f   \bigg\{ \frac{320}{3} \bigg\}
\nonumber\\&       
       - C_F^2 C_A   \bigg\{   \frac{1952}{3} - 96 \zeta_2 \bigg\}       
       + C_F^3   \bigg\{ 456 + 384 \zeta_2 \bigg\}\Bigg] 
+
   \bm{\mathcal{D}^{2}_{\overline{x}'}\mathcal{L}^{1}_{\overline{z}'} }\Bigg[
        C_F^2 C_A   \bigg\{ \frac{880}{3} \bigg\}
       - C_F^2 n_f   \bigg\{   \frac{160}{3} \bigg\}\Bigg] 
\nonumber\\&
+
   \bm{\mathcal{D}^{2}_{\overline{z}'}\mathcal{L}^{1}_{\overline{x}'} }\Bigg[
        C_F^2 C_A   \bigg\{ \frac{880}{3} \bigg\}
       - C_F^2 n_f   \bigg\{   \frac{160}{3} \bigg\}
       + C_F^3   \bigg\{ 384 \bigg\}\Bigg] 
-
   \bm{\mathcal{D}^{2}_{\overline{x}'}\mathcal{L}^{2}_{\overline{z}'} }\Bigg[
        C_F^3   \bigg\{   240 \bigg\}\Bigg] 
-
   \bm{\mathcal{D}^{2}_{\overline{z}'}\mathcal{L}^{2}_{\overline{x}'}} \Bigg[
        C_F^3   \bigg\{   240 \bigg\}\Bigg] 
\nonumber\\&
+
   \bm{\delta_{\overline{x}'}\mathcal{L}^{3}_{\overline{z}'} }\Bigg[
        C_F n_f C_A   \bigg\{ \frac{176}{27} \bigg\}       
       - C_F C_A^2   \bigg\{   \frac{484}{27} \bigg\}
       - C_F n_f^2   \bigg\{   \frac{16}{27} \bigg\}
       - C_F^2 C_A   \bigg\{   \frac{3316}{27} - 32 \zeta_2 \bigg\}
       + C_F^2 n_f   \bigg\{ \frac{400}{27} \bigg\}
\nonumber\\&       
       + C_F^3   \bigg\{ 124 + 128 \zeta_2 \bigg\}\Bigg] 
+
   \bm{\delta_{\overline{z}'}\mathcal{L}^{3}_{\overline{x}'} }\Bigg[
        C_F n_f C_A   \bigg\{ \frac{176}{27} \bigg\}
       - C_F C_A^2   \bigg\{   \frac{484}{27} \bigg\}
       - C_F n_f^2   \bigg\{   \frac{16}{27} \bigg\}
       + C_F^2 n_f   \bigg\{ \frac{1136}{27} \bigg\}
\nonumber\\&       
       - C_F^2 C_A   \bigg\{   \frac{6284}{27} - 32 \zeta_2 \bigg\}
       + C_F^3   \bigg\{ 132 + 128 \zeta_2 \bigg\}\Bigg] 
+
   \bm{\mathcal{D}^{3}_{\overline{x}'}\mathcal{L}^{0}_{\overline{z}'} }\Bigg[
        C_F^2 C_A   \bigg\{ \frac{880}{9} \bigg\}
       - C_F^2 n_f   \bigg\{   \frac{160}{9} \bigg\}
       + C_F^3   \bigg\{ 32 \bigg\}\Bigg] 
\nonumber\\&
+
   \bm{\mathcal{D}^{3}_{\overline{z}'}\mathcal{L}^{0}_{\overline{x}'} }\Bigg[
        C_F^2 C_A   \bigg\{ \frac{880}{9} \bigg\}
       - C_F^2 n_f   \bigg\{  \frac{160}{9} \bigg\}
       + C_F^3   \bigg\{ 96 \bigg\}\Bigg] 
-
   \bm{\mathcal{D}^{3}_{\overline{x}'}\mathcal{L}^{1}_{\overline{z}'}} \Bigg[
        C_F^3   \bigg\{   160 \bigg\}\Bigg] 
-
   \bm{\mathcal{D}^{3}_{\overline{z}'}\mathcal{L}^{1}_{\overline{x}'} }\Bigg[
        C_F^3   \bigg\{   160 \bigg\}\Bigg] 
\nonumber\\&
+
   \bm{\delta_{\overline{x}'}\mathcal{L}^{4}_{\overline{z}'} }\Bigg[
        C_F^2 C_A   \bigg\{ \frac{220}{9} \bigg\}
       - C_F^2 n_f   \bigg\{   \frac{40}{9} \bigg\}
       - C_F^3   \bigg\{   16 \bigg\}\Bigg] 
+
   \bm{\delta_{\overline{z}'}\mathcal{L}^{4}_{\overline{x}'} }\Bigg[
        C_F^2 C_A   \bigg\{ \frac{220}{9} \bigg\}
       - C_F^2 n_f   \bigg\{   \frac{40}{9} \bigg\}
       + C_F^3   \bigg\{ 48 \bigg\}\Bigg] 
\nonumber\\&
-
   \bm{\mathcal{D}^{4}_{\overline{x}'}\mathcal{L}^{0}_{\overline{z}'} }\Bigg[
        C_F^3   \bigg\{   40 \bigg\}\Bigg] 
-
   \bm{\mathcal{D}^{4}_{\overline{z}'}\mathcal{L}^{0}_{\overline{x}'}} \Bigg[
        C_F^3   \bigg\{   40 \bigg\}\Bigg] 
-
   \bm{\delta_{\overline{x}'}\mathcal{L}^{5}_{\overline{z}'} }\Bigg[
        C_F^3   \bigg\{   8 \bigg\}\Bigg] 
-
   \bm{\delta_{\overline{z}'}\mathcal{L}^{5}_{\overline{x}'} }\Bigg[
        C_F^3   \bigg\{   8 \bigg\}\Bigg] 
\, .
\end{align}

%
In analogy to the third order, many terms at the fourth order can be predicted from the knowledge of $\Psi_{d}^{q,(1)}$, $\Psi_{d}^{q,(2)}$ and other higher order results ($A_{3}^{q}, C_{3}^{q}, \tilde{C}_{3}^{q}, D_{3}^{q},\tilde{D}_{3}^{q}$) and the relation $\overline{\varphi}_{q,z',3}^{(3)}=-\overline{\varphi}_{q,x',3}^{(3)} = -16/9 \beta_0 C_F^2$.    
For example, we can predict the coefficients of $\{\delta_{\overline{x}'}\mathcal{L}^{i}_{\overline{z}'},\delta_{\overline{z}'}\mathcal{L}^{i}_{\overline{x}'}\}$ for $i=7,6,5$ and $\{\mathcal{D}_{\overline{x}'}^{i}\mathcal{L}^{j}_{\overline{z}'},\mathcal{D}_{\overline{z}'}^{i}\mathcal{L}^{j}_{\overline{x}'}\}$ for $i+j=6,5,4$ 
and in addition the following terms:$\{\mathcal{D}_{\overline{x}'}^{i}\mathcal{L}^{j}_{\overline{z}'},\mathcal{D}_{\overline{z}'}^{i}\mathcal{L}^{j}_{\overline{x}'}\}$ for $i+j=3$ with constraint $i \geq 2$.
However, with the present information available we cannot predict $\{\delta_{\overline{x}'}\mathcal{L}^{i}_{\overline{z}'},\delta_{\overline{z}'}\mathcal{L}^{i}_{\overline{x}'}\}$ for $i=4,3,2,1,0$  and $\{\mathcal{D}_{\overline{x}'}^{i} \mathcal{L}_{\overline{z}'}^{j},\mathcal{D}_{\overline{z}'}^{i}\mathcal{L}_{\overline{x}'}^{j}\}$ for $i=1,0$ and $j=2,1,0$.  
They require  explicit third  and  fourth order results for $\overline{\varphi}_{q,\xi,3}^{(i)}$ for $i=0,1,2$ and $\overline{\varphi}_{q,\xi,4}^{(i)}$ for $i=0,1,2,3,4$ where $\xi = x', z'$. 
The partial results at ${\mathcal O} (a_s^4)$ are presented below: 
\begin{align}
(\Delta)\mathcal{C}^{(4),\text{NSV}}_{J,cc} &= 
 \bm{  \mathcal{D}^{0}_{\overline{x}'}\mathcal{L}^{4}_{\overline{z}'} }\Bigg[
        C_F^2 n_f C_A   \bigg\{ \frac{3520}{27} \bigg\}
       - C_F^2 C_A^2   \bigg\{   \frac{9680}{27} \bigg\}
       - C_F^3 C_A   \bigg\{   \frac{22144}{27} - 240 \zeta_2 \bigg\}
       + C_F^4   \bigg\{ 600 + 960 \zeta_2 \bigg\}
\nonumber\\&
       - C_F^2 n_f^2   \bigg\{   \frac{320}{27} \bigg\}       
       + C_F^3 n_f   \bigg\{ \frac{2848}{27} \bigg\}\Bigg] 
+
  \bm{ \mathcal{D}^{0}_{\overline{z}'}\mathcal{L}^{4}_{\overline{x}'} }\Bigg[
        C_F^2 n_f C_A   \bigg\{ \frac{3520}{27} \bigg\}   
       - C_F^2 C_A^2   \bigg\{   \frac{9680}{27} \bigg\}
       - C_F^2 n_f^2   \bigg\{   \frac{320}{27} \bigg\}
\nonumber\\&
       - C_F^3 C_A   \bigg\{   \frac{55136}{27} - 240 \zeta_2 \bigg\}
       + C_F^3 n_f   \bigg\{ \frac{9632}{27} \bigg\}
       + C_F^4   \bigg\{ 680 + 960 \zeta_2 \bigg\}\Bigg] 
+
   \bm{\mathcal{D}^{0}_{\overline{x}'}\mathcal{L}^{5}_{\overline{z}'}} \Bigg[
        C_F^3 C_A   \bigg\{ \frac{616}{3} \bigg\}
\nonumber\\&
       - C_F^3 n_f   \bigg\{   \frac{112}{3} \bigg\}
       - C_F^4   \bigg\{   64 \bigg\}\Bigg] 
+
   \bm{\mathcal{D}^{0}_{\overline{z}'}\mathcal{L}^{5}_{\overline{x}'} }\Bigg[
        C_F^3 C_A   \bigg\{ \frac{616}{3} \bigg\}
       - C_F^3 n_f   \bigg\{   \frac{112}{3} \bigg\}
       + C_F^4   \bigg\{ 256 \bigg\}\Bigg] 
\nonumber\\&
-
   \bm{\mathcal{D}^{0}_{\overline{x}'}\mathcal{L}^{6}_{\overline{z}'} }\Bigg[
        C_F^4   \bigg\{   \frac{112}{3} \bigg\}\Bigg] 
-
   \bm{\mathcal{D}^{0}_{\overline{z}'}\mathcal{L}^{6}_{\overline{x}'} }\Bigg[
        C_F^4   \bigg\{   \frac{112}{3} \bigg\}\Bigg] 
+
   \bm{\mathcal{D}^{1}_{\overline{x}'}\mathcal{L}^{3}_{\overline{z}'} }\Bigg[
        C_F^2 n_f C_A   \bigg\{ \frac{14080}{27} \bigg\}
       - C_F^2 C_A^2   \bigg\{   \frac{38720}{27} \bigg\}
\nonumber\\&
       - C_F^3 C_A   \bigg\{   \frac{94336}{27} - 960 \zeta_2 \bigg\}
       - C_F^2 n_f^2   \bigg\{   \frac{1280}{27} \bigg\}
       + C_F^3 n_f   \bigg\{ \frac{12832}{27} \bigg\}     
       + C_F^4   \bigg\{ 2320 + 3840 \zeta_2 \bigg\}\Bigg] 
\nonumber\\&  
+
 \bm{  \mathcal{D}^{1}_{\overline{z}'}\mathcal{L}^{3}_{\overline{x}'} }\Bigg[
        C_F^2 n_f C_A   \bigg\{ \frac{14080}{27} \bigg\}       
       - C_F^2 C_A^2   \bigg\{   \frac{38720}{27} \bigg\}
       - C_F^2 n_f^2   \bigg\{   \frac{1280}{27} \bigg\}       
       - C_F^3 C_A   \bigg\{   \frac{214784}{27} - 960 \zeta_2 \bigg\}
\nonumber\\&       
       + C_F^3 n_f   \bigg\{ \frac{37088}{27} \bigg\}
       + C_F^4   \bigg\{ 2800 + 3840 \zeta_2 \bigg\}\Bigg] 
+
   \bm{\mathcal{D}^{1}_{\overline{x}'}\mathcal{L}^{4}_{\overline{z}'}} \Bigg[
        C_F^3 C_A   \bigg\{ \frac{3080}{3} \bigg\}
       - C_F^3 n_f   \bigg\{   \frac{560}{3} \bigg\}     
       - C_F^4   \bigg\{   224 \bigg\}\Bigg] 
\nonumber\\&  
+
   \bm{\mathcal{D}^{1}_{\overline{z}'}\mathcal{L}^{4}_{\overline{x}'} }\Bigg[
        C_F^3 C_A   \bigg\{ \frac{3080}{3} \bigg\}
       - C_F^3 n_f   \bigg\{   \frac{560}{3} \bigg\}
       + C_F^4   \bigg\{ 1184 \bigg\}\Bigg] 
-
   \bm{\mathcal{D}^{1}_{\overline{x}'}\mathcal{L}^{5}_{\overline{z}'} }\Bigg[
        C_F^4   \bigg\{   224 \bigg\}\Bigg] 
-
   \bm{\mathcal{D}^{1}_{\overline{z}'}\mathcal{L}^{5}_{\overline{x}'} }\Bigg[
        C_F^4   \bigg\{   224 \bigg\}\Bigg] 
\nonumber\\&
+
   \bm{\mathcal{D}^{2}_{\overline{x}'}\mathcal{L}^{0}_{\overline{z}'} }\Bigg[
        C_F n_f C_A^2   \bigg\{ \frac{7324}{9} - \frac{352}{3} \zeta_2 \bigg\}
       - C_F C_A^3   \bigg\{   \frac{43648}{27} - \frac{968}{3} \zeta_2 \bigg\}       
       - C_F n_f^2 C_A   \bigg\{   \frac{1144}{9} - \frac{32}{3} \zeta_2 \bigg\}
\nonumber\\&       
       - C_F^2 C_A^2   \bigg\{   \frac{758188}{81} - \frac{8800}{3} \zeta_3 - \frac{61360}{9} \zeta_2 + \frac{2592}{5} 
         \zeta_2^2 \bigg\}              
       + C_F n_f^3   \bigg\{ \frac{160}{27} \bigg\}
       - C_F^2 n_f^2   \bigg\{   \frac{12184}{81} - \frac{1216}{9} \zeta_2 \bigg\}
\nonumber\\&
       + C_F^2 n_f C_A   \bigg\{ \frac{210284}{81} - \frac{448}{3} \zeta_3 - \frac{17216}{9} \zeta_2 \bigg\}
       + C_F^3 C_A   \bigg\{ \frac{70810}{9} + \frac{22688}{3} \zeta_3 + \frac{33872}{3} \zeta_2 - \frac{13152}{5} \zeta_2^2
          \bigg\}       
\nonumber\\&       
       - C_F^3 n_f   \bigg\{   \frac{7132}{9} + \frac{5696}{3} \zeta_3 + \frac{5696}{3} \zeta_2 \bigg\}
       - C_F^4   \bigg\{   1950 - 1696 \zeta_3 + 6336 \zeta_2 + \frac{9408}{5} \zeta_2^2 \bigg\}\Bigg] 
\nonumber\\&
+
   \bm{\mathcal{D}^{2}_{\overline{z}'}\mathcal{L}^{0}_{\overline{x}'} }\Bigg[
        C_F n_f C_A^2   \bigg\{ \frac{13132}{9} - \frac{352}{3} \zeta_2 \bigg\}
       - C_F C_A^3   \bigg\{   \frac{75592}{27} - \frac{968}{3} \zeta_2 \bigg\}      
       - C_F n_f^2 C_A   \bigg\{   \frac{2200}{9} - \frac{32}{3} \zeta_2 \bigg\}
\nonumber\\&        
       + C_F n_f^3   \bigg\{ \frac{352}{27} \bigg\}
       - C_F^2 C_A^2   \bigg\{   \frac{974548}{81} - \frac{8800}{3} \zeta_3 - \frac{70336}{9} \zeta_2 + \frac{2592}{5} \zeta_2^2 \bigg\}
       - C_F^2 n_f^2   \bigg\{   \frac{23128}{81} - \frac{1216}{9} \zeta_2 \bigg\}
\nonumber\\&       
       + C_F^2 n_f C_A   \bigg\{ \frac{316412}{81} - \frac{448}{3} \zeta_3 - \frac{18848}{9} \zeta_2 \bigg\}    
       + C_F^3 C_A   \bigg\{ 12466 + \frac{24992}{3} \zeta_3 + \frac{55168}{3} \zeta_2 - \frac{13152}{5} \zeta_2^2 \bigg\}
\nonumber\\&          
       - C_F^3 n_f   \bigg\{   1484 + \frac{5696}{3} \zeta_3 + \frac{10144}{3} \zeta_2 \bigg\}    
       - C_F^4   \bigg\{   4182 - 8864 \zeta_3 + 8736 \zeta_2 + \frac{9408}{5} \zeta_2^2 \bigg\}\Bigg] 
\nonumber\\& 
+
   \bm{\mathcal{D}^{2}_{\overline{x}'}\mathcal{L}^{1}_{\overline{z}'}} \Bigg[
        C_F C_A^3   \bigg\{ \frac{5324}{9} \bigg\}
       - C_F n_f C_A^2   \bigg\{   \frac{968}{3} \bigg\}
       + C_F n_f^2 C_A   \bigg\{ \frac{176}{3} \bigg\}
       - C_F n_f^3   \bigg\{   \frac{32}{9} \bigg\}
       + C_F^2 n_f^2   \bigg\{ \frac{1696}{9} \bigg\}       
\nonumber\\&        
       + C_F^2 C_A^2   \bigg\{ \frac{74644}{9} - 1760 \zeta_2 \bigg\}
       - C_F^2 n_f C_A   \bigg\{   \frac{23816}{9} - 320 \zeta_2 \bigg\}
       + C_F^3 C_A   \bigg\{ \frac{6964}{9} - 3360 \zeta_3 - \frac{34016}{3} \zeta_2 \bigg\}
\nonumber\\&             
       - C_F^3 n_f   \bigg\{   \frac{184}{9} - \frac{6080}{3} \zeta_2 \bigg\}
       - C_F^4   \bigg\{   8960 \zeta_3 + 768 \zeta_2 \bigg\}\Bigg] 
+
   \bm{\mathcal{D}^{2}_{\overline{z}'}\mathcal{L}^{1}_{\overline{x}'} }\Bigg[
        C_F C_A^3   \bigg\{ \frac{5324}{9} \bigg\}
       - C_F n_f C_A^2   \bigg\{   \frac{968}{3} \bigg\}
\nonumber\\&          
       + C_F n_f^2 C_A   \bigg\{ \frac{176}{3} \bigg\}
       - C_F n_f^3   \bigg\{   \frac{32}{9} \bigg\}
       + C_F^2 C_A^2   \bigg\{ \frac{111868}{9} - 1760 \zeta_2 \bigg\}
       - C_F^2 n_f C_A   \bigg\{   \frac{39032}{9} - 320 \zeta_2 \bigg\}
\nonumber\\&          
       + C_F^3 C_A   \bigg\{ \frac{58252}{9} - 3360 \zeta_3 - \frac{40352}{3} \zeta_2 \bigg\}
       - C_F^4   \bigg\{   6144 + 8960 \zeta_3 + 9792 \zeta_2 \bigg\}
        - C_F^3 n_f   \bigg\{   \frac{9352}{9} - \frac{6080}{3} \zeta_2 \bigg\}
\nonumber\\&      
       + C_F^2 n_f^2   \bigg\{ \frac{3232}{9} \bigg\}  \Bigg] 
+
   \bm{\mathcal{D}^{2}_{\overline{x}'}\mathcal{L}^{2}_{\overline{z}'} }\Bigg[
        C_F^2 n_f C_A   \bigg\{ \frac{7040}{9} \bigg\}   
       - C_F^2 C_A^2   \bigg\{   \frac{19360}{9} \bigg\}
       - C_F^3 C_A   \bigg\{   5688 - 1440 \zeta_2 \bigg\}
\nonumber\\&       
       - C_F^2 n_f^2   \bigg\{   \frac{640}{9} \bigg\}
       + C_F^3 n_f   \bigg\{ 816 \bigg\}
       + C_F^4   \bigg\{ 3360 + 5760 \zeta_2 \bigg\}\Bigg] 
+
   \bm{\mathcal{D}^{2}_{\overline{z}'}\mathcal{L}^{2}_{\overline{x}'} }\Bigg[
        C_F^2 n_f C_A   \bigg\{ \frac{7040}{9} \bigg\}   
       - C_F^2 n_f^2   \bigg\{   \frac{640}{9} \bigg\}
\nonumber\\&              
       - C_F^2 C_A^2   \bigg\{   \frac{19360}{9} \bigg\}
       - C_F^3 C_A   \bigg\{  \frac{34456}{3} - 1440 \zeta_2 \bigg\}
       + C_F^3 n_f   \bigg\{ \frac{5872}{3} \bigg\}
       + C_F^4   \bigg\{ 4320 + 5760 \zeta_2 \bigg\}\Bigg] 
\nonumber\\&       
+
   \bm{\mathcal{D}^{2}_{\overline{x}'}\mathcal{L}^{3}_{\overline{z}'} }\Bigg[
        C_F^3 C_A   \bigg\{ \frac{6160}{3} \bigg\}
       - C_F^3 n_f   \bigg\{   \frac{1120}{3} \bigg\}
       - C_F^4   \bigg\{   224 \bigg\}\Bigg] 
+
   \bm{\mathcal{D}^{2}_{\overline{z}'}\mathcal{L}^{3}_{\overline{x}'} }\Bigg[
        C_F^3 C_A   \bigg\{ \frac{6160}{3} \bigg\}
       - C_F^3 n_f   \bigg\{   \frac{1120}{3} \bigg\}
\nonumber\\&       
       + C_F^4   \bigg\{ 2144 \bigg\}\Bigg] 
-
   \bm{\mathcal{D}^{2}_{\overline{x}'}\mathcal{L}^{4}_{\overline{z}'} }\Bigg[
        C_F^4   \bigg\{   560 \bigg\}\Bigg] 
-
   \bm{\mathcal{D}^{2}_{\overline{z}'}\mathcal{L}^{4}_{\overline{x}'} }\Bigg[
        C_F^4   \bigg\{   560 \bigg\}\Bigg] 
+
   \bm{\mathcal{D}^{3}_{\overline{x}'}\mathcal{L}^{0}_{\overline{z}'} }\Bigg[
         C_F^2 C_A^2   \bigg\{ \frac{75832}{27} - \frac{1760}{3} \zeta_2 \bigg\}
\nonumber\\&           
       +C_F C_A^3   \bigg\{ \frac{5324}{27} \bigg\}
       - C_F n_f C_A^2   \bigg\{   \frac{968}{9} \bigg\}
       + C_F n_f^2 C_A   \bigg\{ \frac{176}{9} \bigg\}
       - C_F n_f^3   \bigg\{   \frac{32}{27} \bigg\}
       - C_F^2 n_f C_A   \bigg\{   \frac{25088}{27} - \frac{320}{3} \zeta_2 \bigg\}
\nonumber\\&            
       + C_F^3 C_A   \bigg\{ \frac{30880}{27} - 1120 \zeta_3 - \frac{35744}{9} \zeta_2 \bigg\}
       - C_F^4   \bigg\{   512 + \frac{8960}{3} \zeta_3 + 704 \zeta_2 \bigg\}
       - C_F^3 n_f   \bigg\{   \frac{3904}{27} - \frac{6080}{9} \zeta_2 \bigg\}       
\nonumber\\&            
       + C_F^2 n_f^2   \bigg\{ \frac{1888}{27} \bigg\}\Bigg] 
+
   \bm{\mathcal{D}^{3}_{\overline{z}'}\mathcal{L}^{0}_{\overline{x}'} }\Bigg[
        C_F C_A^3   \bigg\{ \frac{5324}{27} \bigg\}
       - C_F n_f C_A^2   \bigg\{   \frac{968}{9} \bigg\}
       + C_F n_f^2 C_A   \bigg\{ \frac{176}{9} \bigg\}
       - C_F n_f^3   \bigg\{   \frac{32}{27} \bigg\}
\nonumber\\&            
       + C_F^2 C_A^2   \bigg\{ \frac{110680}{27} - \frac{1760}{3} \zeta_2 \bigg\}
       - C_F^2 n_f C_A   \bigg\{   \frac{37760}{27} - \frac{320}{3} \zeta_2 \bigg\}
       + C_F^3 C_A   \bigg\{ \frac{34336}{27} - 1120 \zeta_3 - \frac{38624}{9} \zeta_2 \bigg\}       
\nonumber\\&          
       + C_F^2 n_f^2   \bigg\{ \frac{3040}{27} \bigg\}
       - C_F^3 n_f   \bigg\{   \frac{5632}{27} - \frac{6080}{9} \zeta_2 \bigg\}
       - C_F^4   \bigg\{   1536 + \frac{8960}{3} \zeta_3 + 2816 \zeta_2 \bigg\}\Bigg] 
\nonumber\\&     
+
   \bm{\mathcal{D}^{3}_{\overline{x}'}\mathcal{L}^{1}_{\overline{z}'} }\Bigg[
        C_F^2 n_f C_A   \bigg\{ \frac{14080}{27} \bigg\}
       - C_F^2 C_A^2   \bigg\{   \frac{38720}{27} \bigg\}
       - C_F^3 C_A   \bigg\{   \frac{37808}{9} - 960 \zeta_2 \bigg\}
       + C_F^4   \bigg\{ 2160 + 3840 \zeta_2 \bigg\}
\nonumber\\&       
       - C_F^2 n_f^2   \bigg\{   \frac{1280}{27} \bigg\}
       + C_F^3 n_f   \bigg\{ \frac{5696}{9} \bigg\}
       \Bigg] 
+
   \bm{\mathcal{D}^{3}_{\overline{z}'}\mathcal{L}^{1}_{\overline{x}'} }\Bigg[
        C_F^2 n_f C_A   \bigg\{ \frac{14080}{27} \bigg\}
       - C_F^2 C_A^2   \bigg\{   \frac{38720}{27} \bigg\}
       - C_F^2 n_f^2   \bigg\{   \frac{1280}{27} \bigg\}
\nonumber\\&       
       - C_F^3 C_A   \bigg\{   7248 - 960 \zeta_2 \bigg\}
       + C_F^3 n_f   \bigg\{ 1216 \bigg\}
       + C_F^4   \bigg\{ 2960 + 3840 \zeta_2 \bigg\}\Bigg] 
+
   \bm{\mathcal{D}^{3}_{\overline{x}'}\mathcal{L}^{2}_{\overline{z}'} }\Bigg[
        C_F^3 C_A   \bigg\{ \frac{6160}{3} \bigg\}
\nonumber\\&         
       - C_F^3 n_f   \bigg\{   \frac{1120}{3} \bigg\}
       + C_F^4   \bigg\{ 32 \bigg\}\Bigg] 
+
   \bm{\mathcal{D}^{3}_{\overline{z}'}\mathcal{L}^{2}_{\overline{x}'} }\Bigg[
        C_F^3 C_A   \bigg\{ \frac{6160}{3} \bigg\}
       - C_F^3 n_f   \bigg\{   \frac{1120}{3} \bigg\}
       + C_F^4   \bigg\{ 1888 \bigg\}\Bigg] 
\nonumber\\&  
-
   \bm{\mathcal{D}^{3}_{\overline{x}'}\mathcal{L}^{3}_{\overline{z}'} }\Bigg[
        C_F^4   \bigg\{   \frac{2240}{3} \bigg\}\Bigg] 
-
   \bm{\mathcal{D}^{3}_{\overline{z}'}\mathcal{L}^{3}_{\overline{x}'} }\Bigg[
        C_F^4   \bigg\{   \frac{2240}{3} \bigg\}\Bigg] 
+
   \bm{\mathcal{D}^{4}_{\overline{x}'}\mathcal{L}^{0}_{\overline{z}'} }\Bigg[
        C_F^2 n_f C_A   \bigg\{ \frac{3520}{27} \bigg\}   
       - C_F^3 C_A   \bigg\{   \frac{3560}{3} - 240 \zeta_2 \bigg\}
\nonumber\\&        
       - C_F^2 C_A^2   \bigg\{   \frac{9680}{27} \bigg\}       
       - C_F^2 n_f^2   \bigg\{   \frac{320}{27} \bigg\}
       + C_F^3 n_f   \bigg\{ \frac{560}{3} \bigg\}
       + C_F^4   \bigg\{ 520 + 960 \zeta_2 \bigg\}\Bigg] 
+
   \bm{\mathcal{D}^{4}_{\overline{z}'}\mathcal{L}^{0}_{\overline{x}'} }\Bigg[
        C_F^2 n_f C_A   \bigg\{ \frac{3520}{27} \bigg\}   
\nonumber\\&  
       - C_F^2 C_A^2   \bigg\{   \frac{9680}{27} \bigg\}
       - C_F^2 n_f^2   \bigg\{   \frac{320}{27} \bigg\}
       - C_F^3 C_A   \bigg\{   \frac{15080}{9} - 240 \zeta_2 \bigg\}
       + C_F^3 n_f   \bigg\{ \frac{2480}{9} \bigg\}
       + C_F^4   \bigg\{ 760 + 960 \zeta_2 \bigg\}\Bigg] 
\nonumber\\&  
+
   \bm{\mathcal{D}^{4}_{\overline{x}'}\mathcal{L}^{1}_{\overline{z}'} }\Bigg[
        C_F^3 C_A   \bigg\{ \frac{3080}{3} \bigg\}
       - C_F^3 n_f   \bigg\{  \frac{560}{3} \bigg\}
       + C_F^4   \bigg\{ 160 \bigg\}\Bigg] 
-
   \bm{\mathcal{D}^{4}_{\overline{x}'}\mathcal{L}^{2}_{\overline{z}'} }\Bigg[
        C_F^4   \bigg\{   560 \bigg\}\Bigg] 
-
   \bm{\mathcal{D}^{4}_{\overline{z}'}\mathcal{L}^{2}_{\overline{x}'} }\Bigg[
        C_F^4   \bigg\{   560 \bigg\}\Bigg] 
\nonumber\\&  
+
   \bm{\mathcal{D}^{4}_{\overline{z}'}\mathcal{L}^{1}_{\overline{x}'} }\Bigg[
        C_F^3 C_A   \bigg\{ \frac{3080}{3} \bigg\}
       - C_F^3 n_f   \bigg\{   \frac{560}{3} \bigg\}
       + C_F^4   \bigg\{ 800 \bigg\}\Bigg] 
+
   \bm{\delta_{\overline{x}'}\mathcal{L}^{5}_{\overline{z}'} }\Bigg[
        C_F^2 n_f C_A   \bigg\{ \frac{704}{27} \bigg\}
       - C_F^2 C_A^2   \bigg\{   \frac{1936}{27} \bigg\}
\nonumber\\&         
       - C_F^2 n_f^2   \bigg\{   \frac{64}{27} \bigg\}
       - C_F^3 C_A   \bigg\{   \frac{4220}{27} - 48 \zeta_2 \bigg\}
       + C_F^3 n_f   \bigg\{ \frac{512}{27} \bigg\}
       + C_F^4   \bigg\{ 124 + 192 \zeta_2 \bigg\}\Bigg] 
\nonumber\\&
+
  \bm{ \delta_{\overline{z}'}\mathcal{L}^{5}_{\overline{x}'} }\Bigg[
        C_F^2 n_f C_A   \bigg\{ \frac{704}{27} \bigg\}       
       - C_F^2 C_A^2   \bigg\{   \frac{1936}{27} \bigg\}
       - C_F^2 n_f^2   \bigg\{   \frac{64}{27} \bigg\}
       - C_F^3 C_A   \bigg\{   \frac{11236}{27} - 48 \zeta_2 \bigg\}
       + C_F^3 n_f   \bigg\{ \frac{1984}{27} \bigg\}
\nonumber\\&       
       + C_F^4   \bigg\{ 132 + 192 \zeta_2 \bigg\}\Bigg] 
+
   \bm{\mathcal{D}^{5}_{\overline{x}'}\mathcal{L}^{0}_{\overline{z}'} }\Bigg[
        C_F^3 C_A   \bigg\{ \frac{616}{3} \bigg\}
       - C_F^3 n_f   \bigg\{   \frac{112}{3} \bigg\}
       + C_F^4   \bigg\{ 64 \bigg\}\Bigg] 
-
   \bm{\mathcal{D}^{5}_{\overline{x}'}\mathcal{L}^{1}_{\overline{z}'} }\Bigg[
        C_F^4   \bigg\{   224 \bigg\}\Bigg] 
\nonumber\\&
-
   \bm{\mathcal{D}^{5}_{\overline{z}'}\mathcal{L}^{1}_{\overline{x}'} }\Bigg[
        C_F^4   \bigg\{   224 \bigg\}\Bigg] 
+
   \bm{\mathcal{D}^{5}_{\overline{z}'}\mathcal{L}^{0}_{\overline{x}'} }\Bigg[
        C_F^3 C_A   \bigg\{ \frac{616}{3} \bigg\}
       - C_F^3 n_f   \bigg\{   \frac{112}{3} \bigg\}
       + C_F^4   \bigg\{ 128 \bigg\}\Bigg]   
-
   \bm{\mathcal{D}^{6}_{\overline{x}'}\mathcal{L}^{0}_{\overline{z}'} }\Bigg[
        C_F^4   \bigg\{   \frac{112}{3} \bigg\}\Bigg] 
\nonumber\\&
-
   \bm{\mathcal{D}^{6}_{\overline{z}'}\mathcal{L}^{0}_{\overline{x}'} }\Bigg[
        C_F^4   \bigg\{   \frac{112}{3} \bigg\}\Bigg]   
-
   \bm{\delta_{\overline{x}'}\mathcal{L}^{7}_{\overline{z}'} }\Bigg[
        C_F^4   \bigg\{   \frac{16}{3} \bigg\}\Bigg] 
+
   \bm{\delta_{\overline{x}'}\mathcal{L}^{6}_{\overline{z}'} }\Bigg[
        C_F^3 C_A   \bigg\{ \frac{308}{9} \bigg\}
       - C_F^3 n_f   \bigg\{   \frac{56}{9} \bigg\}
       - C_F^4   \bigg\{   \frac{40}{3} \bigg\}\Bigg] 
\nonumber\\&
-
   \bm{\delta_{\overline{z}'}\mathcal{L}^{7}_{\overline{x}'} }\Bigg[
        C_F^4   \bigg\{   \frac{16}{3} \bigg\}\Bigg]
+
   \bm{\delta_{\overline{z}'}\mathcal{L}^{6}_{\overline{x}'} }\Bigg[
        C_F^3 C_A   \bigg\{ \frac{308}{9} \bigg\}
       - C_F^3 n_f   \bigg\{   \frac{56}{9} \bigg\}
       + C_F^4   \bigg\{ \frac{136}{3} \bigg\}\Bigg] 
\, .
\end{align}


\begin{table}[ht!]
\begin{center}
\begin{small}
\begin{tabular}{|p{1.94cm}|p{1.94cm}|p{1.94cm}|p{2.27cm}||p{1.95cm}|p{2.13cm}|p{4.55cm}|}
 \hline
 \multicolumn{4}{|c||}{GIVEN } & \multicolumn{3}{c|}{PREDICTIONS FOR SV}\\
 \hline
 \hline
 \rowcolor{lightgray}
 $\Psi_{d}^{q,(1)}$ & $\Psi_d^{q,(2)}$ &$\Psi_d^{q,(3)}$&$\Psi_d^{q,(n)}$&  
\quad
$ (\Delta){\cal C}_{J,cc}^{(2)} $&
\quad
$ (\Delta){\cal C}_{J,cc}^{(3)} $&
\quad \quad \quad $ (\Delta){\cal C}_{J,cc}^{(i)} $\\
 \hline
$ \delta_{\overline{x}'} \delta_{\overline{z}'},{\cal D}^j_{\overline{x}'}{\cal D}^k_{\overline{z}'}$ 
${\cal D}^{j'}_{\overline{x}'}\delta_{\overline{z}'},{\cal D}^{ j'}_{\overline{z}'} \delta_{\overline{x}'}$
{\scriptsize $j+k=0$
  $j' = 1,0$}
 &
  & &   &  
  ${\cal D}^j_{\overline{x}'}{\cal D}^k_{\overline{z}'}$,
  ${\cal D}^{j'}_{\overline{x}'} \delta_{\overline{z}'},{\cal D}^{j'}_{\overline{z}'} \delta_{\overline{x}'}$
  {\scriptsize $j+k =2,1$
  $j' =3,2 $}
  &
  ${\cal D}^j_{\overline{x}'}{\cal D}^k_{\overline{z}'}$,
  ${\cal D}^{j'}_{\overline{x}'} \delta_{\overline{z}'},{\cal D}^{j'}_{\overline{z}'} \delta_{\overline{x}'}$
  {\scriptsize $j+k =4,3$
  $j' =5,4 $}
  &
  ${\cal D}^j_{\overline{x}'}{\cal D}^k_{\overline{z}'}, {\cal D}^{j'}_{\overline{x}'} \delta_{\overline{z}'}, {\cal D}^{j'}_{\overline{z}'} \delta_{\overline{x}'}$        
  
  {\scriptsize $j+k =2i-2,2i-3$
  
  $j' =2i-1,2i-2 $}\\  
 \hline
  & 
$ \delta_{\overline{x}'} \delta_{\overline{z}'},{\cal D}^j_{\overline{x}'}{\cal D}^k_{\overline{z}'}$ 
${\cal D}^{j'}_{\overline{x}'}\delta_{\overline{z}'},{\cal D}^{ j'}_{\overline{z}'} \delta_{\overline{x}'}$
{\scriptsize $j+k=2,1,0$
  $j' = 3,...,0$}
    &   
  &&&
  ${\cal D}^j_{\overline{x}'}{\cal D}^k_{\overline{z}'}$,
  ${\cal D}^{j'}_{\overline{x}'} \delta_{\overline{z}'},{\cal D}^{j'}_{\overline{z}'} \delta_{\overline{x}'}$
  {\scriptsize $j+k =4,3,2,1$
  $j' =5,4,3,2 $}
  & 
  ${\cal D}^j_{\overline{x}'}{\cal D}^k_{\overline{z}'}, {\cal D}^{j'}_{\overline{x}'} \delta_{\overline{z}'}, {\cal D}^{j'}_{\overline{z}'} \delta_{\overline{x}'}$   
  
{\scriptsize $j+k =2i -2,...,2i-4,2i-5$

$j' = 2i-1,...,2i-3,2i-4$}\\
  \hline
& &
$ \delta_{\overline{x}'} \delta_{\overline{z}'},{\cal D}^j_{\overline{x}'}{\cal D}^k_{\overline{z}'}$ 
${\cal D}^{j'}_{\overline{x}'}\delta_{\overline{z}'},{\cal D}^{ j'}_{\overline{z}'} \delta_{\overline{x}'}$
    {\scriptsize $j+k=4,...,0$
    $j' = 5,...,0$}
    & & & & 
  ${\cal D}^j_{\overline{x}'}{\cal D}^k_{\overline{z}'}, {\cal D}^{j'}_{\overline{x}'} \delta_{\overline{z}'}, {\cal D}^{j'}_{\overline{z}'} \delta_{\overline{x}'}$   
  
{\scriptsize $j+k = 2i-2,...,2i-5,...,2i-7$

$j'=2i-1,...,2i-4,...,2i-6$}\\
  \hline
 & & &  
$ \delta_{\overline{x}'} \delta_{\overline{z}'},{\cal D}^j_{\overline{x}'}{\cal D}^k_{\overline{z}'}$ 
${\cal D}^{j'}_{\overline{x}'}\delta_{\overline{z}'},{\cal D}^{ j'}_{\overline{z}'} \delta_{\overline{x}'}$
{\scriptsize $j+k=2n-2,...,0$
    $j' = 2n-1,...,0$}
 &&&
${\cal D}^j_{\overline{x}'}{\cal D}^k_{\overline{z}'}, {\cal D}^{j'}_{\overline{x}'} \delta_{\overline{z}'}, {\cal D}^{j'}_{\overline{z}'} \delta_{\overline{x}'}$

{\scriptsize $j+k$=$2i-2$,...,$2i-n-2$,...,$2i-2n-1$

$j'=2i-1$,...,$2i-n-1$,...,$2i-2n$}
\\
  \hline

\end{tabular}
\end{small}
\end{center}
\caption{Towers of SV Distributions ($\delta_{\overline{x}'},\delta_{\overline{z}'},\mathcal{D}_{\overline{x}'},\mathcal{D}_{\overline{z}'}$) that can be predicted for $(\Delta){\cal C}_{J,cc}^{(i)}$ for $c =q,\overline{q} $ given $\Psi_d^{q,(n)}$,
$j,k~\geq 0$ and $\delta_{\overline{\xi}} =\delta(1-\xi)$, $\mathcal{D}_{\overline{\xi}}^{j} = \Big[\frac{\ln^{j}(1-\xi)}{1-\xi}\Big]_{+}$ with $\xi = x',z'$.}
\label{TableSVz}
\end{table}

\begin{table}[htb!]
\begin{center}
\begin{small}
\begin{tabular}{|p{1.89cm}|p{1.89cm}|p{1.89cm}|p{2.23cm}||p{2.1cm}|p{2.1cm}|p{4.6cm}|}
 \hline
 \multicolumn{4}{|c||}{GIVEN, SV +} & \multicolumn{3}{c|}{PREDICTIONS FOR NSV}\\
 \hline
 \hline
 \rowcolor{lightgray}
 $\Psi_d^{q,(1)}$ & $\Psi_d^{q,(2)}$ &$\Psi_d^{q,(3)}$&$\Psi_d^{q,(n)}$&  
  \quad
$ (\Delta){\cal C}_{J,cc}^{(2)} $&
 \quad
$ (\Delta){\cal C}_{J,cc}^{(3)} $&
\quad \quad \quad $ (\Delta){\cal C}_{J,cc}^{(i)} $\\
 \hline
${\cal L}^{j'}_{\overline{x}'}\delta_{\overline{z}'},{\cal L}^{j'}_{\overline{z}'}\delta_{\overline{x}'}$
${\cal D}^j_{\overline{x}'}{\cal L}^k_{\overline{z}'}$,
${\cal D}^j_{\overline{z}'}{\cal L}^k_{\overline{x}'}$

 {\scriptsize $j+k=0$

 $j' = 1,0$} 
&
  & &   &  
${\cal D}^j_{\overline{x}'}{\cal L}^k_{\overline{z}'},{\cal D}^j_{\overline{z}'}{\cal L}^k_{\overline{x}'}$
${\cal L}^{j'}_{\overline{x}'}\delta_{\overline{z}'},{\cal L}^{j'}_{\overline{z}'}\delta_{\overline{x}'}$
 {\scriptsize $j+k=2$

 $j'=3$} 
  &
  ${\cal D}^j_{\overline{x}'}{\cal L}^k_{\overline{z}'},{\cal D}^j_{\overline{z}'}{\cal L}^k_{\overline{x}'}$
${\cal L}^{j'}_{\overline{x}'}\delta_{\overline{z}'},{\cal L}^{j'}_{\overline{z}'}\delta_{\overline{x}'}$
 {\scriptsize $j+k=4$

 $j'=5$} 
  &
${\cal D}^j_{\overline{x}'}{\cal L}^k_{\overline{z}'},{\cal D}^j_{\overline{z}'}{\cal L}^k_{\overline{x}'}, {\cal L}^{j'}_{\overline{x}'}\delta_{\overline{z}'},{\cal L}^{j'}_{\overline{z}'}\delta_{\overline{x}'}$

 {\scriptsize $j+k=2i-2$

 $j'=2i-1$} \\  
 \hline
  & 
${\cal L}^{j'}_{\overline{x}'}\delta_{\overline{z}'},{\cal L}^{j'}_{\overline{z}'}\delta_{\overline{x}'}$
${\cal D}^j_{\overline{x}'}{\cal L}^k_{\overline{z}'}$,
${\cal D}^j_{\overline{z}'}{\cal L}^k_{\overline{x}'}$
 
 {\scriptsize $j+k=2,1,0$

 $j'$ = $3,...,0$} 
    &   
  &&&
${\cal D}^j_{\overline{x}'}{\cal L}^k_{\overline{z}'},{\cal D}^j_{\overline{z}'}{\cal L}^k_{\overline{x}'}$
${\cal L}^{j'}_{\overline{x}'}\delta_{\overline{z}'},{\cal L}^{j'}_{\overline{z}'}\delta_{\overline{x}'}$
 {\scriptsize $j+k=4,3,2$

 $j'=5,4$} 
  & 
${\cal D}^j_{\overline{x}'}{\cal L}^k_{\overline{z}'},{\cal D}^j_{\overline{z}'}{\cal L}^k_{\overline{x}'}, {\cal L}^{j'}_{\overline{x}'}\delta_{\overline{z}'},{\cal L}^{j'}_{\overline{z}'}\delta_{\overline{x}'}$

 {\scriptsize $j+k=2i-2,2i-3,2i-4$

 $j'=2i-1,2i-2$} \\
  \hline
& &
${\cal L}^{j'}_{\overline{x}'}\delta_{\overline{z}'},{\cal L}^{j'}_{\overline{z}'}\delta_{\overline{x}'}$
${\cal D}^j_{\overline{x}'}{\cal L}^k_{\overline{z}'}$,
${\cal D}^j_{\overline{z}'}{\cal L}^k_{\overline{x}'}$

 {\scriptsize $j+k=4,...,0$

 $j'$ = $5,...,0$} 
    & & & & 
${\cal D}^j_{\overline{x}'}{\cal L}^k_{\overline{z}'},{\cal D}^j_{\overline{z}'}{\cal L}^k_{\overline{x}'}, {\cal L}^{j'}_{\overline{x}'}\delta_{\overline{z}'},{\cal L}^{j'}_{\overline{z}'}\delta_{\overline{x}'}$

 {\scriptsize $j+k=2i-2,...,2i-5,2i-6$

 $j'=2i-1,...,2i-3$}  \\
  \hline
 & & &  
${\cal L}^{j'}_{\overline{x}'}\delta_{\overline{z}'},{\cal L}^{j'}_{\overline{z}'}\delta_{\overline{x}'}$
${\cal D}^j_{\overline{x}'}{\cal L}^k_{\overline{z}'},{\cal D}^j_{\overline{z}'}{\cal L}^k_{\overline{x}'}$

 {\scriptsize $j+k=2n-2,...,0$

 $j'$ = $2n-1,...,0$} 
 &&&
 ${\cal D}^j_{\overline{x}'}{\cal L}^k_{\overline{z}'},{\cal D}^j_{\overline{z}'}{\cal L}^k_{\overline{x}'}, {\cal L}^{j'}_{\overline{x}'}\delta_{\overline{z}'},{\cal L}^{j'}_{\overline{z}'}\delta_{\overline{x}'}$

 {\scriptsize $j+k$=$2i-2$,...,${2i-n-2}$,...,${2i-2n}$

 $j'=2i-1,...,2i-n$}\\
  \hline
\end{tabular}
\end{small}
\end{center}
\caption{Towers of NSV logarithms that can be predicted for $(\Delta){\cal C}_{J,cc}^{(i)}$ for $c=q\overline{q}$ given $\Psi_d^{q,(n)}$, $j,k~\geq 0$ and $\delta_{\overline{\xi}} =\delta(1-\xi)$, $\mathcal{D}_{\overline{\xi}}^{j} = \Big[\frac{\ln^{j}(1-\xi)}{1-\xi}\Big]_{+}$ and $\mathcal{L}_{\overline{\xi}}^{j} = \ln^{j}(1-\xi)$ with $\xi = x',z'$.
}
\label{TableNSVz}
\end{table}
\end{widetext}


\clearpage

\section{Resummation}
\label{sec:resum}

The formalism for the SV and NSV terms applied in the previous section is useful for obtaining results in $(x',z')$ space at a given order $a_s^n$, provided
that the exponent $\Psi^q_d$ is known to order $a_s^m$, where $m=1, \cdots, n-1$. 
It is impractical, though, for studying the numerical impact at all orders in $a_s$, because it involves iterated convolutions among distributions as well as among distributions and powers of $\ln(1-\xi)$ terms, which become difficult to evaluate at higher orders in $a_s$. 
All-order resummation of our result in eq.~\eqref{eq:MasterF} is therefore tcarried out in Mellin space spanned by two complex numbers $N_i,i=1,2$, taking double Mellin moments with respect to $N_i,i=1,2$ of CF given in eq.~\eqref{eq:MasterF}. 
In Mellin space, the convolution ${\cal C}$ of the exponential in eq.~\eqref{eq:MasterF} becomes an exponential of the double Mellin moment of the exponent $\Psi^q_d$, i.e., 
\begin{widetext}
\begin{eqnarray}
\label{eq:resum}
(\Delta) {\cal C}_{J,qq}^{\vec N}&=&{\cal N} \int_0^1 dx' {x'}^{N_1-1} \int_0^1 dz' {z'}^{N_2-1} (\Delta) {\cal C}^{\rm{SV+NSV}}_{qq}(x',z') \nonumber\\
&=& 
\exp\left({\cal N}\int_0^1 dx' {x'}^{N_1-1}     \int_0^1 dz' {z'}^{N_2-1} \Psi^q_d(x',z')\right)
\nonumber\\
&=& \exp \Big(G_{d,q}^{\vec N}\Big)\, ,
\end{eqnarray}
\end{widetext}
where we have suppressed the dependence on the scales $Q^2$ and $\mu_F^2$.
In Mellin $\vec N$-space, $\vec N=(N_1,N_2)$, 
the threshold limits of the $(x',z')$ space , namely $\{x',z'\} \rightarrow \{1,1\}$,
correspond to taking both $\{N_1,N_2\}\rightarrow \{\infty,\infty \}$ and 
the limits on $N_1$ and $N_2$ are applied simultaneously.
The symbol ${\cal N}$ in eq.~\eqref{eq:resum} indicates that we retain only the $(N_1,N_2)$ space versions of SV and NSV terms,  which are denoted as $\overline {\rm{SV}}$ and $\overline {\rm{NSV}}$, respectively. 
Specifically, $\overline {\rm {SV}}$ terms are of the form $\ln^i(N_1) \ln^j(N_2)$ 
with $i,j=1,\cdots ,\infty$.
$\overline {\rm {NSV}}$ terms, on the other hand, include corrections of the type $1/N_i$.
They are of the form 
$\ln^i(N_1)/N_1 \ln^j(N_2)$ where $i=0,\cdots,\infty, j=1,\cdots,\infty$ or of the form 
$\ln^i(N_1) \ln^j(N_2)/N_2$ where $i=1,\cdots,\infty,j=0,\cdots,\infty$.
All other terms are classified as $\overline {\rm {BNSV}}$.  
We use the overline notations to distiguish the $\vec N$-space approximations for those directly obtained in $(x',z')$ space, because the inverse Mellin transform of
$\overline {\rm {SV}}$ terms in $(x',z')$ space yields SV terms as well as NSV and BNSV terms. 
Similarly, the inverse Mellin transform of $\overline {\rm {NSV}}$ terms will produce
NSV and BNSV terms in $(x',z')$ space.
The computation $G_{d,q}^{\vec N}$ is described in the App.~\ref{DetailMM}. 

The Mellin moments of eq.~\eqref{eq:SVintrep} contain terms of the type 
$\omega = a_s(\mu_R^2) \beta_0 \ln(N_1 N_2)$ as well as single logarithms 
$\omega_i = a_s(\mu_R^2) \beta_0 \ln(N_i)$, 
where large logarithms $\ln(N_i)$ and the strong coupling $a_s(\mu_R^2)$ at moderate values 
combine into an effective expansion parameter of ${\cal O}(1)$, spoiling the reliability of perturbative predictions, if truncated at specific order in $a_s$. 
A systematic reorganization of the $\vec N$-space in the exponent sums all the terms containing $\omega$, $\omega_i$ etc using the resummed expression for strong coupling $a_s$ as described in App.~\ref{asresum}.

By rewriting $G_{d,q}^{\vec N} = \overline G_{d,q}^{\vec N} + G_{d,q}^{\vec N = \vec 1}$ and defining $\tilde g_{d,0}^q = \exp(G_{d,q}^{\vec N = \vec 1})$, we obtain
\begin{align}
\label{DeltaN}
(\Delta) {\cal C}_{J,qq}^{\vec N}   
&=  \tilde g^q_{d,0} (Q^2,\mu_F^2) \exp\left(
\overline G_{d,q}^{\vec N} (Q^2,\mu_F^2) 
\right)\, .
\end{align} 
We expand these $\vec{N}$-independent constants as $\tilde g^q_{d,0}$ = $\sum\limits_{i=0}^\infty a_s^i(\mu_R^2) ~\tilde g^q_{d,0,i}(Q^2,\mu_F^2,\mu_R^2)$, so that the resummed result takes the following form,
\begin{widetext}
\begin{align}\label{GNexp}
\overline G_{d,q}^{\vec N} =& 
g^q_{d,1}(\omega) \ln{N_1} 
+ \sum\limits_{i=0}^\infty a_s^i \Bigg (  \frac{1}{2} g^q_{d,i+2}(\omega)  +  \frac{1}{N_1} \overline{g}^q_{d,i}(\omega)   \Bigg)
+ \frac{1}{N_1} \Bigg(h^q_{d,0}(\omega,N_1)  + \sum\limits_{i=1}^\infty a_s^i h^q_{d,i}(\omega,\omega_{1},N_1) \Bigg)
\nonumber \\ &
+ (N_1 \leftrightarrow N_2 , \omega_1 \leftrightarrow \omega_2 )\, ,
\end{align}
where
\begin{align}
\label{hg}
h^q_{d,0}(\omega,N_l) &= h^q_{d,0,0}(\omega) + h^q_{d,0,1}(\omega) \ln N_l\,,
\nonumber \\ 
    h^q_{d,i}(\omega,\omega_l,N_l)& = \sum_{k=0}^{i-1} h^q_{d,i,k}(\omega,\omega_l)~ \ln^k N_l 
    + \tilde{h}^q_{d,i,i}(\omega,\omega_l)~ \ln^i N_l\,.
\end{align}
\end{widetext}
The functions $g_{d,i}^q$ and $\overline g_{d,i}^q$ are derived from $\Phi^{\text{SV}}_q$, as given in eq.~\eqref{eq:SVintrep}, and $h^q_{d,0}$ and $h^q_{d,i}$ originate from $\Phi^{\text{NSV}}_q$,
as given in eq.~\eqref{eq:NSVintrep}. These functions are listed in App.~\ref{SVNSVResumExp}.




\section{Physical Evolution Kernel}
\label{sec:pek}

The ingredients of SIDIS structure functions, CFs as well as the PDFs and FFs are defined in a factorization scheme,  
but structure functions themselves, being physical observables are scheme independent.
This inherent scheme independence allows to derive a scheme-invariant combination of CFs and splitting functions by constructing a system of differential equations 
among the structure functions~\cite{Furmanski:1981cw,Catani:1996sc,Blumlein:2000wh,vanNeerven:2001pe}. 
Such physical evolution equations have been employed previously in refs.~\cite{Grunberg:2009yi,Grunberg:2009am,Grunberg:2009vs} to obtain certain NSV terms for the CFs of inclusive DIS. 
With an ansatz based on the known results for the CFs and the AP kernels from fixed order computations, 
predictions for the second and third orders CFs of structure functions in inclusive DIS were made~\cite{Grunberg:2009vs}, finding agreement for some of the color factors.
A systematic construction of physical evolution kernels (PEK), relying on the observation about the logarithmic structure of, namely the single-logarithmic enhancement in $(1-z)$ at large $z$ to all orders, 
has been performed for the CFs of inclusive DIS, semi-inclusive $e^+ e^-$ annihilation and the DY process 
in \cite{Moch:2009hr} (also see refs.~\cite{deFlorian:2014vta, Das:2020adl}).
By conjecturing that this behavior holds true at every order in $a_s$, constraints on the structure of the corresponding leading $\ln(1-z)$ terms in the kernel could be derived and predictions for power suppressed 
contributions, including NSV logarithms at higher orders in $a_s$, were obtained, 
in agreement with known third-order results for DY and Higgs production \cite{Anastasiou:2015vya, Duhr:2020seh}.

Following ref.~\cite{Moch:2009hr}, we have established physical evolution equations for the structure functions $F_1$, $F_2$, and $g_1$ to analyze the PEK in the SV and NSV limits. 
In these limits, we have found that the evolution equations simplify significantly, with only diagonal splitting functions and diagonal CFs contributing to the PEK.
Setting $\mu_F^2=Q^2$ for convenience, we find
\begin{widetext}
\begin{align}
Q^2\frac{d}{d Q^2}(g_1)F_{J}  &= \Bigg\{\frac{1}{2}(\Delta)P_{q\leftarrow q}(a_s) +\frac{1}{2}\tilde{P}_{q\leftarrow q} (a_s)+\beta(a_s)\frac{d(\Delta){\cal C}_{J,qq}(a_s) }{d a_s} \otimes \Big((\Delta){\cal C}_{J,qq} (a_s)\Big)^{-1}\Bigg\}\otimes (g_1)F_{J} \nonumber\\
& = (\Delta)K_{J}~\otimes(g_1)F_{J}\nonumber\\
& = \sum_{l=0}^\infty a_s^{l+1}(\Delta)K_{J,l}~\otimes (g_1)F_{J} \, ,
\label{PEKCF}    
\end{align}
where $a_s=a_s(Q^2)$. 
Note that the kernels $K_{J,l}$ in the differential equations contain scheme-dependent CFs and splitting functions. 
However, because of the scheme invariance of the structure functions, the kernels also have to be invariant under scheme transformations.
Using eq.~\eqref{CFexp}, and expanding the QCD $\beta$ function and the splitting functions in powers of $a_s$, we get, 
\begin{align}
\label{eq:PEKi}
K_{0} &= \frac{1}{2}P^{q}_{0} +\frac{1}{2}\tilde{P}^{q}_{0} \, ,\nonumber\\
K_{1} &= \frac{1}{2}P^{q}_{1} +\frac{1}{2}\tilde{P}^{q}_{1} -\beta_0 {\cal C}_{q}^{(1)}\, ,\nonumber\\
K_{2} &= \frac{1}{2}P^{q}_{2} +\frac{1}{2}\tilde{P}^{q}_{2} -\beta_1 {\cal C}_{q}^{(1)} - 2 \beta_0 {\cal C}_{q}^{(2)}+ \beta_0 {\cal C}_{q}^{(1)}\otimes^{2} \, ,  \nonumber\\
K_{3} &= \frac{1}{2}P^{q}_{3} +\frac{1}{2}\tilde{P}^{q}_{3} -\beta_2 {\cal C}_{q}^{(1)} - 2 \beta_1 {\cal C}_{q}^{(2)}- 3 \beta_0 {\cal C}_{q}^{(3)} +  3 \beta_0 {\cal C}_{q}^{(2)} \otimes {\cal C}_{q}^{(1)}
+ \beta_1 {\cal C}_{q}^{(1)}\otimes^2 - \beta_0 {\cal C}_{q}^{(1)}\otimes^3    \, ,\nonumber\\
K_{4} &= \frac{1}{2}P^{q}_{4} +\frac{1}{2}\tilde{P}^{q}_{4} -\beta_3 {\cal C}_{q}^{(1)}- 
 2 \beta_2 {\cal C}_{q}^{(2)} - 3 \beta_1 {\cal C}_{q}^{(3)} - 4 \beta_0 {\cal C}_{q}^{(4)} 
 + 4 \beta_0  {\cal C}_{q}^{(3)}\otimes{\cal C}_{q}^{(1)}  + 
 2 \beta_0 {\cal C}_{q}^{(2)}\otimes^{2}
 \nonumber\\
&  + 3 \beta_1  {\cal C}_{q}^{(2)}\otimes{\cal C}_{q}^{(1)} - 4 \beta_0 {\cal C}_{q}^{(2)}\otimes {\cal C}_{q}^{(1)}\otimes^{2}     
 + \beta_2 {\cal C}_{q}^{(1)}\otimes^{2} - \beta_1 {\cal C}_{q}^{(1)}\otimes^{3}+ \beta_0 {\cal C}_{q}^{(1)}\otimes^{4}\, .
\end{align}
In the above, we have suppressed $\Delta$ and the index $J$, and used a short-hand notation ${\cal C}^{(l)}_{q}$ for 
${\cal C}^{(l)}_{J,qq}$ and ${\cal C}^{(l)}_{q}\otimes^{i}$, for the $(i-1)$-fold convolution of ${\cal C}_{q}^{(l)}$ with itself, 
i.e., ${\cal C}_{q}^{(1)}\otimes^{2}= {\cal C}_{q}^{(1)}\otimes{\cal C}_{q}^{(1)}$ etc.
Following the approach of ref.~\cite{Moch:2009hr} and using CFs known to 
second order~\cite{Bonino:2024qbh,Bonino:2024wgg,Goyal:2024tmo,Goyal:2023zdi,Goyal:2024emo,Bonino:2025qta} in $a_s$, we obtain $K_{l}$ for $l=0,1,2$ in ${\vec N}$-space, as 
%
\begin{align}
K_{0}&=     
    \bm{L_{1}^{0}L_{2}^{0} }\Bigg[
        C_F   \bigg\{ 6 - 8 \gamma_{E} \bigg\}\Bigg]
-
   \bm{\frac{L_{1}^{0}L_{2}^{0}}{N_1}} \Bigg[
        C_F   \bigg\{   2 \bigg\}\Bigg]
-
   \bm{\frac{L_{1}^{0}L_{2}^{0}}{N_2}} \Bigg[
        C_F   \bigg\{   2 \bigg\}\Bigg]
-
   \bm{L_{1}^{0}L_{2}^{1}} \Bigg[
        C_F   \bigg\{   4 \bigg\}\Bigg]
-
  \bm{L_{1}^{1}L_{2}^{0}} \Bigg[
        C_F   \bigg\{   4 \bigg\}\Bigg] 
\nonumber\\
%
%
 K_{1}&=    
\bm{L_{1}^{0}L_{2}^{0}} \Bigg[
       - C_F n_f   \bigg\{   \frac{34}{3} + \frac{8}{3} \zeta_2 - \frac{80}{9} \gamma_{E} - \frac{16}{3} \gamma_{E}^2   \bigg\}
       + C_F C_A   \bigg\{ \frac{193}{3} - 24 \zeta_3 + \frac{44}{3} \zeta_2 - \frac{536}{9} \gamma_{E} + 16 \gamma_{E} \zeta_2
          - \frac{88}{3} \gamma_{E}^2 \bigg\}   
\nonumber\\&  
      + C_F^2   \bigg\{ 3 + 48 \zeta_3 - 24 \zeta_2 \bigg\}
       \Bigg] 
+
   \bm{\frac{L_{1}^{0}L_{2}^{0}}{N_1}} \Bigg[
        C_F n_f   \bigg\{ \frac{56}{9} + \frac{8}{3} \gamma_{E} \bigg\}    
       - C_F C_A   \bigg\{   \frac{332}{9} - 4 \zeta_2 + \frac{44}{3} \gamma_{E} \bigg\}
       + C_F^2   \bigg\{ 12 - 16 \gamma_{E} \bigg\}\Bigg] 
\nonumber\\&
+
   \bm{\frac{L_{1}^{0}L_{2}^{0}}{N_2}} \Bigg[
        C_F n_f   \bigg\{ \frac{8}{9} + \frac{8}{3} \gamma_{E} \bigg\}        
       - C_F C_A   \bigg\{   \frac{68}{9} - 4 \zeta_2 + \frac{44}{3} \gamma_{E}\bigg\}
       - C_F^2   \bigg\{   12 - 16 \gamma_{E} \bigg\}\Bigg] 
+
   \bm{L_{1}^{0}L_{2}^{1}} \Bigg[
        C_F n_f   \bigg\{ \frac{40}{9} + \frac{16}{3} \gamma_{E} \bigg\}        
\nonumber\\&
       + C_F C_A   \bigg\{  - \frac{268}{9} + 8 \zeta_2 - \frac{88}{3} \gamma_{E} \bigg\}\Bigg] 
+
   \bm{\frac{L_{1}^{0}L_{2}^{1}}{N_2}} \Bigg[
        C_F n_f   \bigg\{ \frac{4}{3} \bigg\}
       - C_F C_A   \bigg\{   \frac{22}{3} \bigg\}
       + C_F^2   \bigg\{ 16 \bigg\}\Bigg]        
\nonumber\\&
+
   \bm{\frac{L_{1}^{0}L_{2}^{1}}{N_1}} \Bigg[
        C_F n_f   \bigg\{ \frac{4}{3} \bigg\}
       - C_F C_A   \bigg\{   \frac{22}{3} \bigg\}\Bigg] 
+
   \bm{L_{1}^{0}L_{2}^{2}} \Bigg[
        C_F n_f   \bigg\{ \frac{4}{3} \bigg\}
       - C_F C_A   \bigg\{   \frac{22}{3} \bigg\}\Bigg] 
+
   \bm{L_{1}^{2}L_{2}^{0}} \Bigg[
        C_F n_f   \bigg\{ \frac{4}{3} \bigg\}
       - C_F C_A   \bigg\{   \frac{22}{3} \bigg\}\Bigg] 
\nonumber\\&
+   \bm{L_{1}^{1}L_{2}^{0}} \Bigg[
        C_F n_f   \bigg\{ \frac{40}{9} + \frac{16}{3} \gamma_{E} \bigg\}
       - C_F C_A   \bigg\{   \frac{268}{9} - 8 \zeta_2 + \frac{88}{3} \gamma_{E} \bigg\}\Bigg] 
+
   \bm{L_{1}^{1}L_{2}^{1}} \Bigg[
        C_F n_f   \bigg\{ \frac{8}{3} \bigg\}
       - C_F C_A   \bigg\{   \frac{44}{3} \bigg\}\Bigg] 
\nonumber\\&
+
   \bm{\frac{L_{1}^{1}L_{2}^{0}}{N_1}} \Bigg[
        C_F n_f   \bigg\{ \frac{4}{3} \bigg\}
       - C_F C_A   \bigg\{   \frac{22}{3} \bigg\}
       - C_F^2   \bigg\{   16 \bigg\}\Bigg] 
+
   \bm{\frac{L_{1}^{1}L_{2}^{0}}{N_2}} \Bigg[
        C_F n_f   \bigg\{ \frac{4}{3} \bigg\}
       - C_F C_A   \bigg\{   \frac{22}{3} \bigg\}\Bigg] 
\nonumber\\
%
%
%
K_{2}&=
   \bm{L_{1}^{0}L_{2}^{0}} \Bigg[
     C_F n_f^2   \bigg\{ \frac{220}{9} - \frac{64}{27} \zeta_3 + \frac{304}{27} \zeta_2 - \frac{800}{81} \gamma_{E} - \frac{320}{27}
          \gamma_{E}^2 - \frac{128}{27} \gamma_{E}^3 \bigg\}
       - C_F C_A n_f   \bigg\{   \frac{3052}{9} - \frac{3440}{27} \zeta_3 + \frac{4184}{27} \zeta_2 
\nonumber\\&       
       - \frac{136}{15} 
         \zeta_2^2 - \frac{16408}{81} \gamma_{E} + \frac{320}{9} \gamma_{E} \zeta_2 - \frac{4624}{27} \gamma_{E}^2 + \frac{64}{3} 
         \gamma_{E}^2 \zeta_2 - \frac{1408}{27} \gamma_{E}^3 \bigg\}
       + C_F C_A^2   \bigg\{ \frac{3082}{3} + 80 \zeta_5 - \frac{22600}{27} \zeta_3 
\nonumber\\&       
       +\frac{13708}{27} \zeta_2 - \frac{676}{15} \zeta_2^2 - \frac{62012}{81} \gamma_{E} + 352 \gamma_{E} \zeta_3 + \frac{2144}{9} \gamma_{E} \zeta_2 - \frac{352}{5} \gamma_{E} \zeta_2^2 - \frac{14240}{27} \gamma_{E}^2        
       + \frac{352}{3} \gamma_{E}^2 \zeta_2 - \frac{3872}{27} 
         \gamma_{E}^3 \bigg\}
\nonumber\\&
       - C_F^2 n_f   \bigg\{   \frac{235}{3} + \frac{512}{3} \zeta_3 - \frac{296}{3} \zeta_2 - \frac{112}{15} \zeta_2^2 - 
         \frac{220}{3} \gamma_{E} + 64 \gamma_{E} \zeta_3 - 16 \gamma_{E}^2 \bigg\}
       + C_F^2 C_A   \bigg\{ \frac{232}{3} + 240 \zeta_5 + \frac{3008}{3} \zeta_3 
\nonumber\\&       
       - \frac{2096}{3} \zeta_2 + 32 
         \zeta_2 \zeta_3 - \frac{8}{3} \zeta_2^2 \bigg\}
       + C_F^3   \bigg\{ 29 - 480 \zeta_5 + 136 \zeta_3 + 36 \zeta_2 - 64 \zeta_2 \zeta_3 + \frac{576}{5} \zeta_2^2 \bigg\}\Bigg] 
\nonumber\\&
+
   \bm{\frac{L_{1}^{0}L_{2}^{0}}{N_1}} \Bigg[
        C_F C_A n_f   \bigg\{ \frac{14002}{81} - \frac{176}{9} \zeta_2 + \frac{5120}{27} \gamma_{E} - \frac{32}{3} \gamma_{E} \zeta_2
          + \frac{352}{9} \gamma_{E}^2 \bigg\}
       - C_F n_f^2   \bigg\{   \frac{1208}{81} + \frac{448}{27} \gamma_{E} + \frac{32}{9} \gamma_{E}^2 \bigg\}
\nonumber\\&
       - C_F C_A^2   \bigg\{   \frac{36059}{81} - 88 \zeta_3 - \frac{1064}{9} \zeta_2 + \frac{88}{5} \zeta_2^2 + 
         \frac{13852}{27} \gamma_{E} - \frac{176}{3} \gamma_{E} \zeta_2 + \frac{968}{9} \gamma_{E}^2 \bigg\}
       + C_F^3   \bigg\{ 6 + 96 \zeta_3 - 48 \zeta_2 \bigg\}
\nonumber\\&
       + C_F^2 C_A   \bigg\{ \frac{302}{3} - 48 \zeta_3 + 64 \zeta_2 - \frac{1220}{9} \gamma_{E} + 64 \gamma_{E} \zeta_2
          - \frac{616}{3} \gamma_{E}^2 \bigg\}
       + C_F^2 n_f   \bigg\{ \frac{47}{3} - 16 \zeta_3 - 16 \zeta_2 + \frac{224}{9} \gamma_{E} + \frac{112}{3} \gamma_{E}^2 \bigg\}   \Bigg] 
\nonumber\\&       
 +
   \bm{\frac{L_{1}^{0}L_{2}^{0}}{N_2}} \Bigg[
        C_F n_f^2   \bigg\{ \frac{40}{81} - \frac{64}{27} \gamma_{E} - \frac{32}{9} \gamma_{E}^2 \bigg\}
       + C_F C_A n_f   \bigg\{ \frac{3442}{81} - \frac{80}{9} \zeta_2 + \frac{1616}{27} \gamma_{E} - \frac{32}{3} \gamma_{E} \zeta_2 + 
         \frac{352}{9} \gamma_{E}^2 \bigg\}
\nonumber\\&
       - C_F C_A^2   \bigg\{   \frac{20267}{81} - 88 \zeta_3 - \frac{536}{9} \zeta_2 + \frac{88}{5} \zeta_2^2 + 
         \frac{6196}{27} \gamma_{E} - \frac{176}{3} \gamma_{E} \zeta_2 + \frac{968}{9} \gamma_{E}^2 \bigg\}
       - C_F^3   \bigg\{   6 + 96 \zeta_3 - 48 \zeta_2 \bigg\}
\nonumber\\&
       - C_F^2 C_A   \bigg\{   \frac{302}{3} - 48 \zeta_3 - 112 \zeta_2 - \frac{1220}{9} \gamma_{E} + 64 \gamma_{E} 
         \zeta_2 - \frac{616}{3} \gamma_{E}^2 \bigg\}
       + C_F^2 n_f   \bigg\{ 29 - 16 \zeta_3 - 16 \zeta_2 - \frac{80}{9} \gamma_{E} - \frac{112}{3} \gamma_{E}^2 \bigg\}\Bigg] 
\nonumber\\&
+
   \bm{L_{1}^{0}L_{2}^{1}} \Bigg[
        C_F C_A n_f   \bigg\{ \frac{8204}{81} - \frac{160}{9} \zeta_2 + \frac{4624}{27} \gamma_{E} - \frac{64}{3} \gamma_{E} \zeta_2 + \frac{704}{9} \gamma_{E}^2 \bigg\}
       - C_F n_f^2   \bigg\{   \frac{400}{81} + \frac{320}{27} \gamma_{E} + \frac{64}{9} \gamma_{E}^2 \bigg\}
\nonumber\\&
       - C_F C_A^2   \bigg\{   \frac{31006}{81} - 176 \zeta_3 - \frac{1072}{9} \zeta_2 + \frac{176}{5} \zeta_2^2 + 
         \frac{14240}{27} \gamma_{E} - \frac{352}{3} \gamma_{E} \zeta_2 + \frac{1936}{9} \gamma_{E}^2 \bigg\}
       + C_F^2 n_f   \bigg\{ \frac{110}{3} - 32 \zeta_3 + 16 \gamma_{E} \bigg\}\Bigg] 
\nonumber\\&
+
   \bm{\frac{L_{1}^{0}L_{2}^{1}}{N_1}} \Bigg[
      C_F C_A n_f   \bigg\{ \frac{2740}{27} - \frac{16}{3} \zeta_2 + \frac{352}{9} \gamma_{E} \bigg\}
      - C_F n_f^2   \bigg\{   \frac{224}{27} + \frac{32}{9} \gamma_{E} \bigg\}
       - C_F C_A^2   \bigg\{   \frac{7916}{27} - \frac{88}{3} \zeta_2 + \frac{968}{9} \gamma_{E} \bigg\}
\nonumber\\&       
       - C_F^2 n_f   \bigg\{   12 - \frac{64}{3} \gamma_{E} \bigg\}
       + C_F^2 C_A   \bigg\{ 88 - \frac{352}{3} \gamma_{E} \bigg\}\Bigg] 
-
   \bm{\frac{L_{1}^{0}L_{2}^{1}}{N_2}} \Bigg[
        C_F n_f^2   \bigg\{   \frac{32}{27} + \frac{32}{9} \gamma_{E} \bigg\}
       + C_F^2 n_f   \bigg\{   \frac{260}{9} + \frac{160}{3} \gamma_{E} \bigg\}  
\nonumber\\&
        - C_F C_A n_f   \bigg\{ \frac{988}{27} - \frac{16}{3} \zeta_2 + \frac{352}{9} \gamma_{E} \bigg\} 
       + C_F C_A^2   \bigg\{   \frac{4088}{27} - \frac{88}{3} \zeta_2 + \frac{968}{9} \gamma_{E} \bigg\}
       - C_F^2 C_A   \bigg\{ \frac{2012}{9} - 64 \zeta_2 
       + \frac{880}{3} \gamma_{E} \bigg\}\Bigg] 
\nonumber\\&       
+
   \bm{L_{1}^{0}L_{2}^{2}} \Bigg[
         C_F C_A n_f   \bigg\{ \frac{1156}{27} - \frac{16}{3} \zeta_2 + \frac{352}{9} \gamma_{E} \bigg\} 
       - C_F n_f^2   \bigg\{   \frac{80}{27} + \frac{32}{9} \gamma_{E} \bigg\}
       - C_F C_A^2   \bigg\{   \frac{3560}{27} - \frac{88}{3} \zeta_2 + \frac{968}{9} \gamma_{E} \bigg\}
\nonumber\\&       
       + C_F^2 n_f   \bigg\{ 4 \bigg\}\Bigg] 
+
   \bm{\frac{L_{1}^{0}L_{2}^{2}}{N_1}} \Bigg[
        C_F C_A n_f   \bigg\{ \frac{88}{9} \bigg\}
       - C_F n_f^2   \bigg\{  \frac{8}{9} \bigg\}
       - C_F C_A^2   \bigg\{   \frac{242}{9} \bigg\}\Bigg] 
+
   \bm{\frac{L_{1}^{0}L_{2}^{2}}{N_2}} \Bigg[
        C_F C_A n_f   \bigg\{ \frac{88}{9} \bigg\}
       - C_F n_f^2   \bigg\{  \frac{8}{9} \bigg\}
 \nonumber\\&
       - C_F C_A^2   \bigg\{  \frac{242}{9} \bigg\} 
       - C_F^2 n_f   \bigg\{  16 \bigg\}
       + C_F^2 C_A   \bigg\{ 88 \bigg\}\Bigg] 
+
   \bm{L_{1}^{0}L_{2}^{3}} \Bigg[
        C_F C_A n_f   \bigg\{ \frac{176}{27} \bigg\}
       - C_F n_f^2   \bigg\{   \frac{16}{27} \bigg\}
       - C_F C_A^2   \bigg\{   \frac{484}{27} \bigg\}\Bigg] 
\nonumber\\&
+
   \bm{L_{1}^{1}L_{2}^{0}} \Bigg[
        C_F C_A n_f   \bigg\{ \frac{8204}{81} - \frac{160}{9} \zeta_2 + \frac{4624}{27} \gamma_{E} - \frac{64}{3} \gamma_{E} \zeta_2 + \frac{704}{9} \gamma_{E}^2 \bigg\} 
       - C_F n_f^2   \bigg\{   \frac{400}{81} + \frac{320}{27} \gamma_{E} + \frac{64}{9} \gamma_{E}^2 \bigg\}
\nonumber\\&       
       - C_F C_A^2   \bigg\{   \frac{31006}{81} - 176 \zeta_3 - \frac{1072}{9} \zeta_2 + \frac{176}{5} \zeta_2^2 + 
         \frac{14240}{27} \gamma_{E} - \frac{352}{3} \gamma_{E} \zeta_2 + \frac{1936}{9} \gamma_{E}^2 \bigg\}
       + C_F^2 n_f   \bigg\{ \frac{110}{3} - 32 \zeta_3 + 16 \gamma_{E} \bigg\}\Bigg] 
\nonumber\\&
+
   \bm{\frac{L_{1}^{1}L_{2}^{0}}{N_1}} \Bigg[
        C_F C_A n_f   \bigg\{ \frac{2380}{27} - \frac{16}{3} \zeta_2 + \frac{352}{9} \gamma_{E} \bigg\}
       - C_F n_f^2   \bigg\{   \frac{224}{27} + \frac{32}{9} \gamma_{E} \bigg\}
       - C_F C_A^2   \bigg\{   \frac{5936}{27} - \frac{88}{3} \zeta_2 + \frac{968}{9} \gamma_{E} \bigg\}
\nonumber\\&       
       + C_F^2 n_f   \bigg\{ \frac{332}{9} + \frac{160}{3} \gamma_{E} \bigg\}
       - C_F^2 C_A   \bigg\{   \frac{2012}{9} - 64 \zeta_2 + \frac{880}{3} \gamma_{E} \bigg\}\Bigg] 
+
   \bm{\frac{L_{1}^{1}L_{2}^{0}}{N_2}} \Bigg[
        C_F C_A n_f   \bigg\{ \frac{628}{27} - \frac{16}{3} \zeta_2 + \frac{352}{9} \gamma_{E} \bigg\}
\nonumber\\&
       - C_F n_f^2   \bigg\{   \frac{32}{27} + \frac{32}{9} \gamma_{E} \bigg\}
       - C_F C_A^2   \bigg\{   \frac{2108}{27} - \frac{88}{3} \zeta_2 + \frac{968}{9} \gamma_{E} \bigg\}
       + C_F^2 n_f   \bigg\{ 20 - \frac{64}{3} \gamma_{E} \bigg\}
       - C_F^2 C_A   \bigg\{   88 - \frac{352}{3} \gamma_{E} \bigg\}\Bigg] 
\nonumber\\&       
+
   \bm{L_{1}^{1}L_{2}^{1}} \Bigg[
        C_F C_A n_f   \bigg\{ \frac{2312}{27} - \frac{32}{3} \zeta_2 + \frac{704}{9} \gamma_{E} \bigg\}
       - C_F n_f^2   \bigg\{   \frac{160}{27} + \frac{64}{9} \gamma_{E} \bigg\}
       - C_F C_A^2   \bigg\{   \frac{7120}{27} - \frac{176}{3} \zeta_2 + \frac{1936}{9} \gamma_{E} \bigg\}
\nonumber\\&       
       + C_F^2 n_f   \bigg\{ 8 \bigg\}\Bigg] 
+
   \bm{\frac{L_{1}^{1}L_{2}^{1}}{N_1}} \Bigg[
        C_F C_A n_f   \bigg\{ \frac{176}{9} \bigg\}
       - C_F n_f^2   \bigg\{   \frac{16}{9} \bigg\}
       - C_F C_A^2   \bigg\{   \frac{484}{9} \bigg\}
       + C_F^2 n_f   \bigg\{ \frac{64}{3} \bigg\}
       - C_F^2 C_A   \bigg\{   \frac{352}{3} \bigg\}\Bigg] 
\nonumber\\&
+
   \bm{\frac{L_{1}^{1}L_{2}^{1}}{N_2}} \Bigg[
        C_F C_A n_f   \bigg\{ \frac{176}{9} \bigg\}
        - C_F n_f^2   \bigg\{   \frac{16}{9} \bigg\}       
       - C_F C_A^2   \bigg\{   \frac{484}{9} \bigg\}
       - C_F^2 n_f   \bigg\{   \frac{64}{3} \bigg\}
       + C_F^2 C_A   \bigg\{ \frac{352}{3} \bigg\}\Bigg] 
\nonumber\\&
+
   \bm{L_{1}^{1}L_{2}^{2}} \Bigg[
        C_F C_A n_f   \bigg\{ \frac{176}{9} \bigg\}
       - C_F n_f^2   \bigg\{   \frac{16}{9} \bigg\}
       - C_F C_A^2   \bigg\{   \frac{484}{9} \bigg\}\Bigg] 
+
   \bm{L_{1}^{2}L_{2}^{0}} \Bigg[
        C_F C_A n_f   \bigg\{ \frac{1156}{27} - \frac{16}{3} \zeta_2 + \frac{352}{9} \gamma_{E} \bigg\}
\nonumber\\&
       - C_F n_f^2   \bigg\{   \frac{80}{27} + \frac{32}{9} \gamma_{E} \bigg\}
       - C_F C_A^2   \bigg\{   \frac{3560}{27} - \frac{88}{3} \zeta_2 + \frac{968}{9} \gamma_{E} \bigg\}
       + C_F^2 n_f   \bigg\{ 4 \bigg\}\Bigg] 
+
   \bm{\frac{L_{1}^{2}L_{2}^{0}}{N_1}} \Bigg[
        C_F C_A n_f   \bigg\{ \frac{88}{9} \bigg\}
       - C_F n_f^2   \bigg\{   \frac{8}{9} \bigg\} 
\nonumber\\&       
       - C_F C_A^2   \bigg\{   \frac{242}{9} \bigg\}
       + C_F^2 n_f   \bigg\{ 16 \bigg\}
       - C_F^2 C_A   \bigg\{   88 \bigg\}\Bigg] 
+
   \bm{\frac{L_{1}^{2}L_{2}^{0}}{N_2}} \Bigg[
        C_F C_A n_f   \bigg\{ \frac{88}{9} \bigg\}
       - C_F n_f^2   \bigg\{   \frac{8}{9} \bigg\}
       - C_F C_A^2   \bigg\{   \frac{242}{9} \bigg\}\Bigg] 
\nonumber\\&
+
   \bm{L_{1}^{2}L_{2}^{1}} \Bigg[
        C_F C_A n_f   \bigg\{ \frac{176}{9} \bigg\}
       - C_F n_f^2   \bigg\{   \frac{16}{9} \bigg\}
       - C_F C_A^2   \bigg\{   \frac{484}{9} \bigg\}\Bigg] 
+
   \bm{L_{1}^{3}L_{2}^{0}} \Bigg[
       C_F C_A n_f   \bigg\{ \frac{176}{27} \bigg\}
       - C_F n_f^2   \bigg\{   \frac{16}{27} \bigg\}
\nonumber\\&       
       - C_F C_A^2   \bigg\{   \frac{484}{27} \bigg\}\Bigg] 
\, ,
\label{PEKN}
\end{align}
    
\end{widetext}
where $\bm{L_j^i} = \ln^{i}(N_j)$, for $j=1,2$.  From the above results for $K_i,i=0,1,2$, we observe that
\begin{itemize}
\item In the SV part of $K_{J,l}$, the highest power of logarithms $N_i$ is constrained, for example for the logarithms $\ln^i(N_1)\ln^j(N_2)$, we find $0 \leq i+j \leq l+1 $.
\item In the NSV part of  $K_{J,l}$, the power of highest logarithms is constrained, for example, for the logarithms $\frac{1}{N_1}\ln^i(N_1)\ln^j(N_2)$ or $\frac{1}{N_2}\ln^i(N_1)\ln^j(N_2)$,  we have the limit on $i,j$ as $0 \leq i+j \leq l $.
\end{itemize}
Thus, also the SIDIS PEK shows a single-logarithmic enhancement, both in $(x',z')$ space as well as in $\vec N$ space,  similar to what was observed in ref.~\cite{Moch:2009hr} in the case of PEKs from inclusive cross-sections. 
If we conjecture that this logarithm structure persists at every order in $a_s$ for $K_{J,l}$, then we can determine certain SV and NSV logarithms at a given order $l$ for ${\cal C}_{q}^{(l)}$ from the knowledge of AP kernels and CFs from previous order in $a_s$, i.e., results for all orders up to $a_s^k$ where $k < l$.   

The validation of the SIDIS PEK predictions against the known second-order CFs can be readily performed. 
We use $K_0$ and $K_1$, i.e., the PEKs known from the NLO CFs and NLO splitting functions.   
Then, for SV terms, the maximum sum of logarithmic powers is $(i+j)|_{\rm max}=3$ 
for terms of the form $\ln^i(N_1)\ln^j(N_2)$.
For NSV terms, the constraint is $(i+j)|_{\rm max}= 2$ for logarithms of the form
$\frac{1}{N_1}\ln^i(N_1)\ln^j(N_2)$ or $\frac{1}{N_2}\ln^i(N_1)\ln^j(N_2)$.
By substituting explicit results for $P_2^q, \tilde P_2^q$ and  $\mathcal{C}_q^{(1)}$ 
in $K_2$ of eq.~\eqref{eq:PEKi}, and applying the general structure of SV and NSV terms 
specified in eq.~\eqref{eq:svnsvbnsv}, 
we impose the condition that all terms with $i+j>3$ for SV and $i+j>2$ for NSV must vanish. 
This process yields a set of 13 equations relating the second-order 
coefficients $\mathcal{C}_J^{\xi_1\xi_2,(2)}$ in eq.~\eqref{eq:svnsvbnsv}.
The solutions to these equations provide the NNLO coefficients for terms such as 
$\delta_{\overline{x}'}\{\mathcal{D}^{i}_{\overline{z}'},\mathcal{L}^{i}_{\overline{z}'}\}$ with $i=3$ and $\{\mathcal{D}^{i}_{\overline{x}'}\mathcal{D}^{j}_{\overline{z}'}, \mathcal{D}^{i}_{\overline{x}'}\mathcal{L}^{j}_{\overline{z}'}\}$ with $i+j=2 $, along with similar terms where ${x}'\leftrightarrow {z}'$.
We find that these predictions are in complete agreement with the results obtained from explicit computations \cite{Bonino:2024qbh,Bonino:2024wgg,Goyal:2023zdi,Goyal:2024tmo,Goyal:2024emo,Bonino:2025qta}.

Next, by using known NNLO results, we can predict the third and fourth-order results for CFs, i.e., classes of logarithms for ${\cal C}_{q}^{(3)}$ and ${\cal C}_{q}^{(4)}$.
For SV terms in $K_{3}$, we consider logarithms of the form 
$\ln^i(N_1)\ln^j(N_2)$ with the constraint  $(i+j)|_{\rm max}=4$, and for NSV logarithms
$\frac{1}{N_1}\ln^i(N_1)\ln^j(N_2)$ or $\frac{1}{N_2}\ln^i(N_1)\ln^j(N_2)$, 
the constraint is $(i+j)|_{\rm max}= 3$. 
By imposing that there are no SV logarithms for $i+j>4$ and no NSV logarithms exist for $i+j>3$,
we obtain a system of 35 linear equations that relate the coefficients $C_J^{\xi_1\xi_2,(3)}$.  Solving for these coefficients, we can predict 35 different logarithms at N$^3$LO level.
For example, at third order, for the CF ${\cal C}_{q}^{(3)}$ in $(x',z')$-space, 
we can predict terms such as
$\delta_{\overline{x}'}\{\mathcal{D}^{i}_{\overline{z}'},\mathcal{L}^{i}_{\overline{z}'}\}$ for $i=4,5$ and $\{\mathcal{D}^{i}_{\overline{x}'}\mathcal{D}^{j}_{\overline{z}'}, \mathcal{D}^{i}_{\overline{x}'}\mathcal{L}^{j}_{\overline{z}'}\}$ 
for $i+j=3,4$ and similarly for ${x}'\leftrightarrow {z}'$.
We have found that all these predictions are in complete agreement with those obtained using the resummation approach.

Similarly, we can determine the coefficients of highest logarithms in ${\cal C}_{q}^{(4)}$ by observing that the SV logarithms of the form 
$\ln^i(N_1)\ln^j(N_2)$ are constrained by $(i+j)|_{\rm max}=5$ and for NSV logarithms, 
$\frac{1}{N_1}\ln^i(N_1)\ln^j(N_2)$ or $\frac{1}{N_2}\ln^i(N_1)\ln^j(N_2)$, 
the constraint is $(i+j)|_{\rm max}= 4$. 
These constraints yield a system of 66 equations, with which we can predict 47 SV and NSV logarithms in ${\cal C}_{q}^{(4)}$, namely terms of the form $\delta_{\overline{x}'}\{\mathcal{D}^{i}_{\overline{z}'},\mathcal{L}^{i}_{\overline{z}'}\}$ for $i=6,7$ and $\{\mathcal{D}^{i}_{\overline{x}'}\mathcal{D}^{j}_{\overline{z}'}, \mathcal{D}^{i}_{\overline{x}'}\mathcal{L}^{j}_{\overline{z}'}\}$ for $i+j=5,6$, including analogous terms where ${x}'\leftrightarrow {z}'$.
These predictions are also in agreement with the ones obtained via the resummation approach.
However, the remaining 19 equations require information from ${\cal C}_{q}^{(3)}$ and hence we cannot predict them using this current approach.
%



\section{Phenomenology}
\label{sec:pheno}
We are now in the position to illustrate the impact of resummed SV and NSV contributions on the structure functions $F_1$ and $g_1$ for the EIC at a center-of-mass energy of $\sqrt{s}$ = 140 GeV. 
To that end, we consider the logarithmic accuracies  $\overline{\text{LL}}$,  $\overline{\text{NLL}}$ and  $\overline{\text{NNLL}}$.
As shown in the first row of Tab.~\ref{Table:resumgh},
the resummed result at $\overline{\text{LL}}$ accuracy requires the resummed exponents $\{\tilde g^q_{d,0,0}, g^q_{d,1}, \overline g^q_{d,0}$, $h^q_{d,0}\}$.  
They systematically resum SV logarithms of the form $\{ \ln^i N_1 \ln^j N_2\}$ 
(where $i+j=2n$) and NSV logarithms like 
$\{{\ln^i N_1 \over N_1} \ln^j N_2, {\ln^i N_2 \over N_2} \ln^j N_1\}$ 
(where $i+j=2n-1$) at $n$th order in $a_s$, i.e., $a_s^n$ for all $n>1$:
\begin{widetext}
\begin{align}
\label{LLb}
(\Delta) {\cal C}_{J,qq,\vec N}^{\overline{\rm{LL}}}
&= \Big(\tilde g^q_{d,0,0}\Big)\exp \bigg[ g^q_{d,1}(\omega) \ln N_1 +  \frac{1}{N_1} \bigg(\overline{g}^q_{d,0}(\omega) + h^q_{d,0}(\omega,N_1)\bigg) + (N_1 \leftrightarrow N_2)\bigg] \,.
\end{align}
Similarly, for the resummed result at $\overline{\text{NLL}}$ accuracy, as detailed
in the second row of Tab.~\ref{Table:resumgh}, we require
the additional resummation exponents $\{\tilde g^q_{d,0,1},g^q_{d,2},\overline g^q_{d,1}, h^q_{d,1}\}$ alongside the $\overline{\text{LL}}$ exponents. 
This suffices to sum up the next-to-leading SV logarithms 
and next-to-highest NSV logarithms to all orders.
Specifically, this includes SV logarithms of the form $\{\ln^i N_1  \ln^j N_2\}$ 
(where $i+j=\{2n-1,2n-2\}$) 
and NSV logarithms of the form 
$\{{\ln^i N_1 \over N_1} \ln^j N_2, {\ln^i N_2 \over N_2} \ln^j N_1\}$ 
(where $i+j=2n-2$ with $n>2$), i.e.,
\begin{align}
\label{NLLb}
(\Delta) {\cal C}_{J,qq,\vec N}^{\overline{\rm{NLL}}} 
 &= \Big(\tilde g^q_{d,0,0}+a_s ~\tilde g^q_{d,0,1}\Big)\exp \bigg[ g^q_{d,1}(\omega) \ln N_1 + \frac{1}{2} g^q_{d,2}(\omega) + \frac{1}{N_1}\bigg( \overline{g}^q_{d,0}(\omega) +a_s~\overline{g}^q_{d,1}(\omega)+ h^q_{d,0}(\omega,N_1) \nonumber\\
&+ a_s~h^q_{d,1}(\omega,\omega_1,N_1) \bigg) + (N_1 \leftrightarrow N_2, \omega_1 \leftrightarrow \omega_2) \bigg] \, .
\end{align}
%
\begin{table*}[ht!]
\centering
\begin{small}
{\renewcommand{\arraystretch}{2.2}
\begin{tabular}{|P{1.85cm} P{0.05cm} ||P{2.5cm} |P{2.55cm} |P{2.6cm} |P{1.2cm} |P{3.55cm} |P{2.0cm}|}
 \hline
 \multicolumn{1}{|c}{\textbf{GIVEN}} & & \multicolumn{4}{c}{\textbf{PREDICTIONS - SV and NSV Logarithms}} & & \textbf{Logarithmic Accuracy} \\
 \cline{1-1}\cline{2-7}
  \textbf{Resummed Exponents Up To} & & $ (\Delta){\cal C}_{J,qq,\vec N}^{(2)} $ & $ (\Delta){\cal C}_{J,qq,\vec N}^{(3)} $ & $ (\Delta){\cal C}_{J,qq,\vec N}^{(4)} $ & $\cdots$ & $ (\Delta){\cal C}_{J,qq,\vec N}^{(n)} $ & \\
 \hline
 \hline
 $\tilde g^q_{d,0,0}, g^q_{d,1},$
 $\bar{g}^q_{d,0},h^q_{d,0}$ 
 &  &  
$L_1^i L_2^j{|}_{i+j=4}$
$L_{N_1}^i L_2^j{|}_{i+j=3}$
$L_1^i L_{N_2}^j{|}_{i+j=3}$
 & 
$L_1^i L_2^j{|}_{i+j=6}$
$L_{N_1}^i L_2^j{|}_{i+j=5}$
$L_1^i L_{N_2}^j{|}_{i+j=5}$
 & 
$L_1^i L_2^j{|}_{i+j=8}$
$L_{N_1}^i L_2^j{|}_{i+j=7}$
$L_1^i L_{N_2}^j{|}_{i+j=7}$
 & 
 $\cdots$ 
 & 
$L_1^i L_2^j{|}_{i+j=2n}$
$L_{N_1}^i L_2^j{|}_{i+j=2n-1}$
$L_1^i L_{N_2}^j{|}_{i+j=2n-1}$
 &  
 $\overline{\textbf{LL}}$ 
 \\
 \hline
 $\tilde g^q_{d,0,1}, g^q_{d,2}$ 
$\bar{g}^q_{d,1},h^q_{d,1}$
 &  &  & 
$L_1^i L_2^j{|}_{i+j=6,5,4}$
$L_{N_1}^i L_2^j{|}_{i+j=5,4}$
$L_1^i L_{N_2}^j{|}_{i+j=5,4}$
 & 
$L_1^i L_2^j{|}_{i+j=8,7,6}$
$L_{N_1}^i L_2^j{|}_{i+j=7,6}$
$L_1^i L_{N_2}^j{|}_{i+j=7,6}$
 & 
 $\cdots$ & 
$L_1^i L_2^j{|}_{i+j=2n,...,2n-2}$
$L_{N_1}^i L_2^j{|}_{i+j=2n-1,2n-2}$
$L_1^i L_{N_2}^j{|}_{i+j=2n-1,2n-2}$
 & 
  $\overline{\textbf{NLL}}$ 
 \\
 \hline
 $\tilde g^q_{d,0,2}, g^q_{d,3}$
 $\bar{g}^q_{d,2},h^q_{d,2}$
 &  &  &  & 
$L_1^i L_2^j{|}_{i+j=8,...,4}$
$L_{N_1}^i L_2^j{|}_{i+j=7,...,5}$
$L_1^i L_{N_2}^j{|}_{i+j=7,...,5}$ 
 & 
 $\cdots$ 
 & 
$L_1^i L_2^j{|}_{i+j=2n,...,2n-4}$
$L_{N_1}^i L_2^j{|}_{i+j=2n-1,...,2n-3}$
$L_1^i L_{N_2}^j{|}_{i+j=2n-1,...,2n-3}$
 & 
  $\overline{\textbf{NNLL}}$ 
 \\
 \hline
\end{tabular}}
\end{small}
\caption{The set of resummed exponents required to predict the tower of NSV logarithm in $ {\cal C}_{J,qq,\vec N}^{(n)} $, $L_1^{i}=\ln^{i}(N_1)$, $L_{N_1}^{i}=\frac{\ln^i(N_1)}{N_1} $, $L_{2}^j = \ln^{j}(N_2)$, $L^j_{N_2}= \frac{\ln^j(N_2)}{N_2} $.}\label{Table:resumgh}
\end{table*}


According to the third row of Tab.~\ref{Table:resumgh}, 
we include the resummation exponents 
$\{\tilde g^q_{d,0,2},g^q_{d,3},\overline g^q_{d,2}, h^q_{d,2}\}$
alongside the $\overline{\rm{LL}}$ and $\overline{\rm{NLL}}$ terms to achieve $\overline{\rm{NNLL}}$ accuracy.
This set resums SV logarithms of 
the form $\{\ln^i N_1  \ln^j N_2\}$ (where $i+j=\{2n-3,2n-4\}$) and 
NSV logarithms of the form $\{ {\ln^i N_1 \over N_1} \ln^j N_2, {\ln^i N_2 \over N_2} \ln^j N_1\}$ 
with $i+j=2n-3, n>3$.
\begin{align}
\label{NNLLb}
(\Delta) {\cal C}_{J,qq,\vec N}^{\overline{\rm{NNLL}}}
&= \Big(\tilde g^q_{d,0,0}+a_s ~\tilde g^q_{d,0,1}+a_s^2 ~\tilde g^q_{d,0,2}\Big)\exp \bigg[ g^q_{d,1}(\omega) \ln N_1 + \frac{1}{2} \bigg(g^q_{d,2}(\omega)+a_s ~g^q_{d,3}(\omega)\bigg) + \frac{1}{N_1}\bigg( \overline{g}^q_{d,0}(\omega)+a_s~\overline{g}^q_{d,1}(\omega)\nonumber\\&+a_s^2~\overline{g}^q_{d,2}(\omega)
+ h^q_{d,0}(\omega,N_1) + a_s~h^q_{d,1}(\omega,\omega_1,N_1)+a_s^2~h^q_{d,2}(\omega,\omega_1,N_1) \bigg) + (N_1 \leftrightarrow N_2, \omega_1 \leftrightarrow \omega_2)\bigg] \,.
\end{align}
The resummed results are matched to the respective fixed-order results as follows:
\begin{align}
(g_1){F}_{J}^{\rm N^nLO+\overline{\rm{N^nLL}}}
&= (g_1){F}_{J}^{\rm N^nLO} 
\nonumber\\ &
+ \sum_{cc}
\, \int_{c_{1} - i\infty}^{c_1 + i\infty} \frac{d N_{1}}{2\pi i}
 \int_{c_{2} - i\infty}^{c_2 + i\infty} \frac{d N_{2}}{2\pi i} 
~ 
(\Delta)f_{c,N_1}(\mu_F^2) D_{c,N_2}(\mu_F^2) 
\times \bigg(
(\Delta){\cal C}_{J,cc,\vec N}^{\overline{\rm{N^nLL}}}
- (\Delta){\cal C}_{J,cc,\vec N}\bigg|_{\text{tr}^n} \bigg) \,.
\label{match}
\end{align}
\end{widetext}

The first term in eq.~(\ref{match}) represents the fixed-order contributions up to
N$^n$LO accuracy. The second term corresponds to the resummed result 
at $\overline{\text{N}^\text{n}\text{LL}}$ accuracy.
This term is derived by subtracting its fixed-order expansion, truncated up to order $a_s^n$, 
from the full resummed result, which precisely avoids double-counting of threshold logarithms. 
Therefore, the latter encapsulates the all-order SV and NSV resummed contributions
in perturbation theory starting from $a^{n+1}_s$ onward.
The subscript `tr$^n$' denotes the $n^{\rm th}$-order `truncated' part of the resummed result
which is already contained in the fixed-order expressions.
The CFs required for $(g_1)F_J^{\rm{N^{n}LO}}$ in the first term are known up to NNLO \cite{Altarelli:1979kv,Goyal:2023zdi,Bonino:2024qbh,Goyal:2024tmo,Bonino:2024wgg,Goyal:2024emo,Bonino:2025qta}.

For the unpolarized structure function ($F_1$), we have employed the \texttt{ABMP16} PDF sets~\cite{Alekhin:2017kpj}
and \texttt{NNFF10PIp} FF sets~\cite{Bertone:2017tyb} at their respective orders,
except for the LO where we use NLO PDF set.
For the polarized structure function ($g_1$), we have used the polarized \texttt{BDSSV24NLO} PDFs at LL and NLL accuracies, and \texttt{BDSSV24NNLO} PDFs at NNLL accuracy. 
The strong coupling $a_s$ is taken as provided through the \texttt{LHAPDF} interface~\cite{Buckley:2014ana} with
$n_f$ = 3 active massless quark flavors throughout. 
We have used an in-house \texttt{FORTRAN} code to perform the double Mellin inversion for the resummed contributions, and we have adopted the so-called minimal prescription \cite{Catani:1996yz} to handle the Landau pole in the Mellin inversion routines. 

\begin{widetext}
\begin{center}
\begin{figure}[ht]
\includegraphics[width=0.92\textwidth]{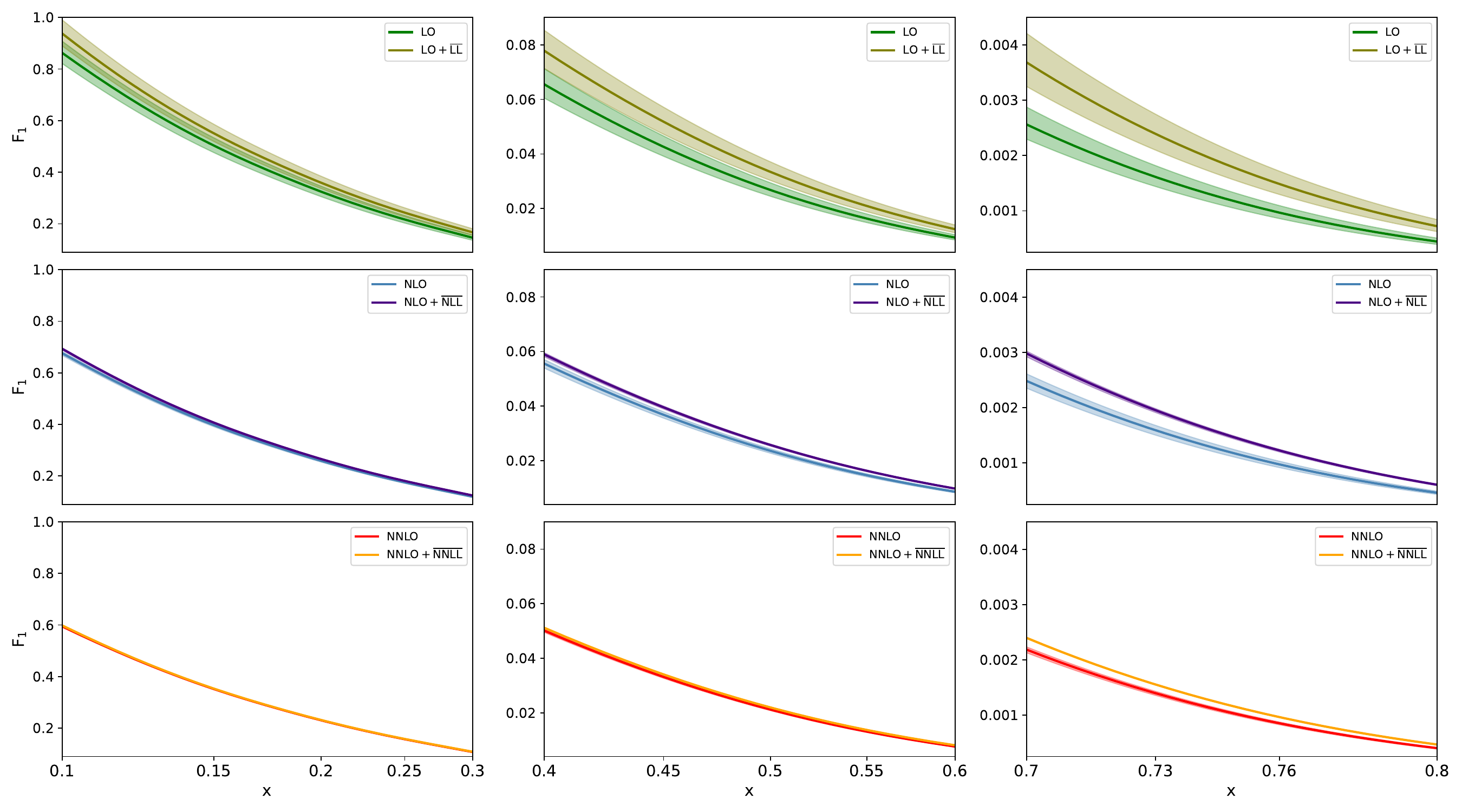}
\caption{Seven-point scale variation of the structure function $F_1$ as a function of $x$ (log-scale) for the EIC at $\sqrt{s}=140$ GeV.}    
\label{fig:Fx}
\end{figure}
\begin{figure}[t]
\includegraphics[width=0.92\textwidth]{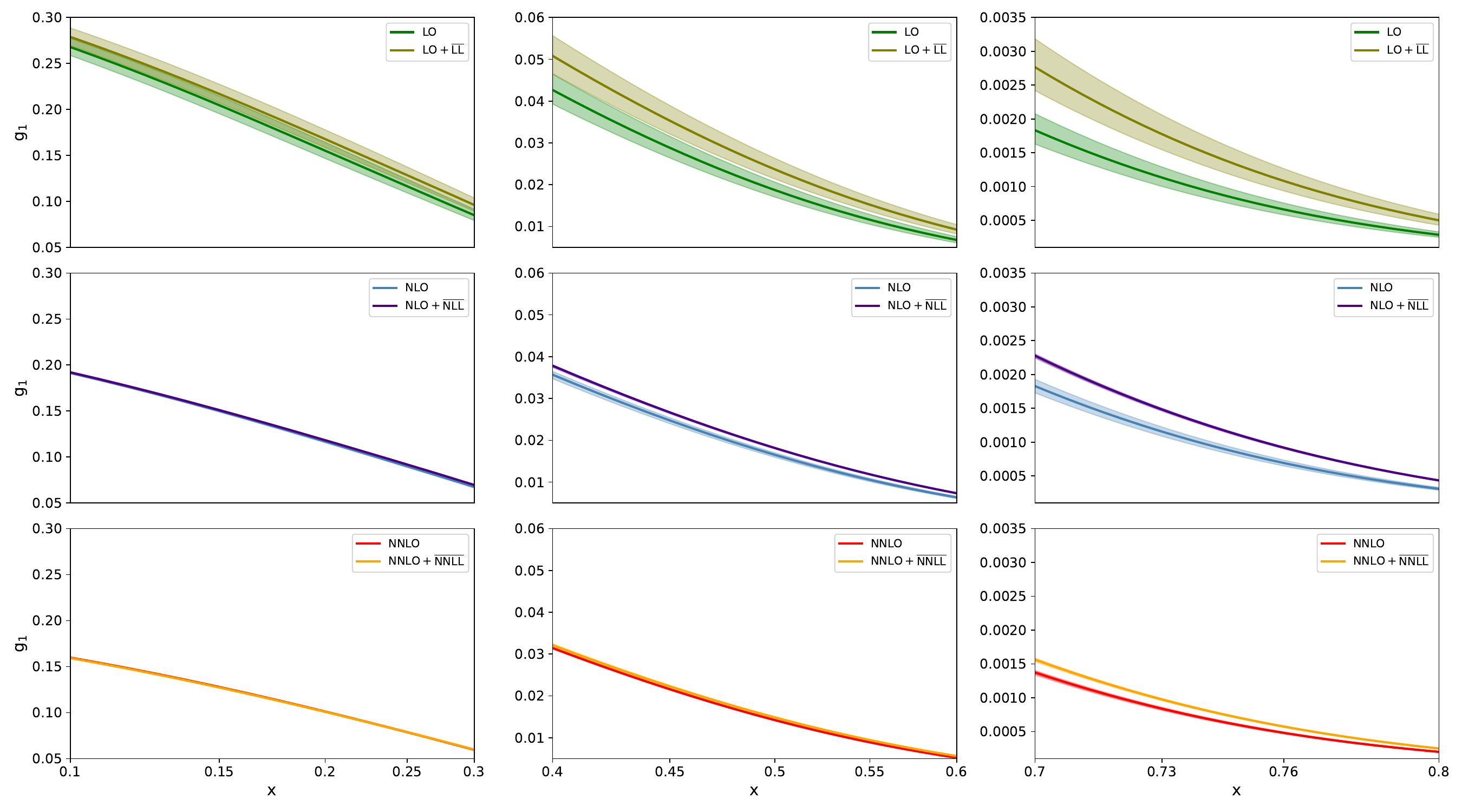}
\caption{Seven-point scale variation of structure function $g_1$ as a function $x$(log-scale) for the EIC at $\sqrt{s}=140$ GeV.}
\label{fig:gx}
\end{figure}
\begin{figure}[t]
\includegraphics[width=0.92\textwidth]{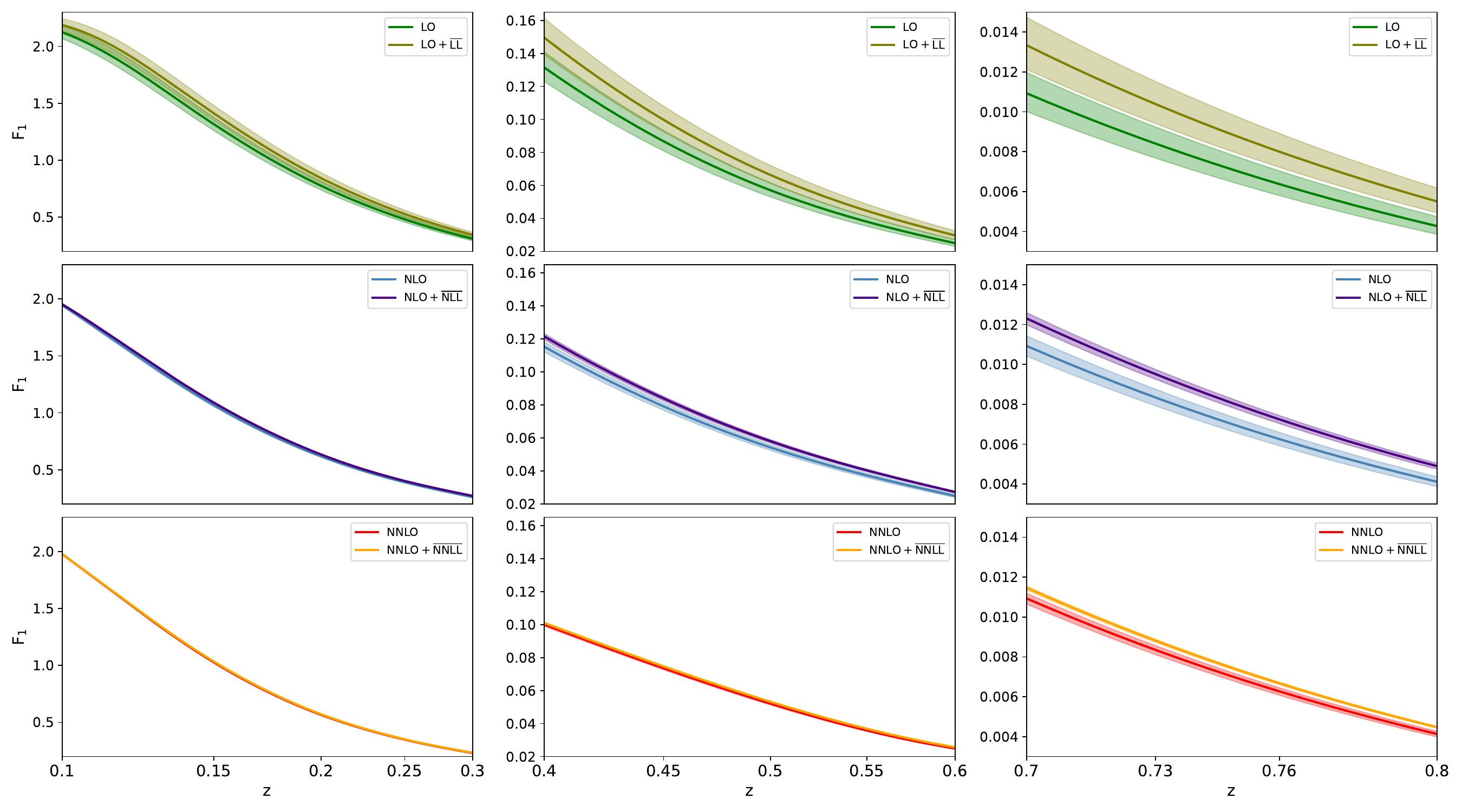}
\caption{Seven-point scale variation of the structure function $F_1$ as a function of $z$ (log-scale) for the EIC at $\sqrt{s}=140$ GeV.}
\label{fig:Fz}
\end{figure}    
\begin{figure}[t]
\includegraphics[width=0.92\textwidth]{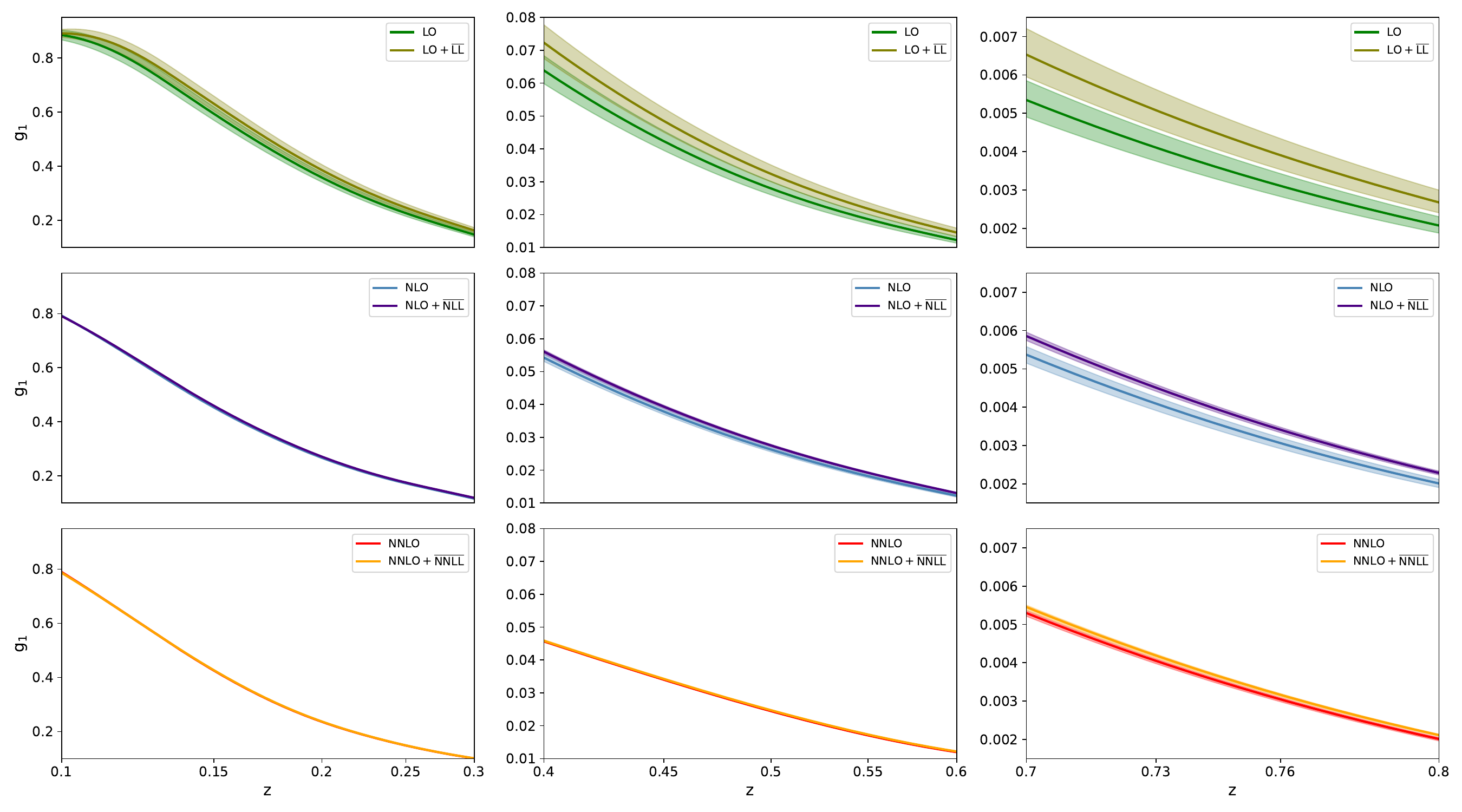}
\caption{Seven-point scale variation of structure function $g_1$ for  $z$(log-scale) for the EIC at $\sqrt{s}=140$ GeV.}
\label{fig:gz}
\end{figure}
\begin{figure}[t]
\includegraphics[width=0.96\textwidth]{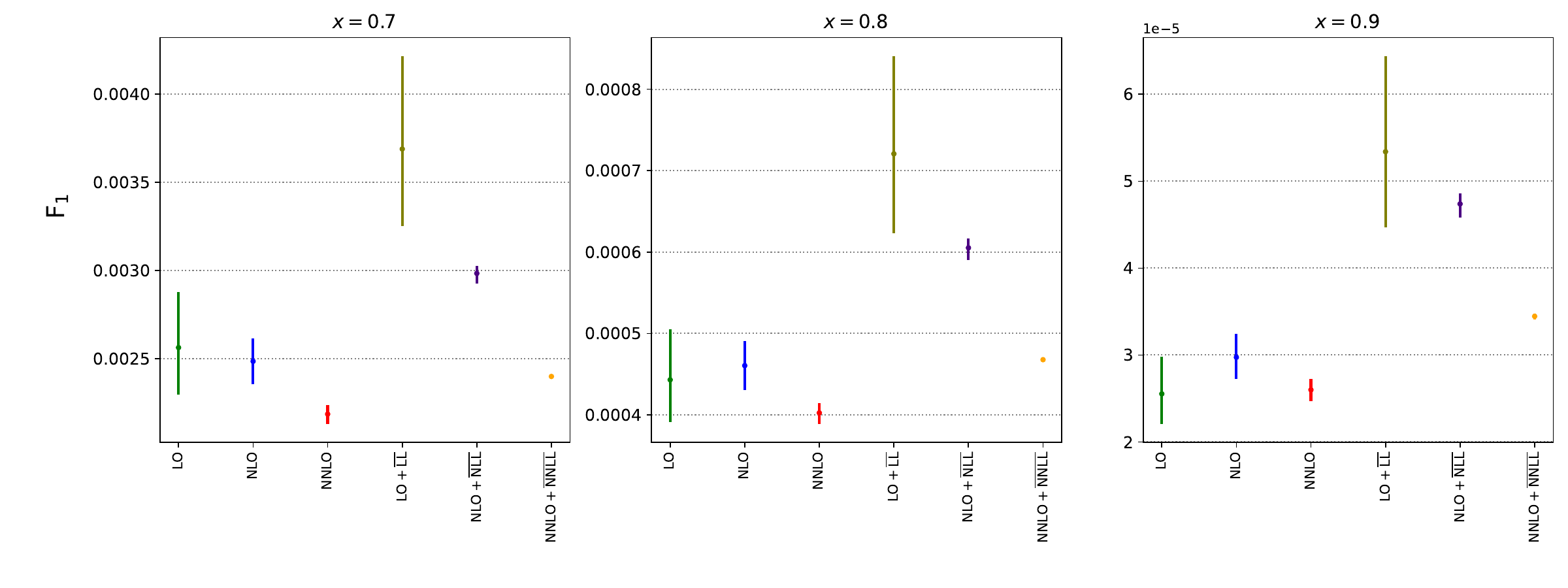}
\caption{Contributions from various perturbative orders to
$F_1$ for three different $x$ values for the EIC at $\sqrt{s}=140$ GeV.}
\label{fig:Fx_7_9}
\end{figure}
\begin{figure}[t]
\includegraphics[width=0.9\textwidth]{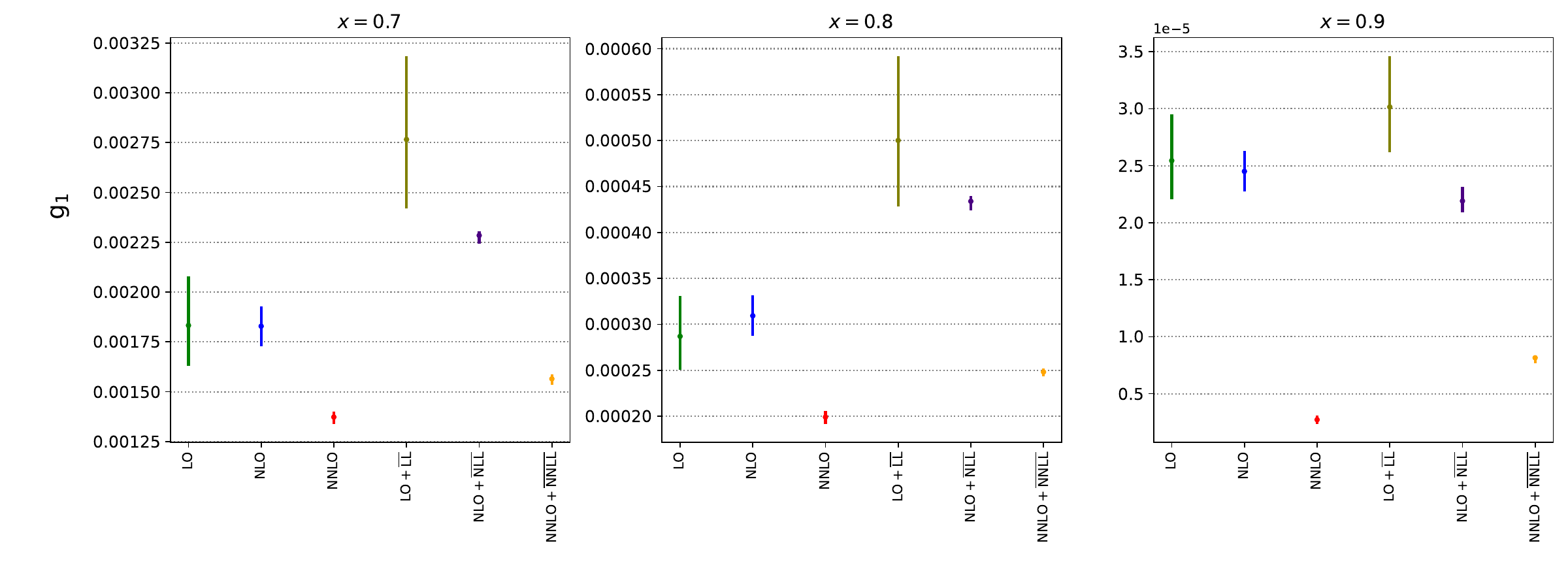}
\caption{
Plots of the structure function $g_1$ against various perturbative orders at different $x$ values , $x$=\{0.7,0.8,0.9\} for the EIC at $\sqrt{s}=140$ GeV. The y-axis of the rightmost plot (for $x = 0.9$) is scaled by $10^5$.} 
\label{fig:gx_7_9}
\end{figure}
\begin{figure}[t]
\includegraphics[width=0.9\textwidth]{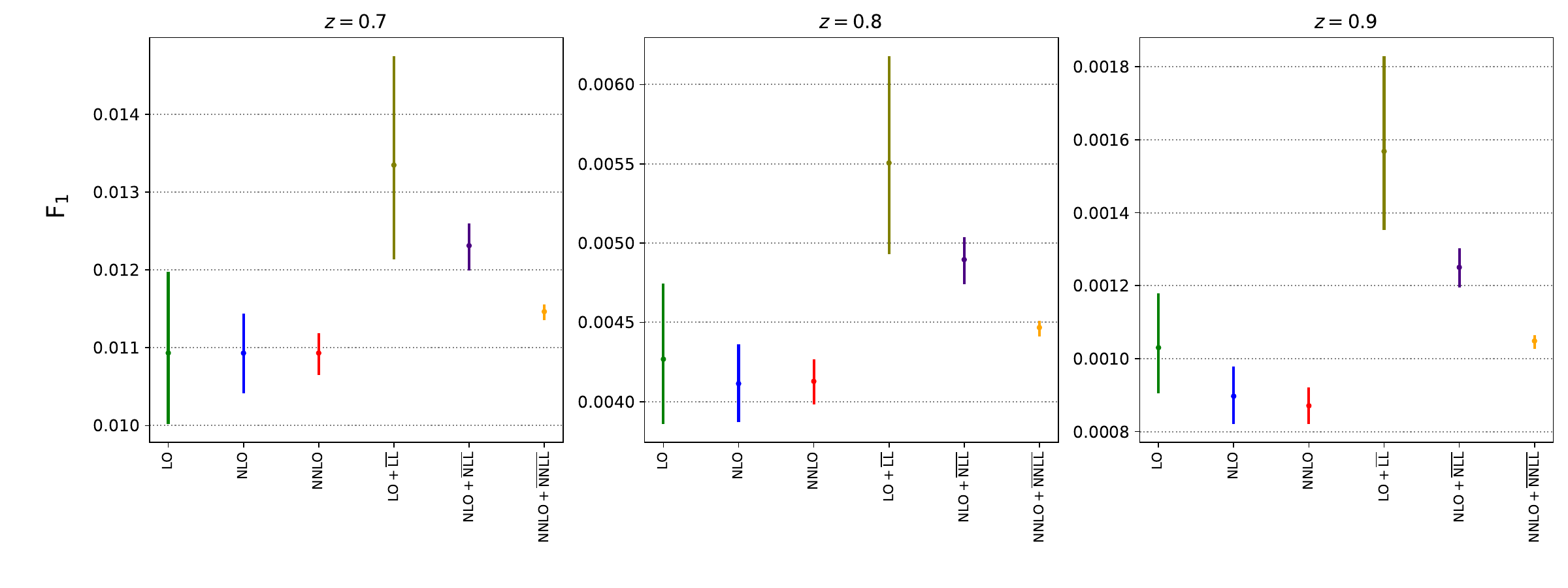}
\includegraphics[width=0.9\textwidth]{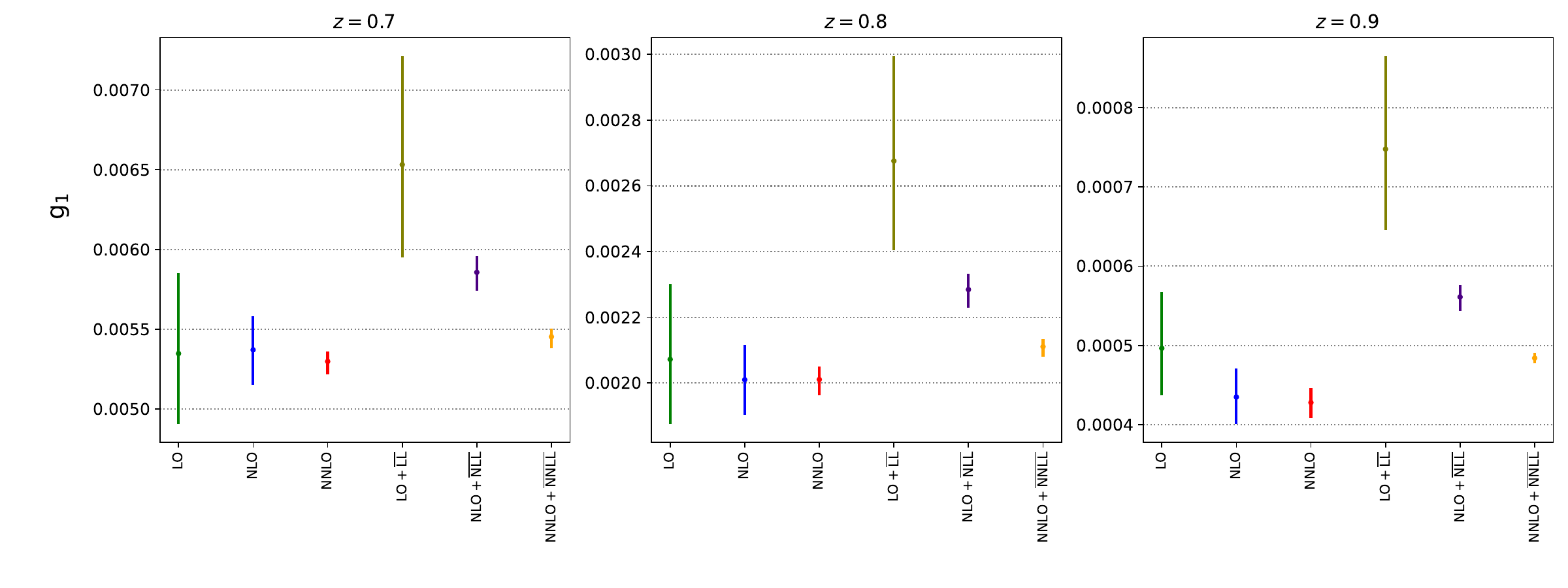}
\caption{Various perturbative contributions to  
$F_1$ (upper) and $g_1$ (lower) at different $z$ values
for the EIC at $\sqrt{s}=140$ GeV.}
\label{fig:Fz_7_9}
\end{figure}
\end{center}
\end{widetext}
In Figs.~\ref{fig:Fx} and \ref{fig:gx}, we present the structure functions $F_1$ and $g_1$ for the EIC 
at $\sqrt{s}= 140$ GeV center-of-mass energy as a function of $x$ after integrating over the kinematic ranges $y \in [0.5, 0.9]$ and $z \in [0.2, 0.85]$. 
The displayed curves correspond to fixed-order QCD predictions up to NNLO accuracy 
and the resummed cross sections to NNLO+$\overline{\text{NNLL}}$ accuracy. 
The associated bands indicates the variation of the $\mu_R^2$ and $\mu_F^2$ scales independently by a factor of two, i.e.\, $\mu_{R}^2$, $\mu_{F}^2$ $\in [Q_{\text{avg}}^2/2, 2 Q_{\text{avg}}^2]$ with the constraint $1/2 \leq \mu_R^2/\mu_F^2 \leq 2$ in the so-called seven-point scale variation, where $Q^2_{avg} = x y_{avg} s$. 

Figs.~\ref{fig:Fz} and \ref{fig:gz} display again $F_1$ and $g_1$ for the EIC at $\sqrt{s}= 140$ GeV, 
now as a function of $z$ after integrating over the kinematic ranges $x \in [0.1, 0.8]$ and $y \in [0.5, 0.9]$, and again showing predictions both at fixed-order NNLO and at NNLO+$\overline{\text{NNLL}}$ accuracies. 
The corresponding bands illustrate the uncertainty from a seven-point scale variation, 
where $\mu_R^2$ and $\mu_F^2$ are independently varied by a factor of two,
i.e. $\mu_{R}^2$, $\mu_{F}^2$ $\in [Q_{\text{avg}}^2/2, 2 Q_{\text{avg}}^2]$, 
where $Q^2_{avg} = x_{avg} y_{avg} s$.

Figs.~\ref{fig:Fx} and \ref{fig:Fz} (\ref{fig:gx} and \ref{fig:gz}) demonstrate an increase of the cross-sections that incorporate the SV+NSV resummed results togetjer with the fixed-order predictions as the logarithmic accuracy improves. 
In addition, we find that the uncertainty band, obtained from variations of the renormalization and factorization scales, shrinks considerably from LO+$\overline{\text{LL}}$ to NNLO+$\overline{\text{NNLL}}$ accuracy. 
This improvement in scale independence is more significant in the large-$x$ or large-$z$ regions e.g. $x > 0.7 $  or $z > 0.7$, due to dominance of the threshold resummation.

The impact of resummation on $F_1$ and $g_1$ is illustrated for three distinct values of $x$ 
($x$=\{0.7, 0.8, 0.9\}) in Figs.~\ref{fig:Fx_7_9} and \ref{fig:gx_7_9}, respectively, when compared to fixed-order perturbative QCD predictions, 
and keeping all other parameters consistent with those in Fig.~\ref{fig:Fx},
e.g., $y$ and $z$ are integrated over the kinematic ranges $y \in [0.5, 0.9]$ and $z \in [0.2, 0.85]$. 
Similarly, in Fig ~\ref{fig:Fz_7_9}, we plot $F_1$ and $g_1$ for three
distinct values of $z$ ($z$=\{0.7, 0.8, 0.9\}) and compare to the different fixed perturbative orders, 
while $y$ and $x$ have been integrated as in the Fig.\ref{fig:Fz}. 
In all these plots (Figs.~\ref{fig:Fx_7_9}--\ref{fig:Fz_7_9}), the vertical lines represent a seven-point $\mu_{R}^2$, $\mu_{F}^2$ scale variation. 
Note, that there is a ignificant scale dependence in the LO+$\overline{\text{LL}}$ predictions in Figs.~\ref{fig:Fx_7_9}--\ref{fig:Fz_7_9}, which arises because the LO contribution
inherently lacks renormalization scale dependence. It then introduced solely by 
the $\overline{\text{LL}}$ terms without counteracting contributions.
These plots clearly demonstrate how, in the large $x$ and/or large $z$ regions, the contribution of resummed terms is significant. 
Furthermore, they illustrate, how resummed predictions substantially reduce the theoretical uncertainties arising from the choice of $\mu_R$ and $\mu_F$.


Finally, in Tabs.~\ref{tab:FG_x} and \ref{tab:FG_z}, we present the explicit results for $F_1$ at a few values of $x$ and $z$ after integrating $y$ and $z$ or $x$, respectively, and setting $\mu_R^2=\mu_F^2=Q^2_{\rm avg}$  as was done for the Figs.~\ref{fig:Fx} and \ref{fig:Fz}. 
A similar table can be constructed for the polarized case; however, we omit it  for brevity.

%
%
\onecolumngrid

\begin{table*}[!hbt]
\vspace*{2mm}
\begin{tabular}{|p{1.0cm}|p{2.5cm}|p{2.5cm}|p{2.5cm}||p{2.5cm}|p{2.5cm}|p{2.5cm}|}
\hline
\centering
  & \multicolumn{5}{c}{$F_1$} & \\
\cline{2-4} \cline{5-7}
\centering
  $x$ &\vspace{0.015cm} LO & \vspace{0.015cm} NLO &\vspace{0.015cm} NNLO &\vspace{0.015cm} LO+$\overline{\text{LL}}$ & \vspace{0.015cm} NLO+$\overline{\text{NLL}}$ &\vspace{0.015cm} 
 NNLO+$\overline{\text{NNLL}}$\\
\hline
 $0.15$  & 
 $0.5044^{+5.971\%}_{-5.389\%}$ & 
 $0.3980^{+1.237\%}_{-1.557\%}$ &
 $0.3526^{+0.570\%}_{-0.700\%}$ &
 $0.5518^{+6.393\%}_{-5.740\%}$ & 
 $0.4077^{+0.715\%}_{-0.934\%}$ &
 $0.3541^{+0.475\%}_{-0.357\%}$       \\
 \hline
 $0.25$   & 
 $0.2168^{+7.160\%}_{-6.375\%}$ & 
 $0.1751^{+1.758\%}_{-2.104\%}$ &
 $0.1569^{+0.719\%}_{-0.899\%}$ & 
 $0.2433^{+7.698\%}_{-6.809\%}$ & 
 $0.1807^{+0.919\%}_{-1.189\%}$ & 
 $0.1579^{+0.484\%}_{-0.397\%}$
        \\
  \hline
$0.45$ & 
$0.0426^{+9.307\%}_{-8.104\%}$ & 
$0.0368^{+2.984\%}_{-3.301\%}$ & 
$0.0331^{+1.180\%}_{-1.452\%}$ & 
$0.0518^{+10.294\%}_{-8.862\%}$ & 
$0.0395^{+1.121\%}_{-1.492\%}$ & 
$0.0341^{+0.244\%}_{-0.360\%}$   
    \\
  \hline
 $0.6$   & 
 $0.0094^{+10.991\%}_{-9.400\%}$ & 
 $0.0087^{+4.157\%}_{-4.360\%}$ & 
 $0.0077^{+1.738\%}_{-2.044\%}$ & 
 $0.0125^{+12.509\%}_{-10.521\%}$ & 
 $0.0098^{+1.269\%}_{-1.740\%}$ & 
 $0.0082^{+0.123\%}_{-0.347\%}$
       \\
  \hline
 $0.75$   & 
 $0.0012^{+13.053\%}_{-10.968\%}$ & 
 $0.0012^{+5.773\%}_{-5.787\%}$  &
 $0.0010^{+2.618\%}_{-2.957\%}$  & 
 $0.0017^{+15.374\%}_{-12.620\%}$ &
 $0.0014^{+1.604\%}_{-2.160\%}$  &
 $0.0011^{+0.349\%}_{-0.465\%}$   
         \\
  \hline\hline
\end{tabular}
\caption{
Values of resummed structure function $F_1$ at various orders in comparison to the
fixed order results at the central scale $\mu_R^2$ = $\mu_F^2$ = $Q^2_{\text{avg}}$ as function of $x$ (see text for integration ranges of $y$, $z$).}
\label{tab:FG_x}
\end{table*}
\clearpage

\begin{table*}[!hbt]
\vspace*{2mm}
\begin{tabular}{|p{1.0cm}|p{2.5cm}|p{2.5cm}|p{2.5cm}||p{2.5cm}|p{2.5cm}|p{2.5cm}|}
\hline
\centering
  & \multicolumn{5}{c}{$F_1$} & \\
\cline{2-4} \cline{5-7}
\centering
  $z$ &\vspace{0.015cm} LO & \vspace{0.015cm} NLO &\vspace{0.015cm} NNLO &\vspace{0.015cm} LO+$\overline{\text{LL}}$ & \vspace{0.015cm} NLO+$\overline{\text{NLL}}$ &\vspace{0.015cm} 
 NNLO+$\overline{\text{NNLL}}$\\
\hline
 $0.15$  & 
 $1.3176^{+4.645\%}_{-4.259\%}$ & 
 $1.0682^{+1.051\%}_{-1.263\%}$ &
 $1.0282^{+0.441\%}_{-0.554\%}$ &
 $1.4127^{+4.927\%}_{-4.497\%}$ & 
 $1.0908^{+0.522\%}_{-0.669\%}$ &
 $1.0350^{+0.214\%}_{-0.176\%}$       \\
 \hline
 $0.25$   & 
 $0.4817^{+5.822\%}_{-5.261\%}$ & 
 $0.3909^{+1.479\%}_{-1.712\%}$ &
 $0.3476^{+0.609\%}_{-0.744\%}$ & 
 $0.5286^{+6.257\%}_{-5.619\%}$ & 
 $0.4020^{+0.759\%}_{-0.950\%}$ & 
 $0.3507^{+0.270\%}_{-0.304\%}$
        \\
  \hline
$0.45$ & 
$0.0866^{+7.581\%}_{-6.726\%}$ & 
$0.0790^{+2.765\%}_{-2.945\%}$ & 
$0.0741^{+1.096\%}_{-1.291\%}$ & 
$0.0997^{+8.307\%}_{-7.299\%}$ & 
$0.0838^{+1.494\%}_{-1.715\%}$ & 
$0.0752^{+0.439\%}_{-0.540\%}$   
    \\
  \hline
 $0.6$   & 
 $0.0251^{+8.644\%}_{-7.598\%}$ & 
 $0.0251^{+3.797\%}_{-3.905\%}$ & 
 $0.0247^{+1.790\%}_{-2.003\%}$ & 
 $0.0299^{+9.544\%}_{-8.297\%}$ & 
 $0.0275^{+1.972\%}_{-2.208\%}$ & 
 $0.0255^{+0.680\%}_{-0.817\%}$
       \\
  \hline
 $0.75$   & 
 $0.0070^{+10.248\%}_{-8.889\%}$ & 
 $0.0069^{+5.253\%}_{-5.220\%}$  &
 $0.0069^{+2.755\%}_{-2.980\%}$  & 
 $0.0087^{+11.255\%}_{-9.656\%}$ &
 $0.0079^{+2.588\%}_{-2.828\%}$  &
 $0.0073^{+0.877\%}_{-1.093\%}$    
       \\
  \hline\hline  
\end{tabular}
\caption{
Values of resummed structure function $F_1$ at various orders in comparison to the
fixed order results at the central scale $\mu_R^2$ = $\mu_F^2$ = $Q^2_{\text{avg}}$ as function of $z$ (see text for integration ranges of $x$, $y$).}
\label{tab:FG_z}
\end{table*}

\twocolumngrid




\section{Conclusions}
\label{sec:concl}  
Threshold-enhanced logarithms, present in the fixed order predictions, need to be resummed in order to obtain reliable phenomenological predictions in the relavant kinematic range.
SIDIS is characterized by the scaling variables $x'$, or $z'$, which tned to unity in threshold region. 
Fixed-order perturbative predictions of the CFs are then dominated by the plus  and delta function distributions (SV terms) as well as $\text{log}^j(1-x')$ or $\text{log}^j(1-z')$  (NSV terms) with $j=0,1,2,\cdots$.
This is a situation similar to the case of rapidity distributions in the DY process. 
Based on the insights there, we have developed a formalism, using collinear factorisation and renormalisation group equations, to achieve the resummation of these threshold logarithms (SV as well as NSV) both in $(x',z')$ space as well as in the space of the conjugate Mellin variables $\vec N =(N_1,N_2)$. 
In Mellin space, these threshold logarithms show up as $\log^i(N_1)\log^j(N_2)$ for SV as well as $\log^i(N_1)\log^j(N_2)/N_2$, $\log^i(N_1)\log^j(N_2)/N_1$ where $\{i,j\} = 0,1,2,\cdots$ and they can be systematically  resummed to all orders in $a_s$.  

The compact expressions for the resummed results encapsulate both SV and NSV logarithms in $\vec N$-space and allow for phenomenological studies with EIC kinematics.
Upon expansion in $a_s$ the resummed result generates predictions for the full SV and partial NSV terms for the CFs $(\Delta)\mathcal{C}^{(3),\text{SV}}_{J,qq}$,
$(\Delta)\mathcal{C}^{(3),\text{NSV}}_{J,qq}$ 
 third order in $(x',z')$ space. 
 At fourth order, using the recently obtained results for four-loop quark virtual and eikonal anomalous dimensions $B_4^q$ and $f_4^q$, 
 we have predicted all SV terms in $(\Delta)\mathcal{C}^{(4),\text{SV}}_{J,qq}$ except  the coefficient of $\delta_{\overline{x}'}\delta_{\overline{z}'}$
The derivation of the NSV terms has been corrobrated by an independent approach that uses PEKs for the structure functions, and 
we have confirmed our predictions both at third and fourth order.  Finally, we have demonstrated the numerical importance of the resummed threshold logarithms for structure functions $F_1$ and $g_1$ at EIC energies. 
We have shown, that at each logarithmic order the resummed contributions are larger than the corresponding fixed order ones. 
Moreover, including these logarithms to all orders through resummation reduces the dependence on the renormalisation and factorisation scales and hence improves the reliability of our predictions. 


\noindent\hrulefill
\section*{Acknowledgements}
This work has been supported through a joint Indo-German research grant by
the Department of Science and Technology (DST/INT/DFG/P-03/2021/dtd.12.11.21). 
S.M. acknowledges the ERC Advanced Grant 101095857 {\it Conformal-EIC}.
N.R. acknowledges the SERB-SRG under Grant No. SRG/2023/000591.
In addition, we also thank the computer administrative unit of IMSc for their help and support.

\appendix
\begin{widetext}

\section{Coefficients in $\Gamma_{\hat{F}_{q}}$:$K_d^{q}$ and ${G}_d^{q}$}\label{FKG}
The coefficients $ K_d^q$ to fourth order in $\hat{a}_s$ are given below:
\begin{align}
K_{d}^{q}(\mu_R^2,\varepsilon) =\sum_{i=0}^{\infty}\hat{a}_s^{i}\bigg(\frac{\mu_R^2}{\mu^2}\bigg)^{i\frac{\varepsilon} {2}}S_{\varepsilon}^i K^{q(i)}_{d}(\varepsilon)
\, ,
\end{align}
where
\begin{align}
\label{eq:App-SolnKFF}
{K}^{q(1)}_{d}(\varepsilon) &= \frac{1}{\varepsilon} \Bigg\{- 2 A^q_{
                       1}\Bigg\}\,,
\nonumber\\
{K}^{q(2)}_{d}(\varepsilon) &= \frac{1}{\varepsilon^2} \Bigg\{ 2 \beta_0
                       A^q_{1} \Bigg\} + \frac{1}{\varepsilon} \Bigg\{-
                       A^q_{ 2}\Bigg\}\,,
\nonumber\\
{K}^{q(3)}_{d}(\varepsilon) &= \frac{1}{\varepsilon^3} \Bigg\{ -\frac{8
                       }{3} \beta_0^2 A^q_{1} \Bigg\} +
                       \frac{1}{\varepsilon^2} \Bigg\{ \frac{2}{3}  \beta_1
                       A^q_{1} + \frac{8}{3}  \beta_0 A^q_{2}
                       \Bigg\} + \frac{1}{\varepsilon} \Bigg\{ -\frac{2
                       }{3} A^q_{ 3} \Bigg\}\,,
\nonumber\\
{K}^{q(4)}_{d}(\varepsilon) &= \frac{1}{\varepsilon^4} \Bigg\{ 4 \beta_0^3
                       A^q_{1} \Bigg\} + \frac{1}{\varepsilon^3}
                       \Bigg\{ -\frac{8}{3} \beta_0 \beta_1 A^q_{1}
                       -6 \beta_0^2 A^q_{2} \Bigg\} +
                       \frac{1}{\varepsilon^{2}} \Bigg\{ \frac{1}{3} \beta_2
                       A^q_{1} + \beta_1 A^q_{2} + 3 \beta_0
                       A^q_{3} \Bigg\}
\nonumber\\
&+ \frac{1}{\varepsilon} \Bigg\{-
                       \frac{1}{2} A^q_{4} \Bigg\}
\, .
\end{align}
The constants $A^q_i$ are the $a_s^i$ coefficients of the cusp anomalous dimension $A^q$:
\begin{equation}
    A^q(a_s(\mu_R^2)) = \sum_{i} ~a_s^i(\mu_R^2)~ A^q_i \,.
\end{equation}
The finite terms $G^{q}_d$ can be written as
\begin{align}
G_{d}^{q}(\hat{a}_s,Q^2,\mu_R^2,\varepsilon) = G_{d}^{q}(a_s(Q^2),\varepsilon)  + \sum_{i=1}^{\infty}\hat{a}_s^i\bigg(\frac{Q^2}{\mu^2}\bigg)^{i\frac{\varepsilon}{2}}S_{\varepsilon}^i~K_{d}^{q,(i)}(\varepsilon) - K_{d}^{q}(\mu_R^2,\varepsilon) 
\, .
\end{align}
The first terms is expanded as: $G_{d}^{q}(a_s(Q^2),\varepsilon) = \sum_{i=1}^{\infty} a_s^i(Q^2){G}_{d,i}^{q}(\varepsilon)$,  where 
\begin{align}\label{eq:gsv}
{G}^{q}_{d,i}(\varepsilon)=2B^q_i+ f_i^q+  \chi_{d,i}^q +
\sum_{j=1}^\infty \varepsilon^j  {\cal G}^{q,(j)}_{d,i}\,,
\end{align}
with
\begin{align}
  \label{eq:gsvi}
\chi^{q}_{d,1} &= 0\, ,
  \chi^{q}_{d,2} = - 2 \beta_{0}  {\cal G}^{q,(1)}_{d,1}\, ,
\chi^{q}_{d,3} = - 2 \beta_{1} {\cal G}^{q,(1)}_{d,1} - 2
        \beta_{0} \left({\cal G}^{q,(1)}_{d,2}  + 2 \beta_{0}{\cal G}^{q,(2)}_{d,1}\right)\, ,
                \nonumber\\
\chi^{q}_{d,4} &= - 2 \beta_{2} {\cal G}^{q,(1)}_{d,1} - 2
        \beta_{1} \left( {\cal G}^{q,(1)}_{d,2}  + 4 \beta_{0}{\cal G}^{q,(2)}_{d,1}\right)
        - 2\beta_{0} 
    \left( {\cal G}^{q,(1)}_{d,3}  + 2 \beta_{0} {\cal G}^{q,(2)}_{d,2} + 4 \beta_{0}^2 {\cal G}^{q,(3)}_{d,1}\right)
                \,.
\end{align}
Next, the results for ${\cal G}^{q,(j)}_{d,i}$ are given below 
%
\begin{align}
{\cal G}^{q,(1)}_{d,1} &= C_F   \Bigg\{  - 8 + \zeta_2 \Bigg\}\, ,\quad\quad
{\cal G}^{q,(2)}_{d,1} =C_F   \Bigg\{ 8 - \frac{7}{3} \zeta_3 - \frac{3}{4}  \zeta_2 \Bigg\}\, ,\quad \quad
{\cal G}^{q,(3)}_{d,1} =C_F   \Bigg\{  - 8 + \frac{7}{4} \zeta_3 + \zeta_2 + \frac{47}{80} \zeta_2^2 \Bigg\} \, ,\nonumber\\
{\cal G}^{q,(4)}_{d,1} &= C_F   \Bigg\{ 8 - \frac{31}{20} \zeta_5 -            \frac{7}{3} \zeta_3 - \zeta_2 + \frac{7}{24} \zeta_2    \zeta_3 - \frac{141}{320} \zeta_2^2 \Bigg\}\, ,\nonumber\\
{\cal G}^{q,(5)}_{d,1} &= C_F   \Bigg\{  - 8 + \frac{93}{80} \zeta_5 +         \frac{7}{3} \zeta_3 - \frac{49}{144} \zeta_3^2 + \zeta_2 -    \frac{7}{32}
          \zeta_2 \zeta_3 + \frac{47}{80} \zeta_2^2 +         \frac{949}{4480} \zeta_2^3 \Bigg\} \, ,\nonumber\\
{\cal G}^{q,(1)}_{d,2} &=C_F n_f   \Bigg\{ \frac{5813}{162} - \frac{8}{3}     \zeta_3 + \frac{37}{9} \zeta_2 \Bigg\}
       - C_F C_A   \Bigg\{   \frac{70165}{324} -             \frac{260}{3} \zeta_3 + \frac{575}{18} \zeta_2 - \frac{88}{5} \zeta_2^2 \Bigg\}
       - C_F^2   \Bigg\{   \frac{1}{4} + 60 \zeta_3 - 58     \zeta_2 + \frac{88}{5} \zeta_2^2 \Bigg\} \, ,\nonumber\\
{\cal G}^{q,(2)}_{d,2} &= C_F n_f   \Bigg\{  - \frac{129389}{1944} +           \frac{301}{27} \zeta_3 - \frac{425}{54} \zeta_2 +             \frac{7}{12} \zeta_2^2
          \Bigg\}
       - C_F^2   \Bigg\{   \frac{109}{16} - 12 \zeta_5 - 184 \zeta_3 + \frac{437}{4} \zeta_2 + 28 \zeta_2
         \zeta_3 - \frac{108}{5} \zeta_2^2 \Bigg\}
\nonumber\\&
         + C_F C_A   \Bigg\{ \frac{1547797}{3888} - 51          \zeta_5 - \frac{12479}{54} \zeta_3 + \frac{7297}{108} \zeta_2
          + \frac{89}{3} \zeta_2 \zeta_3 - \frac{653}{24}     \zeta_2^2 \Bigg\}\, ,
\nonumber\\
{\cal G}^{q,(3)}_{d,2} &=C_F n_f   \Bigg\{ \frac{2628821}{23328} - 6          \zeta_5 - \frac{4085}{162} \zeta_3 + \frac{8405}{648} \zeta_2
          - \frac{11}{6} \zeta_2 \zeta_3 - \frac{1873}{720}   \zeta_2^2 \Bigg\}
       - C_F C_A   \Bigg\{   \frac{31174909}{46656} - 105    \zeta_5 - \frac{154405}{324} \zeta_3 
\nonumber\\&       
       + \frac{569}{6}
         \zeta_3^2 + \frac{155701}{1296} \zeta_2 +            \frac{239}{12} \zeta_2 \zeta_3 - \frac{100907}{1440}
         \zeta_2^2 + \frac{809}{140} \zeta_2^3 \Bigg\}
       + C_F^2   \Bigg\{ \frac{1287}{64} - 9 \zeta_5 -        \frac{1807}{4} \zeta_3 + 134 \zeta_3^2 + \frac{2991}{16}
         \zeta_2 
\nonumber\\&         
         + 13 \zeta_2 \zeta_3 - \frac{127}{2}         \zeta_2^2 + \frac{858}{35} \zeta_2^3 \Bigg\}\, ,\nonumber\\
{\cal G}^{q,(4)}_{d,2} &=C_F n_f   \Bigg\{  - \frac{50947325}{279936} +       \frac{3109}{180} \zeta_5 + \frac{91985}{1944} \zeta_3 -
         \frac{289}{108} \zeta_3^2 - \frac{160493}{7776}      \zeta_2 + \frac{791}{216} \zeta_2 \zeta_3 + \frac{12781}{2160} \zeta_2^2 + \frac{953}{480} \zeta_2^3 \Bigg\}
\nonumber\\&       
       + C_F C_A   \Bigg\{ \frac{599567077}{559872} + 93      \zeta_7 - \frac{96503}{360} \zeta_5 - \frac{3643249}{3888} \zeta_3 + \frac{32339}{216} \zeta_3^2 +         \frac{3110773}{15552} \zeta_2 + \frac{497}{20} \zeta_2
         \zeta_5 + \frac{16739}{432} \zeta_2 \zeta_3 
\nonumber\\&         
         -        \frac{619829}{4320} \zeta_2^2 + \frac{7103}{120}
         \zeta_2^2 \zeta_3 - \frac{7073}{1344} \zeta_2^3      \Bigg\}
       - C_F^2   \Bigg\{   \frac{10901}{256} + \frac{675}{2} \zeta_7 - 99 \zeta_5 - \frac{50219}{48} \zeta_3 + \frac{373}{2} \zeta_3^2 + \frac{19405}{64} \zeta_2 + 3 \zeta_2   \zeta_5 
\nonumber\\&       
       + \frac{235}{6} \zeta_2 \zeta_3 -
         \frac{24453}{160} \zeta_2^2 + \frac{2597}{30}        \zeta_2^2 \zeta_3 + \frac{123}{4} \zeta_2^3 \Bigg\}\, ,\nonumber\\   
{\cal G}^{q,(1)}_{d,3} &= C_F N_{4} n_{fv}   \Bigg\{ 12 - 80 \zeta_5 + 14 \zeta_3 + 30 \zeta_2 -    \frac{6}{5} \zeta_2^2 \Bigg\}
       - C_F n_f^2   \Bigg\{   \frac{258445}{2187} - \frac{536}{81}   \zeta_3 + \frac{3466}{81} \zeta_2 + \frac{40}{9}
         \zeta_2^2 \Bigg\}
\nonumber\\&       
       - C_F C_A^2   \Bigg\{   \frac{48902713}{8748} - \frac{688}{3}         \zeta_5 - \frac{85883}{18} \zeta_3 + \frac{1136}{3}
         \zeta_3^2 + \frac{1083305}{486} \zeta_2 - \frac{1786}{9} \zeta_2     \zeta_3 - \frac{37271}{90} \zeta_2^2
          + \frac{6152}{63} \zeta_2^3 \Bigg\}         
\nonumber\\&       
       + C_F^2 C_A   \Bigg\{ \frac{230}{3} - \frac{3020}{3} \zeta_5 -         \frac{23402}{9} \zeta_3 + 296 \zeta_3^2 +
         \frac{55499}{18} \zeta_2 - \frac{3448}{3} \zeta_2 \zeta_3 +          \frac{2432}{45} \zeta_2^2 - \frac{15448}{105}
         \zeta_2^3 \Bigg\}
\nonumber\\&    
       - C_F^3   \Bigg\{   \frac{1527}{4} - 1992 \zeta_5 + 2130 \zeta_3 - 48 \zeta_3^2 + 206 \zeta_2
          - 840 \zeta_2 \zeta_3 + 534 \zeta_2^2 - \frac{21584}{105} \zeta_2^3 \Bigg\}           
\nonumber\\&       
       + C_F C_A n_f   \Bigg\{ \frac{3702974}{2187} - 72 \zeta_5 -            \frac{68660}{81} \zeta_3 + \frac{155008}{243}
         \zeta_2 + \frac{392}{9} \zeta_2 \zeta_3 - \frac{1298}{45} \zeta_2^2  \Bigg\}
\nonumber\\&   
       + C_F^2 n_f   \Bigg\{ \frac{73271}{162} - \frac{368}{3} \zeta_5 +      \frac{19700}{27} \zeta_3 - \frac{7541}{18} \zeta_2
          - \frac{152}{3} \zeta_2 \zeta_3 - \frac{704}{45} \zeta_2^2 \Bigg\}
\, .
\end{align}

\section{Coefficients in $\Phi^{\text{SV}}_{q}$: $\overline K^q_{d}$ and $\overline{G}^q_d$}
\label{AppendixKbGb}
Here we present the SV coefficients in $\Phi_{q}$.
We expand $\overline K_{d}^{q}(\mu_R^2,x',z',\varepsilon) =\delta(1-x')\delta(1-z')\sum_{i=0}^{\infty}\hat{a}_s^{i}\bigg(\frac{\mu_R^2}{\mu^2}\bigg)^{i\frac{\varepsilon} {2}}S_{\varepsilon}^i \overline K^{q(i)}_{d}(\varepsilon)$  to fourth order in $a_s$, where 
%
\begin{align}
\label{eq:App-SolnK}
\overline{K}^{q(1)}_{d}(\varepsilon) &= \frac{1}{\varepsilon} \Bigg\{ 2 A^q_{
                       1}\Bigg\}\,,
\nonumber\\
\overline{K}^{q(2)}_{d}(\varepsilon) &= \frac{1}{\varepsilon^2} \Bigg\{ -2 \beta_0
                       A^q_{1} \Bigg\} + \frac{1}{\varepsilon} \Bigg\{
                       A^q_{ 2}\Bigg\}\,,
\nonumber\\
\overline{K}^{q(3)}_{d}(\varepsilon) &= \frac{1}{\varepsilon^3} \Bigg\{ \frac{8
                       }{3} \beta_0^2 A^q_{1} \Bigg\} +
                       \frac{1}{\varepsilon^2} \Bigg\{ -\frac{2}{3}  \beta_1
                       A^q_{1} - \frac{8}{3}  \beta_0 A^q_{2}
                       \Bigg\} + \frac{1}{\varepsilon} \Bigg\{ \frac{2
                       }{3} A^q_{ 3} \Bigg\}\,,
\nonumber\\
\overline{K}^{q(4)}_{d}(\varepsilon) &= \frac{1}{\varepsilon^4} \Bigg\{ -4 \beta_0^3
                       A^q_{1} \Bigg\} + \frac{1}{\varepsilon^3}
                       \Bigg\{ \frac{8}{3} \beta_0 \beta_1 A^q_{1}
                       +6 \beta_0^2 A^q_{2} \Bigg\} +
                       \frac{1}{\varepsilon^{2}} \Bigg\{ -\frac{1}{3} \beta_2
                       A^q_{1} - \beta_1 A^q_{2} - 3 \beta_0
                       A^q_{3} \Bigg\}
\nonumber\\
&+ \frac{1}{\varepsilon} \Bigg\{
                       \frac{1}{2} A^q_{4} \Bigg\}
\, .
\end{align}
%
The finite terms $\overline{G}_{d,\text{SV}}^{q(i)}(\varepsilon) $ are related to its renormalised counter parts $\overline {\cal G}_{d,i}^q(\varepsilon)$ in the following way:
\begin{align}
\sum_{i=1}^\infty \hat{a}_s^i \left( \frac{Q^2 (1-x')(1-z')}{\mu^2} \right)^{i \frac{\varepsilon}{2}} S^i_{\varepsilon}~ \overline G_{d,\text{SV}}^{q,(i)}(\varepsilon)
 = \sum_{i=1}^\infty a_s^i \left( Q^2 (1-x')(1-z') \right) \overline{{\cal G}}^{q}_{d,i}(\varepsilon) \,.
\label{Gbar1}
\end{align}
we find,
\begin{eqnarray}
\overline { G}^{q(1)}_{d,\text{SV}}(\varepsilon)&=&\overline {\cal G}_{d,1}^{q}(\varepsilon)
\nonumber\\
\overline { G}^{q(2)}_{d,\text{SV}}(\varepsilon)&=&{1\over \varepsilon} \Bigg(
                  - 2 \beta_0  \overline {\cal G}_{d,1}^{q}(\varepsilon)\Bigg)
                  +\overline {\cal G}_{d,2}^{q}(\varepsilon)
\nonumber\\
\overline { G}^{q(3)}_{d,\text{SV}}(\varepsilon)&=&  {1\over \varepsilon^2} \Bigg(
                    4 \beta_0^2 \overline {\cal G}_{d,1}^{q}(\varepsilon)\Bigg)
                  +{1\over \varepsilon} \Bigg(
                  - \beta_1 \overline {\cal G}_{d,1}^{q}(\varepsilon)
                  -4\beta_0 \overline {\cal G}_{d,2}^{q}(\varepsilon)\Bigg)
                  +\overline {\cal G}_{d,3}^{q}(\varepsilon)
\nonumber\\
\overline {G}^{q(4)}_{d,\text{SV}}(\varepsilon)&=& {1 \over \varepsilon^3} \Bigg(
                    -8 \beta_0^3 \overline {\cal G}_{d,1}^{q}(\varepsilon)\Bigg)
                  +{1 \over \varepsilon^2} \Bigg(
                    {16 \over 3} \beta_0 \beta_1 \overline {\cal G}_{d,1}^{q}(\varepsilon)
                    +12\beta_0^2 \overline {\cal G}_{d,2}^{q}(\varepsilon)\Bigg)
\nonumber\\
&&                  +{1 \over \varepsilon} \Bigg(  -{2 \over 3} \beta_2 \overline {\cal G}_{d,1}^{q}(\varepsilon)
              -2 \beta_1 \overline {\cal G}_{d,2}^{q}(\varepsilon)
                    -6 \beta_0 \overline {\cal G}_{d,3}^{q}(\varepsilon)\Bigg)
                  +\overline {\cal G}_{d,4}^{q}(\varepsilon)\,.
\end{eqnarray}

\section{Details of Mellin moment}\label{DetailMM}
Here we describe how to obtain the Mellin moment of $\mathrm{\Psi}_{d}^q$ given in eq. (\ref{IntSVpNSV}). We decompose the Mellin moment of $\mathrm{\Psi}_{d}^q$ as 
\begin{align}
G_{d,q}^{\vec N}&={\cal N}\int_{0}^{1}dx'x'^{N_1-1}\int_{0}^{1}dz'z'^{N_2-1}\Psi_{d}^q(x',z') 
\nonumber \\
&= \Psi^{q,\vec N}_{d} +\ln\Big(g_{d,0}^q\big(a_s(\mu_F^2)\big)\Big)  \nonumber\\
&= \Bigg[\bigg(\sum_{i=1}^{6}\psi^d_{i}\bigg) + (N_1\leftrightarrow N_2, L^q\leftrightarrow \tilde{L}^q) \Bigg]+\ln\Big(g_{d,0}^q\big(a_s(\mu_F^2)\big)\Big)\, ,
\end{align}
where the symbol ${\cal N}$ indicates we need to keep only ${\text{SV}}$ and ${\text{NSV}}$ terms. The terms $\psi_{i}^d$ in the above equation are given below:
\begin{align}
\psi^d_{1}(N_1) &=  \int_0^1 dx' x'^{N_1-1}
\int_{\mu_F^2}^{Q^2 (1-x')} \frac{d \lambda^2}{\lambda^2}
\Bigg(\frac{A^{q} \Big(a_s\big(\lambda^2\big)\Big)}{(1-x')_{+}}  + L^q\Big(a_s\big(\lambda^2\big) ,(1-x')\Big)\Bigg)\nonumber\\
\psi^d_{2}(N_1) &=\int_0^1 d x' x'^{N_1-1}
\Bigg(\frac{G^q_{d} \Big(a_s\big(Q^2(1-x')\big)\Big)}{(1-x')_{+}} + \varphi_{d,q,f}^q\Big(a_s\big(Q^2(1-x')\big) ,(1-x')\Big)\Bigg)\nonumber\\
\psi^d_{3}(N_1,N_2) &=\frac{1}{2}\int_0^1 dx' x'^{N_1-1}\int_0^1 dz' z'^{N_2-1}
\Bigg(\frac{A^q \Big(a_s\big(Q^2(1-x')(1-z')\big)\Big)}{(1-x')_{+}(1-z')_{+}} \Bigg)\nonumber\\
\psi^d_{4}(N_1,N_2) &=\frac{1}{2}\int_0^1 dx' x'^{N_1-1}\int_0^1 dz' z'^{N_2-1}~Q^2\frac{d}{d Q^2}
\Bigg(\frac{G^q_d \Big(a_s\big(Q^2(1-x')(1-z')\big)\Big)}{(1-x')_{+}(1-z')_{+}} \Bigg)\nonumber\\
\psi^d_{5}(N_1,N_2) &=\int_0^1 dx' x'^{N_1-1}\int_0^1 dz' z'^{N_2-1}
\Bigg(\frac{L^q \Big(a_s\big(Q^2(1-x')(1-z')\big),(1-z')\Big)}{(1-x')_{+}} \Bigg)\nonumber\\
\psi^d_{6}(N_1,N_2) &= \int_0^1 dx' x'^{N_1-1}\int_0^1 dz' z'^{N_2-1}~
Q^2\frac{d}{dQ^2}\Bigg(\frac{\varphi^{q}_{d,q,f} \Big(a_s\big(Q^2(1-x')(1-z')\big),(1-z')\Big)}{(1-x')_{+}} \Bigg)\nonumber
\, .
\end{align}

Following \cite{Laenen:2008ux}, we replace,
\begin{align}
\label{replaceA}
\int_0^1 d\xi \bigg(\frac{{\xi^{N-1}-1}}{1-\xi}\bigg)& \rightarrow \hat{\Gamma}_A\left(N{\frac{d}{dN}}\right) \int_0^1 \frac{d\xi}{1-\xi} ~\theta\left(1-\frac{1} {N}-\xi \right)
\end{align} 
\begin{align}
\label{replaceB}
\int_0^1 d\xi \xi^{N-1}& \rightarrow \frac{1}{N}\hat{\Gamma}_B\left(N{\frac{d}{dN}}\right) \int_0^1 \frac{d\xi}{1-\xi} ~\theta\left(1-\frac{1} {N}-\xi \right) 
\end{align} 
Here, $\xi$ can be $x'$ or $z'$ and corresponding $N$ can be $N_1$ or $N_2$. 
See App.~\ref{expGAGB} for the expansion of $\hat{\Gamma}_A(N\frac{d}{dN})$ and $\hat{\Gamma}_B(N\frac{d}{dN})$ operators. 
We need to apply $\hat{\Gamma}_{A}$ and $\hat{\Gamma}_{B}$ on the integrals through their expansions in terms of  $N \frac{d}{dN}$.
This can be done only after making an  appropriate change of variables and interchange of the order of the integrals.  
For type-$\psi^d_{1}$,
\begin{align}
  \psi_{1}^d(N)  &=  \int_0^1 d\xi \xi^{N-1}
\int_{\mu_F^2}^{Q^2 (1-\xi)} \frac{d \lambda^2}{\lambda^2}
\Bigg(\frac{F^d_{1} \Big(a_s\big(\lambda^2\big)\Big)}{(1-\xi)_{+}}  + \tilde{F}^d_{1}\Big(a_s\big(\lambda^2\big) ,(1-\xi)\Big)\Bigg)\nonumber\\
&=\int_0^1 d\xi \bigg(\frac{\xi^{N-1}-1}{1-\xi}\bigg)
\int_{\mu_F^2}^{Q^2 (1-\xi)} \frac{d \lambda^2}{\lambda^2}
F^d_{1} \Big(a_s\big(\lambda^2\big)\Big) 
+ \int_0^1 d\xi \xi^{N-1}
\int_{\mu_F^2}^{Q^2 (1-\xi)} \frac{d \lambda^2}{\lambda^2}\tilde{F}^d_{1}\Big(a_s\big(\lambda^2\big) ,(1-\xi)\Big)\nonumber\\
&\underset{{\tiny N\rightarrow\infty}}{=}\hat{\Gamma}_{A}\Big(N\frac{d}{dN}\Big) \int_0^{1-\frac{1}{N}} \frac{d\xi}{1-\xi} 
\int_{\mu_F^2}^{Q^2 (1-\xi)} \frac{d \lambda^2}{\lambda^2}
F^d_{1} \Big(a_s\big(\lambda^2\big)\Big) 
\nonumber\\&
+ \frac{1}{N}\hat{\Gamma}_{B}\Big(N \frac{d}{dN}\Big)\int_0^{1-\frac{1}{N}} \frac{d\xi}{1-\xi}
\int_{\mu_F^2}^{Q^2 (1-\xi)} \frac{d \lambda^2}{\lambda^2}\tilde{F}^d_{1}\Big(a_s\big(\lambda^2\big) ,(1-\xi)\Big)
\, .
\end{align}
To simplify further we set $\mu_F^2 = Q^2$ and perform a change of variables by introducing $\mu^2=Q^2(1-\xi)$
\begin{align}
    \int_0^{1-\frac{1}{N}} \frac{d\xi}{1-\xi} 
\int_{Q^2}^{Q^2 (1-\xi)} \frac{d \lambda^2}{\lambda^2} &\rightarrow
 -\int_{\frac{Q^2}{N}}^{Q^2} \frac{d\mu^2}{\mu^2}\int^{Q^2}_{\mu^2}\frac{d\lambda^2}{\lambda^2}\nonumber
\, ,
 \end{align}
and interchange the order of integrations,
\begin{align}
    \int_{\frac{Q^2}{N}}^{Q^2} \frac{d\mu^2}{\mu^2}\int^{Q^2}_{\mu^2}\frac{d\lambda^2}{\lambda^2} = \int_{\frac{Q^2}{N}}^{Q^2}\frac{d\lambda^2}{\lambda^2}\int^{\lambda^2}_{\frac{Q^2}{N}}\frac{d\mu^2}{\mu^2}\nonumber
\, .
\end{align}
These manipulations lead to 
\begin{align}
\psi_{1}^d(N) 
&\underset{N\rightarrow \infty}{=}-\hat{\Gamma}_{A}\Big(N\frac{d}{dN}\Big) \int_{\frac{Q^2}{N}}^{Q^2}\frac{d\lambda^2}{\lambda^2}\int^{\lambda^2}_{\frac{Q^2}{N}}\frac{d\mu^2}{\mu^2}F^d_{1} \Big(a_s\big(\lambda^2\big)\Big) 
- \frac{1}{N}\hat{\Gamma}_{B}\Big(N \frac{d}{dN}\Big)\int_{\frac{Q^2}{N}}^{Q^2}\frac{d\lambda^2}{\lambda^2}\int^{\lambda^2}_{\frac{Q^2}{N}}\frac{d\mu^2}{\mu^2}
\tilde{F}^d_{1}\bigg(a_s\big(\lambda^2\big) ,\frac{\mu^2}{Q^2}\bigg)
\end{align}
For type-$\psi^d_{2}$,
\begin{align}
 \psi^d_{2}(N) &=\int_0^1 d\xi \xi^{N-1}
\Bigg(\frac{F^d_{2} \Big(a_s\big(Q^2(1-\xi)\big)\Big)}{(1-\xi)_{+}}  + \tilde{F}^d_{2}\Big(a_s\big(Q^2(1-\xi)\big) ,(1-\xi)\Big)\Bigg)\nonumber\\
&=\int_0^1 d\xi \Bigg(\frac{\xi^{N-1}-1}{1-\xi}\Bigg)
F^d_{2} \Big(a_s\big(Q^2(1-\xi)\big)\Big)  + 
\int_0^1 d\xi \xi^{N-1} \tilde{F}^d_{2}\Big(a_s\big(Q^2(1-\xi)\big) ,(1-\xi)\Big)\nonumber\\
&\underset{N\rightarrow \infty}{=} \hat{\Gamma}_{A}\Big(N\frac{d}{d N}\Big) \int_{0}^{1-\frac{1}{N}}\frac{d\xi}{1-\xi}F^d_{2} \Big(a_s\big(Q^2(1-\xi)\big)\Big) +\frac{1}{N}\hat{\Gamma}_{B}\Big(N\frac{d}{d N}\Big) \int_{0}^{1-\frac{1}{N}}\frac{d\xi}{1-\xi}\tilde{F}^d_{2}\Big(a_s\big(Q^2(1-\xi)\big), (1-\xi)\Big)
\, .
\end{align}
Changing the integration variable from $\xi \rightarrow \lambda^2 = Q^2(1-\xi)$  we obtain
\begin{align}
\psi^d_{2}(N) 
&\underset{N\rightarrow \infty}{=} \hat{\Gamma}_{A}\Big(N\frac{d}{d N}\Big) \int_{\frac{Q^2}{N}}^{Q^2}\frac{d\lambda^2}{\lambda^2} F^d_{2} \Big(a_s\big(\lambda^2\big)\Big) +\frac{1}{N}\hat{\Gamma}_{B}\Big(N\frac{d}{d N}\Big) \int_{\frac{Q^2}{N}}^{Q^2}\frac{d\lambda^2}{\lambda^2}\tilde{F}^d_{2}\bigg(a_s\big(\lambda^2\big), \frac{\lambda^2}{Q^2}\bigg)
\end{align}
Note that in $\psi^d_{4}$ and $\psi^d_{6}$ after replacing $Q^2\frac{d}{dQ^2}$ by $\bigg(\beta\Big(a_s\big(Q^2(1-x')(1-z')\big)\Big) \times \frac{d}{d a_s}\bigg)$, it becomes similar to $\psi^d_{3}$ and $\psi^d_{5}$ respectively.\\

For type-$\psi^d_{3}$,
\begin{align}
\psi^d_{3}(N_1,N_2) &=\int_0^1 dx' x'^{N_1-1}\int_0^1 dz' z'^{N_2-1}
\Bigg(\frac{F^d_{3} \Big(a_s\big(Q^2(1-x')(1-z')\big)\Big)}{(1-x')_{+}(1-z')_{+}} \Bigg)\nonumber\\
&\underset{N_2\rightarrow\infty}{\underset{N_1\rightarrow\infty}{=}}\hat{\Gamma}_A\Big(N_1\frac{d}{dN_1}\Big)\hat{\Gamma}_A\Big(N_2\frac{d}{dN_2}\Big)\int_0^{1-\frac{1}{N_1}} \frac{dx'}{1-x'} \int_0^{1-\frac{1}{N_2}} \frac{dz'}{1-z'} 
F^d_{3} \Big(a_s\big(Q^2(1-x')(1-z')\big)\Big)
\, .
\end{align}
After an appropriate change of variables we get,
\begin{align}
\psi^d_{3}(N_1,N_2) &
\underset{N_2\rightarrow\infty}{\underset{N_1\rightarrow\infty}{=}}
\hat{\Gamma}_A\Big(N_1\frac{d}{dN_1}\Big)\hat{\Gamma}_A\Big(N_2\frac{d}{dN_2}\Big)
\Bigg[
\int_{\frac{Q^2}{N_1N_2}}^{Q^2}\frac{d\lambda^2}{\lambda^2}\int_{\frac{Q^2}{N_1N_2}}^{\lambda^2}\frac{d\mu^2}{\mu^2} F^d_{3} \Big(a_s\big(\lambda^2\big)\Big) 
-\int_{\frac{Q^2}{N_2}}^{Q^2}\frac{d\lambda^2}{\lambda^2}\int_{\frac{Q^2}{N_1N_2}}^{\lambda^2}\frac{d\mu^2}{\mu^2}F^d_{3} \Big(a_s\big(\lambda^2\big)\Big)\nonumber\\
&-\int_{\frac{Q^2}{N_1}}^{Q^2}\frac{d\lambda^2}{\lambda^2}\int_{\frac{Q^2}{N_1}}^{\lambda^2}\frac{d\mu^2}{\mu^2}F^d_{3} \Big(a_s\big(\lambda^2\big)\Big)
+\int_{\frac{Q^2}{N_2}}^{Q^2}\frac{d\lambda^2}{\lambda^2}\int_{\frac{Q^2}{N_1}}^{Q^2}\frac{d\mu^2}{\mu^2}F^d_{3} \Big(a_s\big(\lambda^2\big)\Big)
\Bigg]
\, .
\end{align}
For type-$\psi^d_{5}$,
\begin{align}
\psi^d_{5}(N_1,N_2) &=\int_0^1 dx' x'^{N_1-1}\int_0^1 dz' z'^{N_2-1}
\Bigg(\frac{F^d_{5} \Big(a_s\big(Q^2(1-x')(1-z')\big),(1-z')\Big)}{(1-x')_{+}} \Bigg)\nonumber\\
&\underset{N_2\rightarrow\infty}{\underset{N_1\rightarrow\infty}{=}}\frac{1}{N_2}\hat{\Gamma}_A\Big(N_1\frac{d}{dN_1}\Big)\hat{\Gamma}_B\Big(N_2\frac{d}{dN_2}\Big)\int_0^{1-\frac{1}{N_1}} \frac{dx'}{1-x'} \int_0^{1-\frac{1}{N_2}} \frac{dz'}{1-z'} 
F^d_{5} \Big(a_s\big(Q^2(1-x')(1-z')\big), (1-z')\Big)
\, .
\end{align}
After a change of variables we get
\begin{align}
\psi^d_{5}(N_1,N_2) &\underset{N_2\rightarrow\infty}{\underset{N_1\rightarrow\infty}{=}}\frac{1}{N_2}\hat{\Gamma}_A\Big(N_1\frac{d}{dN_1}\Big)\hat{\Gamma}_B\Big(N_2\frac{d}{dN_2}\Big)
\Bigg[
\int_{\frac{Q^2}{N_1N_2}}^{Q^2}\frac{d\lambda^2}{\lambda^2}\int_{\frac{Q^2}{N_1N_2}}^{\lambda^2}\frac{d\mu^2}{\mu^2} F^d_{5} \bigg(a_s\big(\lambda^2\big),\frac{\lambda^2}{\mu^2 N_2}\bigg) \nonumber\\&
-\int_{\frac{Q^2}{N_2}}^{Q^2}\frac{d\lambda^2}{\lambda^2}\int_{\frac{Q^2}{N_1N_2}}^{\lambda^2}\frac{d\mu^2}{\mu^2}F^d_{5} \bigg(a_s\big(\lambda^2\big),\frac{\lambda^2}{\mu^2N_2}\bigg)
-\int_{\frac{Q^2}{N_1}}^{Q^2}\frac{d\lambda^2}{\lambda^2}\int_{\frac{Q^2}{N_1}}^{\lambda^2}\frac{d\mu^2}{\mu^2}F^d_{5} \bigg(a_s\big(\lambda^2\big),\frac{\mu^2}{Q^2}\bigg)\nonumber\\&
+\int_{\frac{Q^2}{N_2}}^{Q^2}\frac{d\lambda^2}{\lambda^2}\int_{\frac{Q^2}{N_1}}^{Q^2}\frac{d\mu^2}{\mu^2}F^d_{5} \bigg(a_s\big(\lambda^2\big),\frac{\lambda^2}{\mu^2}\bigg)
\Bigg]
\, .
\end{align}
%
For all $\psi_{i}^{d}$, note that the $\mu^2$ integral is easy to perform. 
Before carrying out the $\lambda^2$ integral we apply $\hat{\Gamma}_{A}$ and $\hat{\Gamma}_{B}$ using Leibniz integral rule, for e.g.,\
\begin{align}
N_1\frac{d }{d N_1}\int_{g_{l}(N_1,N_2)}^{g_{u}(N_1,N_2)} \frac{d\lambda^2}{\lambda^2} F\Big(a_s\big(\lambda^2\big),N_1,N_2\Big) &= \frac{d \ln(g_u)}{d \ln N_1} F\Big(a_s\big(g_u\big),N_1,N_2\Big) - \frac{d \ln(g_l)}{d \ln N_1} F\Big(a_s\big(g_l\big),N_1,N_2\Big) \nonumber\\
&+ \int_{g_{l}(N_1,N_2)}^{g_{u}(N_1,N_2)} \frac{d\lambda^2}{\lambda^2} \frac{\partial }{\partial \ln N_1}F\Big(a_s\big(\lambda^2\big),N_1,N_2\Big)\nonumber
\, ,
\end{align}
and generally 
\begin{align}
\left(N_1\frac{d }{d N_1}\right)^n\int_{g_{l}(N_1,N_2)}^{g_{u}(N_1,N_2)} \frac{d\lambda^2}{\lambda^2} F\Big(a_s\big(\lambda^2\big),N_1,N_2\Big) &= \sum_{j=1}^{n} \Bigg[D^{n-j}_{u,1} ~\bigg(\frac{d\ln(g_u)}{d\ln N_1} \times\frac{\partial^{j-1}}{\partial(\ln N_1)^{j-1}}F\Big(a_s\big(\lambda^2\big),N_1,N_2\Big)\Bigg|_{\lambda^2=g_u}\bigg) \nonumber\\
& - D^{n-j}_{l,1}  ~\bigg(\frac{d\ln(g_l)}{d\ln N_1} \times\frac{\partial^{j-1}}{\partial(\ln N_1)^{j-1}}F\Big(a_s\big(\lambda^2\big),N_1,N_2\Big)\Bigg|_{\lambda^2=g_l}\bigg)\Bigg] 
\nonumber\\
&
+\int_{g_{l}(N_1,N_2)}^{g_{u}(N_1,N_2)} \frac{d\lambda^2}{\lambda^2} \frac{\partial^{n}}{\partial (\ln N_1)^n} F\Big(a_s\big(\lambda^2\big),N_1,N_2\Big)
\, ,
 \end{align}
where $D_{k,1}^i$ for $k=l,u$ is defined as
 \begin{align}
D_{k,1}^i &= \left(\frac{\partial}{\partial \ln N_1} - \frac{d \ln(g_k)}{d \ln N_1}\beta\big(a_s(g_k)\big) \frac{\partial}{\partial a_s(g_k)}\right)^i \,,   \nonumber
\end{align}
which is used to deal with $N_1$ dependence appearing in the argument of $a_s\big(g_k(N_1,N_2)\big)$.
The integral over $\lambda^2$ is done after replacing $a_s(\lambda^2)$ by eq.~\eqref{resumas} and changing variables from $\lambda^2 \rightarrow l = 1+\beta_0 a_s(\mu_R^2)\ln(\lambda^2/\mu_R^2)$. 
While substituting the limits we have used $\omega, \omega_1, \omega_2$ variables where, $\omega = a_s(\mu_R^2) \beta_0 \ln(N_1 N_2)$ and $\omega_l = a_s(\mu_R^2) \beta_0 \ln N_l$ for $l=1,2$.

\section{Expansion of $\hat{\Gamma}_{A}(x)$ and $\hat{\Gamma}_{B}(x)$}\label{expGAGB}
In this section we present the expansion of $\hat{\Gamma}_{A}(x)$ and $\hat{\Gamma}_{B}(x)$ used in App.~\ref{DetailMM}.
%
\begin{align}
\hat \Gamma_A(x) &=- \sum_{k=0}^{\infty}\gamma_{k}^{A}(x)^{k}
\, ,
\end{align}
where the coefficients $\gamma_{k}^{A}$ are given as
\begin{align}
\gamma_{k}^{A} &= \frac{\Gamma_{k}(N)}{k!}(-1)^{k}
\, .
\end{align}
For the definition of $\Gamma_{k}(N)$, see eq.~(25) of \cite{Laenen:2008ux}. We get
\begin{align}
\gamma_{0}^A&= 1\,,
\nonumber\\
\gamma_{1}^A&= \EuGa - \frac{1}{2N}\,,
\nonumber \\
\gamma_{2}^A&= \frac{1}{2}\bigg(\gamma_E^2 + \zeta_2\bigg) -\frac{1}{2N}\bigg(1+\EuGa\bigg) \,,
\nonumber \\
\gamma_{3}^A& = \frac{1}{6}\EuGa^3 + \frac{1}{2}\bigg(\EuGa\zeta_2\bigg) + \frac{1}{3}\zeta_3- \frac{1}{4N} \bigg(\EuGa^2 + 2 \EuGa + \zeta_2 \bigg) \,,
\nonumber \\
\gamma_{4}^A& = \frac{1}{24}\gamma_E^4 + \frac{1}{4}\bigg(\gamma_E^2\zeta_2\bigg) + \frac{9}{40}\zeta_2^2 + \frac{1}{3}\bigg(\gamma_E\zeta_3\bigg) - \frac{1}{12N}\bigg( \EuGa^3 + 3 \EuGa^2 + 
      3 \zeta_2 + 3\EuGa\zeta_2  + 2\zeta_3 \bigg)\,,
      \nonumber \\
\gamma_{5}^A&= \frac{1}{120}\gamma_E^5 + \frac{1}{12}\bigg(\gamma_E^3\zeta_2\bigg) + \frac{1}{40}\bigg(9\gamma_E\zeta_2^2\bigg) + 
     \frac{1}{6}\bigg(\gamma_E^2\zeta_3\bigg) + \frac{1}{6}\bigg(\zeta_2
    \zeta_3\bigg) + \frac{1}{5}\zeta_5
    \nonumber \\
&  -\frac{1}{240N} \bigg(  20 \EuGa^3 +5\EuGa^4 + 
      30  \EuGa^2\zeta_2 + 
      27 \zeta_2^2  +40  \zeta3 
    +   20 \EuGa \Big(3 \zeta_2 + 2 \zeta_3 \Big) \bigg) \,,
\nonumber \\
\gamma_{6}^A&= \frac{1}{720}\gamma_E^6 + \frac{1}{48}\bigg(\gamma_E^4\zeta_2\bigg) + \frac{9}{80}\bigg(\gamma_E^2\zeta_2^2\bigg) + 
     \frac{61}{560}\zeta_2^3 + \frac{1}{18}\bigg(\gamma_E^3\zeta_3\bigg) + \frac{1}{6}\bigg(\gamma_E\zeta_2\zeta_3\bigg) 
     \nonumber\\&
     + 
     \frac{1}{18}\zeta_3^2 + \frac{1}{5}\gamma_E\zeta_5
-\frac{1}{240N}\bigg(
5 \EuGa^4 + \EuGa^5 +10 \EuGa^3\zeta_2            + 27 \zeta_2^2 
 + 20 \zeta_2\zeta_3 + 10 \EuGa^2 \Big(3 \zeta_2 + 2\zeta_3\Big)
 \nonumber\\&
 +\EuGa \Big( 27 \zeta_2^2 + 40 \zeta_3 \Big)
 +24 \zeta_5\bigg)\,,
\nonumber \\
\gamma_{7}^A&= \frac{1}{5040}\gamma_E^7 + \frac{1}{240}\bigg(\gamma_E^5\zeta_2\bigg) + \frac{3}{80}\bigg(\gamma_E^3\zeta_2^2\bigg) + 
     \frac{61}{560}\bigg(\gamma_E\zeta_2^3\bigg) + \frac{1}{72}\bigg(\gamma_E^4\zeta_3\bigg) \nonumber\\&+ \frac{1}{12}\bigg(\gamma_E^2\zeta_2\zeta_3\bigg) + 
     \frac{3}{40}\bigg(\zeta_2^2\zeta_3\bigg) + \frac{1}{18}\bigg(\gamma_E\zeta_3^2\bigg) + \frac{1}{10}\bigg(\gamma_E^2\zeta_5\bigg) + 
     \frac{1}{10}\bigg(\zeta_2\zeta_5\bigg) + \frac{1}{7}\zeta_7
   \nonumber \\  
&-\frac{1}{10080N}\bigg(
42 \EuGa^5 +7  \EuGa^6  + 105\EuGa^4\zeta_2 +
      549\zeta_2^3  
     + 840\zeta_2\zeta_3 
     +140 \EuGa^3 \Big(3 \zeta_2 + 2\zeta_3\Big)
     \nonumber \\
    & + 21 \EuGa^2 \Big(27 \zeta_2^2 + 40\zeta_3 \Big) + 56 \Big(5 \zeta_3^2 + 18\zeta_5 \Big)
     + 42 \EuGa \Big(27 \zeta_2^2  + 20\zeta_2\zeta_3 + 24\zeta_5\Big) \bigg)\,,\nonumber\\
  \gamma_{8}^{A} &=
  \frac{1}{40320}\EuGa^8+\frac{1}{1440}\bigg(\EuGa^6 \zeta_2\bigg)+\frac{1}{360}\bigg(\EuGa^5 \zeta_3\bigg)+\frac{3}{320}
   \bigg(\EuGa^4 \zeta_2^2\bigg)+\EuGa^3 \left(\frac{\zeta_2 \zeta_3}{36}+\frac{\zeta_5}{30}\right)+\EuGa^2 \left(\frac{61 \zeta_2^3}{1120}+\frac{\zeta_3^2}{36}\right)
   \nonumber\\&
   +\EuGa
   \left(\frac{3}{40} \zeta_3 \zeta_2^2+\frac{\zeta_5 \zeta_2}{10}+\frac{\zeta_7}{7}\right)+\frac{1261 }{22400}\bigg(\zeta_2^4 \bigg)+\frac{1}{36} \bigg(\zeta_2 \zeta_3^2\bigg)+\frac{1}{15}\bigg(\zeta_3 \zeta_5\bigg)-\frac{1}{10080 N}\bigg( 
   \EuGa^7+7 \EuGa^6
   \nonumber\\
   &+21 \EuGa^5 \zeta_2
   +35\EuGa^4 \Big(3 \zeta_2+2 \zeta_3\Big)+7\EuGa^3 \Big(27 \zeta_2^2+40 \zeta_3\Big)+21\EuGa^2 \Big(27 \zeta_2^2+20 \zeta_3 \zeta_2+24 \zeta_5\Big)\nonumber\\
   &
   +\EuGa \Big(549 \zeta_2^3+840 \zeta_3 \zeta_2+280 \zeta_3^2+1008
   \zeta_5\Big)+549 \zeta_2^3+280 \zeta_3^2+378 \zeta_2^2 \zeta_3+504 \zeta_2 \zeta_5+720 \zeta_7
   \bigg)
\, .
\end{align}
Similarly $\hat{\Gamma}_{B}(x)$ is given by,
\begin{align}
\hat{\Gamma}_B(x) &=  \sum_{k=0}^{\infty}\gamma^{B}_{k}(x)^{k+1}
\, ,
\end{align}
where $\gamma_{k}^{B}$ are given as
\begin{align}
\gamma^{B}_{k} &= \frac{\Gamma^{k}(1)}{k!}(-1)^{k}
\, .
\end{align}
Here, $\Gamma^k(1)$
is the $k$th derivative of the Euler gamma function. We find explicitly
\begin{align}
\gamma_{0}^B &= 1, \nonumber\\
 \gamma_{1}^B &= \gamma_E, \nonumber\\
 \gamma_{2}^B &= \frac{1}{2}\bigg(\gamma_E^2 + \zeta_2\bigg), \nonumber\\
 \gamma_{3}^B &= \frac{1}{6}\gamma_E^3 + \frac{1}{2}\bigg(\gamma_E\zeta_2\bigg) + \frac{1}{3}\zeta_3, \nonumber\\
 \gamma_{4}^B &= \frac{1}{24}\gamma_E^4 + \frac{1}{4}\bigg(\gamma_E^2\zeta_2\bigg) + \frac{9}{40}\zeta_2^2 + \frac{1}{3}\bigg(\gamma_E\zeta_3\bigg), \nonumber\\
 \gamma_{5}^B &= \frac{1}{120}\gamma_E^5 + \frac{1}{12}\bigg(\gamma_E^3\zeta_2\bigg) + \frac{1}{40}\bigg(9\gamma_E\zeta_2^2\bigg) + 
     \frac{1}{6}\bigg(\gamma_E^2\zeta_3\bigg) + \frac{1}{6}\bigg(\zeta_2
    \zeta_3\bigg) + \frac{1}{5}\zeta_5, \nonumber\\
 \gamma_{6}^B &= \frac{1}{720}\gamma_E^6 + \frac{1}{48}\bigg(\gamma_E^4\zeta_2\bigg) + \frac{9}{80}\bigg(\gamma_E^2\zeta_2^2\bigg) + 
     \frac{61}{560}\zeta_2^3 + \frac{1}{18}\bigg(\gamma_E^3\zeta_3\bigg) + \frac{1}{6}\bigg(\gamma_E\zeta_2\zeta_3\bigg)
     + 
     \frac{1}{18}\zeta_3^2 + \frac{1}{5}\gamma_E\zeta_5,\nonumber\\
\gamma_{7}^B &= \frac{1}{5040}\gamma_E^7 + \frac{1}{240}\bigg(\gamma_E^5\zeta_2\bigg) + \frac{3}{80}\bigg(\gamma_E^3\zeta_2^2\bigg) + 
     \frac{61}{560}\bigg(\gamma_E\zeta_2^3\bigg) + \frac{1}{72}\bigg(\gamma_E^4\zeta_3\bigg) + \frac{1}{12}\bigg(\gamma_E^2\zeta_2\zeta_3\bigg) + 
     \frac{3}{40}\bigg(\zeta_2^2\zeta_3\bigg)
     \nonumber\\& 
     + \frac{1}{18}\bigg(\gamma_E\zeta_3^2\bigg) + \frac{1}{10}\bigg(\gamma_E^2\zeta_5\bigg) + 
     \frac{1}{10}\bigg(\zeta_2\zeta_5\bigg) + \frac{1}{7}\zeta_7\, ,\nonumber\\
\gamma_{8}^B &=
\frac{1}{40320}\EuGa^8+\frac{1}{1440}\bigg(\EuGa^6 \zeta_2\bigg)+\frac{1}{360}\bigg(\EuGa^5
    \zeta_3\bigg)+\frac{3 }{320}\bigg(\EuGa^4 \zeta_2^2\bigg)+\frac{1}{36} \bigg(\EuGa^3 \zeta_2
    \zeta_3\bigg)+\frac{1}{30}\bigg(\EuGa^3 \zeta_5\bigg)+\frac{61 }{1120}\bigg(\EuGa^2 \zeta_2^3\bigg)
    \nonumber\\&
    +\frac{1}{36}\bigg(\EuGa^2
    \zeta_3^2\bigg)+\frac{3}{40} \bigg(\EuGa \zeta_2^2 \zeta_3\bigg)+\frac{1}{10}\bigg(\EuGa \zeta_2
    \zeta_5\bigg)+\frac{1}{7}\bigg(\EuGa \zeta_7\bigg)+\frac{1261 }{22400}\zeta_2^4+\frac{1}{36}\bigg(\zeta_2
    \zeta_3^2 \bigg)+\frac{1}{15}\bigg(\zeta_3 \zeta_5\bigg)
\, .    
\end{align}

\section{Resummed Coupling constant}\label{asresum}
For performing the integration over $\lambda^2$ used in App.~\ref{DetailMM}, we replace $a_s(\lambda^2)$ by
\begin{align}
\label{resumas}
a_s(\lambda^2) =&\bigg({a_s(\mu_R^2) \over l}\bigg) \Bigg[1-{a_s(\mu_R^2)\over l} 
\Bigg\{\frac{\beta_1}{\beta_0}  \log(l) \Bigg\}
   + \bigg({a_s(\mu_R^2) \over l}\bigg)^2 \Bigg\{{\beta_1^2\over \beta_0^2} (\log^2(l)-\log(l)
+l-1)-{\beta_2\over \beta_0} (l-1) \Bigg\} 
\nonumber \\ &
   + \bigg({a_s(\mu_R^2)\over l}\bigg)^3 \Bigg\{ {\beta_1^3\over \beta_0^3} \bigg(2 (1-l) \log(l) + {5\over 2} \log^2(l)
- \log^3(l) -{1\over 2} + l - {1\over 2} l^2\bigg)
+{\beta_3 \over 2 \beta_0} (1-l^2) 
 \nonumber \\ &            
+ {\beta_1 \beta_2 \over \beta_0^2} \Big(2 l \log(l)
- 3 \log(l) - l (1-l)\Big)\Bigg\} 
 \nonumber \\ &  
+ \left(\frac{a_s(\mu_R^2)}{l}\right)^4 
\Bigg\{ \frac{\beta_1^4}{\beta_0^4} \bigg(\frac{7}{6} - 2l + \frac{l^2}{2} + \frac{l^3}{3} 
+ 4 \log(l) - 5l \log(l)
 + l^2 \log(l) - \frac{3}{2} \log(l)^2 + 3l \log(l)^2 
 \nonumber \\ &
 - \frac{13}{3} \log(l)^3 + \log(l)^4 \bigg)
 +\frac{\beta_4}{3\beta_0}  (1 - l^3) 
 + \bigg(\frac{\beta_2}{\beta_0}\bigg)^2 \Big(\frac{5}{3} - 3l + l^2 + \frac{l^3}{3}\Big) 
 \nonumber \\ &
+ \frac{\beta_1  \beta_3}{\beta_0^2} \Big(-\frac{1}{6} - \frac{l^2}{2} + \frac{2}{3} l^3 - 2 \log(l) + l^2 \log(l)\Big) 
+ \frac{\beta_1^2  \beta_2}{\beta_0^3} \Big(-3 + 5l - l^2 - l^3 - 3 \log(l) + 5l \log(l) 
 \nonumber \\ &
- 2 l^2 \log(l) + 6 \log(l)^2 - 3l \log(l)^2\Big) \Bigg\}\Bigg]
\, ,
\end{align}
where $l = 1 + \beta_0 a_s(\mu_R^2) \ln(\lambda^2/\mu_R^2)$ and $\beta_i$ are the coefficients of QCD $\beta$ function
$\beta(a_s) =-\sum\limits_{i=0}^\infty a_s^{i+2} \beta_i$, known to five loops.
\section{QCD $\beta$ functions} \label{App}
\begin{align}
  \beta_0&={11 \over 3 } C_A - {2 \over 3 } n_f \, ,
           \nonumber \\[0.5ex]
  \beta_1&={34 \over 3 } C_A^2- 2 n_f C_F -{10 \over 3} n_f C_A \, ,
           \nonumber \\[0.5ex]
  \beta_2&={2857 \over 54} C_A^3 
           -{1415 \over 54} C_A^2 n_f
           +{79 \over 54} C_A n_f^2
           +{11 \over 9} C_F n_f^2
           -{205 \over 18} C_F C_A n_f
           + C_F^2 n_f 
\end{align}

\section{Anomalous dimensions} \label{App:ano}
\begin{align}
    f^{q}_{1} &= 0 \, ,\nonumber \\
    f^{q}_{2} &=   
       C_F C_A   \Bigg\{ \frac{808}{27} - 28 \zeta_3 -   \frac{22}{3} \zeta_2 \Bigg\}      
     - C_F n_f   \Bigg\{   \frac{112}{27} - \frac{4}{3}  \zeta_2 \Bigg\}
 \, ,\nonumber \\
    f^{q}_{3} &=        
      - C_F^2 n_f   \Bigg\{   \frac{1711}{27} -  \frac{304}{9} \zeta_3 - 4 \zeta_2 - \frac{32}{5} \zeta_2^2    \Bigg\}
      + C_F C_A^2   \Bigg\{ \frac{136781}{729} + 192         \zeta_5 - \frac{1316}{3} \zeta_3 - \frac{12650}{81} \zeta_2 +
         \frac{176}{3} \zeta_2 \zeta_3 + \frac{352}{5}        \zeta_2^2 \Bigg\}\nonumber\\
      &- C_F C_A n_f   \Bigg\{   \frac{11842}{729} -         \frac{728}{27} \zeta_3 - \frac{2828}{81} \zeta_2 +   \frac{96}{5}   \zeta_2^2 \Bigg\}
      - C_F n_f^2   \Bigg\{   \frac{2080}{729} -          \frac{112}{27} \zeta_3 + \frac{40}{27} \zeta_2 \Bigg\}
\end{align}

\begin{align}
  A^{q}_{1} &= 4C_F \, ,\nonumber\\
  A^{q}_{2} &=    C_A C_F \Bigg\{
      \frac{268}{9} - 8 \zeta_2\Bigg\}
    - C_F n_f \Bigg\{ \frac{40}{9}\Bigg\} \, ,\nonumber \\
  A^{q}_{3} &=  C_F C_A^2   \Bigg\{ \frac{490}{3} + \frac{88}{3} \zeta_3 - \frac{1072}{9} \zeta_2 + \frac{176}{5} \zeta_2^2 \Bigg\} 
  - C_F^2 n_f   \Bigg\{  \frac{110}{3} - 32 \zeta_3 \Bigg\} 
       - C_F C_A n_f   \Bigg\{  \frac{836}{27} + \frac{112}{3} \zeta_3 - \frac{160}{9} \zeta_2 \Bigg\}\nonumber \\
        & - C_F n_f^2   \Bigg\{   \frac{16}{27} \Bigg\}\, ,\nonumber\\
A^{q}_{4} &=
        \frac{d^{abcd}_Fd^{abcd}_A}{N_c}  \Bigg\{ \frac{3520}{3} \zeta_5 + \frac{128}{3} \zeta_3 - 384 \zeta_3^2 - 128 \zeta_2 - \frac{7936}{35} \zeta_2^3 \Bigg\}
       - n_f\frac{d^{abcd}_Fd^{abcd}_F}{N_c}   \Bigg\{   \frac{1280}{3} \zeta_5 + \frac{256}{3} \zeta_3 - 256 \zeta_2 \Bigg\}
\nonumber\\&       
       - C_F n_f^3   \Bigg\{   \frac{32}{81} - \frac{64}{27} \zeta_3 \Bigg\}
       - C_F C_A^2 n_f   \Bigg\{   \frac{24137}{81} - \frac{2096}{9} \zeta_5 + \frac{23104}{27} \zeta_3 - \frac{20320}{81}
          \zeta_2 - \frac{448}{3} \zeta_2 \zeta_3 + \frac{352}{15} \zeta_2^2 \Bigg\}          
\nonumber\\&           
       + C_F C_A^3   \Bigg\{ \frac{84278}{81} - \frac{3608}{9} \zeta_5 + \frac{20944}{27} \zeta_3 - 16 \zeta_3^2 -
         \frac{88400}{81} \zeta_2 - \frac{352}{3} \zeta_2 \zeta_3 + \frac{3608}{5} \zeta_2^2 - \frac{20032}{105}
         \zeta_2^3 \Bigg\}
\nonumber\\&          
       + C_F C_A n_f^2   \Bigg\{ \frac{923}{81} + \frac{2240}{27} \zeta_3 - \frac{608}{81} \zeta_2 - \frac{224}{15} \zeta_2^2
          \Bigg\}
          + C_F^2 n_f^2   \Bigg\{ \frac{2392}{81} - \frac{640}{9} \zeta_3 + \frac{64}{5} \zeta_2^2 \Bigg\}
\nonumber\\& 
       - C_F^2 C_A n_f   \Bigg\{   \frac{34066}{81} - 160 \zeta_5 - \frac{3712}{9} \zeta_3 - \frac{440}{3} \zeta_2
          + 128 \zeta_2 \zeta_3 + \frac{352}{5} \zeta_2^2 \Bigg\}
       + C_F^3 n_f   \Bigg\{ \frac{572}{9} - 320 \zeta_5 + \frac{592}{3} \zeta_3 \Bigg\}
\end{align}
\begin{align}
  B^{q}_{1} &= 3 C_F\, ,\nonumber\\
  B^{q}_{2} &= 
        C_F^2   \Bigg\{ \frac{3}{2} + 24 \zeta_3 - 12 \zeta_2 \Bigg\}
       + C_F C_A   \Bigg\{ \frac{17}{6} - 12 \zeta_3 + \frac{44}{3} \zeta_2 \Bigg\}
       - C_F n_f   \Bigg\{  \frac{1}{3} + \frac{8}{3} \zeta_2 \Bigg\}\, ,
 \nonumber\\
  B^{q}_{3} &=        
     C_F^3   \Bigg\{ \frac{29}{2} - 240 \zeta_5 + 68 \zeta_3 + 18 \zeta_2 - 32 \zeta_2 \zeta_3 +
         \frac{288}{5} \zeta_2^2 \Bigg\}
   - C_F^2 n_f   \Bigg\{   23 + \frac{136}{3} \zeta_3 - \frac{20}{3} \zeta_2 - \frac{232}{15} \zeta_2^2 \Bigg\}\nonumber \\
    &+ C_F^2 C_A   \Bigg\{ \frac{151}{4} + 120 \zeta_5 + \frac{844}{3} \zeta_3 - \frac{410}{3} \zeta_2 + 16 \zeta_2 \zeta_3 - \frac{988}{15} \zeta_2^2 \Bigg\}
    - C_F C_A^2   \Bigg\{   \frac{1657}{36} - 40 \zeta_5 + \frac{1552}{9} \zeta_3 - \frac{4496}{27} \zeta_2 + 2\zeta_2^2 \Bigg\}\nonumber\\
    &+ C_F C_A n_f   \Bigg\{ 20 + \frac{200}{9} \zeta_3 - \frac{1336}{27} \zeta_2 + \frac{4}{5} \zeta_2^2 \Bigg\}  
    - C_F n_f^2   \Bigg\{   \frac{17}{9} + \frac{16}{9} \zeta_3 - \frac{80}{27} \zeta_2 \Bigg\}
\end{align}

\begin{align}
  C^{q}_{1} &= 0\, , \nonumber\\
  C^{q}_{2} &= 16C_F^2 \, , \nonumber\\
  C^{q}_{3} &= 
        C_F^2 C_A   \Bigg\{ \frac{2144}{9} - 64 \zeta_2 \Bigg\}
       - C_F^2 n_f   \Bigg\{   \frac{320}{9} \Bigg\}
\end{align}

\begin{align}
  \tilde C^{q}_{1} &= 0\, ,\nonumber\\
  \tilde C^{q}_{2} &= -16C_F^2 \, ,\nonumber\\
  \tilde C^{q}_{3} &= 
              - C_F^2 C_A   \Bigg\{   \frac{2144}{9} - 64 \zeta_2 \Bigg\}
              +C_F^2 n_f   \Bigg\{ \frac{320}{9} \Bigg\}
       
\end{align}

\begin{align}
  D^{q}_{1} &= -4C_F \, ,\nonumber\\
  D^{q}_{2} &= C_F^2   \Bigg\{ 12 \Bigg\} 
       - C_F C_A   \Bigg\{   \frac{400}{9} - 8 \zeta_2 \Bigg\} + C_F n_f   \Bigg\{ \frac{64}{9} \Bigg\}
        \, ,
 \nonumber\\
  D^{q}_{3} &=
        C_F^3   \Bigg\{ 6 + 96 \zeta_3 - 48 \zeta_2 \Bigg\}
       + C_F^2 C_A   \Bigg\{ \frac{302}{3} - 48 \zeta_3 + \frac{104}{3} \zeta_2 \Bigg\}
       + C_F^2 n_f   \Bigg\{ 30 - 32 \zeta_3 - \frac{32}{3} \zeta_2 \Bigg\}\nonumber\\
       &- C_F C_A^2   \Bigg\{   \frac{8582}{27} + \frac{88}{3} \zeta_3 - \frac{1336}{9} \zeta_2 + \frac{176}{5} \zeta_2^2  \Bigg\}
       + C_F C_A n_f   \Bigg\{ \frac{724}{9} + \frac{112}{3} \zeta_3 - \frac{208}{9} \zeta_2 \Bigg\}
       - C_F n_f^2   \Bigg\{   \frac{64}{27} \Bigg\}
\end{align}

\begin{align}
  \tilde D^{q}_{1} &= -4C_F\, ,\nonumber\\
  \tilde D^{q}_{2} &=  
       - C_F^2   \Bigg\{   12 \Bigg\}
       - C_F C_A   \Bigg\{   \frac{400}{9} - 8 \zeta_2 \Bigg\}
       + C_F n_f   \Bigg\{ \frac{64}{9} \Bigg\}\, ,
 \nonumber\\
  \tilde D^{q}_{3} &=       
       - C_F^3   \Bigg\{   6 + 96 \zeta_3 - 48 \zeta_2 \Bigg\}
       - C_F^2 C_A   \Bigg\{   \frac{302}{3} - 48 \zeta_3 + \frac{104}{3} \zeta_2 \Bigg\}
       + C_F^2 n_f   \Bigg\{ \frac{178}{3} - 32 \zeta_3 + \frac{32}{3} \zeta_2 \Bigg\}\nonumber\\
       &- C_F C_A^2   \Bigg\{   \frac{8582}{27} + \frac{88}{3} \zeta_3 - \frac{1336}{9} \zeta_2 + \frac{176}{5} \zeta_2^2   \Bigg\}
       + C_F C_A n_f   \Bigg\{ \frac{724}{9} + \frac{112}{3} \zeta_3 - \frac{208}{9} \zeta_2 \Bigg\}
       - C_F n_f^2   \Bigg\{   \frac{64}{27} \Bigg\}
\end{align}

\section{Resummation exponents}\label{SVNSVResumExp}

In this Appendix we present resummation exponents used in $\Psi_{d}^{q}$ given in eq.~(\ref{IntSVpNSV}) and exponents used in the $N$-space result in eqs.~\eqref{DeltaN} and \eqref{GNexp}. 
First, we present the process dependent $g_{d,0}^{q} = \sum_{i=0} a_s^{i}(\mu_R^2)~ g_{d,0,i}^{q}$ given in eq.~\eqref{IntSVpNSV}.
\begin{align}
g_{d,0,0}^{q} &= 1 \, ,\nonumber\\
g_{d,0,1}^{q} &= C_F   \Bigg\{  - 16 + 6 \Big(L_{qr} - L_{fr}\Big) \Bigg\}\, ,\nonumber\\
g_{d,0,2}^{q} &= 
  C_F^2   \Bigg\{ \frac{511}{4} 
     + 18 \Big(L_{qr}^2 + L_{fr}^2\Big)  
    + \Big(48 \zeta_3  - 24 \zeta_2- 93 \Big)\Big(L_{qr} - L_{fr}\Big)- 36 L_{fr} L_{qr}  - 60 \zeta_3  + 58 \zeta_2  - \frac{88}{5} \zeta_2^2   \Bigg\} \nonumber\\
 &- C_F C_A   \Bigg\{   \frac{1535}{12}  + 11 
     \Big(L_{qr}^2 - L_{fr}^2\Big)  + \Big(24 \zeta_3 - \frac{88}{3} \zeta_2\Big)\Big(L_{qr} -  L_{fr}\Big) - \frac{193}{3} L_{qr}+ \frac{17}{3} L_{fr}  -
     \frac{172}{3} \zeta_3 + \frac{538}{9} \zeta_2  - \frac{68}{5}  \zeta_2^2 \Bigg\}\nonumber\\
 &+ C_F n_f   \Bigg\{ \frac{127}{6}  + 2 
       \Big(L_{qr}^2 - L_{fr}^2\Big) - \frac{16}{3} \zeta_2 \Big( L_{qr} - L_{fr}\Big) - \frac{34}{3}  L_{qr}+ \frac{2}{3} L_{fr}  +  \frac{8}{3} \zeta_3
        + \frac{76}{9} \zeta_2  \Bigg\}  
\end{align}

The $\vec N$-independent $\tilde g_{d,0}^{q} = \sum_{i=0} a_s^{i}(\mu_R^2)~ \tilde g_{d,0,i}^{q}$ given in eq.~(\ref{DeltaN}) are found to be as follows,

\begin{align}
\tilde g_{d,0,0}^q &= 1\, , \nonumber \\    
\tilde g_{d,0,1}^q &= C_F \Bigg\{  - 16 + \Big(6 - 8 \gamma_{E}\Big) \Big(L_{qr}- L_{fr}\Big) + 8 \gamma_{E}^2   + 4
          \zeta_2 \Bigg\}\, ,\nonumber \\    
\tilde g_{d,0,2}^q &=
        C_F^2   \Bigg\{ \frac{511}{4} + \Big(18 - 48 \gamma_{E} + 32 \gamma_{E}^2\Big) \Big(L_{qr}^2 + L_{fr}^2\Big) - \Big(93 - 128 \gamma_{E} -48 \gamma_{E}^2 + 64 \gamma_{E}^3 -48 \zeta_3 + 32 \zeta_2 \gamma_{E}\Big) \Big(L_{qr} - L_{fr}\Big)   \nonumber \\
         & - \Big(36  - 96  \gamma_{E} + 64 \gamma_{E}^2\Big)L_{fr} L_{qr} -128 \gamma_{E}^2 + 32
         \gamma_{E}^4      - 60 \zeta_3 - 6   \zeta_2  + 32 \zeta_2 \gamma_{E}^2 - \frac{48}{5} \zeta_2^2 \Bigg\}\nonumber \\
         & 
- C_F C_A   \Bigg\{   \frac{1535}{12} + \Big(11 - \frac{44}{3} \gamma_{E} \Big)\Big(L_{qr}^2 -L_{fr}^2\Big)+    \Big(\frac{536}{9}  \gamma_{E}- 16 \zeta_2 \gamma_{E} + 24 \zeta_3\Big) \Big(L_{qr} -  L_{fr} \Big) - \frac{193}{3} L_{qr} + \frac{17}{3} L_{fr} \nonumber\\
         &  + \frac{88}{3} \gamma_{E}^2  L_{qr}  - \frac{44}{3} \zeta_2\Big(L_{qr} -2 L_{fr}\Big) - \frac{176}{9}
         \gamma_{E}^3- \frac{536}{9} \gamma_{E}^2 - \frac{1616}{27} \gamma_{E} - \frac{604}{9} \zeta_3  + 56 \zeta_3 \gamma_{E}  + 30 \zeta_2  + 16 \zeta_2 \gamma_{E}^2  - \frac{28}{5} \zeta_2^2 \Bigg\}\nonumber\\
&+C_F n_f   \Bigg\{ \frac{127}{6}  + \Big(2-
         \frac{8}{3} \gamma_{E}\Big) \Big(L_{qr}^2 -  L_{fr}^2\Big) + \frac{80}{9} \gamma_{E} \Big(L_{qr} -  L_{fr}\Big)  + \frac{16}{3} \gamma_{E}^2 L_{qr} - \frac{34}{3} L_{qr}+ \frac{2}{3}  L_{fr}  \nonumber\\
         & - \frac{8}{3} \zeta_2 \Big(L_{qr} -2L_{fr}\Big)- \frac{32}{9}        \gamma_{E}^3 - \frac{80}{9} \gamma_{E}^2- \frac{224}{27} \gamma_{E} + \frac{8}{9} \zeta_3 + 4     \zeta_2
           \Bigg\}
\, .
\end{align}

The SV resummation exponents  $g_{d,i}^{q}(\omega)$ given in eq.~\eqref{GNexp} are presented below.
\begin{align}
g_{d,1}^{q}(\omega) &=  
     C_F \frac{1}{\beta_{0}}   \Bigg\{ 4 +  \frac{4}{\omega} L_{\omega} -  4L_{\omega} \Bigg\} \, ,
 \nonumber\\
g_{d,2}^{q}(\omega) &= 
        C_F \frac{\beta_{1}}{\beta_{0}^3}   \Bigg\{ 4 L_{\omega}  + 2 L_{\omega}^2  + 4 \omega  \Bigg\}
       - C_F \frac{1}{\beta_{0}}   \Bigg\{   8 \gamma_{E} L_{\omega} - 4 \omega L_{fr} - 4 L_{qr} L_{\omega} \Bigg\}
       + C_F n_f \frac{1}{\beta_{0}^2}   \Bigg\{ \frac{40}{9} L_{\omega} + \frac{40}{9} \omega \Bigg\}\nonumber \\
       &    - C_F C_A \frac{1}{\beta_{0}^2}   \Bigg\{   \frac{268}{9} L_{\omega} + \frac{268}{9} \omega - 8 \zeta_2 L_{\omega} - 8 \zeta_2     \omega \Bigg\} \, ,
 \nonumber\\
g_{d,3}^{q}(\omega) &=       
     C_F \frac{1}{(1-\omega)} \frac{\beta_{1}^2}{\beta_{0}^4}   \Bigg\{ 2 L_{\omega}^2  + 4 \omega L_{\omega}  + 2 \omega^2 \Bigg\}
    + C_F \frac{\omega}{(1-\omega)} \frac{\beta_{2}}{\beta_{0}^3}   \Bigg\{ 4 - 2 \omega \Bigg\} 
    + C_F \frac{\beta_{2}}{\beta_{0}^3}   \Bigg\{ 4 L_{\omega}  \Bigg\}\nonumber\\
   &- C_F \frac{1}{(1-\omega)}\frac{\beta_{1}}{\beta_{0}^2}   \Bigg\{   8 \gamma_{E} L_{\omega}  - 4 L_{qr} L_{\omega}  + 8 \omega \gamma_{E}  - 4 \omega L_{qr}  \Bigg\}
   +  C_F \frac{\omega}{(1-\omega)}  \Bigg\{ 4  \zeta_2 + 8  \gamma_{E}^2 - 8  L_{qr} \gamma_{E} + 2
         \omega L_{qr}^2 \Bigg\}\nonumber\\
 & + C_F   \Bigg\{ 2 \omega \Big(L_{qr}^2 - L_{fr}^2\Big) \Bigg\}
 + C_F n_f \frac{1}{\beta_{0}}   \Bigg\{ \frac{40}{9} \omega \Big(L_{qr} -  L_{fr}\Big) \Bigg\}
 + C_F n_f \frac{1}{(1-\omega)} \frac{\beta_{1}}{\beta_{0}^3} \Bigg\{ \frac{40}{9} L_{\omega} + \frac{40}{9} \omega  +    \frac{20}{9} \omega^2   \Bigg\}  \nonumber\\
   &-  C_F n_f \frac{\omega}{(1-\omega)}\frac{1}{\beta_{0}}   \Bigg\{  \frac{112}{27}  - \frac{8}{3}  \zeta_2 + \frac{80}{9}      \gamma_{E} - \frac{40}{9} \omega L_{qr} \Bigg\} -C_F n_f^2 \frac{\omega^2}{(1-\omega)}\frac{1}{\beta_{0}^2}   \Bigg\{   \frac{8}{27}  \Bigg\} \nonumber\\
    &
   -C_F C_A \frac{1}{(1-\omega)}\frac{\beta_{1}}{\beta_{0}^3}   \Bigg\{   \frac{268}{9} L_{\omega}  - 8 \zeta_2 L_{\omega}            +\frac{268}{9} \omega  - 8 \omega \zeta_2 + \frac{134}{9} \omega^2  - 4 \omega^2 \zeta_2  \Bigg\}\nonumber\\
    &
   +  C_F C_A \frac{\omega}{(1-\omega)}\frac{1}{\beta_{0}}   \Bigg\{ \frac{808}{27}  - 28  \zeta_3 - \frac{44}{3}  \zeta_2 + \frac{536}{9}  \gamma_{E} - 16  \gamma_{E} \zeta_2 - \frac{268}{9} \omega L_{qr} + 8  \omega L_{qr} \zeta_2 \Bigg\}\nonumber\\
    &-  C_F C_A n_f \frac{\omega^2}{(1-\omega)}\frac{1}{\beta_{0}^2}   \Bigg\{  \frac{418}{27}  + \frac{56}{3}  \zeta_3 -\frac{80}{9} \zeta_2 \Bigg\}
    +  C_F C_A^2 \frac{ \omega^2}{(1-\omega)} \frac{1}{\beta_{0}^2}   \Bigg\{ \frac{245}{3} + \frac{44}{3}  \zeta_3 -             \frac{536}{9}\zeta_2 + \frac{88}{5}  \zeta_2^2 \Bigg\}\nonumber\\
    &-  C_F^2 n_f \frac{ \omega^2}{(1-\omega)}\frac{1}{\beta_{0}^2}   \Bigg\{   \frac{55}{3} - 16 \zeta_3 \Bigg\}
     - C_F C_A \frac{1}{\beta_{0}}   \Bigg\{   \bigg(\frac{268}{9} \omega - 8 \omega \zeta_2\bigg) \Big(L_{qr}- L_{fr}\Big) \Bigg\}
\, .
\end{align}
The NSV resummation exponents  $\overline g_{d,i}^{q}(\omega)$ given in eq.~\eqref{GNexp} read
\begin{align}
\overline{g}_{d,0}^{q}(\omega) &=  
     C_F \frac{1}{\beta_{0}}   \Bigg\{ 2 L_{\omega} \Bigg\} \, ,\nonumber\\
\overline{g}_{d,1}^{q}(\omega) &=        
         C_F \frac{1}{(1-\omega)} \frac{\beta_{1}}{\beta_{0}^2}   \Bigg\{ 2 L_{\omega}  + 2 \omega  \Bigg\}
       - C_F \frac{1}{(1-\omega)}   \Bigg\{   2 + 4 \gamma_{E} - 2 L_{qr}  \Bigg\}
       - C_F   \Bigg\{ 2 \Big( L_{fr} \Big)  \Bigg\}\nonumber\\
       &+ C_F n_f \frac{\omega}{(1-\omega)} \frac{1}{\beta_{0}}   \Bigg\{ \frac{20}{9}  \Bigg\}
       - C_F C_A \frac{\omega}{(1-\omega)} \frac{1}{\beta_{0}}   \Bigg\{   \frac{134}{9}  - 4 \zeta_2  \Bigg\} \, ,\nonumber\\       
\overline{g}_{d,2}^{q}(\omega) &=       
        C_F \frac{1}{(1-\omega)^2} \frac{\beta_{1}^2}{\beta_{0}^3}   \Bigg\{   \omega^2  - L_{\omega}^2  \Bigg\}
       - C_F \frac{\omega^2}{(1-\omega)^2} \frac{\beta_{2}}{\beta_{0}^2}   \Bigg\{ 1 \Bigg\}
       + C_F \frac{1}{(1-\omega)^2} \frac{\beta_{1}}{\beta_{0}}   \Bigg\{ \Big(2   + 4 \gamma_{E}   - 2 L_{qr}  \Big)    L_{\omega}  \Bigg\}\nonumber\\
       &- C_F \frac{1}{(1-\omega)^2} \beta_{0}   \Bigg\{   4 \gamma_{E} + 4 \gamma_{E}^2 - \big(1-\omega\big)^2L_{fr}^2  - 2 L_{qr} - 4 L_{qr} \gamma_{E} + L_{qr}^2 + 2 \zeta_2 \Bigg\}\nonumber\\
       &- C_F n_f \frac{1}{(1-\omega)^2} \frac{\beta_{1}}{\beta_{0}^2}   \Bigg\{  \frac{20}{9}  L_{\omega}  
       + \frac{20}{9} \omega    - \frac{10}{9} \omega^2  \Bigg\}
       + C_F n_f \frac{1}{(1-\omega)^2}   \Bigg\{ \frac{116}{27} + \frac{40}{9} \gamma_{E}  -\frac{20}{9} L_{qr} - \frac{4}{3} \zeta_2 \Bigg\}\nonumber\\
       &+ C_F n_f   \Bigg\{ \frac{20}{9} \bigg( L_{fr}\bigg) \Bigg\}
       + C_F n_f^2 \frac{\omega}{(1-\omega)^2} \frac{1}{\beta_{0}}   \Bigg\{ \frac{8}{27}  - \frac{4}{27} \omega \Bigg\}
       + C_F C_A   \Bigg\{  \bigg( \frac{134}{9} - 4 \zeta_2 \bigg) \bigg( L_{fr} \bigg)\Bigg\}\nonumber\\
       &+ C_F C_A \frac{1}{(1-\omega)^2} \frac{\beta_{1}}{\beta_{0}^2}   \Bigg\{ \Big(2 L_{\omega} +2 \omega-\omega^2\Big)\bigg(\frac{67}{9} -2\zeta_2\bigg) \Bigg\}
       + C_F C_A \frac{1}{(1-\omega)^2}   \Bigg\{   \bigg(\frac{134}{9}- 4\zeta_2\bigg) L_{qr}  -\frac{806}{27}\nonumber\\
    &  - \frac{268}{9} \gamma_{E} + 8 \zeta_2 \gamma_{E} + 14 \zeta_3 + \frac{34}{3} \zeta_2 \Bigg\} 
    - C_F C_A^2 \frac{\omega}{(1-\omega)^2} \frac{1}{\beta_{0}}   \Bigg\{ \Big(2-\omega\Big)\bigg( \frac{245}{6} -\frac{268}{9}\zeta_2 +\frac{44}{5}\zeta_2^2+\frac{22}{3}\zeta_3\bigg) \Bigg\}   \nonumber\\
    & + C_F^2 n_f \frac{\omega}{(1-\omega)^2} \frac{1}{\beta_{0}}   \Bigg\{ \Big(2-\omega\Big) \bigg(\frac{55}{6} -8\zeta_3\bigg)\Bigg\}
    + C_F C_A n_f \frac{\omega}{(1-\omega)^2} \frac{1}{\beta_{0}}   \Bigg\{ \Big(2-\omega\Big)\bigg(\frac{209}{27} - \frac{40}{9}\zeta_2 +\frac{28}{3}\zeta_3\bigg) \Bigg\}
\, .
\end{align}
The NSV resummation exponents  $h_{d,i,j}^{q}$ and $\tilde h_{d,i,i}^{q}$ given in eq.~\eqref{hg} are listed next:
\begin{align}
    h_{d,0,0}^{q}(\omega) &= C_F \frac{1}{\beta_{0}}   \Bigg\{  - 4 L_{\omega} \Bigg\} \, ,\nonumber\\
    h_{d,0,1}^{q}(\omega) &= 0\, , \nonumber\\
    h_{d,1,0}^{q}(\omega,\omega_l) &= 
      -  C_F \frac{1}{(1-\omega)} \frac{\beta_{1}}{\beta_{0}^2}   \Bigg\{   4 L_{\omega}  + 4 \omega \Bigg\}
       - C_F \frac{1}{(1-\omega)}   \Bigg\{ (-1)^{l}4 - 8 \gamma_{E} - 4 L_{fr} + 4 \omega L_{fr} + 4 L_{qr} \Bigg\}\nonumber\\
       &- C_F n_f \frac{\omega}{(1-\omega)} \frac{1}{\beta_{0}}   \Bigg\{   \frac{64}{9} \Bigg\}
       + C_F C_A \frac{\omega}{(1-\omega)} \frac{1}{\beta_{0}}   \Bigg\{ \frac{400}{9}  - 8 \zeta_2  \Bigg\}
       + C_F^2 \frac{\omega}{(1-\omega)} \frac{1}{\beta_{0}}   \Bigg\{  (-1)^{l} \Big(12  - 16  \gamma_{E}\Big) \Bigg\}\, ,  \nonumber\\    
    \tilde h_{d,1,1}^{q}(\omega,\omega_l) &= 
         (-1)^l\Bigg[C_F^2 \frac{1}{(1-\omega)^2} \frac{1}{\beta_{0}}   \Bigg\{ 4\omega_l \Bigg\}
       - C_F^2 \frac{\omega}{(1-\omega)} \frac{1}{\beta_{0}}   \Bigg\{  16  \Bigg\} \Bigg]\, ,\nonumber\\
{h_{d,2,0}^{q}(\omega,\omega_1)} &=  {h_{d,2,0}^{q}(\omega,\omega_2)} 
       +\Bigg[
       C_F \frac{1}{(1-\omega)^2}\beta_{0}   \Bigg\{ 16 \gamma_{E} - 8 L_{qr} \Bigg\}
       - C_F \frac{1}{(1-\omega)^2}\frac{\beta_{1}}{\beta_{0}}   \Bigg\{   8 L_{\omega}  \Bigg\} 
\nonumber\\
&       +  C_F C_A \frac{1}{(1-\omega)^2}  \Bigg\{ \frac{128}{9} - 8 \zeta_2 - 20 \gamma_{E} \Bigg\}
       +C_F^2  \frac{1}{(1-\omega)^2} \frac{\beta_{1}}{\beta_{0}^2}   \Bigg\{ \Big(12-16\gamma_{E}\Big)\Big(2L_{\omega} +2\omega-\omega^2\Big)  \Bigg\}\nonumber\\
&      + C_F^2 \frac{1}{(1-\omega)^2}   \Bigg\{ 16 \zeta_2 - 28 \gamma_{E} + 56 \gamma_{E}^2 
       +\Big(24-32\gamma_{E}\Big) L_{qr} \Bigg\}  - C_F n_f \frac{1}{(1-\omega)^2}  \Bigg\{ \frac{104}{9}\Bigg\}  \nonumber\\
& 
  - C_F^2   \Bigg\{ \Big(24-32\gamma_{E}\Big)\Big(L_{fr}\Big) \Bigg\}  
  + C_F^2 n_f \frac{\omega}{(1-\omega)^2} \frac{1}{\beta_{0}}   \Bigg\{ \Big(2-\omega\Big)\bigg( \frac{44}{3} -\frac{320}{9}\gamma_{E}+\frac{32}{3}\zeta_2\bigg) \Bigg\}\nonumber\\
&       -  C_F^2 C_A \frac{\omega}{(1-\omega)^2}\frac{1}{\beta_{0}}   \Bigg\{  \Big(2-\omega\Big)\bigg( \frac{302}{3} -\frac{2144}{9}\gamma_{E} +\frac{104}{3} \zeta_2 +64\gamma_{E}\zeta_{2}-48\zeta_{3}
  \bigg)\Bigg\}\nonumber\\
&      - C_F^3 \frac{\omega}{(1-\omega)^2} \frac{1}{\beta_{0}}   \Bigg\{ \Big(2-\omega\Big)\Big(6-48\zeta_2+96\zeta_3\Big) \Bigg\}
\Bigg] \, ,\nonumber\\
{h_{d,2,0}^{q}(\omega,\omega_2)} &= 
        C_F \frac{1}{(1-\omega)^2} \frac{\beta_{1}^2}{\beta_{0}^3}   \Bigg\{ 2 L_{\omega}^2  - 2 \omega^2  \Bigg\}
       + C_F \frac{\omega^2}{(1-\omega)^2} \frac{\beta_{2}}{\beta_{0}^2}   \Bigg\{ 2   \Bigg\} + C_F \frac{1}{(1-\omega)^2} \frac{\beta_{1}}{\beta_{0}}   \Bigg\{  \Big( 4  - 8 \gamma_{E} + 4 L_{qr}\Big)  L_{\omega} \Bigg\}\nonumber\\&
      - C_F \frac{1}{(1-\omega)^2} \beta_{0}   \Bigg\{ 8 \gamma_{E} - 8 \gamma_{E}^2 - 4 L_{qr} + 8 L_{qr} \gamma_{E} -2 L_{qr}^2      - 4 \zeta_2 \Bigg\}
       - C_F \beta_{0}   \Bigg\{   2 L_{fr}^2 \Bigg\} \nonumber\\
       &+ C_F n_f \frac{1}{(1-\omega)^2} \frac{\beta_{1}}{\beta_{0}^2}   \Bigg\{ \frac{64}{9} L_{\omega}  + \frac{64}{9} \omega  -  \frac{32}{9} \omega^2  \Bigg\} 
        - C_F n_f   \Bigg\{   \frac{64}{9} \bigg(  L_{fr}\bigg) \Bigg\} \nonumber\\&
- C_F n_f \frac{1}{(1-\omega)^2}   \Bigg\{   \frac{52}{27} + \frac{128}{9} \gamma_{E} -\frac{64}{9} L_{qr}- \frac{8}{3} \zeta_2 \Bigg\}   
+ C_F n_f^2 \frac{\omega}{(1-\omega)^2} \frac{1}{\beta_{0}}   \Bigg\{ \frac{64}{27}  - \frac{32}{27} \omega \Bigg\}\nonumber\\&
- C_F C_A \frac{1}{(1-\omega)^2} \frac{\beta_{1}}{\beta_{0}^2}   \Bigg\{  \Big(2L_{\omega}+2\omega-\omega^2\Big)\bigg(\frac{200}{9}-4\zeta_2\bigg)\Bigg\} 
 +C_F C_A   \Bigg\{ \bigg(\frac{400}{9} - 8 \zeta_2 \bigg)\bigg(   L_{fr} \bigg)\Bigg\} \nonumber\\ &
+ C_F C_A \frac{1}{(1-\omega)^2}   \Bigg\{ \frac{1108}{27} + \frac{890}{9} \gamma_{E} -\bigg(\frac{400}{9}-8\zeta_2\bigg)  L_{qr}
 - 28 \zeta_3 - \frac{44}{3} \zeta_2 - 16 \zeta_2 \gamma_{E}  \Bigg\} \nonumber\\&       
- C_F C_A n_f \frac{\omega}{(1-\omega)^2} \frac{1}{\beta_{0}}   \Bigg\{   \Big(2-\omega\Big)\bigg(\frac{362}{9} - \frac{104}{9}\zeta_2+\frac{56}{3}\zeta_3\bigg)\Bigg\}
+ C_F^2   \Bigg\{  \bigg(12- 16 \gamma_{E} \bigg) \bigg(  L_{fr}\bigg) \Bigg\}
\nonumber\\&
- C_F^2 \frac{1}{(1-\omega)^2} \frac{\beta_{1}}{\beta_{0}^2}   \Bigg\{ \Big(6-8\gamma_{E}\Big)\Big(2 L_{\omega}+2 \omega-\omega^2\Big) \Bigg\} 
+ C_F^2 \frac{1}{(1-\omega)^2}   \Bigg\{14 \gamma_{E} - 28 \gamma_{E}^2 -20 \zeta_2\nonumber\\ 
&
-\Big(12-16\gamma_{E}\Big) L_{qr}
 \Bigg\} 
+ C_F C_A^2 \frac{\omega}{(1-\omega)^2} \frac{1}{\beta_{0}}   \Bigg\{ \Big(2-\omega\Big)\bigg(\frac{4291}{27} -\frac{668}{9}\zeta_2 +\frac{88}{5}\zeta_{2}^{2} +\frac{44}{3}\zeta_{3}\bigg) \Bigg\} 
\nonumber\\   & 
+ C_F^2 C_A \frac{\omega}{(1-\omega)^2} \frac{1}{\beta_{0}}   \Bigg\{ \Big(2-\omega\Big)\bigg(  
\frac{151}{3} -\frac{1072}{9}\gamma_{E} +\frac{52}{3}\zeta_{2}+32\gamma_{E}\zeta_{2} -24\zeta_{3}
\bigg) \Bigg\} 
\nonumber\\  &
-C_F^2 n_f \frac{\omega}{(1-\omega)^2} \frac{1}{\beta_{0}} \Bigg\{ \Big(2-\omega\Big)\bigg(\frac{89}{3}  -\frac{160}{9}\gamma_{E} +\frac{16}{3}\zeta_{2}-16\zeta_3  \bigg)
\Bigg\} 
\nonumber\\  &
+ C_F^3 \frac{\omega}{(1-\omega)^2} \frac{1}{\beta_{0}}   \Bigg\{ \Big(2-\omega\Big)\Big(3-24\zeta_2 +48\zeta_3\Big) \Bigg\}\, ,
\nonumber\\
 h_{d,2,1}^{q}(\omega,\omega_l) &= (-1)^{l}\Bigg[
    C_F C_A \frac{1}{(1-\omega)^2}   \Bigg\{ 10 \Bigg\}
     + C_F^2 \frac{1}{(1-\omega)^2} \frac{\beta_{1}}{\beta_{0}^2}   \Bigg\{   16 L_{\omega}  + 16 \omega  - 8 \omega^2  \Bigg\}
       - C_F^2  \Bigg\{ 16 \Big( L_{fr}\Big) \Bigg\}     
     \nonumber\\
    &- C_F^2 \frac{1}{(1-\omega)^2}   \Bigg\{ 10 + 24 \gamma_{E}  - 16 L_{qr}  \Bigg\}+ C_F^2 n_f \frac{\omega}{(1-\omega)^2} \frac{1}{\beta_{0}}   \Bigg\{  \frac{160}{9}\Big( 2-\omega \Big) \Bigg\}
\nonumber\\&
 - C_F^2 C_A \frac{\omega}{(1-\omega)^2} \frac{1}{\beta_{0}}   \Bigg\{ \frac{16}{9}\Big(2-\omega\Big)\Big(67-18\zeta_2\Big)  \Bigg\}
       \Bigg]\, ,
         \nonumber\\
\tilde h_{d,2,2}^{q}(\omega,\omega_l) &= (-1)^{l}\Bigg[  C_F^2 n_f  \frac{1}{(1-\omega)^3} \frac{1}{\beta_{0}}   \Bigg\{ \frac{32}{27} \omega_l \Bigg\}
  - C_F^2 C_A \frac{1}{(1-\omega)^3}\frac{1}{\beta_{0}}   \Bigg\{ \frac{176}{27} \omega_l \Bigg\}\Bigg]
\, ,
\end{align}
where $\gamma_E$ is the Euler-Mascheroni constant, 
$\omega = a_s\beta_{0}\ln\big(N_1N_2\big)$, $\omega_l =a_s\beta_{0}\ln\big(N_l\big)$ for $l = 1,2$, 
$L_{\omega} = \ln\big(1-\omega\big)$ and $L_{qr} = \ln\Big(\frac{Q^2}{\mu_R^2}\Big)$, $L_{fr} = \ln\Big(\frac{\mu_F^2}{\mu_R^2}\Big)$. 
Here $\tilde h_{d,2,2}^{q}(\omega,\omega_l)$ is written using $\varphi^{(3)}_{q,x',3}= - \varphi^{(3)}_{q,z',3}$ and $\varphi^{(3)}_{q,z',3} = -16/9~\beta_{0}C_F^2$, see the main text.

\section{Convolutions}
Relevant formulae for convolutions of distributions ${\delta_{\xi}}$ = $\delta(1-\xi)$, $\mathcal{D}_{\xi,i}$ = $\Big[\frac{\ln^i(1-\xi)}{1-\xi}\Big]_{+}$ and logarithms $\mathcal{L}_{\xi,i}$ = $\ln^i(1-\xi)$ in (N)SV limits are listed below.
\begin{align*}
\bm{\delta_{\xi} \otimes \delta_{\xi}} &= \bm{1} \, ,
  \nonumber \\
\bm{\delta_{\xi} \otimes \mathcal{D}_{\xi,i}} & = \bm{\mathcal{D}_{\xi,i}} \, ,  
  \nonumber \\
\bm{\mathcal{D}_{\xi,0} \otimes \mathcal{D}_{\xi,0}} & =  \bm{\mathcal{D}_{\xi,1}}\Big\{2\Big\} + \bm{\mathcal{L}_{\xi,0}} - \bm{\delta_{\xi}}\Big\{\zeta_2 \Big\} \, ,
\nonumber \\
\bm{\mathcal{D}_{\xi,0} \otimes \mathcal{D}_{\xi,1}}  & = \bm{\mathcal{D}_{\xi,0}}\Big\{-\zeta_2\Big\}  + \bm{\mathcal{D}_{\xi, 2}}\Big\{\frac{3 }{2}\Big\} + \bm{\mathcal{L}_{\xi,1}} + \bm{\delta_{\xi}}\Big\{\zeta_3 \Big\} \, ,
\nonumber \\ 
\bm{\mathcal{D}_{\xi,0} \otimes \mathcal{D}_{\xi,2}} & =  \bm{\mathcal{D}_{\xi,0}} \Big\{2 \zeta_3 \Big\}  - \bm{\mathcal{D}_{\xi,1}} \Big\{2 \zeta_2 \Big\}  + \bm{\mathcal{D}_{\xi,3}} \Big\{\frac{4}{3} \Big\}  + \bm{\mathcal{L}_{\xi,2}} - \bm{\delta_{\xi}} \Big\{\frac{4}{5} \zeta_2^2 \Big\}  \, ,
\nonumber \\ 
\bm{\mathcal{D}_{\xi,0} \otimes \mathcal{D}_{\xi,3}} & =\bm{\mathcal{D}_{\xi,0}}
\Big\{-\frac{12 \zeta_2^2 }{5}\Big\} + \bm {\mathcal{D}_{\xi,1}} \Big\{6 \zeta_3 \Big\} - \bm {\mathcal{D}_{\xi,2}} \Big\{3 \zeta_2 \Big\}  + \bm{\mathcal{D}_{\xi,4}} \Big\{\frac{5}{4} \Big\}  + \bm{\mathcal{L}_{\xi,3}} + \bm{\delta_{\xi}} \Big\{6 \zeta_5\Big\}  \, ,
\nonumber \\ 
\bm{\mathcal{D}_{\xi,0} \otimes \mathcal{D}_{\xi,4}} & = \bm{\mathcal{D}_{\xi,0}}\Big\{24 \zeta_5\Big\}  - \bm{\mathcal{D}_{\xi,1}} \Big\{\frac{48 \zeta_2^2 }{5} \Big\} + \bm{\mathcal{D}_{\xi,2}}\Big\{12 \zeta_3 \Big\}  - \bm{\mathcal{D}_{\xi,3}}\Big\{4 \zeta_2\Big\}  + \bm{\mathcal{D}_{\xi,5}}\Big\{\frac{6}{5}\Big\}  + \bm{\mathcal{L}_{\xi,4}} 
\nonumber \\ &
-  \bm{\delta_{\xi}} \Big\{\frac{192}{35} \zeta_2^3 \Big\}\, ,
\nonumber \\ 
\bm{\mathcal{D}_{\xi,0} \otimes \mathcal{D}_{\xi,5}} & = \bm{\mathcal{D}_{\xi,0}}\Big\{-\frac{192 \zeta_2^3}{7}\Big\}  + \bm{\mathcal{D}_{\xi,1}}\Big\{120 \zeta_5\Big\}  - \bm{\mathcal{D}_{\xi,2}}\Big\{24 \zeta_2^2\Big\}  + \bm{\mathcal{D}_{\xi,3}}\Big\{20 \zeta_3\Big\}  - \bm{\mathcal{D}_{\xi,4}}\Big\{5 \zeta_2\Big\}  + \bm{\mathcal{D}_{\xi,6}}\Big\{\frac{7}{6} \Big\} 
\nonumber \\ &
+ \bm{\mathcal{L}_{\xi,5}} + \bm{\delta_{\xi}}\Big\{120 \zeta_7 \Big\} \, ,
\nonumber \\ 
\bm{\mathcal{D}_{\xi,1} \otimes \mathcal{D}_{\xi,1}} & = \bm{\mathcal{D}_{\xi,0}}\Big\{2 \zeta_3\Big\}  - \bm{\mathcal{D}_{\xi,1}}\Big\{2 \zeta_2\Big\}  + \bm{\mathcal{D}_{\xi,3}} - \bm{\mathcal{L}_{\xi,0}}\Big\{\zeta_2\Big\}  + \bm{\mathcal{L}_{\xi,2}} - \bm{\delta_{\xi}}\Big\{\frac{1}{10} \zeta_2^2\Big\} \, ,
\nonumber \\ 
\bm{\mathcal{D}_{\xi,1} \otimes \mathcal{D}_{\xi,2}} & = \bm{\mathcal{D}_{\xi,0}}\Big\{-\zeta_2^2\Big\}  +\bm{\mathcal{D}_{\xi,1}} \Big\{6 \zeta_3 \Big\}  - \bm{\mathcal{D}_{\xi,2}}\Big\{3 \zeta_2\Big\}  + \bm{\mathcal{D}_{\xi,4}}\Big\{\frac{5}{6}\Big\}  + \bm{\mathcal{L}_{\xi,0}}\Big\{2 \zeta_3\Big\}  - \bm{\mathcal{L}_{\xi,1}}\Big\{2 \zeta_2\Big\}  + \bm{\mathcal{L}_{\xi,3}} 
\nonumber \\ &
+ \bm{\delta_{\xi}}\Big\{4 \zeta_5 -2 \zeta_2 \zeta_3  \Big\} \, ,
\nonumber \\ 
\bm{\mathcal{D}_{\xi,1} \otimes \mathcal{D}_{\xi,3}} & = \bm{\mathcal{D}_{\xi,0}}\Big\{18 \zeta_5 - 6\zeta_2  \zeta_3  \Big\} - \bm{\mathcal{D}_{\xi,1}}\Big\{\frac{27}{5}\zeta_2^2\Big\}  + \bm{\mathcal{D}_{\xi,2}} \Big\{12 \zeta_3\Big\}  - \bm{\mathcal{D}_{\xi,3}} \Big\{4\zeta_2\Big\}  + \bm{\mathcal{D}_{\xi,5}}\Big\{\frac{3}{4}\Big\}  -  \bm{\mathcal{L}_{\xi,0}}\Big\{\frac{12}{5} \zeta_2^2 \Big\} 
\nonumber \\ &
+ \bm{\mathcal{L}_{\xi,1}}\Big\{6 \zeta_3\Big\}   
- \bm{\mathcal{L}_{\xi,2}}\Big\{3 \zeta_2\Big\}  + \bm{\mathcal{L}_{\xi,4} }
+ \bm{\delta_{\xi}} \Big\{ 3 \zeta_3^2-\frac{36 \zeta_2^3}{35} \Big\} \, ,
\nonumber \\ 
\bm{\mathcal{D}_{\xi,1} \otimes \mathcal{D}_{\xi,4}} & = \bm{\mathcal{D}_{\xi,0}}\Big\{12 \zeta_3^2 -\frac{48 \zeta_2^3 }{5}\Big\}   + \bm{\mathcal{D}_{\xi,1}}\Big\{96 \zeta_5 -24 \zeta_2 \zeta_3 \Big\}  + \bm{\mathcal{D}_{\xi,3}}\Big\{20 \zeta_3\Big\}  +  \bm{\mathcal{D}_{\xi,6}}\Big\{\frac{7}{10}\Big\} + \bm{\mathcal{L}_{\xi,0}}\Big\{24 \zeta_5\Big\} 
\nonumber \\ & 
- \bm{\mathcal{D}_{\xi,2}}\Big\{\frac{78}{5} \zeta_2^2\Big\}  - \bm{\mathcal{L}_{\xi,1}}\Big\{\frac{48}{5} \zeta_2^2\Big\}  
+ \bm{\mathcal{L}_{\xi,2}}\Big\{12 \zeta_3\Big\}   - \bm{\mathcal{D}_{\xi,4}}\Big\{5\zeta_2\Big\}  - \bm{\mathcal{L}_{\xi,3}}\Big\{4 \zeta_2\Big\} + \bm{\mathcal{L}_{\xi,5}}  
\nonumber \\ &
- \bm{\delta_{\xi}}\Big\{\frac{48}{5} \zeta_2^2 \zeta_3 + 24 \zeta_2 \zeta_5 - 72 \zeta_7 \Big\} \, ,
\nonumber \\ 
\bm{\mathcal{D}_{\xi,1} \otimes \mathcal{D}_{\xi,5}} & = \bm{\mathcal{D}_{\xi,0}}\Big\{480 \zeta_7 -120 \zeta_5 \zeta_2 - 48 \zeta_2^2 \zeta_3  \Big\} + \bm{\mathcal{D}_{\xi,1}}\Big\{60 \zeta_3^2 -\frac{528}{7} \zeta_2^3 \Big\}  + \bm{\mathcal{D}_{\xi,2}}\Big\{300 \zeta_5-60 \zeta_3 \zeta_2  \Big\} + \bm{\mathcal{D}_{\xi,4}}\Big\{30 \zeta_3 \Big\}  
\nonumber \\ & 
+ \bm{\mathcal{D}_{\xi,7}}\Big\{\frac{2}{3}\Big\}  
 -\bm{\mathcal{L}_{\xi,0}}\Big\{\frac{192}{7} \zeta_2^3\Big\}  + \bm{\mathcal{L}_{\xi,1}}\Big\{ 120 \zeta_5 \Big\} - \bm{\mathcal{D}_{\xi,3}}\Big\{34 \zeta_2^2 \Big\}  - \bm{\mathcal{L}_{\xi,2}}\Big\{24 \zeta_2^2\Big\}  + \bm{\mathcal{L}_{\xi,3}}\Big\{20 \zeta_3 \Big\} 
\nonumber \\ &
- \bm{\mathcal{D}_{\xi,5}}\Big\{6 \zeta_2 \Big\} - \bm{\mathcal{L}_{\xi,4}}\Big\{5 \zeta_2 \Big\} + \bm{\mathcal{L}_{\xi,6}} - \bm{\delta_{\xi}}  \Big\{\frac{144}{7} \zeta_2^4 - 120 \zeta_3 \zeta_5\Big\} \, ,
\nonumber \\
\bm{\mathcal{D}_{\xi,2} \otimes \mathcal{D}_{\xi,2}} & = 
\bm{\mathcal{D}_{\xi,0}}\Big\{16 \zeta_5 - 8 \zeta_2 \zeta_3 \Big\}  + \bm{\mathcal{D}_{\xi,2}}\Big\{12 \zeta_3 \Big\} + \bm{\mathcal{D}_{\xi,5}}\Big\{\frac{2}{3} \Big\} 
-  \bm{\mathcal{D}_{\xi,1}}\Big\{ 4 \zeta_2^2\Big\}  - \bm{\mathcal{L}_{\xi,0}}\Big\{\frac{2}{5} \zeta_2^2\Big\}  + \bm{\mathcal{L}_{\xi,1}} \Big\{ 8 \zeta_3 \Big\}  
\nonumber \\ &
- \bm{\mathcal{D}_{\xi,3}}\Big\{4 \zeta_2 \Big\}- \bm{\mathcal{L}_{\xi,2}} \Big\{4 \zeta_2 \Big\}+ \bm{\mathcal{L}_{\xi,4}} 
+ \bm{\delta_{\xi}}\Big\{ 4 \zeta_3^2 - \frac{46 \zeta_2^3}{35} \Big\}   \, ,
\nonumber \\
\bm{\mathcal{D}_{\xi,2} \otimes \mathcal{D}_{\xi,3}} & =
 \bm{\mathcal{D}_{\xi,0}} \Big\{ 18 \zeta_3^2-6\zeta_2^3 \Big\}  
+ \bm{\mathcal{D}_{\xi,1}}\Big\{84 \zeta_5 -36 \zeta_2 \zeta_3  \Big\}  
- \bm{\mathcal{D}_{\xi,2}}\Big\{\frac{57}{5} \zeta_2^2\Big\}   + \bm{\mathcal{D}_{\xi,3}} \Big\{20 \zeta_3 \Big\}  - \bm{\mathcal{D}_{\xi,4}}\Big\{5 \zeta_2 \Big\}  
\nonumber \\ &
+ \bm{\mathcal{D}_{\xi,6}}\Big\{\frac{7}{12} \Big\}  
-   \bm{\mathcal{L}_{\xi,0}}\Big\{12\zeta_2 \zeta_3 - 24 \zeta_5\Big\}  
- \bm{\mathcal{L}_{\xi,1}}\Big\{6 \zeta_2^2 \Big\}  
+ \bm{\mathcal{L}_{\xi,2}}\Big\{18 \zeta_3 \Big\}  
- \bm{\mathcal{L}_{\xi,3}}\Big\{6 \zeta_2  \Big\} 
+ \bm{\mathcal{L}_{\xi,5} }
\nonumber \\ &
- \bm{\delta_{\xi}} \Big\{6\zeta_2^2 \zeta_3 + 24 \zeta_2 \zeta_5 - 60 \zeta_7\Big\}  \, ,
\nonumber \\
\bm{\mathcal{D}_{\xi,2} \otimes \mathcal{D}_{\xi,4}} & = 
\bm{\mathcal{D}_{\xi,0}}\Big\{384 \zeta_7  -144 \zeta_2  \zeta_5 -\frac{216}{5} \zeta_2^2 \zeta_3  \Big\}  + \bm{\mathcal{D}_{\xi,1}}\Big\{96 \zeta_3^2 -\frac{216}{5}\zeta_2^3   \Big\}
+ \bm{\mathcal{D}_{\xi,2}}\Big\{264 \zeta_5 -96 \zeta_2  \zeta_3 \Big\}  
\nonumber \\ &
+ \bm{\mathcal{D}_{\xi,4}}\Big\{30 \zeta_3 \Big\} 
+ \bm{\mathcal{D}_{\xi,7}}\Big\{\frac{8}{15}  \Big\}
+ \bm{\mathcal{L}_{\xi,0}}\Big\{24 \zeta_3^2 -\frac{288}{35} \zeta_2^3 \Big\}
  + \bm{\mathcal{L}_{\xi,1}}\Big\{144 \zeta_5 -48 \zeta_2  \zeta_3\Big\} 
 - \bm{\mathcal{D}_{\xi,3}}\Big\{\frac{128}{5} \zeta_2^2 \Big\} 
\nonumber \\ &
-  \bm{\mathcal{L}_{\xi,2}}\Big\{\frac{108}{5} \zeta_2^2 \Big\} 
+ \bm{\mathcal{L}_{\xi,3}} \Big\{32 \zeta_3 \Big\}   - \bm{\mathcal{D}_{\xi,5}} \Big\{ 6 \zeta_2 \Big\}   
- \bm{\mathcal{L}_{\xi,4}}\Big\{ 8 \zeta_2  \Big\} + \bm{\mathcal{L}_{\xi,6}} 
+ \bm{\delta_{\xi}}\Big\{ 144 \zeta_3 \zeta_5 - 24 \zeta_2 \zeta_3^2-\frac{2928 \zeta_2^4}{175}  \Big\} \, ,
\nonumber \\
\bm{\mathcal{D}_{\xi,2} \otimes \mathcal{D}_{\xi,5}} & =
\bm{ \mathcal{D}_{\xi,0}}  \Big\{960  \zeta_3 \zeta_5 - \frac{624}{5}\zeta_2^4 - 120 \zeta_2 \zeta_3^2\Big\}  
- \bm{ \mathcal{D}_{\xi,1}} \Big\{312 \zeta_2^2 \zeta_3 + 960 \zeta_2 \zeta_5 - 2880 \zeta_7\Big\}  
+ \bm{ \mathcal{D}_{\xi,2}} \Big\{ 300 \zeta_3^2 -\frac{1284}{7} \zeta_2^3 \Big\}  
\nonumber \\ &
-  \bm{ \mathcal{D}_{\xi,3}} \Big\{200 \zeta_2 \zeta_3  - 640 \zeta_5 \Big\}  
-  \bm{ \mathcal{D}_{\xi,4}}  \Big\{ 49 \zeta_2^2\Big\}
+ \bm{ \mathcal{D}_{\xi,5}} \Big\{ 42 \zeta_3  \Big\}
- \bm{ \mathcal{D}_{\xi,6}}  \Big\{ 7 \zeta_2 \Big\}
+ \bm{ \mathcal{D}_{\xi,8}} \Big\{\frac{1}{2}  \Big\}
\nonumber \\ &
- \bm{\mathcal{L}_{\xi,0}}  \Big\{96 \zeta_2^2 \zeta_3 + 240 \zeta_2 \zeta_5 - 720 \zeta_7\Big\}  
- \bm{\mathcal{L}_{\xi,1}}  \Big\{96 \zeta_2^3 - 120 \zeta_3^2 \Big\} 
-  \bm{\mathcal{L}_{\xi,2}}  \Big\{120\zeta_2 \zeta_3 - 480 \zeta_5 \Big\} 

- \bm{\mathcal{L}_{\xi,3}}  \Big\{52 \zeta_2^2  \Big\}
\nonumber \\ &
+ \bm{\mathcal{L}_{\xi,4}} \Big\{50 \zeta_3  \Big\} 
- \bm{\mathcal{L}_{\xi,5}} \Big\{10 \zeta_2  \Big\} 
+ \bm{\mathcal{L}_{\xi,7}} 
- \bm{\delta_{\xi}} \Big\{96 \zeta_2^3 \zeta_3 + 216 \zeta_2^2 \zeta_5 + 720 \zeta_2 \zeta_7 - 40 \zeta_3^3 -  2240 \zeta_9 \Big\} \, ,
\nonumber \\ 
\bm{\mathcal{D}_{\xi,3} \otimes \mathcal{D}_{\xi,3}} & =\bm{\mathcal{D}_{\xi,0}} \Big\{360 \zeta_7 -36 \zeta_2^2 \zeta_3 - 144 \zeta_2 \zeta_5  \Big\}  - 
\bm{\mathcal{D}_{\xi,1}}  \Big\{ 36\zeta_2^3 - 108 \zeta_3^2 \Big\}  - 
\bm{\mathcal{D}_{\xi,2}}  \Big\{ 108 \zeta_2 \zeta_3 - 252 \zeta_5 \Big\}  - 
\bm{\mathcal{D}_{\xi,3}} \Big\{\frac{114 \zeta_2^2}{5}\Big\}  
\nonumber \\ &
+ \bm{\mathcal{D}_{\xi,4}}\Big\{ 30 \zeta_3\Big\}  - 
\bm{\mathcal{D}_{\xi,5}}\Big\{ 6 \zeta_2  \Big\}
+ \bm{\mathcal{D}_{\xi,7}} \Big\{\frac{1}{2}\Big\}  + 
\bm{\mathcal{L}_{\xi,0}} \Big\{ 36 \zeta_3^2 -\frac{414 \zeta_2^3}{35}  \Big\}  - 
 \bm{\mathcal{L}_{\xi,1}}\Big\{ 72  \zeta_2 \zeta_3 - 144 \zeta_5\Big\}  
\nonumber \\ &
- \bm{\mathcal{L}_{\xi,2}} \Big\{18 \zeta_2^2 \Big\} 
+ \bm{\mathcal{L}_{\xi,2}}\Big\{ 36 \zeta_3\Big\}  
- \bm{\mathcal{L}_{\xi,4}} \Big\{9 \zeta_2\Big\}  + \bm{\mathcal{L}_{\xi,6}} 
+ \bm{\delta_{\xi}}\Big\{ 144 \zeta_3 \zeta_5 - 36 \zeta_2 \zeta_3^2-\frac{4491 \zeta_2^4}{350}   \Big\} \, ,
\nonumber \\ 
\bm{\mathcal{D}_{\xi,3} \otimes \mathcal{D}_{\xi,4}} & = \bm{\mathcal{D}_{\xi,0}} \Big\{ 1008 \zeta_3 \zeta_5- 216 \zeta_2 \zeta_3^2  -\frac{2538 \zeta_2^4}{25}   \Big\}  - 
\bm{\mathcal{D}_{\xi,1}}\Big\{  \frac{1368}{5}  \zeta_2^2 \zeta_3 +  1008  \zeta_2 \zeta_5 -  2592 \zeta_7 \Big\}  
\nonumber \\ &
+ \bm{\mathcal{D}_{\xi,2}} \Big\{360 \zeta_3^2 -\frac{684 \zeta_2^3}{5} \Big\}
- \bm{\mathcal{D}_{\xi,3}}  \Big\{ 240 \zeta_2 \zeta_3 - 600 \zeta_5 \Big\}  - 
\bm{\mathcal{D}_{\xi,4}}\Big\{ 42 \zeta_2^2\Big\}  +  \bm{\mathcal{D}_{\xi,5}} \Big\{42 \zeta_3\Big\} - \bm{\mathcal{D}_{\xi,6}} \Big\{7 \zeta_2\Big\}  
\nonumber \\ &
+  \bm{\mathcal{D}_{\xi,8}} \Big\{\frac{9}{20}\Big\} 
- \bm{\mathcal{L}_{\xi,0}}   \Big\{ 72\zeta_2^2 \zeta_3 + 288 \zeta_2 \zeta_5 - 720 \zeta_7 \Big\}
- \bm{\mathcal{L}_{\xi,1}} \Big\{ 72 \zeta_2^3 - 216  \zeta_3^2 \Big\}  - 
\bm{\mathcal{L}_{\xi,2}}  \Big\{ 216\zeta_2 \zeta_3 - 504 \zeta_5\Big\}  
\nonumber \\ &
- \bm{ \mathcal{L}_{\xi,3}}\Big\{ \frac{228 \zeta_2^2}{5}\Big\} 
+ \bm{\mathcal{L}_{\xi,4}}\Big\{ 60 \zeta_3\Big\}  - \bm{\mathcal{L}_{\xi,5}}\Big\{ 12 \zeta_2 \Big\} + \bm{\mathcal{L}_{\xi,7}}
\nonumber \\ &
- \bm{\delta_{\xi}}\Big\{ 72 \zeta_2^3 \zeta_3 +  \frac{1008}{5} \zeta_2^2 \zeta_5 
+  720 \zeta_2 \zeta_7 -  72 \zeta_3^3 + 2016 \zeta_9 \Big\} \, ,
\nonumber \\ 
\bm{\mathcal{D}_{\xi,3} \otimes \mathcal{D}_{\xi,5}} & = \bm{\mathcal{D}_{\xi,0}} \Big\{480  \zeta_3^3 + 16800 \zeta_9 - 648 \zeta_2^3 \zeta_3 - 1656 \zeta_2^2 \zeta_5 - 5760 \zeta_2 \zeta_7  \Big\}  - \bm{\mathcal{D}_{\xi,1}}  \Big\{ 882 \zeta_2^4 + 1440 \zeta_2 \zeta_3^2 - 7920 \zeta_3 \zeta_5 \Big\}  
\nonumber \\ &
- \bm{\mathcal{D}_{\xi,2}} \Big\{ 1152 \zeta_2^2 \zeta_3 + 3960 \zeta_2 \zeta_5 - 10800 \zeta_7 \Big\}  +  \bm{\mathcal{D}_{\xi,3}} \Big\{ 900 \zeta_3^2 -\frac{2880 \zeta_2^3}{7} \Big\}  - 
\bm{\mathcal{D}_{\xi,4}} \Big\{ 450 \zeta_2 \zeta_3 - 1230 \zeta_5\Big\}  
\nonumber \\ &
- \bm{ \mathcal{D}_{\xi,5}} \Big\{\frac{357 \zeta_2^2}{5} \Big\}
+ \bm{\mathcal{D}_{\xi,6}} \Big\{56 \zeta_3\Big\}  - \bm{\mathcal{D}_{\xi,7}} \Big\{8 \zeta_2\Big\}  + \bm{\mathcal{D}_{\xi,9}}\Big\{\frac{5 }{12}\Big\} + \bm{\mathcal{L}_{\xi,0}} \Big\{ 2160 \zeta_3 \zeta_5  - 360 \zeta_2 \zeta_3^2 -\frac{8784 \zeta_2^4}{35}  \Big\}  
\nonumber \\ &
- \bm{\mathcal{L}_{\xi,1}}  \Big\{ 648 \zeta_2^2 \zeta_3 + 2160 \zeta_2 \zeta_5 - 5760 \zeta_7 \Big\}  - \bm{\mathcal{L}_{\xi,2}}\Big\{324 \zeta_2^3 - 720 \zeta_3^2 \Big\}  - 
\bm{\mathcal{L}_{\xi,3}} \Big\{ 480 \zeta_2 \zeta_3 - 1320 \zeta_5 \Big\}  
\nonumber \\ &
- \bm{\mathcal{L}_{\xi,4}} \Big\{96 \zeta_2^2\Big\}  
+ \bm{\mathcal{L}_{\xi,5}}\Big\{ 90 \zeta_3\Big\}  - \bm{\mathcal{L}_{\xi,6}}\Big\{ 15 \zeta_2 \Big\} + \bm{\mathcal{L}_{\xi,8}} 
\nonumber \\ &
- \bm{ \delta_{\xi}}  \Big\{\frac{20016}{55}  \zeta_2^5 
+ 324 \zeta_2^2 \zeta_3^2 + 2160 \zeta_2 \zeta_3 \zeta_5 -2880  \zeta_5^2 - 5760 \zeta_3 \zeta_7  \Big\}\, ,
\nonumber \\
\bm{\mathcal{D}_{\xi,4} \otimes \mathcal{D}_{\xi,4}} & = \bm{\mathcal{D}_{\xi,0}} \Big\{576 \zeta_3^3 + 161 28 \zeta_9  -576 \zeta_2^3 \zeta_3 -  \frac{8064}{5} \zeta_2^2 \zeta_5 -  5760 \zeta_2 \zeta_7 \Big\} -\bm{\mathcal{D}_{\xi,1}}  \Big\{\frac{20304}{25} \zeta_2^4 + 1728 \zeta_2 \zeta_3^2 - 8064 \zeta_3 \zeta_5\Big\} 
\nonumber \\ &
- \bm{\mathcal{D}_{\xi,2}}\Big\{\frac{5472}{5}  \zeta_2^2 \zeta_3 + 4032 \zeta_2 \zeta_5 - 10368 \zeta_7\Big\}  
- \bm{\mathcal{D}_{\xi,3}} \Big\{\frac{1824}{5} \zeta_2^3 - 960 \zeta_3^2\Big\} 
-  \bm{\mathcal{D}_{\xi,4}} \Big\{480 \zeta_2 \zeta_3 - 1200 \zeta_5\Big\}
\nonumber \\ &
- \bm{\mathcal{D}_{\xi,5}} \Big\{ \frac{336}{5} \zeta_2^2 \Big\}
+\bm{\mathcal{D}_{\xi,6}}  \Big\{56 \zeta_3\Big\} - \bm{\mathcal{D}_{\xi,7}}\Big\{ 8 \zeta_2 \Big\} + \bm{\mathcal{D}_{\xi,9}}\Big\{ \frac{2}{5} \Big\}+ \bm{\mathcal{L}_{\xi,0}} \Big\{ 2304 \zeta_3 \zeta_5  - 576 \zeta_2 \zeta_3^2 -\frac{35928 \zeta_2^4}{175}  \Big\} 
\nonumber \\ &
-   \bm{\mathcal{L}_{\xi,1}}  \Big\{ 576 \zeta_2^2 \zeta_3 + 2304 \zeta_2 \zeta_5 - 5760 \zeta_7\Big\}
- \bm{\mathcal{L}_{\xi,2}} \Big\{ 288\zeta_2^3 - 864 \zeta_3^2\Big\} 
- \bm{\mathcal{L}_{\xi,3}} \Big\{576 \zeta_2 \zeta_3 - 1344 \zeta_5\Big\}
\nonumber \\ &
- \bm{\mathcal{L}_{\xi,4}} \Big\{\frac{456}{5} \zeta_2^2 \Big\} 
+  \bm{\mathcal{L}_{\xi,5}}\Big\{ 96 \zeta_3 \Big\}- \bm{\mathcal{L}_{\xi,6}}\Big\{16 \zeta_2\Big\}  + \bm{\mathcal{L}_{\xi,8}} 
\nonumber \\ &
+  \bm{\delta_{\xi}}\Big\{2880 \zeta_5^2 + 5760 \zeta_3 \zeta_7 - 288 \zeta_2^2 \zeta_3^2 - 2304 \zeta_2 \zeta_3 \zeta_5  -\frac{96552 \zeta_2^5}{275} \Big\}\, ,
\nonumber \\ 
\bm{\mathcal{D}_{\xi,4} \otimes \mathcal{D}_{\xi,5}} & = \bm{\mathcal{D}_{\xi,0}}  \Big\{25920 \zeta_5^2 + 51840 \zeta_3 \zeta_7-2736 \zeta_2^2 \zeta_3^2 - 20160 \zeta_2 \zeta_3 \zeta_5 -\frac{16056}{5}  \zeta_2^5  \Big\} 
- \bm{\mathcal{D}_{\xi,1}}  \Big\{ 5472 \zeta_2^3 \zeta_3 - 4800 \zeta_3^3 
\nonumber \\ &
+ 14688 \zeta_2^2 \zeta_5 + 51840 \zeta_2 \zeta_7 - 147840 \zeta_9 \Big\}  + \bm{\mathcal{D}_{\xi,10}}\Big\{\frac{11}{30}\Big\}  - \bm{\mathcal{D}_{\xi,2}} \Big\{\frac{18972}{5} \zeta_2^4 + 7200 \zeta_2 \zeta_3^2 - 36000 \zeta_3 \zeta_5\Big\} 
\nonumber \\ &
-  \bm{\mathcal{D}_{\xi,3}} \Big\{ 3360 \zeta_2^2 \zeta_3 +  12000 \zeta_2 \zeta_5 
-  31680 \zeta_7\Big\} +  \bm{\mathcal{D}_{\xi,4}} \Big\{ 2100 \zeta_3^2-\frac{6072 \zeta_2^3}{7} \Big\} - \bm{\mathcal{D}_{\xi,5}} \Big\{840 \zeta_2 \zeta_3 - 2184 \zeta_5\Big\}  
\nonumber \\ &
- \bm{\mathcal{D}_{\xi,6}}\Big\{ \frac{518}{5} \zeta_2^2\Big\}  + \bm{\mathcal{D}_{\xi,7}}\Big\{72 \zeta_3\Big\}  
- \bm{\mathcal{D}_{\xi,8}} \Big\{9 \zeta_2 \Big\}
- \bm{\mathcal{L}_{\xi,0}}  \Big\{1440 \zeta_2^3 \zeta_3 + 4032 \zeta_2^2 \zeta_5 + 14400 \zeta_2 \zeta_7 - 1440 \zeta_3^3 
\nonumber \\ &
- 40320 \zeta_9\Big\} 
- \bm{\mathcal{L}_{\xi,1}} \Big\{\frac{10152}{5} \zeta_2^4 + 4320\zeta_2 \zeta_3^2 - 20160 \zeta_3 \zeta_5\Big\}  
- \bm{\mathcal{L}_{\xi,2}}  \Big\{ 2736 \zeta_2^2 \zeta_3 +  10080 \zeta_2 \zeta_5 -  25920 \zeta_7\Big\} 
\nonumber \\ &
- \bm{\mathcal{L}_{\xi,3}}  \Big\{912 \zeta_2^3 - 2400 \zeta_3^2\Big\}  
-  \bm{\mathcal{L}_{\xi,4}}  \Big\{ 1200 \zeta_2 \zeta_3 - 3000 \zeta_5\Big\}
-  \bm{\mathcal{L}_{\xi,5}} \Big\{168 \zeta_2^2 \Big\}
+ \bm{\mathcal{L}_{\xi,6}} \Big\{140 \zeta_3\Big\}  - \bm{\mathcal{L}_{\xi,7}}\Big\{ 20 \zeta_2 \Big\} 
\nonumber \\ &
+ \bm{\mathcal{L}_{\xi,9}} 
+ \bm{\delta_{\xi}} \Big\{120960 \zeta_{11} - \frac{10152}{5} \zeta_2^4 \zeta_3 - 4032 \zeta_2^3 \zeta_5 + 10080 \zeta_3^2 \zeta_5 - 10656 \zeta_2^2 \zeta_7 
-1440\zeta_2 \zeta_3^3 -  40320 \zeta_2  \zeta_9\Big\} \, ,
\nonumber \\
\bm{\mathcal{D}_{\xi,5} \otimes \mathcal{D}_{\xi,5}} & = \bm{\mathcal{D}_{\xi,0}} \Big\{1209600 \zeta_{11}-20304 \zeta_2^4\zeta_3     - 40320 \zeta_2^3 \zeta_5 + 100800 \zeta_3^2 \zeta_5 - 106560 \zeta_2^2 \zeta_7 - 14400 \zeta_2 \zeta_3^3 - 403200 \zeta_2 \zeta_9 \Big\}
\nonumber \\ &
- \bm{\mathcal{D}_{\xi,1}}  \Big\{32112 \zeta_2^5 
+ 27360 \zeta_2^2 \zeta_3^2 
+ 201600 \zeta_2 \zeta_3 \zeta_5 - 259200 \zeta_5^2 - 518400 \zeta_3 \zeta_7 \Big\} 
+ \bm{\mathcal{D}_{\xi,11}} \Big\{\frac{1}{3} \Big\} 
\nonumber \\ &
- \bm{\mathcal{D}_{\xi,2}}  \Big\{27360 \zeta_2^3 \zeta_3 - 24000 \zeta_3^3 + 73440 \zeta_2^2 \zeta_5 + 259200 \zeta_2 \zeta_7 - 739200 \zeta_9\Big\} 
- \bm{\mathcal{D}_{\xi,3}} \Big\{12648 \zeta_2^4 + 24000 \zeta_2 \zeta_3^2 
\nonumber \\ &
- 120000 \zeta_3 \zeta_5\Big\}  
- \bm{\mathcal{D}_{\xi,4}} \Big\{8400 \zeta_2^2 \zeta_3 + 30000 \zeta_2 \zeta_5 - 79200 \zeta_7\Big\}  +  \bm{\mathcal{D}_{\xi,5}} \Big\{ 4200 \zeta_3^2 -\frac{12144 \zeta_2^3}{7} \Big\}
\nonumber \\ &
- \bm{\mathcal{D}_{\xi,6}}  \Big\{1400 \zeta_2 \zeta_3 - 3640 \zeta_5\Big\} 
- \bm{\mathcal{D}_{\xi,7}} \Big\{148 \zeta_2^2\Big\} + \bm{\mathcal{D}_{\xi,8}}\Big\{ 90 \zeta_3 \Big\} - \bm{\mathcal{D}_{\xi,9}} \Big\{10 \zeta_2 \Big\}
- \bm{\mathcal{L}_{\xi,0}} \Big\{ \frac{96552}{11} \zeta_2^5 +  7200  \zeta_2^2 \zeta_3^2 
\nonumber \\ &
+ 57600 \zeta_2 \zeta_3 \zeta_5 
-  72000 \zeta_5^2 - 144000 \zeta_3 \zeta_7 \Big\}  
- \bm{\mathcal{L}_{\xi,1}} \Big\{14400 \zeta_2^3 \zeta_3 + 40320 \zeta_2^2 \zeta_5 + 144000 \zeta_2 \zeta_7 - 14400 \zeta_3^3 
\nonumber \\ &
- 403200 \zeta_9 \Big\} 
+ \bm{\mathcal{L}_{\xi,10}} 
-  \bm{\mathcal{L}_{\xi,2}} \Big\{10152 \zeta_2^4 + 21600 \zeta_2 \zeta_3^2 
- 100800 \zeta_3 \zeta_5\Big\} - \bm{\mathcal{L}_{\xi,3}} \Big\{9120 \zeta_2^2 \zeta_3 + 33600 \zeta_2 \zeta_5 
\nonumber \\ &
- 86400 \zeta_7\Big\}  
- \bm{\mathcal{L}_{\xi,4}} \Big\{2280 \zeta_2^3 - 6000 \zeta_3^2\Big\} 
-  \bm{\mathcal{L}_{\xi,5}}  \Big\{2400 \zeta_2 \zeta_3 - 6000 \zeta_5\Big\}
- \bm{\mathcal{L}_{\xi,6}} \Big\{280 \zeta_2^2\Big\}  +  \bm{\mathcal{L}_{\xi,7}} \Big\{200 \zeta_3\Big\} 
\nonumber \\ &
- \bm{\mathcal{L}_{\xi,8}}\Big\{ 25 \zeta_2 \Big\} 
+  \bm{\delta_{\xi}} \Big\{   3600 \zeta_3^4 + 374400 \zeta_5 \zeta_7 + 403200 \zeta_3 \zeta_9 - 7200 \zeta_2^3 \zeta_3^2 - 40320 \zeta_2^2 \zeta_3 \zeta_5 
-  72000 \zeta_2 \zeta_5^2 
\nonumber \\ &
- 144000 \zeta_2 \zeta_3 \zeta_7 -\frac{509128524 }{35035}\zeta_2^6\Big\} \, ,
\end{align*}

\begin{align*}
\bm{\delta_{\xi} \otimes \mathcal{L}_{\xi,i}} &= \bm{\mathcal{L}_{\xi,i}} \, ,
\nonumber \\
\bm{\mathcal{D}_{\xi,0} \otimes \mathcal{L}_{\xi,0}} & =\bm{\mathcal{L}_{\xi,1}} \, ,
\nonumber \\
\bm{\mathcal{D}_{\xi,0}  \otimes \mathcal{L}_{\xi,1}} & =\bm{\mathcal{L}_{\xi,0}}\Big\{ -\zeta_2 \Big\}  + \bm{\mathcal{L}_{\xi,2}} \, ,
\nonumber \\
\bm{\mathcal{D}_{\xi,0}  \otimes \mathcal{L}_{\xi,2}} & = \bm{\mathcal{L}_{\xi,0}} \Big\{2 \zeta_3\Big\}  - \bm{\mathcal{L}_{\xi,1}} \Big\{2 \zeta_2\Big\}  + \bm{\mathcal{L}_{\xi,3}}\, ,
\nonumber \\
\bm{\mathcal{D}_{\xi,0}  \otimes \mathcal{L}_{\xi,3}} & = \bm{\mathcal{L}_{\xi,0}} \Big\{-\frac{12}{5} \zeta_2^2\Big\}  + \bm{\mathcal{L}_{\xi,1}}\Big\{ 6 \zeta_3\Big\}  - \bm{\mathcal{L}_{\xi,2}}\Big\{ 3 \zeta_2\Big\}  + \bm{\mathcal{L}_{\xi,4}}\, ,
\nonumber \\
\bm{\mathcal{D}_{\xi,0} \otimes \mathcal{L}_{\xi,4}} & = \bm{\mathcal{L}_{\xi,0}}\Big\{ 24 \zeta_5 \Big\} -\bm{\mathcal{L}_{\xi,1}} \Big\{\frac{48 \zeta_2^2}{5} \Big\} +\bm{\mathcal{L}_{\xi,2}} \Big\{ 12 \zeta_3 \Big\} - \bm{\mathcal{L}_{\xi,3}}\Big\{ 4 \zeta_2\Big\}  + \bm{\mathcal{L}_{\xi,5}}\, ,
\nonumber \\
\bm{\mathcal{D}_{\xi,0} \otimes \mathcal{L}_{\xi,5}} & = \bm{\mathcal{L}_{\xi,0}} \Big\{-\frac{192}{7} \zeta_2^3 \Big\} +   \bm{\mathcal{L}_{\xi,1}} \Big\{120 \zeta_5\Big\} -  \bm{\mathcal{L}_{\xi,2}}\Big\{ 24 \zeta_2^2\Big\}  +  \bm{\mathcal{L}_{\xi,3}}\Big\{ 20 \zeta_3 \Big\} -   \bm{\mathcal{L}_{\xi,4}}\Big\{ 5 \zeta_2\Big\} +  \bm{\mathcal{L}_{\xi,6}}\, ,
\nonumber \\
\bm{\mathcal{D}_{\xi,1} \otimes \mathcal{L}_{\xi,0}} & = \bm{\mathcal{L}_{\xi,2}} \Big\{\frac{1}{2} \Big\}\, ,
\nonumber \\
\bm{\mathcal{D}_{\xi,1} \otimes \mathcal{L}_{\xi,1}} & = \bm{\mathcal{L}_{\xi,0}}\Big\{  \zeta_3\Big\}  - \bm{\mathcal{L}_{\xi,1}}\Big\{ \zeta_2\Big\}  + \bm{\mathcal{L}_{\xi,3}}\Big\{  \frac{1}{2} \Big\}\, ,
\nonumber \\
\bm{\mathcal{D}_{\xi,1}  \otimes \mathcal{L}_{\xi,2}} & = \bm{\mathcal{L}_{\xi,0}} \Big\{- \frac{1}{5} \zeta_2^2\Big\}  + \bm{\mathcal{L}_{\xi,1}}\Big\{  4 \zeta_3 \Big\} - \bm{\mathcal{L}_{\xi,2}}\Big\{  2 \zeta_2\Big\}  + \bm{\mathcal{L}_{\xi,4}}\Big\{\frac{1}{2}\Big\} \, ,
\nonumber \\
\bm{\mathcal{D}_{\xi,1} \otimes \mathcal{L}_{\xi,3}} & = \bm{\mathcal{L}_{\xi,0}} \Big\{12 \zeta_5 - 6 \zeta_2 \zeta_3\Big\}  - \bm{\mathcal{L}_{\xi,1}}\Big\{ 3 \zeta_2^2 \Big\} +  \bm{\mathcal{L}_{\xi,2}}\Big\{ 9 \zeta_3 \Big\}   -\bm{\mathcal{L}_{\xi,3}} \Big\{3 \zeta_2\Big\}  + \bm{\mathcal{L}_{\xi,5}}\Big\{ \frac{1}{2}\Big\} \, ,
\nonumber \\
\bm{\mathcal{D}_{\xi,1} \otimes \mathcal{L}_{\xi,4}} & = \bm{\mathcal{L}_{\xi,0}}\Big\{12 \zeta_3^2 -\frac{144 \zeta_2^3}{35}  \Big\}  + \bm{\mathcal{L}_{\xi,1}} \Big\{ 72 \zeta_5 - 24 \zeta_2 \zeta_3 \big\}  - \bm{\mathcal{L}_{\xi,2}}\Big\{ \frac{54 \zeta_2^2}{5}\Big\}  + \bm{\mathcal{L}_{\xi,3}}\Big\{ 16 \zeta_3\Big\}  - \bm{\mathcal{L}_{\xi,4}}\Big\{ 4 \zeta_2\Big\}  
\nonumber \\ &
+ \bm{\mathcal{L}_{\xi,6}}\Big\{\frac{1}{2}\Big\} \, ,
\nonumber \\
\bm{\mathcal{D}_{\xi,1} \otimes \mathcal{L}_{\xi,5}} & = \bm{\mathcal{L}_{\xi,0}}\Big\{360 \zeta_7-48 \zeta_2^2 \zeta_3 - 120 \zeta_2 \zeta_5 \Big\}  + \bm{\mathcal{L}_{\xi,1}} \Big\{ 60 \zeta_3^2 -48 \zeta_2^3 \Big\}  -  \bm{\mathcal{L}_{\xi,2}} \Big\{60\zeta_2 \zeta_3 - 240 \zeta_5\Big\} - \bm{\mathcal{L}_{\xi,3}}\Big\{ 26 \zeta_2^2\Big\}  
\nonumber \\ &
+ \bm{\mathcal{L}_{\xi,4}}\Big\{25 \zeta_3\Big\}  
- \bm{\mathcal{L}_{\xi,5}} \Big\{5 \zeta_2\Big\}  + \bm{\mathcal{L}_{\xi,7}} \Big\{\frac{1}{2} \Big\}\, ,
\nonumber \\
\bm{\mathcal{D}_{\xi,2} \otimes \mathcal{L}_{\xi,0}} & = \bm{\mathcal{L}_{\xi,3}}\Big\{\frac{1}{3}\Big\}\, ,
\nonumber \\
\bm{\mathcal{D}_{\xi,2} \otimes \mathcal{L}_{\xi,1}} & = \bm{\mathcal{L}_{\xi,0}}\Big\{-\frac{4}{5} \zeta_2^2\Big\}  + \bm{\mathcal{L}_{\xi,1}} \Big\{2 \zeta_3\Big\}  - \bm{\mathcal{L}_{\xi,2}}\Big\{ \zeta_2\Big\}  + \bm{\mathcal{L}_{\xi,4}}\Big\{\frac{1}{3}\Big\}\, ,
\nonumber \\
\bm{\mathcal{D}_{\xi,2} \otimes \mathcal{L}_{\xi,2}} & = \bm{\mathcal{L}_{\xi,0}} \Big\{8 \zeta_5 - 4 \zeta_2 \zeta_3 \Big\}  - \bm{\mathcal{L}_{\xi,1}}\Big\{ 2 \zeta_2^2\Big\}  + \bm{\mathcal{L}_{\xi,2}} \Big\{6 \zeta_3 \Big\}  -\bm{\mathcal{L}_{\xi,3}}\Big\{ 2 \zeta_2\Big\}  + \bm{\mathcal{L}_{\xi,5}}\Big\{ \frac{1}{3}\Big\}\, ,
\nonumber \\
\bm{\mathcal{D}_{\xi,2} \otimes \mathcal{L}_{\xi,3}} & = \bm{\mathcal{L}_{\xi,0}}\Big\{12 \zeta_3^2 -\frac{138 \zeta_2^3}{35} \Big\} +\bm{\mathcal{L}_{\xi,1}} \Big\{48  \zeta_5 -24\zeta_2 \zeta_3 \Big\}  - \bm{\mathcal{L}_{\xi,2}}\Big\{ 6 \zeta_2^2 \Big\} + \bm{\mathcal{L}_{\xi,3}} \Big\{12 \zeta_3\Big\}  - \bm{\mathcal{L}_{\xi,4}}\Big\{ 3 \zeta_2 \Big\} + \bm{\mathcal{L}_{\xi,6}}\Big\{ \frac{1}{3}\Big\}\, ,
\nonumber \\
\bm{\mathcal{D}_{\xi,2} \otimes \mathcal{L}_{\xi,4}} & = \bm{\mathcal{L}_{\xi,0}}  \Big\{240 \zeta_7-24\zeta_2^2 \zeta_3 - 96 \zeta_2 \zeta_5\Big\} - \bm{\mathcal{L}_{\xi,1}}  \Big\{24 \zeta_2^3 - 72 \zeta_3^2\Big\} - \bm{\mathcal{L}_{\xi,2}} \Big\{72 \zeta_2 \zeta_3 - 168\zeta_5\Big\}  -  \bm{\mathcal{L}_{\xi,3}}\Big\{\frac{76 \zeta_2^2}{5}\Big\} 
\nonumber \\ &
+ \bm{\mathcal{L}_{\xi,4}} \Big\{20 \zeta_3 \Big\}-  \bm{\mathcal{L}_{\xi,5}} \Big\{4 \zeta_2\Big\} + \bm{\mathcal{L}_{\xi,7}}\Big\{\frac{1}{3}\Big\}\, ,
\nonumber \\
\bm{\mathcal{D}_{\xi,2} \otimes \mathcal{L}_{\xi,5}} & =   \bm{\mathcal{L}_{\xi,0}} \Big\{ 720 \zeta_3 \zeta_5- 120 \zeta_2 \zeta_3^2 -\frac{2928 \zeta_2^4}{35} \Big\} -   \bm{\mathcal{L}_{\xi,1}} \Big\{216 \zeta_2^2 \zeta_3 + 720 \zeta_2 \zeta_5 - 1920 \zeta_7\Big\}  -   \bm{\mathcal{L}_{\xi,2}} \Big\{108 \zeta_2^3 - 240 \zeta_3^2\Big\}  
\nonumber \\ & 
-   \bm{\mathcal{L}_{\xi,3}} \Big\{160 \zeta_2 \zeta_3 - 440 \zeta_5\Big\}  -  \bm{\mathcal{L}_{\xi,4}} \Big\{32 \zeta_2^2\Big\}  +  \bm{\mathcal{L}_{\xi,5}} \Big\{30 \zeta_3 \Big\} -  \bm{\mathcal{L}_{\xi,6}} \Big\{5 \zeta_2\Big\}  +  \bm{\mathcal{L}_{\xi,8}}\Big\{\frac{1}{3}\Big\}\, ,
\nonumber \\
\bm{\mathcal{D}_{\xi,3} \otimes \mathcal{L}_{\xi,0}} & = \bm{\mathcal{L}_{\xi,4}}\Big\{\frac{1}{4}\Big\}\, , 
\nonumber \\
\bm{\mathcal{D}_{\xi,3} \otimes \mathcal{L}_{\xi,1}} & = \bm{\mathcal{L}_{\xi,0}}\Big\{ 6 \zeta_5\Big\}  -\bm{\mathcal{L}_{\xi,1}} \Big\{\frac{12 \zeta_2^2}{5}\Big\}  +\bm{\mathcal{L}_{\xi,2}}\Big\{  3 \zeta_3\Big\}  - \bm{\mathcal{L}_{\xi,3}}\Big\{\zeta_2\Big\}  + \bm{\mathcal{L}_{\xi,5}}\Big\{\frac{1}{4}\Big\}\, ,
\nonumber \\
\bm{\mathcal{D}_{\xi,3} \otimes \mathcal{L}_{\xi,2}} & = \bm{\mathcal{L}_{\xi,0}}\Big\{6 \zeta_3^2-\frac{72 \zeta_2^3}{35} \Big\}  -  \bm{\mathcal{L}_{\xi,1}}\Big\{12\zeta_2 \zeta_3 - 36 \zeta_5\Big\}  - \bm{\mathcal{L}_{\xi,2}}\Big\{ \frac{27 \zeta_2^2}{5}\Big\}  +\bm{\mathcal{L}_{\xi,3}}\Big\{ 8 \zeta_3\Big\}  - \bm{\mathcal{L}_{\xi,4}} \Big\{2 \zeta_2\Big\}  +\bm{\mathcal{L}_{\xi,6}}\Big\{ \frac{1}{4}\Big\}\, ,
\nonumber \\
\bm{\mathcal{D}_{\xi,3} \otimes \mathcal{L}_{\xi,3}} & = \bm{\mathcal{L}_{\xi,0} }\Big\{ 180 \zeta_7-18\zeta_2^2 \zeta_3 - 72 \zeta_2 \zeta_5 \Big\} - \bm{\mathcal{L}_{\xi,1}} \Big\{18\zeta_2^3 - 54 \zeta_3^2\Big\}  - \bm{\mathcal{L}_{\xi,2}}\Big\{54 \zeta_2 \zeta_3 - 126 \zeta_5\Big\}  -  \bm{\mathcal{L}_{\xi,3}}\Big\{\frac{57 \zeta_2^2}{5}\Big\} 
\nonumber \\ &
+ \bm{\mathcal{L}_{\xi,4}}\Big\{ 15 \zeta_3 \Big\} 
 - \bm{\mathcal{L}_{\xi,5}}\Big\{ 3 \zeta_2\Big\}  +\bm{\mathcal{L}_{\xi,7}}\Big\{ \frac{1}{4}\Big\}\, ,
\nonumber \\
\bm{\mathcal{D}_{\xi,3} \otimes \mathcal{L}_{\xi,4}} & = \bm{\mathcal{L}_{\xi,0}}\Big\{576 \zeta_3 \zeta_5- 144 \zeta_2 \zeta_3^2 -\frac{8982 \zeta_2^4}{175} \Big\}  -  \bm{\mathcal{L}_{\xi,1}}\Big\{144\zeta_2^2 \zeta_3 + 576 \zeta_2 \zeta_5 - 1440 \zeta_7 \Big\}  -  \bm{\mathcal{L}_{\xi,2}}  \Big\{ 72 \zeta_2^3 - 216 \zeta_3^2\Big\}
\nonumber \\ & 
-  \bm{\mathcal{L}_{\xi,3}} \Big\{144 \zeta_2 \zeta_3 - 336 \zeta_5\Big\}  - \bm{\mathcal{L}_{\xi,4}}\Big\{\frac{114 \zeta_2^2}{5}\Big\}  + \bm{\mathcal{L}_{\xi,5}} \Big\{24 \zeta_3 \Big\} - \bm{\mathcal{L}_{\xi,6}}\Big\{ 4 \zeta_2\Big\}  + \bm{\mathcal{L}_{\xi,8}}\Big\{\frac{1}{4}\Big\}\, ,
\nonumber \\
\bm{\mathcal{D}_{\xi,3} \otimes \mathcal{L}_{\xi,5}} & = \bm{\mathcal{L}_{\xi,0}}\Big\{360 \zeta_3^3 + 10080 \zeta_9-360 \zeta_2^3 \zeta_3 - 1008 \zeta_2^2 \zeta_5 - 3600 \zeta_2 \zeta_7  \Big\}  -\bm{\mathcal{L}_{\xi,1}} \Big\{\frac{2538}{5}  \zeta_2^4 + 1080 \zeta_2 \zeta_3^2 - 5040\zeta_3 \zeta_5\Big\} 
\nonumber \\ & 
- \bm{\mathcal{L}_{\xi,2}} \Big\{684 \zeta_2^2 \zeta_3 + 2520 \zeta_2 \zeta_5 - 6480 \zeta_7\Big\}  +  \bm{\mathcal{L}_{\xi,3}}\Big\{600 \zeta_3^2 -228 \zeta_2^3 \Big\} - \bm{\mathcal{L}_{\xi,4}} \Big\{300 \zeta_2 \zeta_3 - 750 \zeta_5\Big\}  
\nonumber \\ & 
- \bm{\mathcal{L}_{\xi,5}} \Big\{42 \zeta_2^2 \Big\}
+ \bm{\mathcal{L}_{\xi,6}} \Big\{35 \zeta_3\Big\}  - \bm{\mathcal{L}_{\xi,7}}\Big\{ 5 \zeta_2 \Big\} + \bm{\mathcal{L}_{\xi,9}}\Big\{ \frac{1}{4}\Big\}\, ,
\nonumber \\ 
\bm{\mathcal{D}_{\xi,4} \otimes \mathcal{L}_{\xi,0}} & =\bm{\mathcal{L}_{\xi,5}} \Big\{ \frac{1}{5}\Big\}\, ,
\nonumber \\ 
\bm{\mathcal{D}_{\xi,4} \otimes \mathcal{L}_{\xi,1}} & = \bm{\mathcal{L}_{\xi,0}} \Big\{-\frac{192}{35} \zeta_2^3\Big\}  +  \bm{\mathcal{L}_{\xi,1}} \Big\{24 \zeta_5 \Big\}- \bm{\mathcal{L}_{\xi,2}}\Big\{ \frac{24 \zeta_2^2 }{5}\Big\} + \bm{\mathcal{L}_{\xi,3}} \Big\{4 \zeta_3 \Big\} - \bm{\mathcal{L}_{\xi,4}} \Big\{ \zeta_2\Big\}  + \bm{\mathcal{L}_{\xi,6}}\Big\{\frac{1}{5}\Big\}\, ,
\nonumber \\ 
\bm{\mathcal{D}_{\xi,4} \otimes \mathcal{L}_{\xi,2}} & = \bm{\mathcal{L}_{\xi,0}}\Big\{ -\frac{96}{5} \zeta_2^2 \zeta_3 -48 \zeta_2 \zeta_5 +144 \zeta_7 \Big\}  -  \bm{\mathcal{L}_{\xi,1}} \Big\{\frac{96}{5} \zeta_2^3 - 24 \zeta_3^2\Big\} - \bm{\mathcal{L}_{\xi,2}} \Big\{24 \zeta_2 \zeta_3 - 96 \zeta_5\Big\}  
\nonumber \\&  
- \bm{\mathcal{L}_{\xi,3}} \Big\{\frac{52}{5}  \zeta_2^2 \Big\} + \bm{\mathcal{L}_{\xi,4}} \Big\{  10 \zeta_3\Big\} - \bm{\mathcal{L}_{\xi,5}} \Big\{ 2 \zeta_2 \Big\} +  \bm{\mathcal{L}_{\xi,7}} \Big\{\frac{1}{5}\Big\}\, ,
\nonumber \\ 
\bm{\mathcal{D}_{\xi,4} \otimes \mathcal{L}_{\xi,3}} & = \bm{\mathcal{L}_{\xi,0}} \Big\{ 432 \zeta_3 \zeta_5 - 72 \zeta_2 \zeta_3^2 -\frac{8784 \zeta_2^4}{175}  \Big\}  - \bm{ \mathcal{L}_{\xi,1}}  \Big\{\frac{648}{5} \zeta_2^2 \zeta_3 + 432 \zeta_2 \zeta_5 - 1152 \zeta_7 \Big\} - \bm{ \mathcal{L}_{\xi,2}} \Big\{\frac{324}{5}  \zeta_2^3 - 144 \zeta_3^2 \Big\}  
\nonumber \\ &
+  \bm{\mathcal{L}_{\xi,3}}  \Big\{ 264 \zeta_5 -96 \zeta_2 \zeta_3  \Big\}  - \bm{\mathcal{L}_{\xi,4}} \Big\{ \frac{96}{5} \zeta_2^2 \Big\}  + \bm{\mathcal{L}_{\xi,5}} \Big\{  18 \zeta_3 \Big\} - \bm{\mathcal{L}_{\xi,6}} \Big\{ 3 \zeta_2 \Big\}  + \bm{\mathcal{L}_{\xi,8}} \Big\{ \frac{1}{5}  \Big\}\, ,
\nonumber \\ 
\bm{\mathcal{D}_{\xi,4} \otimes \mathcal{L}_{\xi,4}} & = \bm{\mathcal{L}_{\xi,0}}   \Big\{288\zeta_3^3 + 8064 \zeta_9 -288 \zeta_2^3 \zeta_3 - \frac{4032}{5} \zeta_2^2 \zeta_5 - 2880 \zeta_2 \zeta_7 \Big\} 
- \bm{\mathcal{L}_{\xi,1}} \Big\{\frac{10152}{25} \zeta_2^4 +864\zeta_2 \zeta_3^2 - 4032 \zeta_3 \zeta_5\Big\} 
\nonumber \\ &
- \bm{\mathcal{L}_{\xi,2}}\Big\{\frac{2736}{5}\zeta_2^2 \zeta_3 +  2016 \zeta_2 \zeta_5 -  5184 \zeta_7\Big\} 
+ \bm{\mathcal{L}_{\xi,3}} \Big\{480 \zeta_3^2 -\frac{912 \zeta_2^3}{5} \Big\}  - \bm{\mathcal{L}_{\xi,4}}\Big\{240 \zeta_2 \zeta_3 - 600 \zeta_5\Big\} 
\nonumber \\ &
-\bm{\mathcal{L}_{\xi,5}}  \Big\{\frac{168 \zeta_2^2 }{5}\Big\} 
+ \bm{\mathcal{L}_{\xi,6}} \Big\{28 \zeta_3 \Big\} 
- \bm{\mathcal{L}_{\xi,7}} \Big\{4 \zeta_2\Big\}  + \bm{\mathcal{L}_{\xi,9}}\Big\{\frac{1}{5}\Big\}\, ,
\nonumber \\
\bm{\mathcal{D}_{\xi,4} \otimes \mathcal{L}_{\xi,5}} & = \bm{\mathcal{L}_{\xi,0}}   \Big\{ 14400 \zeta_5^2  -\frac{96552}{55} \zeta_2^5 - 1440 \zeta_2^2 \zeta_3^2 - 11520\zeta_2 \zeta_3 \zeta_5+ 22800 \zeta_3 \zeta_7\Big\} 
- \bm{\mathcal{L}_{\xi,1}} \Big\{ 2880 \zeta_2^3 \zeta_3 +  8064 \zeta_2^2 \zeta_5 
\nonumber \\ &
+  28800 \zeta_2 \zeta_7 -  2880 \zeta_3^3 - 80640 \zeta_9\Big\} 
- \bm{\mathcal{L}_{\xi,2}}\Big\{\frac{10152}{5} \zeta_2^4 + 4320 \zeta_2 \zeta_3^2 - 20160 \zeta_3 \zeta_5\Big\}  - \bm{\mathcal{L}_{\xi,3}} \Big\{1824 \zeta_2^2 \zeta_3 
\nonumber \\ & 
+6720 \zeta_2 \zeta_5 - 17280 \zeta_7\Big\}  - \bm{\mathcal{L}_{\xi,4}} \Big\{456 \zeta_2^3 
- 1200 \zeta_3^2\Big\}  - \bm{\mathcal{L}_{\xi,5}} \Big\{480 \zeta_2 \zeta_3 - 1200 \zeta_5\Big\}  - \bm{\mathcal{L}_{\xi,6}}\Big\{ 56 \zeta_2^2\Big\}  
\nonumber \\ &
+ \bm{\mathcal{L}_{\xi,7}} \Big\{40 \zeta_3\Big\}  
- \bm{\mathcal{L}_{\xi,8}}\Big\{ 5 \zeta_2 \Big\} + \bm{\mathcal{L}_{\xi,10}}\Big\{\frac{1}{5}\Big\}\, ,
\nonumber \\
\bm{\mathcal{D}_{\xi,5} \otimes \mathcal{L}_{\xi,0}}  & = \bm{\mathcal{L}_{\xi,6}}\Big\{\frac{1}{6} \Big\}\, ,
\nonumber \\
\bm{\mathcal{D}_{\xi,5} \otimes \mathcal{L}_{\xi,1}}  & = \bm{\mathcal{L}_{\xi,0}}\Big\{ 120 \zeta_7 \Big\} - \bm{\mathcal{L}_{\xi,1}}\Big\{ \frac{192 \zeta_2^3}{7}\Big\}  + \bm{\mathcal{L}_{\xi,2}} \Big\{60 \zeta_5\Big\}  - \bm{\mathcal{L}_{\xi,3}} \Big\{8 \zeta_2^2\Big\}  + \bm{\mathcal{L}_{\xi,4}}\Big\{ 5 \zeta_3\Big\}  - \bm{\mathcal{L}_{\xi,5}} \Big\{\zeta_2\Big\}  + \bm{\mathcal{L}_{\xi,7}} \Big\{\frac{1}{6} \Big\}\, ,
\nonumber \\ 
\bm{\mathcal{D}_{\xi,5} \otimes \mathcal{L}_{\xi,2}}  & = \bm{\mathcal{L}_{\xi,0} } \Big\{240 \zeta_3 \zeta_5-\frac{288}{7} \zeta_2^4 \Big\}  - \bm{\mathcal{L}_{\xi,1}}  \Big\{96 \zeta_2^2 \zeta_3 + 240 \zeta_2 \zeta_5 - 960 \zeta_7\Big\}  + \bm{\mathcal{L}_{\xi,2}}\Big\{60 \zeta_3^2 -\frac{528 \zeta_2^3}{7} \Big\}  
\nonumber \\ &
- \bm{\mathcal{L}_{\xi,3}}\Big\{40\zeta_2 \zeta_3 
- 200 \zeta_5\Big\} - \bm{\mathcal{L}_{\xi,4}} \Big\{17 \zeta_2^2\Big\}  + \bm{\mathcal{L}_{\xi,5}} \Big\{12 \zeta_3 \Big\} - \bm{\mathcal{L}_{\xi,6}} \Big\{2 \zeta_2 \Big\} + \bm{\mathcal{L}_{\xi,8}} \Big\{\frac{1}{6} \Big\}\, ,
\nonumber \\
\bm{\mathcal{D}_{\xi,5} \otimes \mathcal{L}_{\xi,3}} & = \bm{\mathcal{L}_{\xi,0}}\Big\{120\zeta_3^3 + 6720 \zeta_9 -288 \zeta_2^3 \zeta_3 - 648 \zeta_2^2 \zeta_5 -  2160 \zeta_2 \zeta_7 \Big\} - \bm{\mathcal{L}_{\xi,1}} \Big\{\frac{1872}{5} \zeta_2^4 + 360 \zeta_2 \zeta_3^2 - 2880 \zeta_3 \zeta_5\Big\} 
\nonumber \\ &
- \bm{\mathcal{L}_{\xi,2}} \Big\{468 \zeta_2^2 \zeta_3 + 1440 \zeta_2 \zeta_5 
- 4320 \zeta_7\Big\}+ \bm{\mathcal{L}_{\xi,3}}\Big\{300 \zeta_3^2 -\frac{1284 \zeta_2^3}{7} \Big\}  - \bm{\mathcal{L}_{\xi,4}} \Big\{150 \zeta_2 \zeta_3 - 480 \zeta_5\Big\} 
\nonumber \\ &
- \bm{\mathcal{L}_{\xi,5}} \Big\{ \frac{147 \zeta_2^2}{5}\Big\} + \bm{\mathcal{L}_{\xi,6}}\Big\{ 21 \zeta_3\Big\}  - \bm{\mathcal{L}_{\xi,7}} \Big\{3 \zeta_2 \Big\} + \bm{\mathcal{L}_{\xi,9}}\Big\{ \frac{1}{6} \Big\}\, ,
\nonumber \\
\bm{\mathcal{D}_{\xi,5} \otimes \mathcal{L}_{\xi,4}} & = \bm{\mathcal{L}_{\xi,0}} \Big\{11520 \zeta_5^2 + 23040 \zeta_3 \zeta_7-\frac{80064}{55} \zeta_2^5 - 1296 \zeta_2^2 \zeta_3^2 -  8640 \zeta_2 \zeta_3 \zeta_5 \Big\} -  \bm{\mathcal{L}_{\xi,1}} \Big\{2592 \zeta_2^3 \zeta_3 + 6624 \zeta_2^2 \zeta_5 
\nonumber \\ &
+ 23040 \zeta_2 \zeta_7 - 1920\zeta_3^3 - 67200 \zeta_9\Big\}- \bm{\mathcal{L}_{\xi,2}}\Big\{1764 \zeta_2^4 + 2880 \zeta_2 \zeta_3^2 - 15840 \zeta_3 \zeta_5\Big\} - \bm{\mathcal{L}_{\xi,3}} \Big\{1536 \zeta_2^2 \zeta_3 
\nonumber \\ &
+ 5280 \zeta_2 \zeta_5  
- 14400 \zeta_7\Big\} +  \bm{\mathcal{L}_{\xi,4}}\Big\{ 900 \zeta_3^2-\frac{2880 \zeta_2^3}{7} \Big\} +  \bm{\mathcal{L}_{\xi,5}}\Big\{984 \zeta_5 -360 \zeta_2 \zeta_3\Big\} - \bm{\mathcal{L}_{\xi,6}} \Big\{\frac{238 \zeta_2^2}{5}\Big\}  
\nonumber \\ &
+ \bm{\mathcal{L}_{\xi,7}}\Big\{ 32 \zeta_3 \Big\} 
- \bm{\mathcal{L}_{\xi,8}}\Big\{ 4 \zeta_2 \Big\} + \bm{\mathcal{L}_{\xi,10}} \Big\{\frac{1}{6}\Big\}\, , 
\nonumber \\
\bm{\mathcal{D}_{\xi,5} \otimes \mathcal{L}_{\xi,5}} & = \bm{\mathcal{L}_{\xi,0}} \Big\{604800 \zeta_{11} - 10152 \zeta_2^4 \zeta_3 - 20160 \zeta_2^3 \zeta_5 + 50400 \zeta_3^2 \zeta_5 - 53280 \zeta_2^2 \zeta_7 -  7200\zeta_2\zeta_3^3 - 201600 \zeta_2 \zeta_9\Big\}  
\nonumber \\ &
-  \bm{\mathcal{L}_{\xi,1}} \Big\{16056 \zeta_2^5 + 13680 \zeta_2^2 \zeta_3^2 
+ 100800 \zeta_2 \zeta_3 \zeta_5 - 129600 \zeta_5^2 - 259200 \zeta_3 \zeta_7\Big\}  - \bm{\mathcal{L}_{\xi,2}}\Big\{13680 \zeta_2^3 \zeta_3 - 12000 \zeta_3^3 
\nonumber \\ &
+ 36720 \zeta_2^2 \zeta_5 + 129600 \zeta_2 \zeta_7 - 369600 \zeta_9\Big\}   -  \bm{\mathcal{L}_{\xi,3}}\Big\{6324 \zeta_2^4 
+ 12000 \zeta_2 \zeta_3^2 - 60000 \zeta_3 \zeta_5\Big\}  
\nonumber \\ &
-  \bm{\mathcal{L}_{\xi,4}}\Big\{4200 \zeta_2^2 \zeta_3 + 15000 \zeta_2 \zeta_5 - 39600 \zeta_7\Big\} +  \bm{\mathcal{L}_{\xi,5}}\Big\{2100 \zeta_3^2 -\frac{6072 \zeta_2^3}{7} \Big\}  - \bm{\mathcal{L}_{\xi,6}}\Big\{700 \zeta_2 \zeta_3 - 1820 \zeta_5\Big\}  
\nonumber \\ &
- \bm{\mathcal{L}_{\xi,7}} \Big\{74 \zeta_2^2\Big\}  
+  \bm{\mathcal{L}_{\xi,8}}\Big\{ 45 \zeta_3 \Big\}  -\bm{\mathcal{L}_{\xi,9}} \Big\{5 \zeta_2\Big\}   + \bm{\mathcal{L}_{\xi,11}} \Big\{\frac{1}{6}\Big\}  \, .
\end{align*}

\end{widetext}


\if{1=0}
\section{Possible plots}
Dear Sven,  
we have explained our confusion,
please let us know whatyou think:

{\bf {regarding the convergence of the resummed result, it might be illustrative to look at them im N-space first.
I.e. integrate over z and plot as a function of N.
Then we could compare with the results from resummation for inclusive $F_2$, see e.g. Fig.1 in
   https://arxiv.org/pdf/hep-ph/0506288
There we have it in N-space (left) and x-space (right).}}

Let us start with the eq.~(7) :
\begin{widetext}
\begin{eqnarray}\label{eq:StrucCoeff}
\big(g_{1}\big) F_J (x,z,Q^2)
&=& \sum_{a,b = q,\overline{q},g}
\int_x^1 \frac{dx_1}{x_1} (\Delta )f_{a/\text{P}}\left( x_1,\mu_F^2\right)
\int_z^1 \frac{dz_1}{z_1}
  D_{\text{H}'/b}\left( z_1,\mu_F^2\right)
  (\Delta) {\cal C}_{J,ab}\left(\frac{x}{x_1},\frac{z}{z_1},Q^2,\mu_F^2\right)\, ,
\end{eqnarray}
\end{widetext}

If we integrate over $z$ numerically (analytically it is not easy due to complicated polylogs with two variables) , we get

\begin{widetext}
\begin{eqnarray} 
\int_0^1 dz \big(g_{1}\big) F_J (x,z,Q^2)
&=& \sum_{a,b = q,\overline{q},g}
\int_x^1 \frac{dx_1}{x_1} (\Delta )f_{a/\text{P}}\left( x_1,\mu_F^2\right)
(\Delta)\overline {\cal C}^{H'}_{J,a}\left(\frac{x}{x_1}, Q^2,\mu_F^2\right)
\end{eqnarray}
where,
\begin{eqnarray}
(\Delta) \overline {\cal C}^{H'}_{J,a}\left(\frac{x}{x_1},Q^2,\mu_F^2\right) =\int_0^1 dz \int_z^1 \frac{dz_1}{z_1}
D_{\text{H}'/b}\left( z_1,\mu_F^2\right)
  (\Delta) {\cal C}_{J,ab}\left(\frac{x}{x_1},\frac{z}{z_1},Q^2,\mu_F^2\right)\, ,
\end{eqnarray}
\end{widetext}
Because $(\Delta)\overline {\cal C}^{H'}_{J,a}\left(\frac{x}{x_1},\frac{z}{z_1},Q^2,\mu_F^2\right)$
contains information of $D_{H'/b}$ for all $b$ and hence
will not coincide with the DIS structure funtion $g_1(F_J)(x,Q^2)$ unless we sum over $H'$ using their fragmentation function.  Hence somehow after mass factorisation, we loose connection between structure function of SIDIS and structure function of DIS.

Instead of integrating $g_1,F_J$, if we try to integrate the CF  with respect to $z$, again we end up with complicated poly logs with complicated argument.  So both hadron structure function level and parton CF level, it is difficult to evaluate integration over $z$ variable analytically.

{\bf {Another comment concerns the $1/N$ corrections.
When including them (for example for Higgs in Fig.1 in https://arxiv.org/pdf/2004.00563), we found found some numerical sensitivity from those $1/N$ corrections.
That is to say, we should be careful in the way we truncate them to avoid spurious large numerical terms.
We found then, that the $ln(N)$ resummed results and the $ln(N)+1/N$ terms give a nice envelope band for the exact result.}}

About your suggestion:
May be I am missing something:
At NNLO level, we can not obtain
exact result for Double Mellin moment ($N_1,N_2$) of CFs as there are complicated poly logs with complicated arguments.
However we can expand the CFs in x,z space and then calcualte Mellin Moment w.r.t N1,N2 and obtain 
SV ($log N_1, log N_2$) and NSV ($1/N_1  log N_1, 1/N_2 log N_2$)
at third and fourth in $a_s$.
Hoeever we do not have exact
$N_1,N_2$ space result , hence
we can not plot like Fig1 of  the paper https://arxiv.org/pdf/2004.00563)

\fi

\bibliographystyle{apsrev4-1}

\bibliography{main}

\end{document}